\documentclass[12pt,reqno]{ksudiss}
\usepackage{verbatim,epsfig,caption,amstext,amsmath,amssymb,afterpage,graphicx,graphics}

\title{Non-equilibrium hydrodynamics of \\the quark-gluon plasma}
\author{Mohammad Nopoush}{}
\namedegreedate{Mohammad Nopoush, Ph.D., August 2018}
\pages{137}
\advisor{Michael Strickland}
\degrees{B.S., University of Tabriz, Iran, 2006\\
M.S., Amirkabir University of Technology, Iran, 2009\\
Ph.D., Kent State University, USA, 2018
}

\newcommand{\cdate}{
\bigskip
\medskip
\\
	\textsf{\textcircled{c}} Copyright\\
	All rights reserved\\
	Except for previously published materials
}  
\makeindex   
\usepackage{float}
\usepackage{multirow}
\usepackage{graphicx}
\usepackage{amsmath}
\usepackage{amssymb}
\usepackage{tabularx}
\usepackage[colorlinks=false,linktocpage=true]{hyperref}
\usepackage[usenames]{color}
\usepackage{mathtools,slashed}
\usepackage[utf8]{inputenc}
\usepackage{hyperref}
\usepackage{slashed}
\usepackage{dsfont}
\usepackage{lipsum,graphicx,subcaption}
\usepackage{enumitem}
\usepackage{dsfont}

\usepackage{bbm}
\usepackage{dsfont}

\usepackage{tikz}
\usepackage{tikz-3dplot}
\usepackage{tikz-cd}
\usepackage[nottoc]{tocbibind}

\newcommand{\Reals}{\rm I\kern-.19emR}
\newcommand{\Notin}{/\kern-.6em\hbox{$\in$}}
\newcommand{\MM}{\rm I\kern-.19emM}
\newcommand{\Notequiv}{/\kern-.6em\hbox{$\equiv$}}
\newcommand{\Ceals}{\rm I\kern-.5emC}
\newcommand{\nsubset}{/\kern-.6em\hbox{$\subset$}}

\newcommand{\nin}{\backslash \kern-.5em\in}

\DeclareMathOperator{\arcsinh}{arcsinh}

\def\be{\begin{equation}}
\def\ee{\end{equation}}
\def\ba{\begin{eqnarray}}
\def\ea{\end{eqnarray}}

\newcommand{\bsq}{\begin{subequations}}
\newcommand{\esq}{\end{subequations}}

\newcommand{\intdP}{\int\!dP}

\tikzset{
    vector/.style = {
        thick,
        > = stealth',
    },
    axis/.style = {
        very thick,
        > = stealth',
    },
}

\begin{document}
\frontmatter  %
\prefacesection{List of Publications}

{\small
{\bf Journal Papers}
\begin{enumerate}
 \item M. Alqahtani, {\bf M. Nopoush}, M. Strickland, ``Relativistic anisotropic hydrodynamics'', \\ \emph{Progress in Particle and Nuclear Physics} 101 (2018) 204-248,  \href{https://arxiv.org/abs/1712.03282}{arXiv:1712.03282 [nucl-th]}  
 \item {\bf M. Nopoush}, Y. Guo, M. Strickland, ``The static hard-loop gluon propagator to all orders in anisotropy'', \emph{JHEP} 1709 (2017) 063,  \href{https://arxiv.org/abs/1706.08091}{arXiv:1706.08091 [hep-ph]}  
     \item M. Alqahtani, {\bf M. Nopoush},  R. Ryblewski , M. Strickland, ``Anisotropic hydrodynamic modeling of 2.76 TeV Pb-Pb collisions'', \emph{Physical Review C} 96, 044910 (2017),  \href{https://arxiv.org/abs/1705.10191}{arXiv:1705.10191 [nucl-th]}
    \item M. Alqahtani, {\bf M. Nopoush},  R. Ryblewski , M. Strickland, ``3+1d quasiparticle anisotropic hydrodynamics for ultrarelativistic heavy-ion collisions'', \emph{Physical Review Letters} 119, 042301 (2017),  \href{https://arxiv.org/abs/1703.05808}{arXiv:1703.05808 [nucl-th]}
    \item B. Kasmaei, {\bf M. Nopoush}, M. Strickland, ``Quark self-energy in an ellipsoidally anisotropic quark-gluon plasma'',  \emph{Physical Review D} 94, 125001 (2017),  \href{https://arxiv.org/abs/1608.06018}{arXiv:1608.06018 [hep-ph]}    
    \item M. Alqahtani, {\bf M. Nopoush} , M. Strickland, ``Quasiparticle anisotropic hydrodynamics for central collisions'', \emph{Physical Review C} 95, 034906 (2016),  \href{https://arxiv.org/abs/1605.02101}{arXiv:1605.02101 [nucl-th]}     
      \item M. Alqahtani, {\bf M. Nopoush}, R. Ryblewski, M. Strickland ``Quasiparticle equation of state for anisotropic hydrodynamics'', \emph{Physical Review C} 92, 054910 (2015),  \href{https://arxiv.org/abs/1509.02913}{arXiv:1509.02913 [hep-ph]}   
      \item {\bf M. Nopoush}, M. Strickland, R. Ryblewski, D. Bazow, U. Heinz, M. Martinez, ``Leading-order anisotropic hydrodynamics for central collisions'', \emph{Physical Review C} 92, 044912 (2015),  \href{https://arxiv.org/abs/1506.05278}{arXiv:1506.05278 [nucl-th]}   
       \item {\bf M. Nopoush}, R. Ryblewski, M. Strickland, ``Anisotropic hydrodynamics for conformal Gubser flow'', \emph{Physical Review D} 91, 045007 (2015),  \href{https://arxiv.org/abs/1410.6790}{arXiv:1410.6790 [nucl-th]}   
             \item {\bf M. Nopoush}, R. Ryblewski, M. Strickland, ``Bulk viscous evolution within anisotropic hydrodynamics'', \emph{Physical Review C} 90, 014908 (2014),  \href{https://arxiv.org/abs/1405.1355}{arXiv:1405.1355 [hep-ph]}    
 \end{enumerate}}
       
{\bf Conference Proceedings}
\begin{enumerate}
\item M. Alqahtani, D. Almaalol, {\bf M. Nopoush},  R. Ryblewski , M. Strickland, ``Anisotropic hydrodynamic modeling of heavy-ion collisions at
LHC and RHIC'', To be published in \emph{Nuclear Physics A} (2018), \href{https://arxiv.org/pdf/1807.05508.pdf}{arXiv:1807.05508 [hep-ph]}
\item M. Alqahtani, {\bf M. Nopoush},  R. Ryblewski , M. Strickland, ``Quasiparticle anisotropic hydrodynamics for ultrarelativistic heavy-ion collisions'', \emph{Proceedings of science}, 311, 070 (2018) \href{https://arxiv.org/pdf/1711.07416.pdf}{arXiv:1711.07416 [hep-ph]}
 \item {\bf M. Nopoush}, M. Strickland, R. Ryblewski, ``Phenomenological predictions of 3+1d anisotropic hydrodynamics'',  \emph{Journal of Physics: Conference Series} 832 , 012054 (2017),  \href{https://arxiv.org/abs/1610.10055}{arXiv:1610.10055 [nucl-th]}
    \item M. Strickland, {\bf M. Nopoush}, R. Ryblewski, ``Anisotropic hydrodynamics for conformal Gubser flow'', \emph{Nuclear Physics A} 956, 268-271 (2016),  \href{https://arxiv.org/abs/1512.07334}{arXiv:1512.07334 [nucl-th]}  
         \item U. Heinz, D. Bazow, G.S. Denicol, M. Martinez, {\bf M. Nopoush}, J. Noronha, R. Ryblewski, M. Strickland, ``Exact solutions of the Boltzmann equation and optimized hydrodynamic approaches for relativistic heavy-ion collisions'', \emph{Nuclear and Particle Physics Proceedings} 276-278, 193-196 (2016),  \href{https://arxiv.org/abs/1509.05818}{arXiv:1509.05818 [nucl-th]} 
\end{enumerate}
\prefacesection{Acknowledgments}

First of all, I would like to express my sincere gratitude to my advisor Professor Michael Strickland for his continuous support and motivation. His guidance helped me in all the time of research and writing of this dissertation. 
I also thank my collaborators, Dr. Radoslaw Ryblewski, Dr. Mubarak Alqahtani and Babak S. Kasmaei for the hours of motivating discussions.

Last but not the least, I would like to acknowledge my family. I would like to thank my lovely wife, Mahrokh, for her continuous support, patience, and encouragement throughout my life, and my sweet daughter, Ayleen, which has filled my life with love and joy. 

This work is supported by the U.S. Department of Energy, Office of Science, Office of Nuclear Physics under Award No. DE-SC0013470.

\prefacesection{Abstract}
\vspace{0.2in}

Relativistic heavy-ion collision experiments are currently the only controlled way to generate and study matter in the most extreme temperatures ($T \gtrsim 10^{12}$ K). At these extreme temperatures matter undergoes a phase transition to an exotic phase of matter called the quark-gluon plasma (QGP). The QGP is an extremely hot and deconfined phase of matter where sub-nucleonic constituents (quarks and gluons) are asymptotically free. The QGP phase is important for different reasons. First of all, our universe existed in this phase up to approximately $t\sim 10^{-5}$ s  after the Big Bang, before it cools down sufficiently to form any kind of quark bound states. In this regard, studying the QGP provides us with useful information about the dynamics and evolution of the early universe. Secondly, high-energy collisions serve as a microscope with a resolution on the order of $10^{-15}$ m (several orders of magnitude more powerful than the best ever developed electron microscopes). With this fantastic probe, penetrating into the detailed structure of nucleons, and the discovery of new particles and fundamental phases are made possible. The dynamics of the QGP is based on quantum chromodynamics (which governs the interactions of quarks and gluons) and the associated force is ``strong force''. The strong collective behaviors observed experimentally inspired people to use dissipative fluid dynamics to model the dynamics of the medium. 
 The QGP produced in heavy-ion collisions, experiences strong longitudinal expansion at early times which leads to a large momentum-space anisotropy in the local rest frame distribution function. The rapid longitudinal expansion casts doubt on the application of standard viscous hydrodynamics (vHydro) models, which lead to unphysical predictions such as negative pressure, negative one-particle distribution function, and so on \cite{Strickland:2014eua}. Anisotropic hydrodynamics (aHydro) takes into account the strong momentum-space anisotropy in the leading order distribution function in a consistent and systematic way. 
 
 My dissertation is about the formulation and application of anisotropic hydrodynamics as a successful non-equilibrium hydrodynamics model for studying the QGP. For this purpose, I introduce the basic conformal anisotropic hydrodynamics formalism and then explain the ways we included realistic features (bulk degree of freedom \cite{Nopoush:2014pfa}, quasiparticle implementation of realistic equation of state \cite{Borsanyi:2010cj}, more realistic collisional kernel \cite{Alqahtani:2015qja}), to make it a suitable hydrodynamics model for studying the QGP generated in heavy-ion collisions. For verification of our model we have compared the evolution of model parameters predicted by aHydro and vHydro, with exact analytical solution of the Boltzmann equation \cite{Denicol:2014xca}. For this purpose, we have studied the evolution of the system under conformal Gubser flow using the aHydro model. By transforming to de Sitter spacetime (a non-trivial curved coordinate system) we simplified the dynamics to 0+1d spacetime \cite{Nopoush:2014qba}. Comparisons with exact solutions show that aHydro better reproduces the exact solutions than the best available vHydro models \cite{Denicol:2012cn}. However, the system is not conformal and the aHydro needed to be improved to include a realistic prescription for the equation of state which takes care of non-ideal effect in the dynamics. In the framework of finite temperature field theory the equation of state is provided by numerical calculation of QCD partition function using lattice QCD (LQCD), whereas, devising an equation of state for aHydro model is challenging because therein we deal with anisotropic pressures. In the next step of my research, we have designed a novel method for implementing the realistic equation of state (provided by lattice QCD) in the aHydro formalism \cite{Alqahtani:2016rth,Alqahtani:2017jwl}. This model, called the quasiparticle aHydro model, integrates the non-conformal effects in the aHydro model. The non-conformal effects are due to strong interactions of plasma constituents which leads to temperature-dependence of the particles' effective mass in the system. Based on the quasiparticle picture, we have developed the quasiparticle aHydro (aHydroQP) model which has all necessary components for studying the phenomenology of the QGP created in heavy-ion collisions. We have then compared the phenomenological predictions of the aHydroQP model with experimental observations. Comparisons illustrate a high level of consistency between our model and the experimental data \cite{Alqahtani:2017tnq}. The last two chapters are about two applications of the aHydro model to field-theoretical measurables in the QGP. In these chapters, we have calculated the quark self-energy in an anisotropic QGP \cite{Kasmaei:2016apv}. The quark self-energy is important because it encodes the way quarks gain interactional mass while in the hot QGP. I also have presented the calculation of gluon self-energy in hard loop approximation in an anisotropic QGP \cite{Nopoush:2017zbu}.  The gluon self-energy is important since it is related to heavy-quark potential and heavy quarkonium suppression. Heavy quarkonia bound states, besides theoretical importance, serve as a thermometer for the QGP \cite{Mocsy:2008eg}.

\startthesis
\chapter{\bf Theoretical Background}    
\label{chap:intro}
\setcounter{figure}{0}
\setcounter{table}{0}
\setcounter{equation}{0}



\section{The standard model}
\begin{figure}[t!]
\hspace{-0.5cm}
\includegraphics[width=.99\linewidth]{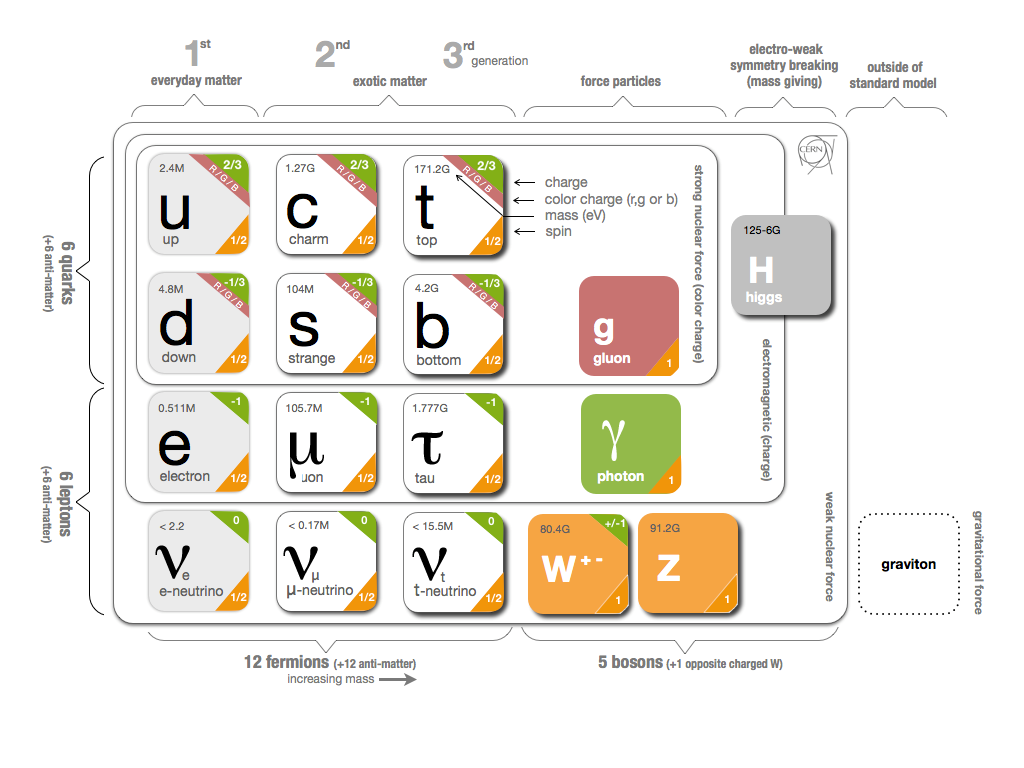}
\vspace{-0.8cm}
\caption{Periodic table of particles/force carriers indicated by the standard model \cite{standardTable}.}
\label{fig:standardmodel}
\end{figure}

There are four fundamental forces in nature. Sorted by strength, they are {\it gravity}, {\it weak}, {\it electromagnetism}, and {\it strong forces}. (i) Gravity is the attractive interaction between all massive objects. Over the years people used Newtonian mechanics and general relativity to study gravity. However, some newly developed theories (e.g. quantum gravity) are trying to approach gravity in a modern way aiming to its unification with other three forces. (ii)
The weak force is a mechanism of interactions that is effective in very small ranges ($\sim 0.01$ fm). It underlies some forms of radioactivity, governs the decay of unstable subnucleon particles, and initiates the nuclear fusion reaction. (iii) The electromagnetic (EM) force describes the interaction between the electrically-charged particles. Maxwell's equations form a complete set of equations for studying EM interactions in classical and relativistic settings. The {\it quantum electrodynamics} (QED) is a modern quantum field theory (QFT) developed to study electromagnetic forces in quantum ranges. (iv) The strong force governs the interactions of quarks and gluons, and holds the nuclei of atoms together. {\it Quantum chromodynamics} (QCD) developed in 1960s and 70s to study strong interactions. At the time of writing this dissertation standard model predict 38 elementary particles which are briefly listed in Fig.~\ref{fig:standardmodel}.
 According to the standard model, all matter in the universe is made of three generations of quarks (that is six quarks and six anti-quarks), three generations of leptons (that is three leptons, their respected neutrinos and their anti particles), four force mediators (which come in 13 types), and finally one Higgs boson. This totally adds up to 12+12+13+1=38. The quark generations are (i) up and down (ii) strange and charm (iii) top and bottom (aka beauty). The lepton generations are (i) electron and electron neutrino (ii) muon and muon neutrino (iii) tau and tau neutrino. 
 
 According to QCD, each quark comes with an additional degree of freedom, called {\it color}, which was originally introduced to resolve the contradictions encountered with the Pauli exclusion principle. 
 
 The fundamental bosons in the standard model are, the photon ($\gamma$) which is the mediator of electromagnetic interactions, the gluon ($g$) which is the mediator of strong interactions (and comes in 8 color states), the $W^{\pm}$ and $Z$ bosons which are the mediators of weak interaction, graviton ($G$) which is the conjectured mediator of gravitational force, and the Higgs boson ($H^0$) which is the fundamental quantized excitation of the Higgs field. The interaction of elementary particles with the Higgs field provides their bare mass (except for photons and gluons which are massless). 
 
 Since the inception of the standard model in the 1970s, many of its predictions have been solidly confirmed by  experiment, e.g. top quark (1995), tau neutrino (2000), Higgs boson (2012). However, there are some questions and ambiguities which have not yet been addressed properly in this model. For example, the theory does not explain baryon asymmetry; it is unable to include a full theory of gravity which accounts for expanding universe; it has no explanation for neutrino oscillation which results from the fact that neutrinos have a small but non-vanishing mass; and so on.
\section{High-energy nuclear physics}
High-energy nuclear physics focuses on studying the behaviors of nuclear matter at extremely high energies. Technically, this field is interdisciplinary which connects particle physics, nuclear physics, astrophysics, and cosmology. The goal of high-energy nuclear physics is to study nuclear  matter at the most extreme energies using finite temperature and density QCD.
{\it Ultrarelativistic heavy-ion collision} (URHIC) experiments are designed and developed to help to understand the behavior of nuclear matter at extremely high energies.  This is important because it provides us with the highest resolution probe for investigating the most fundamental sub-nucleonic structures. It actually helps to understand the dynamics of quarks, leptons, force carriers, and assessing different field theories. More interestingly, the dynamics experienced by nuclear matter throughout experiment closely mimics the condition of the early universe up to a few microseconds after Big Bang when the universe cooled down sufficiently so that hadrons could form. Therefore, studying matter under extreme conditions using the URHIC experiments provides us with helpful clues about the dynamics, structure, and evolution of the early universe. In this regard, the {\it quark-gluon plasma} (QGP) created in URHICs, in correspondence to the Big Bang, is sometimes called the Little Bang. This correspondence persuaded people to set up some connections among the two Bangs, e.g. photon and dilepton pairs generated by QGP form the analogue of the {\it Cosmic Microwave Background} in cosmology and so on. However, there is a fundamental difference between the two Bangs, that is the expansion rates which are different by a factor of the Planck mass ($m_{\rm Pl}=1.2\times 10^{19}$ GeV), due to different governing mechanisms in the dynamics of two systems, i.e. in the Little Bang the strong force is the dominant interaction while in the Big Bang the weak interaction plays an equally important role.
\begin{figure}[t!]
\hspace{-0.9cm}
\centering
\includegraphics[width=.95\linewidth]{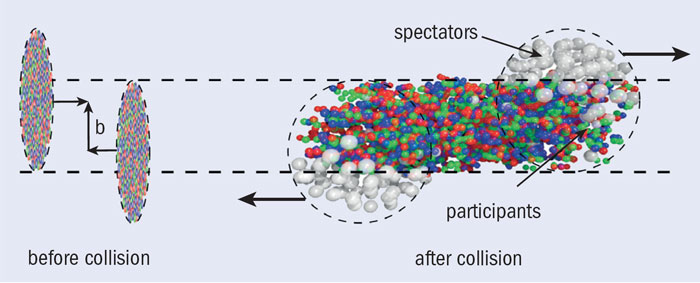}
\caption{The heavy nuclei colliding with impact parameter b. The spectators continue unaffected, while in the participant zone particle production takes place \cite{toia}}.
\label{fig:collision}
\end{figure}
 
 Etymologically, the term ``ultrarelativistic'' denotes that the kinetic energy of the particles being collided significantly exceeds the particles' rest energy. Collisions of heavy nuclei are interesting because one is able to generate a high multiplicity of partons through the collisions, which helps to generate the highest energy densities ever created in a laboratory setting. The resulting collective phenomena can drive the system into a fluid-like phase with rich physics underlying, i.e. QGP \cite{Pasechnik:2016wkt}. On the other hand, theoretical interpretation of the results requires the development of models based on finite temperature QCD.
\subsection{Historical background of heavy-ion collision experiments}
Relativistic heavy-ion experiments started in the mid-1970s with a set of experiments at Bevalac at the Lawrence Berkeley National Laboratory and the Intersecting Storage Rings collider at CERN. These experiments hoped to create a droplet of QGP through the high-energy collision of ions. Starting the investigations with beams of light ions, the signals were not much different from proton-proton collisions, indicating no signature of QGP. However, experiments with heavier ions observed  new flow patterns. In 1982-84, employing beams of niobium ($^{93}$Nb) and later gold ($^{197}\!$Au) at the Bevalac, at fixed-target energies from 200 MeV to 2 GeV per nucleon, some forms of collective phenomena in hadron emission were observed. Nowadays, there are mainly two leading experimental facilities in the world which study matter using URHIC: the RHIC at Brookhaven National Laboratory and the Large Hadron Collider (LHC) at CERN. At CERN, the accelerator complex has a better mass separation at the early stages of acceleration. Therefore, lead ($^{208}$Pb) which is a spherical and has heavier nucleus was selected for the heavy ion program at the SPS and it has also been used for the LHC program. For the program at the RHIC, gold ions were chosen, because gold has only one stable isotope, $^{197}\!$Au, and 18 different radioisotopes. The choice of spherical nuclei helps the collision geometry to be simpler.  
Currently, gold is considered the heaviest monoisotopic element (formerly $^{209}$Bi held this position but has been found to be slightly radioactive \cite{cern:2010}). The mechanism of operation of URHIC experiments is as follows. The intended element (i.e. Au, Pb, Cu, U and so on) is heated up to vaporize. Then, in several steps which consist of a chain of accelerating loops with different sizes and functions they get ultrarelativistic accelerations. In the early years of heavy-ion collision experiments, RHIC, as the most powerful particle collider, was the dominant apparatus for the new discoveries in connection with the QGP. However, since 2010 the LHC now achieves the highest possible collision energy. These days, RHIC is running collisions with different targets and energies and has been mostly working on beam energy scan which is related to finite chemical potential experiments. RHIC is planned to be upgraded to be launched as electron-ion collider (EIC) to better understand the 3d wavefunctions of hadrons and nuclei. Another important designated mission of EIC is to explore glue as the fundamental building block of matter. 
\subsection{Physics of heavy-ion collisions}
During URHIC, matter passes through a few stages, ending up with an ensemble of hadrons, leptons, and photons, eventually flying freely to the detectors. The physics relevant to each stage is explained briefly in the followings (see also left panel of Fig.~\ref{fig:gluon-sat}). 
\begin{itemize}
\item Color-glass condensate: Prior to the collisions, due to ultra-relativistic energies, incoming nuclei are Lorentz contracted pancakes. In addition, one finds that nuclei at high resolution (high energy collisions) are an extremely dense system of gluons which are ``saturated'' and each of which carries only a small fraction of longitudinal momentum of the nuclei, Fig.~\ref{fig:gluon-sat}. By the uncertainty principle, these compact gluons carry large transverse momenta, which based on asymptotic freedom of QCD form a weakly coupled system. This dense strongly-interacting but weakly-coupled gluonic system generated at high energies is universal, with similar properties regardless of the type of colliding nuclei. This phase is known as color-glass condensate (CGC) and is interpreted in terms of classical fields obeying the Yang-Mills equations \cite{McLerran:1993ka,Gelis:2010nm}. The ``color'' refers to the color-charged property of gluons as carriers of strong force. The word "glass" is refers to the materials that are disordered and behave like solids on short time scales and like a liquids on long enough time scales.
\begin{figure}[t]
\centering
\hspace{-0.5cm}
\includegraphics[width=.72\linewidth]{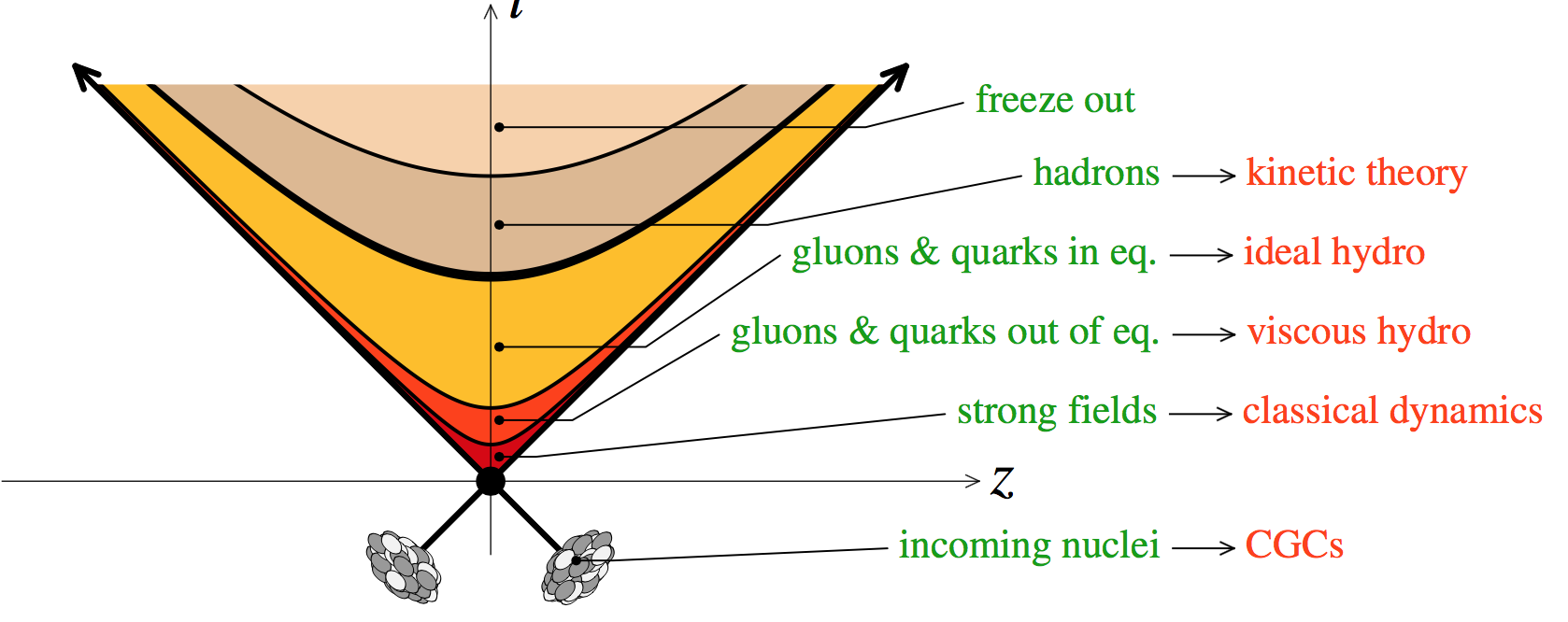}
\hspace{-1cm}
\includegraphics[width=.3\linewidth]{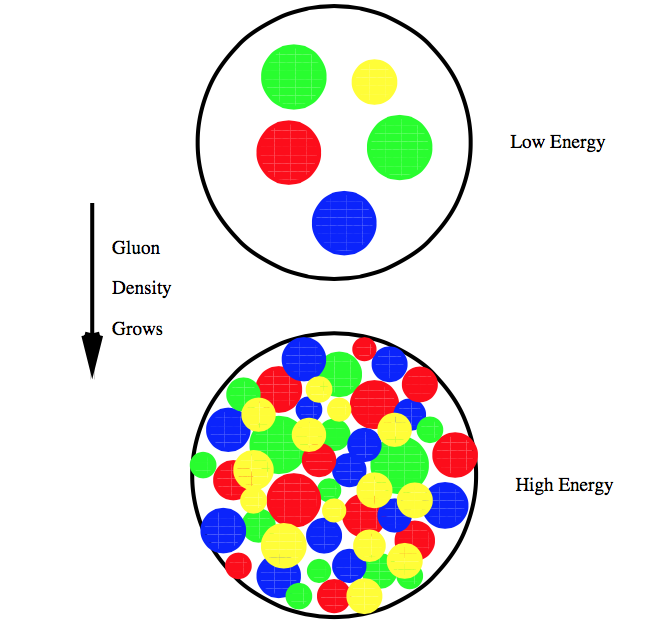}
\caption{Left: Schematic representation of HIC system as a function of time $t$ and longitudinal coordinate $z$ \cite{Iancu:2012xa}. Right: Saturation of gluons in the accelerating nuclei before the collision \cite{McLerran:2001sr}.}
\label{fig:gluon-sat}
\end{figure}
\item Early dynamics after the collision: Once the nuclei collide, mutual interactions start to develop. Among the multiple interactions, there are few hard processes (transverse momenta Q  $\sim$ 10 GeV) which involve large momentum transfers and are basically the main mechanisms for generating high-energy jets, direct photons, heavy quarks, and vector bosons. At $\tau \sim$ 0.2 fm/c the interactions are of the `semi-hard' nature (Q $\sim$ 1 GeV) and this is the time when most of the bulk of partonic constituents of the colliding nuclei, including the highly occupied gluon content of CGC, are released. At this stage, the liberated partons form a relatively dense and non-equilibrium phase which is called Glasma \cite{Gelis:2009wh}. The transition between a Glasma and a thermalized QGP is continuous. As the Glasma expands it interacts with itself and produces additional partons \cite{Lappi:2006fp}.

\item QGP: If there were no interaction, the partons would propagate freely for some time and then hadronize and fly out to the detectors. However, the data from URHIC experiments show signs of collective phenomena throughout the lifetime of the QGP, which is the evidence of strongly interacting matter. In other words, only strong interactions would be able to drive the system toward thermalization rapidly. Such a result seems inconsistent with perturbation theory presumptions of a weakly coupled system after the impact, however, it is possible to understand QGP thermalization using a perturbative treatment of the strong interaction. In any case, the outcome of this thermalization is the generation of a near-equilibrium QGP. 
\item Hadronization and freezeout: At about $\tau\sim 10$ fm/c when the local effective temperature falls below a certain limit ($T\sim 150-180$ MeV), the partons become confined within colorless hadrons. At later times, the system is in the form of interacting hot hadron gas. At about $\tau\sim 20\,{\rm fm/c}$ hadrons stop interacting strongly and colliding inelastically when they undergo chemical freezeout, i.e. the particles number stay conserved. Chemical freezeout is later followed by a kinetic freezeout where the hadrons stop having any kind of collisions, i.e. particles' momenta stay constant. After this, the hadrons fly freely to the detector.
\end{itemize}

\subsection{Quark-gluon plasma}
By definition, the QGP is a hot, dense, and strongly-interacting state of matter in which partons, e.g. quarks and gluons, are deconfined and free to move around (as opposed to the normal hadronic state where partons are confined in colorless hadrons held together by the strong force). Since there is a large population of color charge carriers, the QGP possesses large-range color charge conductivity. 

In order to illustrate the properties of the QGP, putting aside the procedure for making a QGP for the moment, I present a thought experiment which makes more sense for pedagogical purposes. At low temperatures, the system consists of a gas of colorless hadron states, which are the eigenstates of the QCD Hamiltonian at zero temperature. By increasing the temperature, hadrons interact strongly. For temperatures of order $T\sim$ 150 MeV and higher, the hadron interactions are so intense that hadronic states do not present a favorable quantum basis to describe the properties of the medium any more. In this condition the matter transforms into QGP, where quarks and gluons are the degrees of freedom.

In general there are two recipes to generate QGP: (i) At extremely high densities: By squeezing a large number of baryons into a small volume, the baryons wave functions start to overlap at a certain critical baryon density, $\rho_c$ and dissolve into a system of degenerate quark matter. The magnitude of $\rho_c$ for this purpose must typically be several times the nuclear saturation density ($\rho_{\rm 0}=0.16\,\, {\rm fm}^{-3}$). Such conditions are accessible in compact stars, i.e. white
dwarfs, neutron stars, quark stars, perhaps black holes. The white dwarfs are
made entirely of electrons and nuclei, while neutron stars are mainly made of a liquid of neutrons, with some protons and electrons. If the density at the center of neutron
stars reaches 5-10 $\rho_{\rm 0}$, the neutrons will possibly melt into cold quark matter. Secondly, at extremely high temperatures: Producing QGP at high temperatures do not even need the nuclear matter and one can heat up the QCD vacuum. At low temperatures hadrons are thermally excited from the vacuum. At higher temperatures, the hadrons start to overlap and then at the critical temperature ($T\gtrsim 150$ MeV) they melt into quark and gluon degrees of freedom. URHICs are the only way to generate QGP in the laboratory and they rely on this mechanism. 
Moreover, Friedman's solution of Einstein's equation suggests that the universe has experienced an expansion from a singularity at time zero. This scenario has, also, been confirmed by Hubble's law for the red shift of distant galaxies. Extrapolating the observed properties of the universe backward in time, the universe becomes hotter and denser, crossing the QCD phase transition  ($\sim 200$ MeV) at about $\sim 10^{-5} s$ after its inception. This phase transition preceded by electroweak phase transition at $\sim 200$ GeV.

The QGP is a strongly interacting phase of matter with the quarks and gluons existing in deconfined state. Similar to an electromagnetic plasma, the QGP is color conducting and, therefore, possesses color charge screening. The deconfinement of hadronic states xis a direct consequence of the running coupling constant and color screening. Phenomenologically, the QGP's behavior can be explained by hydrodynamic equations with a small specific shear viscosity.  On the other hand, calculations based on perturbation theory and string
theory indicate that the mean free path of quarks and gluons in the QGP may be comparable to the average interparticle spacing. Hence the QGP is like a liquid with the lowest measured value of shear viscosity, as long as one focuses on its flow properties. The  argument that the QGP at high temperature is weakly coupled is subtle. QGP is a multi-scale system and depending on the scale under consideration it might demonstrate weakly- or strongly-coupled behavior. In a QGP (and generally any physical plasma), there exist long wavelength modes which are strongly coupled, and also short wavelength modes which are weakly coupled and dominate the thermodynamics. Concerning the short wavelength modes, e.g. thermodynamics functions, asymptotic freedom implies that the coupling constant at high energies ($T\gg \Lambda_{\rm QCD}$) is weak, turning the system into a nearly ideal gas of quarks and gluons. On the other hand, the QGP reveals a strongly-coupled character which does not seem to be related to a large value of the coupling constant. For instance, even in the limit $T\gg \Lambda_{\rm QCD}$ the QGP contains a non-perturbative sector of static magnetic color fields, which is strongly coupled, but it does not contribute significantly to thermodynamic properties of the plasma.

\section{Overview of QCD}
\subsection{QCD phase transition and its diagram}
\begin{figure}[t!]
\hspace{-0.5cm}
\centering
\includegraphics[width=.9\linewidth]{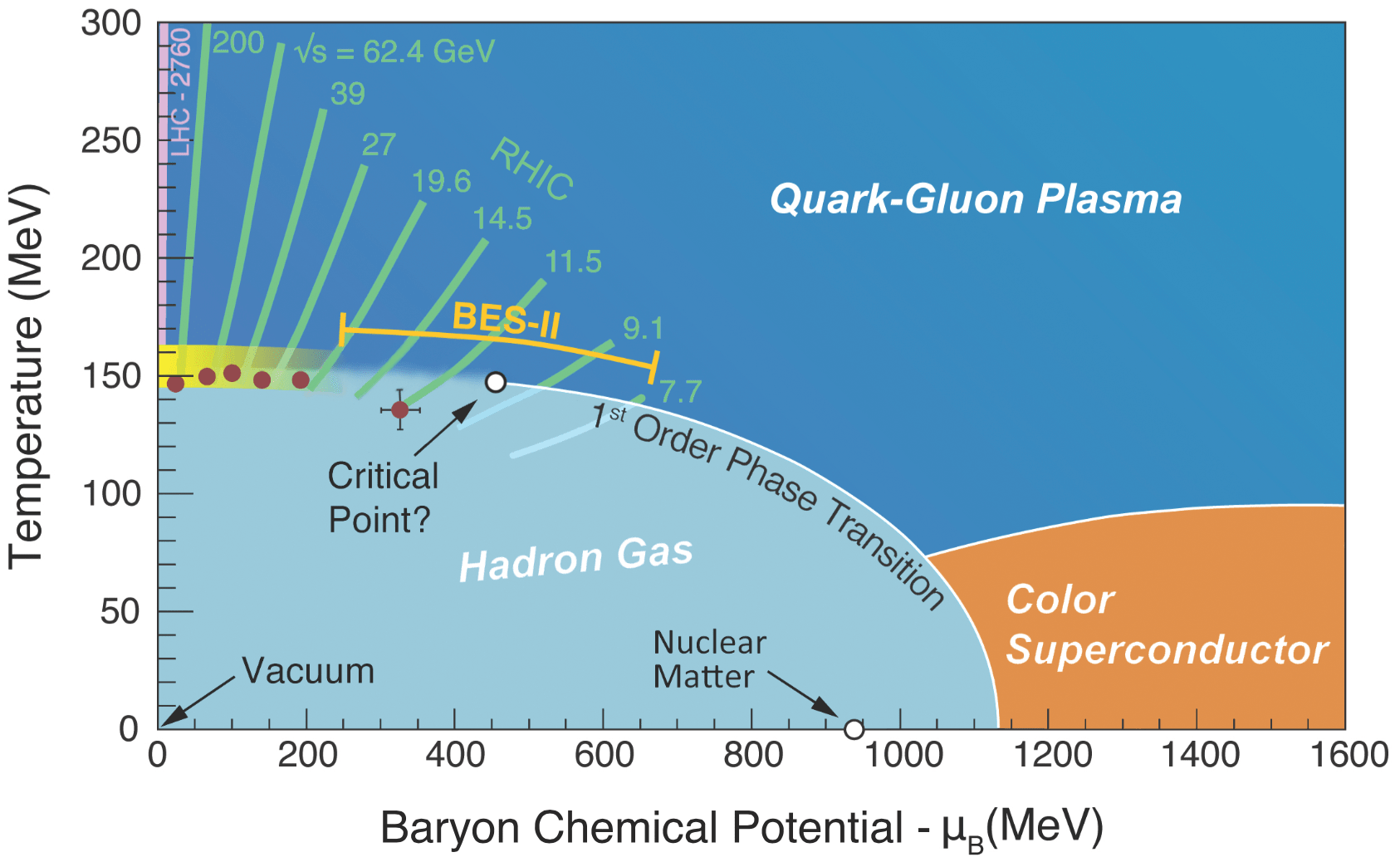}
\caption{QCD phase diagram as a function of temperature and baryon chemical potential. For sufficiently high temperatures one finds matter in QGP form \cite{Friese:2006za}.}
\label{fig:phase}
\end{figure}

One of the most crucial missions of URHIC experiments is to help reconstructing the QCD phase diagram Fig.~\ref{fig:phase}. The QCD phase diagram illustrates different states of nuclear matter as a function temperature and baryon chemical potential. The baryon chemical potential in this context is a measure of the imbalance between quarks and anti-quarks. The point $T=\mu=0$ corresponds to the QCD vacuum. Ordinary nuclear matter lives at nearly vanishing temperature and small chemical potential ($\mu=\mu_{\rm q}\approx 310$ MeV) region.
 Increasing chemical potential keeping the temperature low 
 leads to a strongly correlated dense nuclear matter phase. This phase is naturally observed in neutron stars. By further compressing the system, it transitions to a quark matter phase at an unknown critical chemical potential $\mu_c$. At higher densities, we end up with ultra high density phase which is expected to be color super conducting \cite{Alford:2007xm,Rischke:2003mt}. 

 On the other hand, by heating the system up at vanishing chemical potential, first a strongly interacting gas of hadrons is generated. With further heating, collisions between hadrons become stronger until the energies become sufficient to overcome the hadron binding energy and, then, the system transforms into the QGP phase. The nature of transition between hadronic gas and QGP depends on the value of the chemical potential at the transition. For large values of $\mu$, the transition is expected to be of first order. At some critical chemical potential $\mu=\mu_{\rm cp}$, the system undergoes the second order phase transition at a critical point $T_{\rm cp}$. For smaller values of $\mu$, lattice QCD studies have shown that is a crossover \cite{Borsanyi:2010cj,Bazavov:2009zn}. A crossover is a smooth transition between two phases (herein between QGP and hadronic matter) which occurs in a region of temperatures (instead of a single point). During a crossover, thermodynamical quantities change quickly, however, the free energy and its derivatives to all orders are continuous, thus the crossover is also called an $\infty$-order phase transition. This means that the value of the thermodynamics parameters (e.g. energy density, pressure, etc) change dramatically in a narrow range of temperature, however, there are
no discontinuities in the thermodynamical quantities (no discontinuities in the derivatives of system's free energy).
\begin{figure}[t]
\centering
\hspace{-0.5cm}
\includegraphics[width=.7\linewidth]{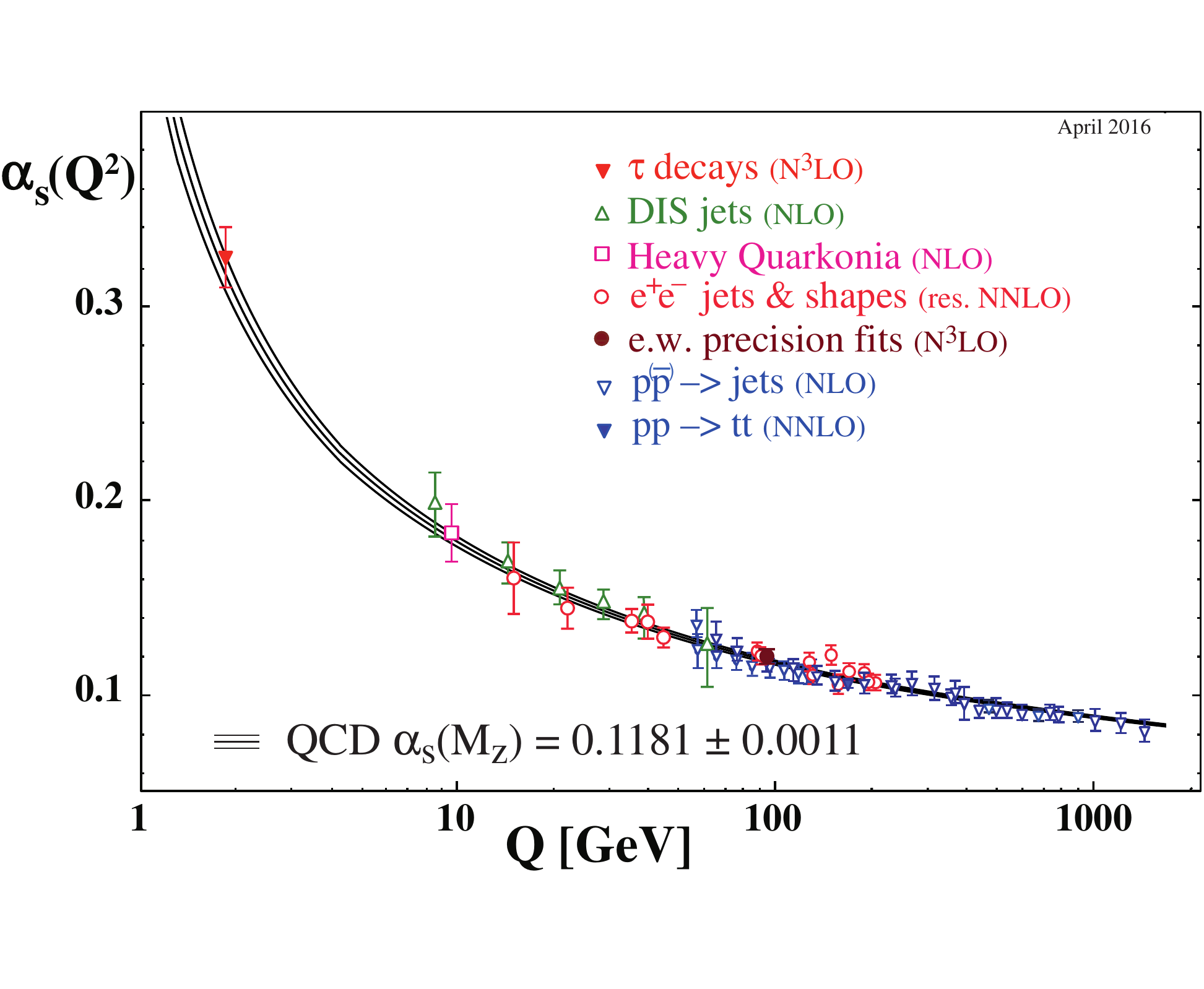}
\caption{Left: Sketch of the QCD phase diagram in the $m_{ud}-m_s$ plane. This plot is known as Columbia plot \cite{deForcrand:2007rq}.}
\label{fig:col}
\end{figure}
In order to better understand the QCD equation of state, I briefly review the theoretical calculations based on lattice QCD (LQCD)  with vanishing chemical potential. For a system of quarks with massless u and d and other quark species being infinitely massive, one expects a phase transition of second order \cite{Brown:1990ev,deForcrand:2007rq,Heller:2006ub}, see Fig.~\ref{fig:col}.
 In the case of massless u, d, and s quarks with the rest quarks being infinitely massive the transition would be of the first order. Also, when all quarks species are infinitely massive we have again a first order phase transition. However, for a system with the realistic quark species masses ($m_u\simeq m_d\simeq 10$ MeV and $m_s\simeq 120$ MeV) it is expected to be a rapid crossover. For this last case, use of `critical point' terminology is meaningless, although its concept is used to define a pseudo-critical transition temperature by searching for the point where the chiral susceptibility is maximal, i.e. the points with maximum fluctuation of the chiral order parameter. By increasing the chemical potential, the crossover turns to a first order phase transition at $\mu_{\rm cp}$. Due to the fermion sign problem \cite{Philipsen:2005mj}, which results from a complex fermionic determinant in the path integral, it is currently not possible to find $\mu_{\rm cp}$ using LQCD. There are some techniques which can be used which are (i) Taylor expanding the fermionic determinant around $\mu=0$ (ii) introducing purely imaginary chemical potential, or (iii) re-weighting methods where we treat non-zero chemical potential as an observable rather than being a part of the integration measure. It is interesting to know the location of critical point because this is where the correlation length diverges potentially. 
There are several groups around the world that are trying to obtain the high-energy equation of state by numerically solving QCD lagrangian on lattice. 
\begin{figure}[t!]
\hspace{-0.5cm}
\centering
\includegraphics[width=.7\linewidth]{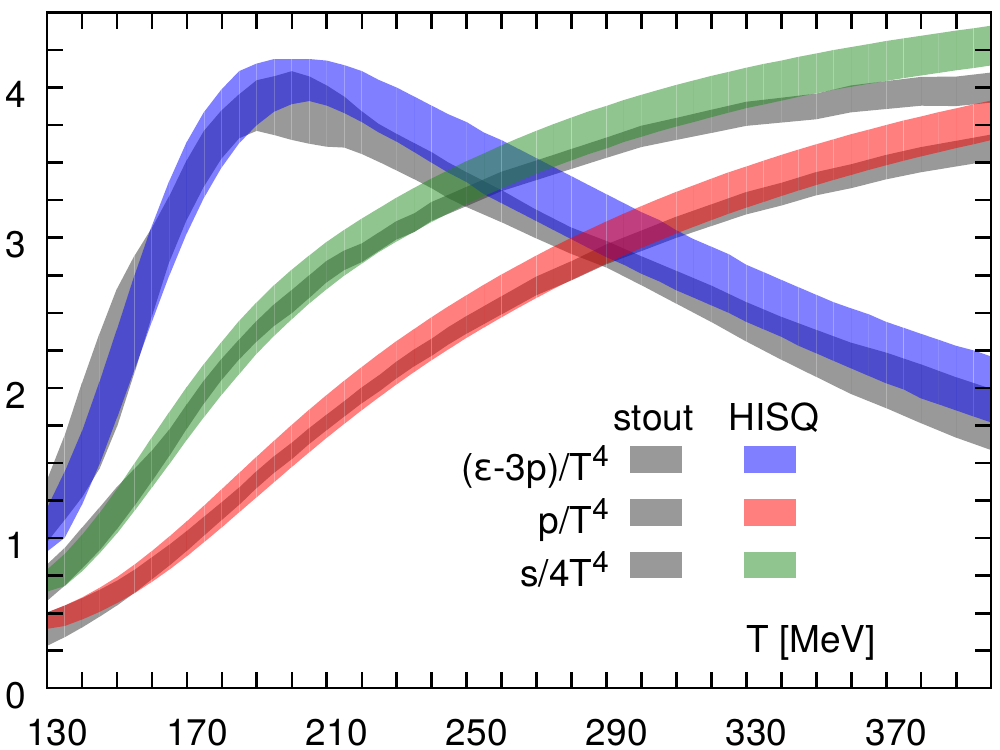}
\caption{The comparison of the HPQCD and WB collaborations
for the trace anomaly, the pressure, and the entropy density \cite{Bazavov:2014pvz}.}
\label{fig:eos-lattice}
\end{figure}
Lattice QCD calculations are important since they provide the only first principle way to calculate the thermal properties of in-medium hadrons and the EoS
of QCD matter, etc.  Lattice calculations of the QCD EoS were first performed in 1980 \cite{Engels:1980ty}. For recent reviews see for instance Refs. \cite{Petreczky:2012rq,Philipsen:2012nu,DeTar:2009ef}.

Studying the QCD-based EoS in vicinity of QCD phase transition requires nonperturbative  techniques. This is originated from the fact that in such regions coupling constant is fairly large (g $\sim 2$) and we deal with a strongly coupled system. Below the phase transition,  hadron  resonance  gas  models
(HRG)  for  the  equation  of
state are quite successful \cite{BraunMunzinger:2003zd}. This provides the LQCD calculation with a boundary value problem at the phase transition where both approaches should smoothly merge. In Fig.~\ref{fig:eos-lattice}, I provide the comparisons between two most successful LQCD results obtained for (2+1)-flavor QCD. The results are obtained from HPQCD collaboration \cite{Follana:2006rc} and Wuppertal-Budapest collaboration \cite{Borsanyi:2010cj}. This Figure compares the trace anomaly, pressure, and entropy density all scaled by their Stephan-Boltzmann limit. This figure also indicates the rapid crossover in the pressure between 150-250 MeV. The comparison shows a high level of consistency between two approaches.  
 
 \subsection{Heavy quarks}

Finding the evidence of a phase transition to a QGP is one of the essential goals of URHICs. Therefore, people are interested in the particles with cleanest trace of early time dynamics of the QGP. For this purpose, it is preferable to consider the particles that have undergone the least amount of interactions with plasma and therefore carry a good memory of the early stage dynamics. In principle, the external probes, e.g. ultra fast laser pulses and particle beams, are completely ruled out due to the short lifetime of the QGP. Also, particles produced late in the evolution of the system are not a suitable choice, because due to their strong (late time) interactions with the late time hadron gas their ``memory'' of the information they carry is quite erased. Among the particles that are produced at early times and also are not significantly affected by final-state interactions are lepton pairs and heavy quarkonia. The former one is hard to disentangle since it is technically hard to precisely distinguish the dileptons produced at early stages of the collision from the ones generated after freezeout due to interactions of hadrons. However, heavy quarkonia measurements and analysis is more practical and plausible. The bound states of heavy quarks (quarkonia) do not significantly dissolve in a QGP and they can exist as bound states in QGP up to rather large temperatures. For this reason, these bound states are considered as interesting probes for measuring the temperature of the QGP. Because heavy quark bound states may not completely melt in QGP, their formation and breakup are affected by Debye screening\footnote{Debye screening is the process of shielding the long-range interactions in the QGP which is similar to electromagnetic screening in plasmas. As a result of this screening, the color-charged partons feel a weaker color force at long distances.} in QGP, e.g. $J/\Psi$ suppression and enhancement \cite{Matsui:1986dk,Thews:2000rj}.

\subsection{Finite-temperature field theory}
 In order to study the dynamics and evolution of heavy quark bound states in QGP more accurately people use the finite temperature field theory to calculate the heavy quark potential. In general, we have two formulations for finite-temperature field theory  which is based on the way we treat the time variable in the dynamics. (i) The imaginary time formalism (ITF): ITF is based on quantum statistical mechanics and is applicable only to systems in thermal equilibrium. In this formulation, an imaginary time variable is introduced through $\tau=it$. Using the new variable, under some enforced periodic boundary conditions, the time evolution operator defined in the Feynman path integral framework looks  like the statistical mechanics partition function. This manipulation helps us to take advantage of QFT path integral for finite-temperature QFT calculations. As mentioned above, the fact that ITF deals with the thermodynamic partition function, means that it can only be applied to equilibrium systems. One of the disadvantages of ITF is that the time variable is traded for the temperature and hence the time coordinate is not accessible in the dynamics. This issue is resolved in the real-time formalisms. (ii) Real time formalisms (RTF): In RTF, the time coordinate is maintained as a real variable alongside the temperature. There are different types of RTF. One useful RTF is ``thermo-field dynamics" (TFD) which is another good candidate for studying thermal equilibrium systems \cite{landsman1987real,chu1994unified}. In this framework, one tries to define the ensemble average of an observable as its expectation value with respect to a thermal vacuum state (in contrast to zero temperature vacuum in QFT). This exercise demands working in a fictitious Hilbert space with two times as many as ordinary Hilbert space degrees of freedom. This leads to a doubling of the number of system's degrees of freedom which itself causes the fundamental functions, e.g. propagators, self-energies, to become 2$\times$2 matrices in contrast to a single function in ITF. Another important RTF is the ``Schwinger-Keldysh''(closed path) formalism. This framework is the only well-developed formalism for non-equilibrium thermal systems. In this formalism, without loss of generality one assumes that the system exists in an equilibrium state at the distant past in time ($t_0\rightarrow -\infty$) and it falls out of equilibrium during the evolution along the real-time contour. However, generally the system never reaches equilibrium at any finite time after $t_0$, so, in order to define a physically sound ensemble average for observables one needs to deform the contour and close it back at $t=-\infty$. The action of folding the contour leads to doubling degrees of freedom, which, similar to TFD formalism, results in 2$\times$2 matrices for fundamental functions. In following chapters, I provide an explicit computation of the heavy-quark potential for an anisotropic QGP using the Schwinger-Keldysh formalism.
\subsection{Asymptotic freedom}
\begin{figure}[t!]
\hspace{-1.5cm}
\centering
\includegraphics[width=.8\linewidth]{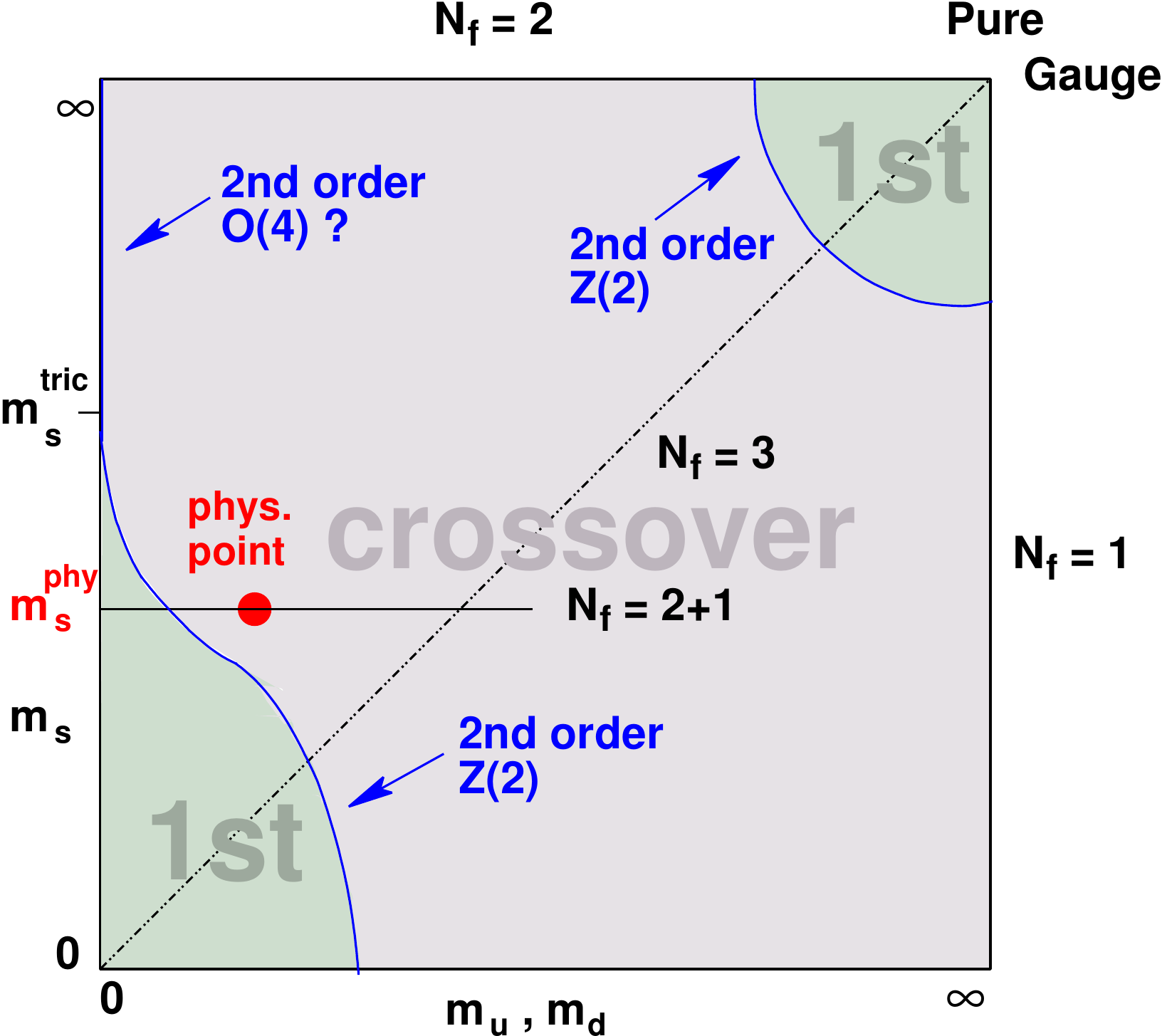}
\caption{The figure shows strong coupling constant as a function of energy. Different theoretical and experimental data illustrate the asymptotic freedom at high energy limit. The figure is taken from \cite{Tanabashi:2018}.}
\label{fig:asy}
\end{figure}

One of the distinguishing properties of gluons compared to other force carriers is that gluons carry color charge and, besides interacting with quarks, interact with themselves as well. This property turns QCD into a more complex QFT with unique features, e.g. asymptotic freedom and infrared confinement. 
QCD predicts that the strength of the force between quarks changes with distance (or momentum) in a particular calculable way.
That being said, quarks interact weakly at small scales (high energies), allowing perturbative calculations to be implemented for studying the QGP as high energies. On the other hand, at low energies the interactions becomes strong, leading to the confinement of quarks and gluons within hadrons, i.e. see Fig.~\ref{fig:asy}.
In general, the variation of a physical coupling constant under changes of scale can be understood qualitatively as the mutual interactions between virtual particles and the fields \cite{peskin2018introduction}.
QCD deals with virtual quark-antiquark pairs, which tend to screen the color charge (the same as electrons and ions screen electrostatic potential in EM plasma). However, the gluons which carry color charges do screen the color charge in a different way. The color fields tend to weaken due to quark screening whereas they tend to enhance due to gluon screening (sometimes is called anti-screening). By getting closer to the quarks as we decrease the length scale (or increase resolution of the experiment) one finds out that gluons anti-screening effect dominates and color field strength weakens, allowing for perturbative computations. Therefore, QCD is known as an asymptotically free theory.
\section{Hydrodynamics of quark-gluon plasma}
One of the most remarkable properties of the QGP which has been verified by experiment is its collective behavior. For instance, observation of flow harmonics in the hadronic spectrum indicates that the QGP carries a memory of initial geometry of the targets, e.g. impact parameter and event-by-event fluctuation. In other words, the QGP is able to exhibit collective effects during the evolution. This inspired the idea of using the relativistic hydrodynamics to describe the dynamics and evolution of the QGP. This idea was soon verified when ideal and viscous hydrodynamics successfully reproduced  the URHIC experimental observations. This great and somehow unexpected triumph tempted people to establish new formalisms for relativistic hydrodynamics in order to increase its accuracy and range of validity in the context of heavy-ion collisions. 

\subsection{Early thermalization puzzle}
Another fantastic feature of hydrodynamics is its ability to incorporate the phase transition in the system self-consistently, only using a realistic equation of state. Using the modern hydrodynamics approaches \cite{Denicol:2012cn} one is able to describe the evolution of QGP down to a fraction of 1 fm/c. On the other hand, in principle, it is usually assumed that hydrodynamics demands local equilibration which implies that the QGP system must undergo the fast thermalization (of order of a fraction of $1$ fm/c). Based on this speculation, researchers suggested several theoretical models to describe the fast thermalization of the QGP. (i) Some people tried to model that by including the collisions of all type in parton-cascade model \cite{Geiger:1991nj}, e.g. binary collision, gluon radiation (1$\rightarrow 2$, $2\rightarrow 2$, $2\rightarrow 3$, $3\rightarrow 3$, etc) (ii) some used bottom-up thermalization \cite{Baier:2000sb} which occurs at weak coupling limit and is dominated by soft-enhanced scattering. In this case, people assume that the initial partons are produced by hard collisions with the initial state of the collision being described by QCD saturation mechanisms, e.g. CGC. (iii) Some studied the role of instabilities which are effective in an anisotropic quark-gluon plasma and are more important than the collisions in the weak coupling limit \cite{Arnold:2003rq}. (iv) Another approach is the equilibration without any secondary interactions, e.g.. Schwinger mechanism in the strong color-fields for thermalization of transverse momentum and Hawking-Unruh effect for overall thermalization \cite{Kharzeev:2005iz}. (v) Considering the thermalization as an effect of chaotic dynamics of the non-abelian classical color fields \cite{florkowski2010phenomenology}. All these models decrease the estimations for thermalization time only down to about 2-3 fm/$c$. On one hand, successful predictions of experimental observations using viscous hydrodynamics and, on the other hand, failure of theoretical models to describe the fast themalization, led to a big puzzle, i.e.``early thermalization puzzle'' \cite{Florkowski:2012mz}. 

\subsection{Range of validity of hydrodynamics}
In order to deal with early thermalization puzzle, I remind the reader that the hydrodynamics is a classical effective field theory that mimics  properties of underlying microscopic description of evolution of the system toward equilibrium.  However, such an effective description, in practice, is not derived from a microscopic theory. Instead, it is derived based on some general postulates in connection with symmetries and resulting conservation laws, and then connecting to microscopic description by matching the gradient expansion between hydrodynamics and underlying microscopic framework. Baier et al. in \cite{Baier:2007ix} demonstrated that hydrodynamics description can be matched with any microscopic framework up to the second order in the gradient expansion. Using these developments, one concludes that the range of applicability of hydrodynamics is much wider than what it seemed before. It means the hydrodynamics does not need the system to reach a well-defined isotropic equilibrium state in order to accurately describe the dynamics. This important result motivates the application of hydrodynamics to anisotropic, inhomogeneous, or small systems. In fact, the ``early thermalization puzzle'' could be circumvented by replacing ``thermalization'' with ``hydrodynamization'' as the condition when the hydrodynamics formulations are useful \cite{Romatschke:2016hle}. 
\subsection{Hadronization and freezeout}
Due to the hydrodynamic expansion, the QGP cools down and undergoes a phase transition from quark and gluon degrees of freedom to a hot gas of strongly interacting hadrons. In other words, at the lower temperatures quarks and gluons become confined into colorless hadrons. Further expansion leads to another stage namely hadron gas. Such decoupling of hadrons is called hadronization. After that, the hadron gas undergoes freezeout stage which turn a hot strongly interacting phase into a free streaming gas of hadrons as flying out to the detectors. The freezeout takes place in two different steps: (i) chemical freezeout when the inelastic collisions stop happening and chemical potentials freeze, i.e. particle number become conserved (ii) thermal (kinetic) freezeout when any kind of collisions (including elastic and inelastic) cease. The latter happens when $\tau_{\rm coll}\gg\tau_{\rm exp}$, with $\tau_{\rm coll}\sim 1/(\sigma n)$ and $\tau_{\rm exp}\sim \nabla\cdot u$ ($\sigma$ is the collision rate).
In general, the freezeout is a kinematically complicated procedure which occurs as a hierarchy of decouplings of various particles and force carriers happening at different times. In other words, different plasma constituents undergo freezeout at different temperatures leaving the plasma as a partially decoupled medium. The comprehensive analysis of this stage is feasible through kinetic theory. After freezeout, the hadrons behave like a free streaming non-interacting gas which keeps its distribution unchanged. However, the soft region of phase space may still be modified by decays of unstable hadron resonances.
\subsection{Elliptic flow}
Flow is a crucial observable for the studying the dynamics and evolution of the QGP. It provides useful information about the equation of state and the transport properties of matter created in heavy-ion collisions. Different forms of flow are initiated by pressure gradients developed due to the geometry of initial state of the colliding nuclei. The acceleration of expanding partons is maximal along the direction of the largest pressure gradient, i.e., along the short axis of the ellipsoidal QGP. Collective behavior of the system evolves the pressure gradient into different forms of flow.
  This results in an anisotropic azimuthal distribution of the final-state hadrons in lab momenta. In order to study the flow it is common to Fourier analyze the azimuthal distribution of particles in momentum space and compute the Fourier coefficients. 
 The first Fourier coefficient is called directed flow and the second Fourier coefficient of this azimuthal asymmetry is known as elliptic flow. The magnitude of elliptic flow coefficient depends strongly on the friction in the created matter, characterized be the ratio $\eta/s$, where $\eta$ is the shear viscosity and $s$ the entropy density. The elliptic flow coefficient is closely related to eccentricity of the overlap region. Mathematical analysis shows that $v_2$ increases for the events (single collision) with higher impact parameter and eventually turns down for higher centrality classes. However, due to the event-by-event (quantum) fluctuations at the initial state of targets one always measures a finite $v_2$ even for central collisions. On a general basis, these fluctuations also develop the higher flow harmonics ($n>2$) regardless of centrality of the collisions.
\begin{figure}[t!]
\hspace{-0.5cm}
\includegraphics[width=.5\linewidth]{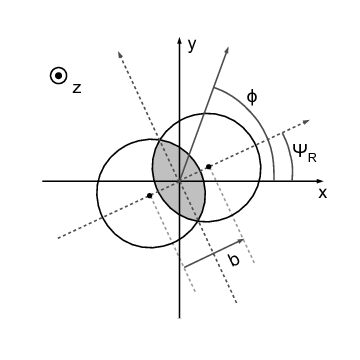}
\includegraphics[width=.5\linewidth]{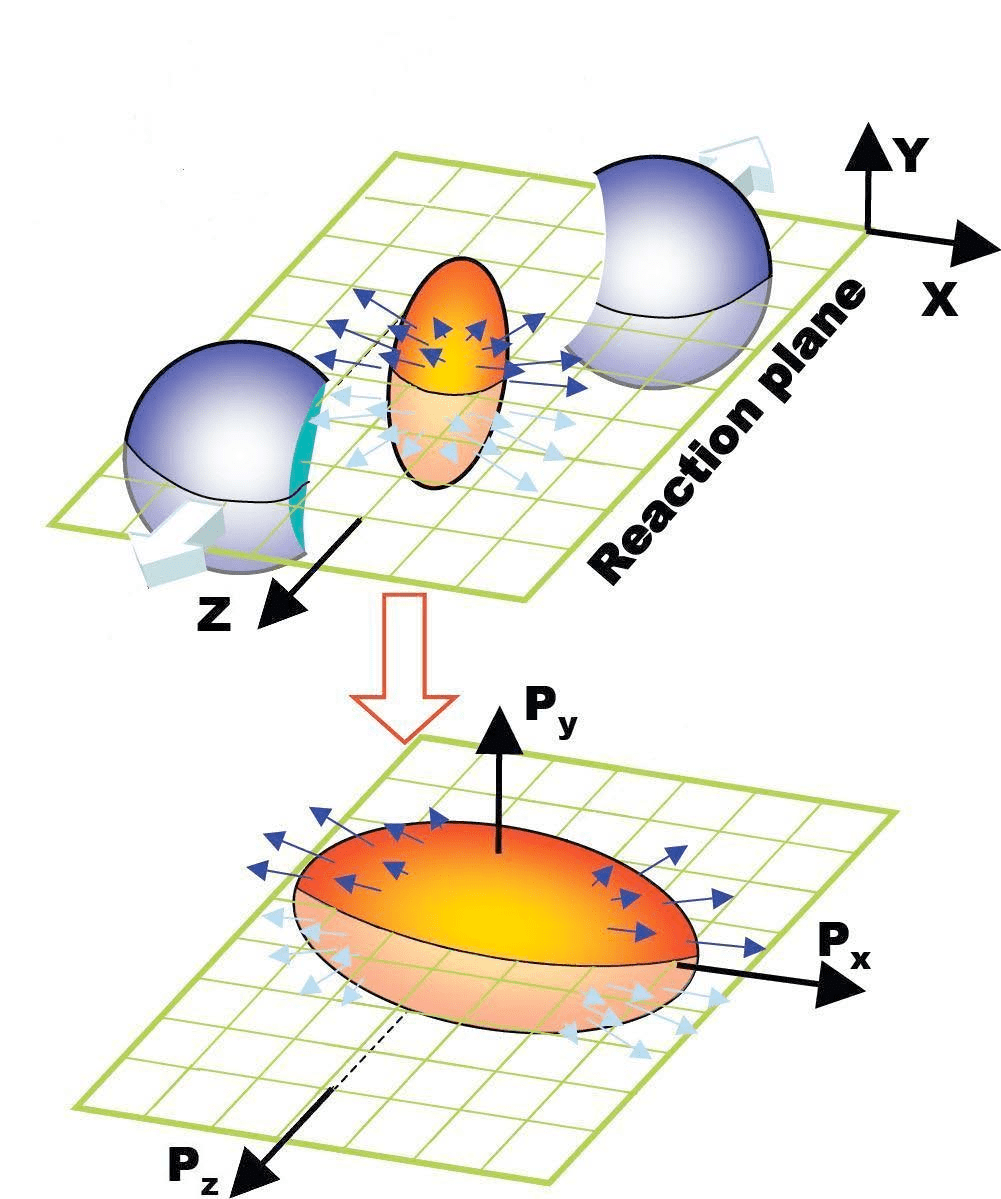}
\caption{The geometry of colliding nuclei is demonstrated. Left: configuration of reaction plane and azimuthal angle of flying partons respect to the lab frame. Right: The pressure gradients transform into an ellipsoidal distribution function in the momentum space.}
\label{fig:collision-geo}
\end{figure}
 One can expand the triple differential invariant distribution of particles emitted in the final state as a Fourier series
over azimuthal angle as
\ba
E\frac{d^3 N}{d^3{\bf p}}=\frac{d^3N}{p_T dp_T d\phi dy}\Bigg(1+ 2\sum_{n=1}^\infty v_n \cos[n(\phi-\Psi_n)]\Bigg)\,.
\ea
In above relation, ${\bf p}$ is the three vector momentum, $p_T$ is the transverse momentum, $\phi$ is the azimuthal angle, $y$ is the rapidity, and $\Psi_{n}$ the reaction plane angle pertained to each harmonic respect to the lab coordinate.
The Fourier coefficients are given by
\ba 
v_n(p_T,y)&=&\langle\cos[n(\phi-\Psi_{R})]\rangle\,. \nonumber 
\ea
In the smooth initial distribution one has vanishing odd coefficients
in Fourier series by symmetry. Also, in this case one has $\Psi_n=\Psi_{R}$. The reaction plane $\Psi_{R}$ is
defined as the plane made by the beam direction and impact parameter vector Fig.~\ref{fig:collision-geo}.
\section{The overview of my dissertation}
In my PhD dissertation, I studied the QGP created in heavy-ion collisions using a non-equilibrium hydrodynamics framework, called anisotropic hydrodynamics. For this purpose, I and my collaborators developed an anisotropic hydrodynamics framework whose details are elaborated in this dissertation. The overview of my dissertation is that, I started from an elementary aHydro framework and extended it to include the realistic aspects such that it is able to study the phenomenology of the QGP. Along the way, we have used two benchmarks for assessing aHydro: exact solutions to the Boltzmann equation and experimental data. First, the exact solutions of Boltzmann equation which are obtained for some special symmetric cases, is used frequently as a reference solution for comparisons between aHydro and other vHydro frameworks. Second, the URHIC experimental data which is used for verification of phenomenological predictions of 3+1d aHydro model. The hadron spectra is one of the observables that can be directly measured in the experiments. 
In different sections of this dissertation, I have reviewed my projects based on the development of aHydro. In the chapter \ref{chap:ahydro}, I comprehensively introduce the conformal anisotropic hydrodynamics formalism and review some basic analytical derivations. In the chapter \ref{chap:bulk}, I discuss the massive (single constant mass) anisotropic hydrodynamics equations in 0+1d, and the way it enhances the formalism by including the bulk viscous degree of freedom and breaking the conformal symmetry of the system. In chapter \ref{chap:gubser}, the solution of anisotropic hydrodynamics equations under conformal Gubser flow is discussed. The idea is to test the accuracy of a given aHydro model for a system with non-trivial transverse expansion. The system with transverse expansion evolves in 1+1d, which through translating to de Sitter coordinate simplifies to 0+1d evolution. In chapter \ref{chap:qp}, I introduce quasiparticle model, as a systematic way of implementing realistic equation of state in aHydro. Then, I have compared the results of our model with the experimental data. Chapters \ref{chap:gluon} and \ref{chap:quark}  are about a QFT study of the QGP possessing an anisotropic distribution function. In chapter \ref{chap:gluon}, I present the steps of calculation of the gluon self-energy for anisotropic QGP using hard-thermal loop summation. In this calculation, I have used spheroidal anisotropic distribution function to all orders in anisotropy. The gluon self-energy is applied to calculation of the heavy-quark potential in the QGP. Chapter \ref{chap:quark} is about calculation of the quark self-energy in a QGP with an ellipsoidal distribution function. In the chapter \ref{chap:summary}, the summary, outlook, and future perspectives are discussed. Finally, I present some appendices containing mathematical identities and special functions used in this dissertation.

\part{Developments of Anisotropic Hydrodynamics}
\chapter{\bf Overview of anisotropic hydrodynamics}
\label{chap:ahydro}
\setcounter{figure}{0}
\setcounter{table}{0}
\setcounter{equation}{0}

\section{Introduction}
Relativistic hydrodynamics has become an important and effective tool for studying the dynamics of the QGP generated in heavy-ion collisions. Different relativistic hydrodynamics formalisms (ideal and viscous hydrodynamics) have been able to reproduce the heavy-ion collision experimental results to different extents. One of the most challenging questions in this regard is the range of applicability of the hydrodynamical models. Canonical viscous hydrodynamic  models (vHydro) are built upon the assumption of proximity to an isotropic equilibrium state in the local rest frame (LRF) of the flow. Based on this assumption, one can perform a perturbative expansion of the one-particle distribution function around an isotropic equilibrium state $f\approx f_0+\delta f$. This, somehow restricting, assumption confines the validity of vHydro to the regions where the system experiences only small deviations from an isotropic equilibrium state. Under extreme conditions, i.e. large $\delta f$, the perturbative correction can overcome the leading order term and causing the whole distribution function $f$ to be dominated by $\delta f$ term. This immediately results in an unconstrained (negative or very large) distribution function and hence nonphysical values for the thermodynamics observables (negative pressure, etc).

In practice, one finds that, at early times after the collisions and in the vicinity of transverse edges of the system throughout all time, the state of the system presents a large momentum-space anisotropy \cite{Strickland:2014eua}. This results from the large longitudinal momentum carried by the colliding ions, which  causes the QGP to experience strong expansion along the beamline right after the collision. This longitudinal expansion results in a large anisotropy in the momentum-space distribution of the QGP. As a result, the vHydro formalism, despite its impressive phenomenological success in describing soft hadronic spectra, gives non-physical predictions where the momentum-space anisotropy is sufficiently large. 

To resolve this issue, let's take a closer look at the perturbative expansion of vHydro. As discussed before, the perturbative expansion becomes invalid when the perturbation is large enough to control the whole expansion. One of the ways to avoid this, is to sum up the perturbative (non-equilibrium) terms in the leading order term by introducing a non-equilibrium distribution function. The resummed distribution function, which incorporates the non-equilibrium anisotropic effects, guarantees positivity in all spacetime regions. 

\section{Conventions and notations}
In this section, I introduce the notation, mathematical identities, and different coordinate systems which are used throughout this document. Unless otherwise indicated, the metric signature is taken to be `mostly minus' (also called: timelike, west coast, or particle physics convention)\footnote{In the chapter 4, where I introduce the anti-de Sitter coordinate, with `mostly plus' metric convention.}. Based on that, the line element in Minkowski space, i.e. $x^\mu=(t,x,y,z)$, is 
\ba
ds^2=\eta_{\mu\nu}dx^\mu dx^\nu=dt^2-dx^2-dy^2-dz^2\,,
\ea
where $t$ is the time, $z$ is the coordinate along the beamline, and $x$ and $y$ denote the coordinate in the transverse plane to the beamline. $\eta^{\mu\nu}\equiv{\rm diag}(1,-1,-1,-1)$ denotes the Minkowski metric while $g^{\mu\nu}$ is reserved for other metric tensors used in this document, e.g. Milne, Cartesian Milne, and anti de Sitter coordinates.
Because sometimes in the calculations I deal with boost-invariant cylindrically-symmetric systems, it is useful to introduce another coordinate called (polar) Milne spacetime coordinate, i.e. $x^\mu=(\tau,r,\phi,\varsigma)$. In this coordinate, $\tau\equiv\sqrt{t^2-z^2}$  denotes the longitudinal proper time and $\varsigma\equiv\tanh^{-1}(z/t)$ specifies the longitudinal spacetime rapidity. Also, in this framework, $\phi\equiv\tan^{-1}(y/x)$ is the azimuthal angle, and $r\equiv\sqrt{x^2+y^2}$ is the radial coordinate, both in the plane transverse to the beamline. There is also a Cartesian Milne coordinate, i.e. $x^\mu=(\tau,x,y,\varsigma)$, which is used for the systems with cylindrical symmetry explicitly broken. 

 The transverse projector $\Delta^{\mu\nu}=\eta^{\mu\nu}-u^\mu u^\nu$ is used to project
four-vectors and/or tensors into the space orthogonal to $u^\mu$. Parentheses and square brackets on indices denote symmetrization and anti-symmetrization, respectively, 
i.e. $A^{(\mu\nu)}\equiv (A^{\mu\nu}+A^{\nu\mu})/2$ and $A^{[\mu\nu]}\equiv (A^{\mu\nu}-A^{\nu\mu})/2$. Angle brackets on indices indicate projection with a four-index transverse
projector, $A^{\langle\mu\nu\rangle}\equiv \Delta^{\mu\nu}_{\alpha\beta} A^{\alpha\beta}$, where $\Delta^{\mu\nu}_{\alpha\beta}\equiv \Delta^{(\mu}_\alpha \Delta^{\nu)}_\beta-\Delta^{\mu\nu}\Delta_{\alpha\beta}/3$ projects out the traceless and
$u^\mu$-transverse component of a rank-two tensor.  

\section{Conformal anisotropic distribution function}
Before introducing the anisotropic hydrodynamics distribution function, let me review some basic definitions in statistical mechanics. `Statistical mechanics' is used to describe systems with a large number ($\sim$ Avogadro's number) of particles. In kinetic theory, the one particle distribution function is defined as a continuous and real-valued function in eight-dimensional phase space (for on-shell particles one deals with seven dimensional phase space). The eight dimensions of phase space consists of four spacetime coordinates, i.e. $x^\mu=(t,{\bf r})$, and four momentum coordinates, i.e. $p^\mu=(p^0,{\bf p})$. The value of this function at fixed time, $t$, specifies the number of particles per unit volume of six-dimensional phase space, $\bf{x}, \bf{p}$. Integrating the distribution function over the three-momentum provides the number density as a function of four dimensional spacetime. The particles follow different statistics based on the spin they carry. For instance, fermions (half-integer spin) follow Fermi-Dirac statistics, bosons (integer spin) follow Bose-Einstein statistic, or classical particles (the effect of spin is ignored as assumed in the classical limits) follow Maxwell-Boltzmann distribution. The isotropic equilibrium distribution function is defined as\footnote{Herein, it is assumed that the chemical potential is zero.}
\ba
f_{\rm iso}(x,p)=f_{\rm iso}(p)=\bigg[\exp\left(\frac{u^\mu p_\mu}{T}\right)+a\bigg]^{-1}\,,
\label{eq:eqdist}
\ea
with $a=0,+1,-1$ corresponding the Maxwell-Boltzmann, Fermi-Dirac, and Bose-Einstein statistics, respectively. As expected from equilibrium distribution, the argument of exponential term in Eq.~(\ref{eq:eqdist}) is isotropic. Note that throughout this dissertation, unless otherwise is indicated, by ignoring the effects of spin statistics I specialize to the Maxwell-Boltzmann statistic, i.e. $a=0$ in Eq.~(\ref{eq:eqdist}) which yields $f_{\rm eq}({\bf p})=\exp(-|{\bf p}|/T)$.

The leading order anisotropic distribution function is parametrized through deforming the argument of exponential in the distribution function into the anisotropic one \cite{Martinez:2012tu}
\ba 
f(x,p)=f_{\rm eq}\bigg(\frac{\sqrt{p_\mu\Xi^{\mu\nu}p_\nu}}{\lambda}\bigg)\,,
\label{eq:anisodist1}
\ea
where $\lambda$ has dimensions of energy and can be identified with temperature only in the isotropic equilibrium limit. Note that in practice $f_{\rm eq}$ need not be a thermal equilibrium distribution. However, throughout this document, I take it to be of thermal equilibrium form, i.e. Eq.~(\ref{eq:eqdist}). In the conformal (massless) case, the rank-2 tensor $\Xi^{\mu\nu}$ specifying the shape of the distribution in the momentum space is defined as 
\ba 
\Xi^{\mu\nu}=u^\mu u^\nu +\xi^{\mu\nu}\,,
\label{eq:xi}
\ea
\begin{figure}[t!]
\hspace{-0.5cm}
\centering
\includegraphics[width=.98\linewidth]{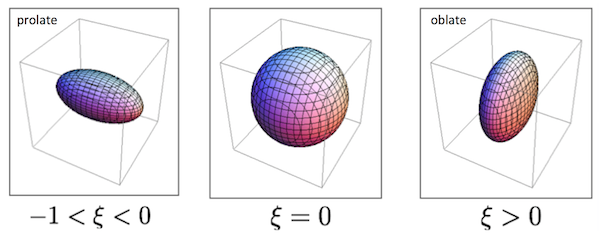}
\caption{This figure depicts the shape of spheroidal distribution function in the momentum space for different values of longitudinal anisotropy parameter \cite{Strickland:2014pga}.}
\label{fig:spheroids}
\end{figure}
where $u^\mu$ is the flow velocity four-vector, and $\xi^{\mu\nu}$ denotes a symmetric traceless anisotropy tensor, i.e. $\xi_x+\xi_y+\xi_z=0$. The quantities $\lambda$, $u^\mu$ and $\xi^{\mu\nu}$ are understood to be functions of spacetime and satisfy the following identities
\ba
u^\mu u_\mu = 1 ;\quad
{\xi^{\mu}}_\mu = 0 ;\quad
u_\mu \xi^{\mu\nu} = 0 \, .
\ea
In the LRF of the flow one has $u^\mu=(1,0,0,0)$ and at the leading order one has $\xi^{\mu\nu}=\rm diag(0,\xi_x,\xi_y,\xi_z)$. 
The three scalar parameters $\xi_i$ encode the degree of anisotropy in the three Cartesian directions in the LRF. In fact, based on tracelessness of $\xi^{\mu\nu}$, the number of independent anisotropy parameters for a conformal system is two. Using (\ref{eq:xi}), the distribution function (\ref{eq:anisodist1}) takes the form 
\ba 
f(x,p)=f_{\rm eq}\bigg(\frac{1}{\lambda}\sqrt{(1+\xi_x)({\bf p}\cdot X)^2+(1+\xi_y)({\bf p}\cdot Y)^2+(1+\xi_z)({\bf p}\cdot Z)^2}\bigg)\,.
\label{eq:anisodist2}
\ea
Based on geometrical considerations, the $\xi_i$ ($i\in x,y,z)$ range as $-1<\xi_i<\infty$. Negative/positive values for $\xi_i$ corresponds to the  stretched/squeezed distributions along $i$-$\rm th$ direction. In the most general case, when all three anisotropy parameters are non-zero, the system is anisotropically deformed in three directions, yielding an ellipsoidal distribution function in the LRF. However, sometimes for simplicity of the dynamics or because of symmetries, e.g. transversally homogeneity, one can ignore the anisotropy in the transverse plane and define a spheroidally anisotropic distribution which is known as Romatschke-Strickland form \cite{Romatschke:2003ms,Romatschke:2004jh} (with a new set of parameters $\Lambda$ and $\xi$) 
\ba 
f(x,p)=f_{\rm eq}\bigg(\frac{1}{\Lambda}\sqrt{{\bf p}^2+\xi({\bf p}\cdot {\bf n})^2}\bigg)\,.
\label{eq:anisodist3}
\ea
In this relation ${\bf n}$ is the unit vector along anisotropy direction.
Fig. \ref{fig:spheroids} demonstrates the spheroidal distribution for different ranges of $\xi$. In order to make the definition of spheroidal distribution function more clear, one can construct a map between the two sets of variables $(\xi_i,\lambda)$ and $(\xi,\Lambda)$. The logic is based on the fact that spheroidal distribution is equivalent to an ellipsoidal distribution when its spheroidal limit is taken, i.e. $\xi_x=\xi_y$. By taking ${\bf n}=\hat{z}$:
\begin{equation}
\begin{split}
\frac{\sqrt{(1+\xi_x)p_\perp^2+(1+\xi_z)p_z^2}}{\lambda}=\frac{\sqrt{{\bf p}^2+\xi p_z^2}}{\Lambda} \quad \Rightarrow\quad
\end{split}
 \begin{cases}
\Lambda = \frac{\lambda}{\sqrt{1+\xi_x}}\,,\\
1+\xi=\frac{1+\xi_z}{1+\xi_x}\,.
 \end{cases}
 \label{eq:map}
\end{equation}

The value $\xi_i=0$ (or $\xi=0$) gives an isotropic distribution distribution, i.e. $\lambda=T$ (or $\Lambda=T$). In this case, $\Xi^{\mu\nu}$ reduces to $\Xi^{\mu\nu}=\rm diag (1,0,0,0)$, which yields the expected isotropic expression, i.e. $\sqrt{p_\mu\Xi^{\mu\nu}p_\nu}=E=|{\bf p}|$.

\section{Spacetime basis vectors}
In order to obtain relativistic hydrodynamic equations of motion of the QGP from the Boltzmann equation, it is useful to define four spacetime four-vectors, which is the LRF are simply the  Minkowski space unit vectors. In the LRF of the fluid, they have the form
\ba
X_{0,\rm LRF}^\mu &\equiv& u^\mu_{\rm LRF}=(1,0,0,0)\,,\nonumber\\
X_{1,\rm LRF}^\mu &\equiv& X^\mu_{\rm LRF}=(0,1,0,0)\,,\nonumber\\
X_{2,\rm LRF}^\mu &\equiv& Y^\mu_{\rm LRF}=(0,0,1,0)\,,\nonumber\\
X_{3,\rm LRF}^\mu &\equiv& Z^\mu_{\rm LRF}=(0,0,0,1)\,.
\label{eq:basisLRF}
\ea

The LRF is defined as a frame constructed at the location of a specific fluid element, which moves with the same velocity as the local fluid element. In this frame the fluid element is always at rest. The set of LRF basis vectors are mutually related. They are orthogonal to each other and they fulfill the normalization condition, which can be summarized as
\ba
(X_{\alpha,\rm LF})_\mu(X_{\beta,\rm LF})^\mu=\eta_{\alpha\beta}\,.
\label{eq:ortho}
\ea
 I point out that one can express the metric tensor itself in terms of these 4-vectors as
\be
\eta^{\mu \nu}= X^\mu_0 X^\nu_0 - \sum_i X^\mu_i X^\nu_i \, ,
\label{eq:gbasis}
\ee
where the sum extends over $i=1,2,3$.  In addition, the standard transverse projection operator which is orthogonal to $X^\mu_0$ can be expressed in terms of the basis (\ref{eq:basisLRF})
\be
\Delta^{\mu \nu} = \eta^{\mu\nu} - X^\mu_0 X^\nu_0 = - \sum_i X^\mu_i X^\nu_i \, ,
\label{eq:transproj}
\ee
which yields $u_\mu \Delta^{\mu \nu} = u_\nu \Delta^{\mu \nu} = 0$ as expected.  Note that the spacelike components of the tensor basis are eigenfunctions of this operator, i.e. $X_{i\mu} \Delta^{\mu \nu} = X^\nu_{i}$.

In order to study the phenomenology of the system one needs to construct Lab Frame (LF) which is the coordinate system sitting in the lab where measurements are performed on the system. Based on the geometry of heavy-ion collisions system, one can define a sequence of Lorentz transformations to connect the two coordinate systems. Let's get started with taking longitudinal ($z$) axis of lab frame along the beamline. First, one can boost the lab frame by $L_z(\vartheta)$ to make it move longitudinally with the same speed as the fluid element. Then by performing a rotation $R_z(\theta)$, one can align the $x$-axis along the LRF $x$-axis. Finally, by boosting as $L_x(\psi)$ transversally along $x$ axis, the transformation to LRF of the fluid is completed. This set of transformations in mathematical language is written as 
\ba
X^\mu_{\alpha,\rm LRF}=L_x(\psi)R_z(\theta)L_z(\vartheta) X^\mu_{\alpha,\rm LF}\,,
\ea
where $\alpha\in \{0,1,2,3\}$. This will yield 
\ba
X^\mu_{\alpha,\rm LF}=L^{-1}_z(\vartheta)R^{-1}_z(\theta)L^{-1}_x(\psi) X^\mu_{\alpha,\rm LRF} \,.
\ea
Using standard definitions for rotation and boost matrices one obtains the LF basis vectors
\ba
X_{0,\rm LF}^\mu &\equiv & u^\mu =(\cosh\psi\cosh\vartheta, \sinh\psi\cos\varphi,\sinh\psi\sin\varphi,\cosh\psi\sinh\vartheta)\,,\nonumber \\
X_{1,\rm LF}^\mu &\equiv &X^\mu =(\sinh\psi\cosh\vartheta, \cosh\psi\cos\varphi,\cosh\psi\sin\varphi,\sinh\psi\sinh\vartheta)\,,\nonumber \\
X_{4,\rm LF}^\mu &\equiv &Y^\mu =(0, -\sin\varphi,\cos\varphi,0)\,,\nonumber \\
X_{3,\rm LF}^\mu &\equiv &Z^\mu =(\sinh\vartheta, 0,0,\cosh\vartheta)\,.
\label{eq:basis-gen}
\ea
The Lorentz transformations do not change the orthogonality and orthonormality of the basis vectors, so the LF basis vectors satisfy the same relations as (\ref{eq:ortho}). 
The four-vectors defined above are completely general and yet no specific assumption for the symmetries of the system is applied.
Introducing another parametrization by using the temporal and transverse components of flow velocity
\ba 
u_0&=&\cosh\psi\, , \\
u_x&=& u_\perp \cos\varphi\, , \\
u_y&=& u_\perp \sin\varphi\, , 
\label{eq:u-par}
\ea
where $u_\perp\equiv \sqrt{u_x^2+u_y^2}=\sqrt{u_0^2-1} = \sinh\psi$, one has
\ba
u^\mu &\equiv& (u_0 \cosh\vartheta,u_x,u_y,u_0 \sinh\vartheta) \, , \nonumber\\
X^\mu &\equiv& \Big(u_\perp\cosh\vartheta,\frac{u_0 u_x}{u_\perp},\frac{u_0 u_y}{u_\perp},u_\perp\sinh\vartheta\Big) , \nonumber \\ 
Y^\mu &\equiv& \Big(0,-\frac{u_y}{u_\perp},\frac{u_x}{u_\perp},0\Big)  , \nonumber \\
Z^\mu &\equiv& (\sinh\vartheta,0,0,\cosh\vartheta ) \, .
\label{eq:4vectors}
\ea

\subsection{Simplification for some symmetrical cases}
For simplicity of calculations sometimes, however, I assume some symmetries for the system.   For future reference, I point out that in the limit that the system is boost invariant one can identify the longitudinal boost as longitudinal spacial rapidity, i.e. $\vartheta=\varsigma$, with
\ba
t&=&\tau\cosh\varsigma\,,\nonumber\\
z&=&\tau\sinh\varsigma\,.
\ea
Moreover, if the system is cylindrically symmetric, the rotation angle is equal to the azimuthal angle of the system, i.e. $\varphi=\phi$, and the transverse boost is equal to the radial flow, i.e. $\psi=\theta_\perp$, such that it is related to the radial flow velocity as $v_\perp=\tanh\theta_\perp$. With these definitions the basis vectors in the LF in the boost-invariant and cylindrically symmetric (1+1d) case is summarized as
\ba
u^\mu &=&(\cosh\theta_\perp\cosh\varsigma, \sinh\theta_\perp\cos\phi,\sinh\theta_\perp\sin\phi,\cosh\theta_\perp\sinh\varsigma)\,,\nonumber \\
X^\mu &=&(\sinh\theta_\perp\cosh\varsigma, \cosh\theta_\perp\cos\phi,\cosh\theta_\perp\sin\phi,\sinh\theta_\perp\sinh\varsigma)\,,\nonumber \\
Y^\mu &=&(0, -\sin\phi,\cos\phi,0)\,,\nonumber \\
Z^\mu &=&(\sinh\varsigma, 0,0,\cosh\varsigma)\,.
\label{eq:basis1+1d}
\ea
For transversally homogenous system, the transverse flow is absent, i.e. $\theta_\perp=0$, and,  
\ba
u^\mu &=& (\cosh\varsigma,0,0, \sinh\varsigma)\,,\nonumber \\
X^\mu &=& (0,\cos\phi,\sin\phi,0)\,,\nonumber \\
Y^\mu &=& (0,-\sin\phi,\cos\phi,0)\,,\nonumber \\
Z^\mu &=& (\sinh\varsigma,0,0,\cosh\varsigma).
\label{eq:basis0+1d}
\ea
Note that in the last case, $X^\mu$ and $Y^\mu$ are simply unit vectors in polar coordinates.

\section{Dynamical equations}
The dynamical equations in anisotropic hydrodynamics, similar to other hydrodynamics approaches, can be derived from the Boltzmann equation. The Boltzmann equation which is the main kinetic theory equation, is a non-linear differential equation which describes the dynamical behavior of a system which is not necessarily in equilibrium.  In a system with a large number of particles it is very hard to analyze the individual positions and momenta of the particles in the fluid. The Boltzmann equation, however, deals with this issue by considering the phase space probability distribution of the particles, i.e. $f(x,p)$. In the absence of external forces, the Boltzmann equation in covariant form is 
\ba
p^\mu\partial_\mu f(x,p)=-{\cal C}[f(x,p)]\,.
\label{eq:boltzmann}
\ea
In this equation, ${\cal C}[f]$ is the collisional kernel. The collisional kernel is a potentially complicated function which encodes all elastic and inelastic collisional interactions which modify the distribution of the particles in the plasma.\footnote{Note that the above equation is valid for a system with massless or constant-mass particles. Later, I will discuss a modified version of Boltzmann equation which governs quasiparticles with varying mass.} Note that the Boltzmann equation in the form of Eq.~(\ref{eq:boltzmann}) is completely general and no specific assumptions for the system's interactions is made. However, as discussed previously, hydrodynamics studies the system's evolution in the near equilibrium and small gradient limit. The typical way to obtain the hydrodynamic equations is to take different moments of the Boltzmann equation with an appropriate prescription for the collisional kernel. For a system which evolves close to equilibrium, one can use a simple formula for the collisional kernel, i.e. relaxation-time approximation (RTA)
\ba 
{\cal C}[f(x,p)]=\frac{p\cdot u}{\tau_{\rm eq}}\Big[f(x,p)-f_{\rm eq}(x,p)\Big]\,,
\label{eq:RTA}
\ea
 where in this relation $\tau_{\rm eq}$ is the relaxation time and $f_{\rm eq}$ is the late-time  equilibrium distribution function. Throughout this dissertation I am going to use RTA for the collisional kernel. By taking the moments I mean multiplying the quantity with four momenta and integrating over all momentum space, where in general integral measure is defined as
\ba
 \int dP \equiv N_{\rm dof}\int \frac{d^4 p}{(2\pi)^4}2\sqrt{-g}\Theta(p^0)(2\pi) \delta(p^\mu p_\mu -m^2)\,.
\ea
In this expression, $N_{\rm dof}$ is number of degrees of freedom and $g\equiv {\rm det} [g^{\mu\nu}]$.\footnote{Note that for Minkowski spacetime $g=-1$} Taking the integral over $p^0$ sends all particles on mass shell, i.e. $p^0\equiv E$ due to Dirac delta function which enforces the mass-shell condition. One has
\ba
\int dP=N_{\rm dof}\int \frac{d^3{\bf p}}{(2\pi)^3}\frac{1}{\sqrt{-g}E} =\tilde{N}\int \frac{d^3{\bf p}}{\sqrt{-g}E}\,,
\label{eq:measure}
\ea
where $\bf{p}$ is the on-shell momentum vector, E is the on-shell energy, and the notation $\tilde{N}\equiv N_{\rm dof}/(2\pi)^3$ is defined for simplicity. The $n^{\rm th}$ moment of quantity ${\cal O}$ is defined as 
\ba
\int dP p^{\mu_1}p^{\mu_2}\dots p^{\mu_n} {\cal O}\,.
\label{eq:nth-moment}
\ea
In above expressions, $N_{\rm dof}$ can be obtained by the assumption that the pressures and energy densities obtained by any hydrodynamics approach, including anisotropic hydrodynamics, should simplify to the Stefan-Boltzmann limit in the ideal case. For a QCD system with quarks and gluons degrees of freedom ideal pressure reads
\ba 
P_{\rm SB}=\frac{\pi^2 T^4}{45}\bigg(N_c^2-1+\frac{7}{4}N_c N_f\bigg)=\frac{\epsilon_{\rm SB}}{3}\,,
\ea
with $N_c$ and $N_f$ being the number of color charges and number of quark flavors in the model, respectively. Throughout this dissertation, one has $N_c=3$ and $N_f=3$.
\section{Bulk variables}
The plasma bulk variables can be obtained by taking different moments of distribution function. In other words, replacing the quantity ${\cal O}$ with distribution function $f(x,p)$ in (\ref{eq:nth-moment}), for any value of $n$ it provides us with a $n^{\rm th}$ order hydrodynamics quantity. The set of $n$-th order quantities is unlimited and in principle one can obtain the quantities of any order. However, by increasing the moment order used, the number of degrees of freedom also increases, and one needs to construct a scheme for obtaining the necessary dynamical equations by taking higher moments of Boltzmann equation. On the other hand, stopping at $n$-th order, the number of degrees of freedom and the number of dynamical equations might not be matched. In this condition, the system will be under- or over-determined. A reasonable prescription is to stop at the lowest possible order (sufficient to investigate the dissipative quantities of the desired order) and select necessary dynamical equations from available ones. The higher order moments encode the system at higher resolution.

Taking the first moment of distribution function provides us with four-current density
\ba 
J^\mu= \int dP\,p^\mu f(x,p)\,,
\label{eq:J-intg}
\ea
and taking the second moment yields `well-known' energy-momentum tensor
\ba
T^{\mu\nu} =\int dP\, p^\mu p^\nu f(x,p)\,.
\label{eq:T-intg}
\ea
Taking higher order moments provides some less-known tensor quantities. However, for future reference, I introduce ${\cal I}^{\mu\nu\lambda}$ obtained from the third moment of distribution function
\ba
{\cal I}^{\mu\nu\lambda} = \int dP\,p^\mu p^\nu p^\lambda f(x,p)\,.
\label{eq:I-intg}
\ea
The set of basis vectors form a complete set (either in the LRF or LF) and any tensor quantity can be expanded in terms of them. This will helps to simplify the calculation of tensor quantities to only calculation of some scalar expansion coefficients. With a general prescription for distribution function, we have 4 degrees of freedom for $J^\mu$, 10 for $T^{\mu\nu}$, and 20 for ${\cal I}^{\mu\nu\lambda}$, i.e. $J^\mu=n u^\mu +J_x X^\mu +J_y Y^\mu +J_z Z^\mu$ and so on. To obtain these numbers the intrinsic symmetry of above quantities under switching four momenta in integrals (\ref{eq:T-intg}-\ref{eq:I-intg}) is used. However, one can show that the leading order anisotropic distribution function defined at (\ref{eq:anisodist2}) possesses some extra symmetries, that reduces the number of degrees of freedom to 1, 4, and 3 for above tensor quantities, respectively.\footnote{It turns out that ${\cal I}_u$ is not an independent quantity and it can be obtained from ${\cal I}_i$'s.} 
\ba
 J^\mu &=&n \,u^\mu\,,  \nonumber \\
 T^{\mu\nu}&=&\epsilon\, u^\mu u^\mu+P_x X^\mu X^\nu+P_y Y^\mu Y^\nu+P_z Z^\mu Z^\nu\,,\nonumber \\
 {\cal I}^{\mu\nu\lambda} &=&  {\cal I}_u \, u^\mu u^\nu u^\lambda+\,  {\cal I}_x \left[ u^\mu X^\nu X^\lambda +X^\mu u^\nu X^\lambda + X^\mu X^\nu  u^\lambda \right]
\nonumber \\ 
&&\hspace{1.85cm}+ x\rightarrow y + x \rightarrow z\,.
\label{eq:expand}
\ea
Above, $n$ is the number density, $\epsilon$ is the energy density, $P_i$ are the components of pressure, and ${\cal I}$'s are the second order quantities which encode the dissipative properties of the system. 
Taking appropriate projections provides us with their components.

 To conclude the discussion of this chapter, I present the ideal (non-conformal) equilibrium thermodynamics quantities for a massless and massive system. The equilibrium distribution function for a classical gas is defined through a Maxwell-Boltzmann distribution 
\ba
f(p)=\exp\!\bigg(\!\!\!-\frac{E}{T}\bigg) \,.
\ea
The on-shell energy reads $E=|{\bf p}|$ for massless and $E=\sqrt{{\bf p}^2+m^2}$ for massive case. Using the expanded forms of $J^\mu$ and $T^{\mu\nu}$ and Eqs.~(\ref{eq:J-intg}) and (\ref{eq:T-intg}), one finds
\begin{eqnarray}
\epsilon_{\rm eq}(T,m) &=& 4 \pi \tilde{N} T^4 \, \hat{m}_{\rm eq}^2
 \Big[ 3 K_{2}\left( \hat{m}_{\rm eq} \right) + \hat{m}_{\rm eq} K_{1} \left( \hat{m}_{\rm eq} \right) \Big] \, , \nonumber  \\
P_{\rm eq}(T,m) &=& 4 \pi \tilde{N} T^4 \, \hat{m}_{\rm eq}^2 K_2\left( \hat{m}_{\rm eq}\right) \,, \nonumber \\
n_{\rm eq}(T,m) &=& 4 \pi \tilde{N} T^3 \, \hat{m}_{\rm eq}^2 K_2\left( \hat{m}_{\rm eq}\right) \,  ,
\label{eq:bulk-eq}
\end{eqnarray}
where $\hat{m}_{\rm eq}\equiv m/T$ and $K_n(x)$ are the modified Bessel function of the second kind. In the massless case these equations simplify to 
\ba
 \epsilon_{\rm eq}(T) &=& 24 \pi \tilde{N}T^4=3 P_{\rm eq}(T)\,,\nonumber \\
 n_{\rm eq}(T)&=& 8\pi \tilde{N}T^3\,.
 \label{eq:bulk-eq2}
\ea

\section{Anisotropic hydrodynamics and viscous hydrodynamics}
As discussed before, viscous hydrodynamics frameworks are able to reliably study small deviations from an isotropic equilibrium state. On the other hand, aHydro is able to provide an accurate description of the system even when the momentum space anisotropy is large. Therefore, one expects aHydro simplifies to vHydro in the limit of small anisotropy. Substituting (\ref{eq:anisodist2}) in (\ref{eq:T-intg}) and expanding for small $\xi_i$ one  has \cite{Tinti:2013vba}
\ba
 T^{\mu\nu}\simeq T^{\mu\nu}_{\rm eq}-\frac{32\pi\tilde{N}T^4}{4}(\xi_x X^\mu X^\nu+\xi_y Y^\mu Y^\nu+\xi_z Z^\mu Z^\nu)\,.
\ea
Comparing to Eq.~(\ref{eq:expand}), the corrections to the pressure components, being components of shear tensor in the vHydro framework, are linearly proportional to anisotropy parameters. 
\section{Realistic equation of state}
\label{sec:eos}
Generally, hydrodynamics models can be used to study the collective flow to different degrees, depending on how close to equilibrium the system is and how precise hydrodynamics formalism deals with viscous effects. For both ideal and viscous hydrodynamics, one needs to use an equation of state (EoS) obtained specifically for the system under consideration. The EoS which connects different thermodynamics quantities is obtained from theoretical considerations of the microscopic interactions. As discussed before, the governing interaction in QGP system is the strong force which is well studied in QCD. To obtain a realistic version of equation of state suitable for modeling the QGP, one needs to solve QCD dynamical equations for a relevant range of temperatures and chemical potentials. The complexity of the QCD Lagrangian rules out first-principle analytical calculations in this case. However, different groups are working on numerical calculations of the QCD partition function on a four dimensional lattice. In this dissertation, I am going to use a well-known parametrization of QCD lattice data provided by Budapest-Wuppertal collaboration \cite{Borsanyi:2010cj}.

Herein, I consider a system at finite temperature and zero chemical potential. At asymptotically high temperatures, the pressure of a gas of quarks and gluons tends the Stefan-Boltzmann (SB) limit. In what follows I take $N_c=N_f=3$.  At the temperatures probed in heavy-ion collisions there are important corrections to the SB limit and at low temperatures the relevant degrees of freedom change to hadrons.  The standard way to determine the QGP EoS is to use non-perturbative lattice QCD calculations.  For this purpose, I use an analytic parameterization of lattice data for the QCD interaction measure (trace anomaly), $I_{\rm eq} = \epsilon_{\rm eq} - 3 P_{\rm eq}$, taken from the Wuppertal-Budapest collaboration \cite{Borsanyi:2010cj}
\be
\frac{I_{\rm eq}(T)}{T^4}=\left[\frac{h_0}{1+h_3 t^2}+\frac{f_0\big[\tanh(f_1t+f_2)+1\big]}{1+g_1t+g_2t^2}\right]\exp\!\left(\!-\!\frac{h_1}{t}-\frac{h_2}{t^2}\right) ,
\label{eq:I-func}
\ee
with $t\equiv T/(0.2 \; \rm GeV)$. For $N_f=3$ one has $h_0=0.1396$, $h_1=-0.1800$, $h_2=0.0350$, $f_0=2.76$, $f_1=6.79$, $f_2=-5.29$, $g_1=-0.47$, $g_2=1.04$, and $h_3=0.01$.
\begin{figure}[t]
\hspace{-6mm}
\includegraphics[width=.46\linewidth]{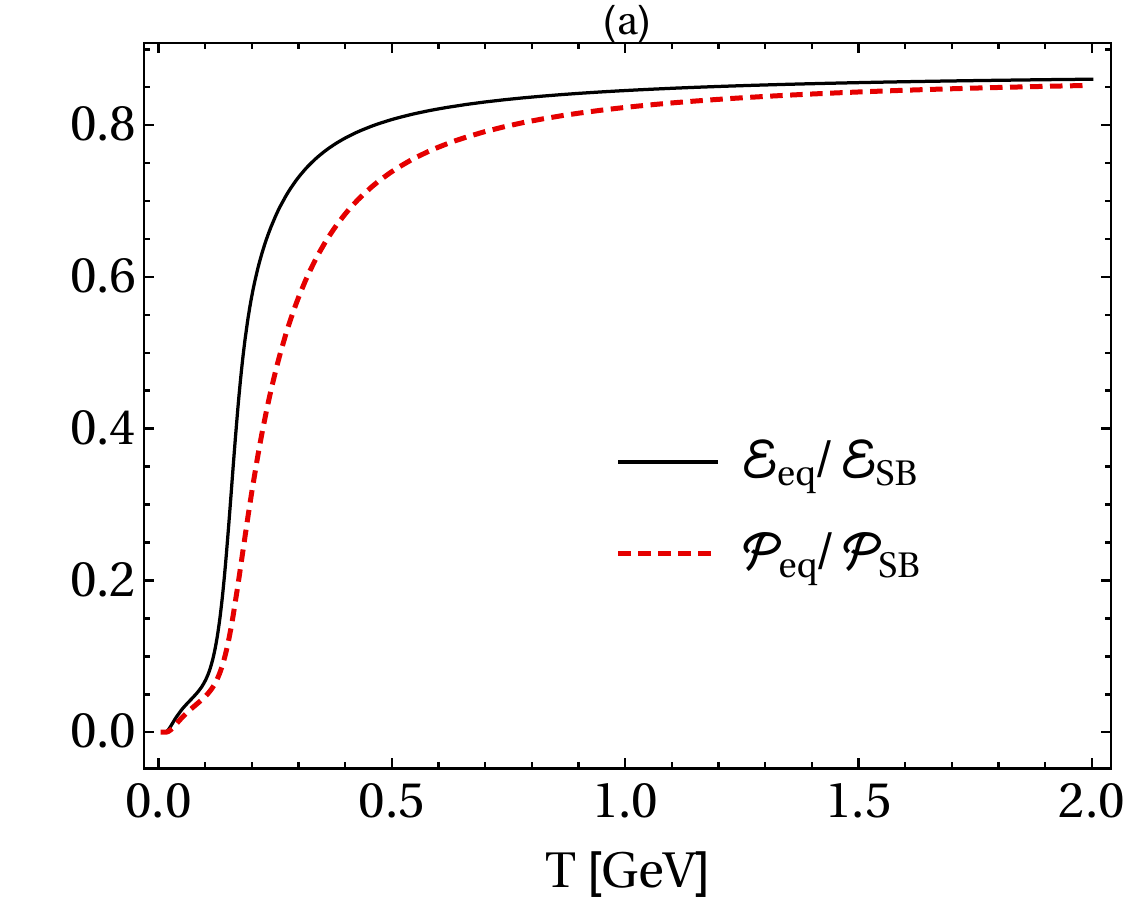}
\includegraphics[width=.48\linewidth]{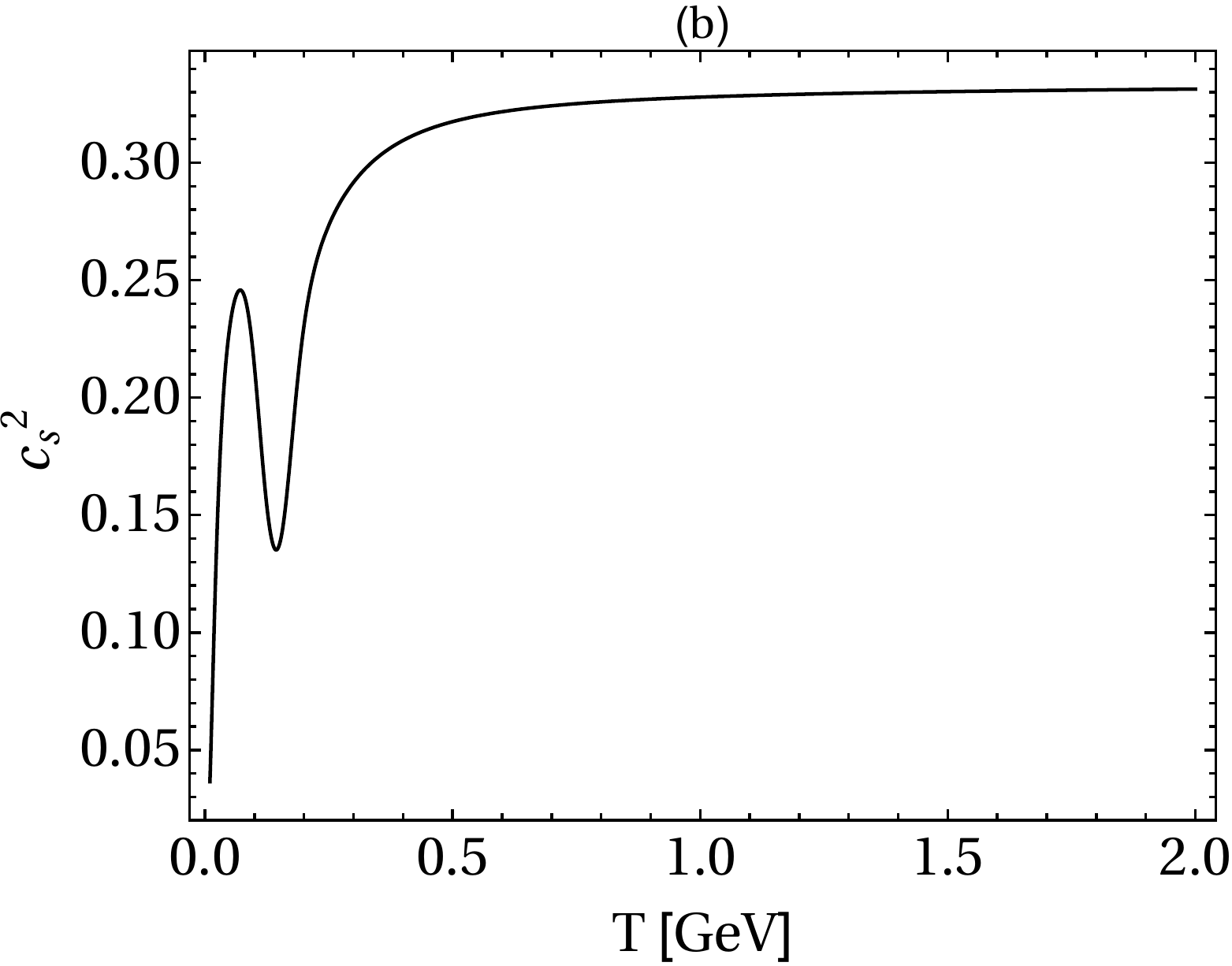}
\caption{panel (a) shows the energy density and pressure scaled by their respective SB limits and panel (b) shows the speed of sound squared both as a function of temperature.}
\label{fig:eos}
\end{figure}
So,
\be
\frac{P_{\rm eq}(T)}{T^4}=\int_0^T \frac{dT}{T}\frac{I_{\rm eq}(T)}{T^4} \, ,
\label{eq:P-func}
\ee
where I have assumed $P_{\rm eq}(T=0)=0$.  Having $P_{\rm eq}(T)$, one can obtain the energy density $\epsilon_{\rm eq}$ using $\epsilon_{\rm eq}(T) = 3 P_{\rm eq}(T) + I_{\rm eq}(T)$.  In the limit $T\rightarrow\infty$, the system tends to the ideal limit as expected.\footnote{In the original parametrization presented in Ref.~\cite{Borsanyi:2010cj} the authors used $h_3=0$, however, as pointed out in Ref.~\cite{Nopoush:2015yga}, choosing $h_3=0$ gives the wrong high temperature limit.} The temperature dependence of the resulting equilibrium energy density, pressure, and speed of sound squared ($c_s^2 = \partial P_{\rm eq}/\partial\epsilon_{\rm eq}$) are shown in the two panels of Fig.~\ref{fig:eos}.

\chapter{\bf Bulk viscosity in anisotropic hydrodynamics}  
\label{chap:bulk}
\setcounter{figure}{0}
\setcounter{table}{0}
\setcounter{equation}{0}
\section{Introduction}
So far, I have discussed the conformal massless aHydro formalism relevant for modeling a massless QGP. In this chapter, I introduce non-conformal aHydro for a massive gas.  Generalizing the model to a massive gas helps to study the bulk viscosity which is an essential component of hydrodynamics. In fact, lack of bulk degree of freedom in hydrodynamics model does not lead to a comprehensive and precise model for studying the QGP. In this chapter, I will discuss how introducing a bulk degree of freedom (DoF) will resolve the deficiencies of the previous non-confomal aHydro models, leading to better consistency with the exact solution of Boltzmann equation for a massive gas.

\section{Massive system and conformal symmetry}
In fluid mechanics, we define two main types of viscosity or internal friction of the fluid. (i) Shear viscosity which measures the fluid resistance to shearing flows. When the adjacent layers of the fluid move with different speeds, they experience resistance against sliding on each other. The fluids with more resistance of this kind, present higher values for shear viscosity. (ii) Bulk (volume) viscosity which plays the role when the compressibility is not negligible. It means, when a compressible fluid is compressed or expanded (without shear) it may exhibit resistance to flow which is called bulk viscosity. Bulk viscosity together with shear viscosity are the main governing dissipative mechanisms in the fluid dynamics. While shear viscosity reflects only molecular motions, in contrast the bulk viscosity reflects the relaxation of both ``rotational'', ``vibrational'' DoFs, and inelastic number changing processes.

Following the discussion of ahydro in the last chapter, the natural way to make the formalism more general is to extend it to a massive plasma. Inclusion of mass in the dynamics provides us with a new scale in the system. On the other hand, bulk viscosity deals with the bulk size of the system and only vanishes when the system is scale-invariant (i.e. there is no resistance against expanding or squeezing). Therefore, including mass leads to non-zero bulk viscosity, requiring the bulk DoF in the dynamics.
The system with zero bulk viscosity is assumed to be conformal, so, including mass breaks the ``conformal symmetry''. The conformal symmetry, a terminology borrowed from the geometry, specifies a type of symmetry in connection with an object or a pattern that is scale invariant (self-similar).

Starting from the conformal aHydro formalism introduced in the last chapter, some people tried to study the massive system by naively extending the anisotropic distribution to a massive one. This approach is followed by Florkowski et al. \cite{Florkowski:2014bba}, where two prescriptions for choosing the equations (i) from zeroth and first (ii) from first and second moments of Boltzmann equation have been tested. However, the numerical results were not promising. The reason is that the naive inclusion of mass in the distribution function without having any specific parameter manifesting the bulk viscosity explicitly is not only physically unreasonable but it is against the standard viscous hydrodynamics prescriptions in this regards. In this work, I have treated this inconsistency by introducing a new DoF in anisotropic distribution function which plays the role of the bulk DoF. Our approach is somewhat similar to the vhydro approach where one decomposes the shear tensor into traceless and traceful contributions. Herein, such a decomposition is performed when parametrizing the anisotropy tensor $\Xi^{\mu\nu}$. In this work, I derive the aHydro equations for a massive boost-invariant and cylindrically symmetric system, i.e. 1+1d, and then simplify to transversally symmetric case, i.e. 0+1d, where I compare with the exact solution of the Boltzmann equation. 
I begin by specifying the basic setup necessary for treating a boost-invariant and cylindrically-symmetric system allowing for an explicit bulk DoF.
\section{Massive ellipsoidal distribution function }
\label{sec:massivedf}
As discussed previously, naively extending the anisotropic distribution function in the conformal massless case, i.e. Eq.~(\ref{eq:anisodist1}), to the massive cases leads to conceptual and physical inconsistencies. To resolve this issue, I extend the previous definition of anisotropy tensor Eq.~(\ref{eq:xi}) to include a new scalar $\Phi$ degree of freedom, i.e. bulk DoF 
\be
\Xi^{\mu\nu} = u^\mu u^\nu + \xi^{\mu\nu} - \Delta^{\mu\nu} \Phi \, .
\label{eq:xi2}
\ee
As other parameters, $\Phi$ is understood to be a function of space and time and obeys ${\Xi^\mu}_\mu = 1 - 3 \Phi$. Using the anisotropic distribution function defined in Eq.~(\ref{eq:anisodist1}) and using (\ref{eq:xi2}) instead of (\ref{eq:xi}) and also considering $p\cdot u=E=\sqrt{{\bf p}^2+m^2}$, with mass $m$ being constant, one has in the LRF 
\be
f(x,p) = f_{\rm eq}\!\left(\frac{1}{\lambda} \sqrt{p_\mu \Xi^{\mu\nu} p_\nu} \right) 
=  f_{\rm eq}\!\left(\frac{1}{\lambda}\sqrt{\sum_i \frac{p_i^2}{\alpha_i^2} + m^2}\right)  \, ,
\label{eq:fform}
\ee
where $i\in \{x,y,z\}$ and the anisotropy parameters is introduced as 
\be
\alpha_i \equiv (1 + \xi_i + \Phi)^{-1/2} \, ,
\label{eq:alphadef}
\ee  
whose range of definition is $0<\alpha_i<+\infty$.  In the isotropic equilibrium limit, where $\xi_i = \Phi = 0$, $\alpha_i =1$, one has $p_\mu \Xi^{\mu\nu} p_\nu = (p \cdot u)^2 = E^2$, as expected. Here we have 3 anisotropy parameters plus $\Phi$, however, the tracelessness of $\xi^{\mu\nu}$ decreases the number of independent variables to three. In practice, I will use the variables $\alpha_i$ as the dynamical parameters and then convert back to the $\xi_i$ and $\Phi$ when necessary. I note that, using Eq.~(\ref{eq:alphadef}) and the tracelessness of the $\xi^{\mu\nu}$ tensor, one has 
\be
\Phi = \frac{1}{3} \sum_i \alpha_i^{-2} - 1 \, .
\ee

\subsection{Simplifying for the spheroidal form}
\label{app:varmap}
Note that, for a transversally homogeneous 0+1d system, one can further simplify the distribution function by using $\xi_x = \xi_y \equiv \xi_\perp$ and transforming to spheroidal form; however, it is more useful to keep the notation general and perform this simplification at the end of the calculation.  In practice, therefore, the general ellipsoidal form (\ref{eq:fform}) for the remainder of this chapter is used. However, for future comparisons one can define a spheroidal distribution function for massive system (including bulk DoF) using the old notations $(\xi,\Lambda)$
\be
f(x,p) = f_{\rm eq}\!\left(\frac{1}{\Lambda}\sqrt{p_\perp^2 + (1+\xi) p_z^2 + (1+\tilde\Phi) m^2 } \right) .
\label{eq:rsf2}
\ee
Sometimes it is useful to make connections between these two notations, similar to the map which is introduced at (\ref{eq:map})
\ba
1 + \xi &=& \frac{\alpha_x^2}{\alpha_z^2} = \frac{1 + \xi_z + \Phi}{1 - \xi_z/2 + \Phi} \, ,
\nonumber \\
\Lambda &=& \alpha_x \lambda = \frac{\lambda}{\sqrt{1 - \xi_z/2 + \Phi}} \, ,
\nonumber \\
1+\tilde\Phi &=& \alpha_x^2 = \frac{1}{1 - \xi_z/2 + \Phi} \, ,
\label{eq:esrel1}
\ea 
from which one can obtain the following relations
\ba
\xi_z &=& \frac{2 \xi}{3(1 + \tilde\Phi)};\quad
\Phi = \frac{\xi - 3\tilde\Phi}{3(1 + \tilde\Phi)};\quad
\lambda = \frac{\Lambda}{\sqrt{1 + \tilde\Phi}} \, .
\label{eq:esrel2}
\ea
Note that for $\Phi \rightarrow 0$ one obtains $\tilde\Phi = \xi/3$ and for $\tilde\Phi \rightarrow 0$ one obtains $\Phi = \xi/3$.

\section{System bulk variables}
\label{sect:bulkvars}
Before deriving the dynamical equations of the system, one needs to determine the thermodynamic variables of the system: number density, energy density, and pressures. To do so, I use Eqs.~(\ref{eq:expand}) to determine the relevant projections of $J^\mu$ and $T^{\mu\nu}$ for each variable. Then, one can take projections of the integral forms of $J^\mu$ and $T^{\mu\nu}$, i.e. Eqs.~(\ref{eq:T-intg}-\ref{eq:I-intg}), being implemented by massive ellipsoidal distribution function defined at (\ref{eq:fform}). As discussed before, in this chapter the variables and equations of motions are primarily derived for the boost-invariant cylindrically symmetric 1+1d case and then specialized to 0+1d case. Hence, the relevant basis vectors would be the ones defined at (\ref{eq:basis1+1d}). Equivalently, the tensor $T^{\mu\nu}$ has a simpler form in the 1+1d case, since $P_x=P_y\equiv P_T$.
The number density is 
\ba
n({\boldsymbol\alpha},m) &=& \int dP E \, 
f_{\rm eq}\!\left(\frac{1}{\lambda}\sqrt{\sum_i \frac{p_i^2}{\alpha_i^2} + m^2}\right) =\alpha \, n_{\rm eq}(\lambda,m)\,,
\label{eq:nequation}
\ea
where $i \in  \{x,y,z\}$ and $\alpha \equiv \prod_i \alpha_i$.
Note that $n_{\rm eq}(\lambda,m)$ follows
 $n_{\rm eq}(T,m)$ specified by Eq.~(\ref{eq:bulk-eq}).  
The energy density and transverse/longitudinal pressures are given by 
\ba
\epsilon &=& \int dP E^2 \, f_{\rm eq}\!\left(\frac{1}{\lambda}\sqrt{\sum_i \frac{p_i^2}{\alpha_i^2} + m^2}\right)
 ={\cal H}_3({\boldsymbol\alpha},\hat{m}) \, \lambda^4\,,  \label{eq:edensint}\\
 P_T &=& \frac{1}{2} \int dP \, (p_x^2+p_y^2) \, f_{\rm eq}\!\left(\frac{1}{\lambda}\sqrt{\sum_i \frac{p_i^2}{\alpha_i^2} + m^2}\right)
= {\cal H}_{3T}({\boldsymbol\alpha},\hat{m}) \, \lambda^4\,, \label{eq:ptint}\\
P_L &=&  \int dP \, p_z^2 \, f_{\rm eq}\!\left(\frac{1}{\lambda}\sqrt{\sum_i \frac{p_i^2}{\alpha_i^2} + m^2}\right)
={\cal H}_{3L}({\boldsymbol\alpha},\hat{m}) \, \lambda^4\,,
\label{eq:plint}
\ea
where $\hat{m} \equiv m/\lambda$. For a transversally homogeneous system one has $\alpha_x = \alpha_y$, which gives 
\ba
\epsilon &=& \tilde{\cal H}_3({\boldsymbol \alpha},\hat{m}) \, \lambda^4 \, ,
\label{eq:edenst}\\
P_T &=& \tilde{\cal H}_{3T}({\boldsymbol \alpha},\hat{m}) \, \lambda^4 \, , \\
P_L &=& \tilde{\cal H}_{3L}({\boldsymbol \alpha},\hat{m}) \, \lambda^4 \, .
\label{eq:plt}
\ea
All functions used above are collected in App.~\ref{subapp:h-functions-1}. 

\section{Moments of the Boltzmann equation}
\label{sect:beqmoments}
To obtain the necessary equations of motion, I take moments of the Boltzmann equation in the relaxation time approximation. Below, I compute the zeroth, first, and second moments of Eq.~(\ref{eq:boltzmann}) using (\ref{eq:basis1+1d}) for a boost-invariant cylindrically symmetric 1+1d system. At the end of each subsection, I specialize to transversally homogeneous 0+1d system.  
\subsection{Zeroth moment}
\label{sect:0mom}
Computing the zeroth moment gives
\be
D_u n + n \theta_u = \frac{1}{\tau_{\rm eq}} ( n_{\rm eq} - n ) \, , 
\label{eq:zeromom1}
\ee
with co-moving derivatives and divergences defined in App.~\ref{app:identities}. For one-dimensional transversally homogeneous expansion one has 
\be
\partial_\tau n +  \frac{n}{\tau} = \frac{1}{\tau_{\rm eq}} ( n_{\rm eq} - n ) \, ,
\label{eq:zeromom2}
\ee
which upon using (\ref{eq:nequation}) gives
\be
\partial_\tau \log \alpha_x^2 \alpha_z
+ \left[ 3 + \hat{m} \frac{K_1(\hat{m})}{K_2(\hat{m})} \right] \, \partial_\tau \log \lambda + \frac{1}{\tau}
=  \frac{1}{\tau_{\rm eq}} \left[ \frac{1}{\alpha_x^2 \alpha_z} \frac{T}{\lambda}\frac{K_2(\hat{m}_{\rm eq})}{K_2(\hat{m})} - 1 \right] .
\label{eq:final0mom}
\ee

\subsection{First moment}
\label{sect:1mom}
The first moment of the Boltzmann equation gives energy-momentum conservation
\be
\partial_\mu T^{\mu\nu} = 0 \, .
\label{eq:firstmom}
\ee
In principle, the right-hand side of this equation (i.e. the first moment of the collisional kernel in RTA) is not trivially vanishing. By forcing it to vanish, one guarantees energy-momentum conservation. This provides us with a new constraint equation which is called  dynamical Landau matching condition (LMC). Using the LMC one can define the temperature in terms of other dynamical variables of the system. It also introduces one way to define the local rest frame of the flow. The first moment of collisional kernel in RTA results in
\be
u_\mu T^{\mu\nu} = u_\mu T^{\mu\nu}_{\rm eq} \, .
\label{eq:dynlandau}
\ee
Herein $T^{\mu \nu}_{\rm eq}$ is the equilibrium energy-momentum tensor  
\ba
T^{\mu \nu}_{\rm eq} = \left( \epsilon_{\rm eq}
+ P_{\rm eq} \right)  u^\mu u^\nu
- P_{\rm eq} \eta^{\mu\nu}  \, ,
\label{eq:TEQ}
\ea
where $\epsilon_{\rm eq}$ and $P_{\rm eq}$ are given by Eqs.~(\ref{eq:bulk-eq}).  
I will return to dynamical Landau matching later.
For a boost-invariant and cylindrically symmetric system the energy-momentum tensor $T^{\mu\nu}$ has the general structure
\be
T^{\mu \nu} = \epsilon u^\mu u^\nu
+ P_x X^\mu X^\nu
+ P_y Y^\mu Y^\nu
+ P_z Z^\mu Z^\nu \, .
\label{eq:TAHg}
\ee
The resulting non-trivial dynamical equations in this case are~\cite{Tinti:2013vba}
\ba
D_u \epsilon + \epsilon \theta_u &=& -\sum_i P_i u_\mu D_i X_i^\mu  \, , \label{eq:1st-1}\\
D_x P_x + P_x \theta_x &=& \epsilon (X_\mu D_u u^\mu) 
+ P_y (X_\mu D_y Y^\mu) + P_z (X_\mu D_z Z^\mu) \, ,
\label{eq:1st-2}
\ea
where the equations can be expanded using the identities listed in App.~\ref{app:identities}. For a transversally homogeneous 0+1d system one can take $\alpha_x = \alpha_y$ and the energy-momentum tensor $T^{\mu\nu}$ has a somewhat simpler structure \cite{Martinez:2012tu}
\be
T^{\mu \nu} = \left( \epsilon
+ P_T \right)  u^\mu u^\nu
- P_T \eta^{\mu\nu}
+\left( P_L - P_T \right) Z^\mu Z^\nu \, .
\label{eq:TAH}
\ee
In this limit, Eq.~(\ref{eq:1st-2}) gives $\partial_r P_x=0$, which is a result of the homogeneity in the transverse plane. However, Eq.~(\ref{eq:1st-1}) gives
\be
\partial_\tau \epsilon = - \frac{\epsilon + P_L}{\tau} \, .
\label{eq:firstmom1D}
\ee
Using Eqs.~(\ref{eq:edenst}) and (\ref{eq:plt}), this becomes explicitly
\be
\left( 4 \tilde{\cal H}_3 - \tilde\Omega_m \right) \partial_\tau \log\lambda
+ \tilde\Omega_T \partial_\tau \log\alpha_x^2
+ \tilde\Omega_L \partial_\tau \log\alpha_z = 
-\frac{1}{\tau} \tilde\Omega_L \, .
\label{eq:final1mom}
\ee
Now lets discuss LMC in more details. Using (\ref{eq:expand}), the Eq.~(\ref{eq:dynlandau}) results in explicit energy conservation equation
\ba
\epsilon({\boldsymbol \xi},\Phi,\hat{m})=\epsilon_{\rm eq}(T)\quad \Rightarrow \quad \tilde{\cal H}_3 \lambda^4 = 4 \pi \tilde{N} T^4 \hat{m}_{\rm eq}^2 
 \Big[ 3 K_{2}\left( \hat{m}_{\rm eq} \right) + \hat{m}_{\rm eq} K_{1} \left( \hat{m}_{\rm eq} \right) \Big]\,,
\label{eq:energycon}
\ea 
which implies that the non-equilibrium energy density is equal to its equilibrium counterpart at all spacetime points during the evolution of the system. 
Equation~(\ref{eq:energycon}) is another constraint equation which can be solved using non-linear root finding algorithms, together with the other dynamical equations.
Note however that instead of using a root solver to enforce the LMC, it is possible to transform this equation into a differential equation by taking a derivative with respect to $\tau$ on the left and right hand sides~\cite{Florkowski:2012as}.
Taking a derivative of Eq.~(\ref{eq:energycon}) and using (\ref{eq:final1mom}) one has
\be
\partial_\tau \log T = - \frac{1}{\tau} \frac{\lambda^4}{T^4} \frac{\tilde\Omega_L}{\tilde\Omega_{\rm eq}} \, ,
\label{eq:coneq}
\ee
with
\be
\tilde\Omega_{\rm eq} \equiv 4 \pi \tilde{N} \hat{m}_{\rm eq}^2 \left[ 12 K_2(\hat{m}_{\rm eq}) + 5 \hat{m}_{\rm eq} K_1(\hat{m}_{\rm eq}) + \hat{m}_{\rm eq}^2 K_0(\hat{m}_{\rm eq}) \right] .
\label{eq:lmatch}
\ee
If one uses this method, one needs to ensure that Eq.~(\ref{eq:coneq}) is satisfied at $\tau=\tau_0$ and then evolve (\ref{eq:lmatch}) with the other dynamical equations.  I will use both methods to check our numerical results, but will primarily use the root-finding method since, in practice, it is slightly more numerically efficient for the case at hand.
\subsection{Second moment}
\label{sect:2mom}
Computing the second moment of the Boltzmann equation, one finds
\be
\partial_\lambda {\cal I}^{\lambda\mu\nu} = \frac{1}{\tau_{\rm eq}} \left(u_\lambda{\cal I}_{\rm eq}^{\lambda\mu\nu} - u_\lambda{\cal I}^{\lambda\mu\nu}\right),
\label{eq:tmom}
\ee
where the ${\cal I}$-tensor is defined in Eqs.~(\ref{eq:I-intg}) and (\ref{eq:expand}). 
Evaluating the necessary integrals using the distribution function (\ref{eq:fform}), one finds
\ba
{\cal I}_u &=& \Big(\sum_i \alpha_i^2\Big) \alpha \, {\cal I}_{\rm eq}(\lambda,m) + \alpha m^2 n_{\rm eq}(\lambda,m) \, , \nonumber \\
{\cal I}_i &=& \alpha \, \alpha_i^2 \, {\cal I}_{\rm eq}(\lambda,m) \, ,
\label{eq:I-funcs}
\ea
with 
\be
{\cal I}_{\rm eq}(\lambda,m) \equiv \frac{4\pi\tilde{N}\lambda^5}{3}  \int_0^\infty \, d\hat{p} \, \hat{p}^4  f_{\rm eq}\!\left(\!\sqrt{\hat{p}^2 + \hat{m}^2}\,\right)=   4 \pi {\tilde N} \lambda^5 \hat{m}^3 K_3(\hat{m})\,.
\label{eq:Ieq}
\ee
Note that, in general, one has
\be
{\cal I}_u - \sum_i {\cal I}_i = \alpha m^2 n_{\rm eq}(\lambda,m) \, ,
\label{eq:thetauID}
\ee
and $\lim_{m \rightarrow 0} {\cal I}_u = \sum_i {\cal I}_i$.
The second moment equation is obtained by expanding Eq.~(\ref{eq:tmom})
\ba
\partial_\lambda {\cal I}^{\lambda \mu \nu} &=&
u^\mu u^\nu D_u {\cal I}_u
+ {\cal I}_u \left[ u^\mu u^\nu \theta_u + 2 u^{(\nu} D_u u^{\mu)} \right]
+ X^\mu X^\nu D_u {\cal I}_x + 2 u^{(\mu} X^{\nu)} D_x {\cal I}_x
\nonumber \\ &&
+ {\cal I}_x \left[ X^\mu X^\nu {\cal I} + 2 X^{(\nu} D_u X^{\mu)} \right] 
\nonumber 
+ 2 {\cal I}_x \left[ u^{(\mu} X^{\nu)} \partial_\alpha X^\alpha + X^{(\nu} D_x u^{\mu)}  + u^{(\mu} D_x X^{\nu)} \right] 
\nonumber \\ &&
+ ( X \rightarrow Z) + (X \rightarrow Z) =\frac{1}{\tau_{\rm eq}} \left(u_\lambda{\cal I}_{\rm eq}^{\lambda\mu\nu} - u_\lambda{\cal I}^{\lambda\mu\nu}\right)\,.
\label{eq:dTheta}
\ea
The above is a tensor equation which contains several scalar equations which can be obtained using different projections (4 diagonal and 6 off-diagonal). Let's begin with the diagonal projections. By taking $uu$-, $XX$-, $YY$-, and $ZZ$-projections one has, respectively,
\ba
D_u {\cal I}_u +  {\cal I}_u \theta_u+ 2 \sum_i {\cal I}_i u_\mu D_i X_i^\mu &=&\frac{1}{\tau_{\rm eq}} ( {\cal I}_{u,\rm eq} - {\cal I}_u ) \, , \label{eq:uu-proj}\\
D_u {\cal I}_i + {\cal I}_i (\theta_u + 2 u_\mu D_i X_i^\mu) 
&=& \frac{1}{\tau_{\rm eq}} ( {\cal I}_{\rm eq} - {\cal I}_i ) \,; \quad i\in \{x,y,z\}\,.
\label{eq:ii-proj}
\ea
Among off-diagonal equations, only $uX$-projection provides us with a useful equation
\ba
D_x{\cal I}_x+{\cal I}_x\theta_x+({\cal I}_x+{\cal I}_u)D_uX^\mu-{\cal I}_yX_\mu D_y Y^\mu-{\cal I}_z X_\mu D_z Z^\mu=0 \, .
\label{eq:ux-proj}
\ea
 The $uY$- and $uZ$-projections provide two equations
\ba
 D_y {\cal I}_y=\partial_\phi {\cal I}_y = 0\,,\\
 D_z {\cal I}_z =\partial_{\varsigma} {\cal I}_z= 0\,,
\ea
which are expressing the cylindrically symmetry and boost invariance of the system, respectively.
Other off-diagonal equations, i.e. $XY$-, $XZ$-, and $YZ$- projections, are trivially satisfied in the 1+1d case. The identities used are listed in App.~\ref{app:identities}.

The set of Eqs.~(\ref{eq:uu-proj})-(\ref{eq:ii-proj}) and (\ref{eq:ux-proj}) are five independent dynamical equations obtained from the second moment~\cite{Tinti:2013vba}, which together with (\ref{eq:zeromom1}) from the zeroth moment, (\ref{eq:1st-1}) and (\ref{eq:1st-2}) from the first moment, and Landau Matching condition (\ref{eq:lmatch}) seem to form a set of nine dynamical equations for the six unknowns ($\alpha_i$, $\lambda$, T, and $\theta_\perp$) necessary to describe boost-invariant cylindrically-symmetric system. However, making use of Eq.~(\ref{eq:thetauID}) one finds that Eq.~(\ref{eq:uu-proj}) is not an independent equation and can be obtained from the three equations (\ref{eq:ii-proj}).
To see this, subtracting Eqs.~(\ref{eq:ii-proj}) from Eq.~(\ref{eq:uu-proj}) and then using (\ref{eq:thetauID}) gives
\be
m^2 \biggl[ D_u n + n \theta_u \biggr] = m^2 \left[ \frac{1}{\tau_{\rm eq}} ( n_{\rm eq} - n ) \right] . 
\ee
For massless systems this is satisfied trivially, however, if $m$ is finite, one has
\be
D_u n + n \theta_u  =  \frac{1}{\tau_{\rm eq}} ( n_{\rm eq} - n )  \, ,
\ee
which is precisely the zeroth moment equation obtained previously (\ref{eq:zeromom1}).  Since the second moment equation for ${\cal I}_u$ is identical to the zeroth moment equation, this leaves us with eight equations for six unknowns.  As demonstrated by Tinti and Florkowski \cite{Tinti:2013vba}, the three equations for ${\cal I}_i$ (\ref{eq:ii-proj}) can be reduced to two equations since the third is guaranteed if the other two are satisfied.  This leaves us with seven equations for six unknowns.  Also, one can follow the prescription of Tinti and Florkowski, which is to disregard the $uX$-projection equation (\ref{eq:ux-proj}). Using this scheme, one then has the same number of equations as unknowns, namely six.  I will return to this issue in the conclusions.

For a 0+1d system, using the identities listed in App.~\ref{app:identities} one can further simplify the equations. The set of Eqs.~(\ref{eq:ii-proj}) provide only two equations for $i=x,z$ (in the 0+1d the system does not distinct between $x$ and $y$ directions). The equation (\ref{eq:ux-proj}) simplifies to $\partial_r {\cal I}_x=0$ which guarantees that the model variables of the system are invariant respect to r (as expected in 0+1d case). Summing up equations from all moments in 0+1d, one obtains 5 equations for 4 ($\alpha_x$, $\alpha_z$, $\lambda$, and $T$) variables.\footnote{Note that in 0+1d case, the system is homogeneous in transverse plane so $\theta_\perp=0$.} In order to match the number of equations and variables, one can choose one of the second moment equations. However, following \cite{Tinti:2013vba} one can subtract one third of the sum of the ${\cal I}_i$ equations from each of the ${\cal I}_i$ equations.\footnote{Following Ref.~\cite{Tinti:2013vba} I also discard the equation implied by the sum of the ${\cal I}_i$ equations.} This logic in 0+1d translates into subtracting $x$ and $z$ components, which yields 
\be
\partial_\tau \log\left( \frac{\alpha_x}{\alpha_z} \right) - \frac{1}{\tau} + \frac{3}{4\tau_{\rm eq}} \frac{\xi_z}{\alpha_x^2\alpha_z} \left( \frac{T}{\lambda} \right)^2 \frac{K_3(\hat{m}_{\rm eq})}{K_3(\hat{m})} = 0 \, ,
\label{eq:final2mom}
\ee
where $\xi_z = \frac{2}{3}(\alpha_z^{-2} - \alpha_x^{-2})$.

\section{0+1d dynamical equations}
Our final set of three dynamical equations describing the evolution of a massive 0+1d system including bulk viscous pressure are given by Eqs.~(\ref{eq:final0mom}), (\ref{eq:final1mom}), (\ref{eq:final2mom}), and (\ref{eq:coneq})
\ba
&& \partial_\tau \log \alpha_x^2 \alpha_z
+ \left[ 3 + \hat{m} \frac{K_1(\hat{m})}{K_2(\hat{m})} \right] \, \partial_\tau \log \lambda + \frac{1}{\tau}
=  \frac{1}{\tau_{\rm eq}} \left[ \frac{1}{\alpha_x^2 \alpha_z} \frac{T}{\lambda}\frac{K_2(\hat{m}_{\rm eq})}{K_2(\hat{m})} - 1 \right] \, ,
\label{eq:final0m}
\\
&&
\left( 4 \tilde{\cal H}_3 - \tilde\Omega_m \right) \partial_\tau \log\lambda
+ \tilde\Omega_T \partial_\tau \log\alpha_x^2
+ \tilde\Omega_L \partial_\tau \log\alpha_z =
-\frac{1}{\tau} \tilde\Omega_L \, ,
\label{eq:final1m}
\\
&&
 \partial_\tau \log\left( \frac{\alpha_x}{\alpha_z} \right) - \frac{1}{\tau} + \frac{3}{4\tau_{\rm eq}} \frac{\xi_z}{\alpha_x^2\alpha_z} \left( \frac{T}{\lambda} \right)^2 \frac{K_3(\hat{m}_{\rm eq})}{K_3(\hat{m})} = 0 \, ,
\label{eq:final2m}
\\
&&
\partial_\tau \log T = - \frac{1}{\tau} \frac{\lambda^4}{T^4} \frac{\tilde\Omega_L}{\tilde\Omega_{\rm eq}} \, .
\ea
These four equations can be used to evolve $\xi_z$, $\Phi$, $\lambda$, and $T$ in proper-time (or alternatively $\alpha_x$, $\alpha_z$, $T$, and $\lambda$). As discussed previously in order to build a more stable algorithm for this problem I solve Eqs.~(\ref{eq:final0m}), (\ref{eq:final1m}), and (\ref{eq:final2m}) and solve Eq.~(\ref{eq:energycon}) using root finding in order to obtain the local effective temperature.  

\begin{figure*}[t]
      \includegraphics[width=.57\linewidth]{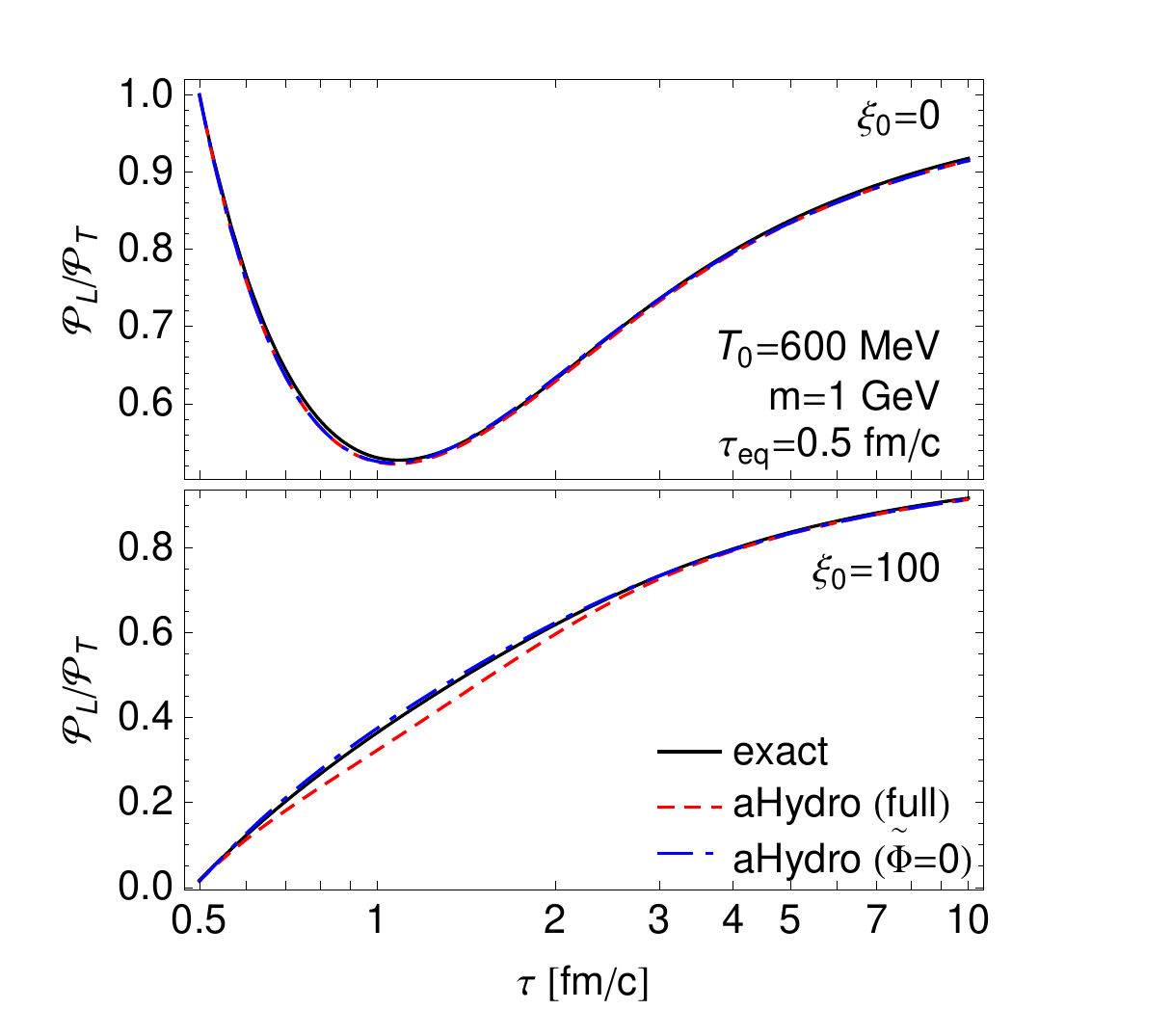}\hspace{-1.2cm}
    \includegraphics[width=.57\linewidth]{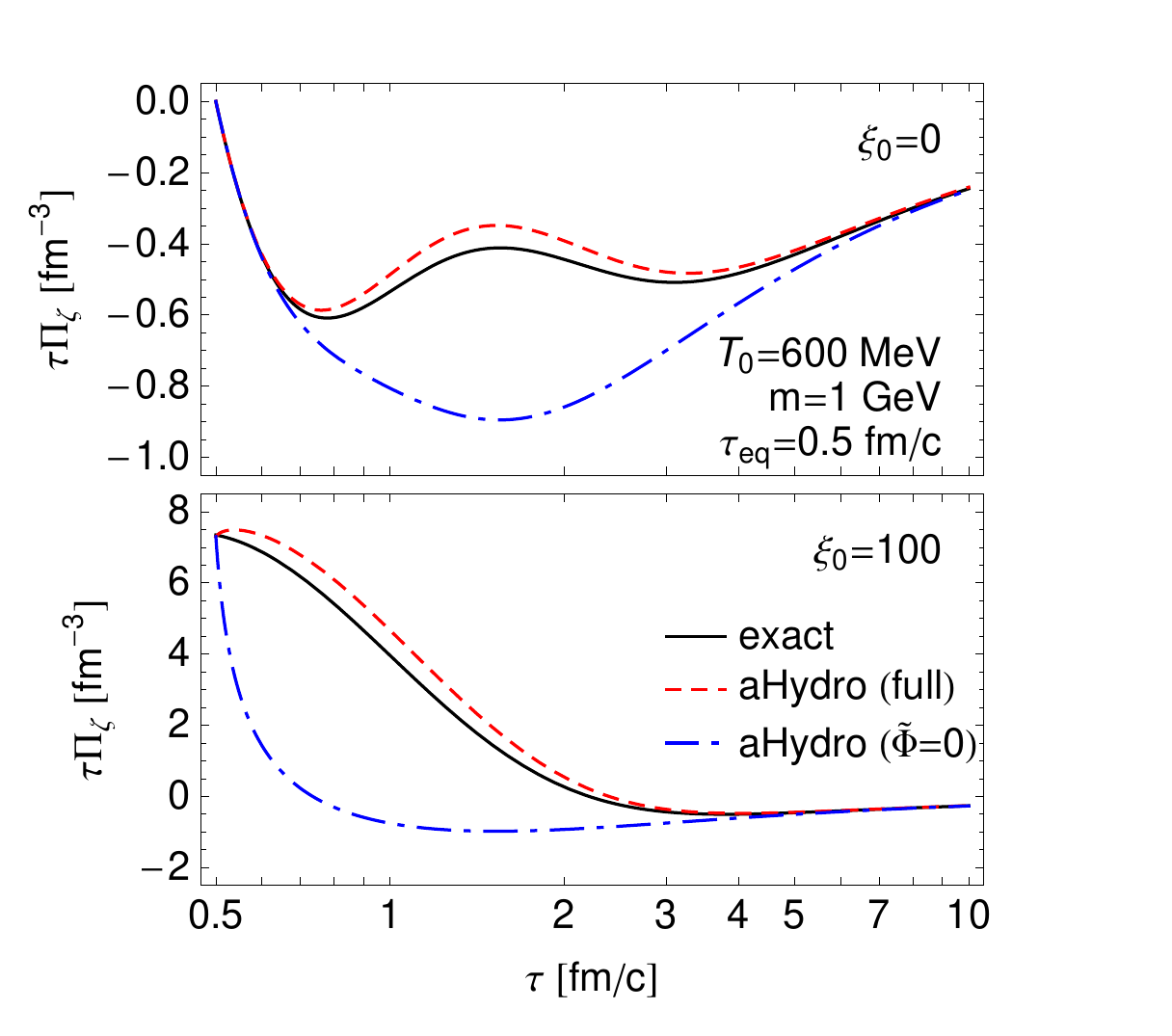}
    \caption{In all figures, the three lines correspond to the exact solution of the Boltzmann equation \cite{Florkowski:2014sfa} (black solid line), the full aHydro equations including the bulk DoF (red dashed line), and the aHydro equations with the ellipsoidal bulk DoF set to zero (blue dot-dashed line).  For all panels, $m=$ 1 GeV, $\tau_0$ = 0.5 fm/c, $\tau_{\rm eq}$ = 0.5 fm/c, and $T_0$ = 600 MeV are used.  In the top panels the initial spheroidal anisotropy parameter $\xi_0=0$ are fixed and in the bottom panels $\xi_0 = 100$ is used. Left: (Color online) Proper-time evolution of $P_L/P_T$.   Right: (Color online) Proper-time evolution of the bulk pressure.}
    \label{fig:aniso1}
\end{figure*}
\section{Numerical results}
\label{sect:results}
I now compare the evolution predicted by Eqs.~(\ref{eq:final0m})-(\ref{eq:final2m}) with the exact solution of the massive Boltzmann equation obtained in Ref.~\cite{Florkowski:2014sfa}.  Instead of evolving the anisotropy parameter $\xi_z$ and bulk parameter $\Phi$, $\alpha_x$ and $\alpha_z$ are evolved numerically.  I fix the initial conditions for $\alpha_x^0$, $\alpha_z^0$, and $T_0=$ 600 MeV at $\tau_0 = 0.5$ fm/c and fix $\lambda_0$ using Eq.~(\ref{eq:energycon}).  I then use Eqs.~(\ref{eq:final0m})-(\ref{eq:final2m}) to evolve $\alpha_{x}$, $\alpha_{y}$, and $\lambda$.  At each step of the numerical integration, Eq.~(\ref{eq:energycon}) is used to self-consistently determine the effective temperature $T$ which appears in the equations of motion or, alternatively, evolve the temperature using Eq.~(\ref{eq:coneq}).

In Fig.~\ref{fig:aniso1} I plot the proper-time evolution of $P_L/P_T$ and the bulk pressure $\Pi_\zeta$, respectively.  The bulk pressure is computed via
\be
\Pi_\zeta(\tau) = \frac{1}{3}
\left[P_L(\tau) + 2 P_T(\tau)
- 3 P_{\rm eq}(\tau) \right] ,
\label{eq:PIkz}
\ee
where $P_{\rm eq}$ is the equilibrium pressure evaluated at the effective temperature $T(\tau)$.  In both figures the three lines correspond to the exact solution of the Boltzmann equation \cite{Florkowski:2014sfa} (black solid line), the full aHydro equations including the bulk DoF (red dashed line), and the aHydro equations with the spheroidal bulk DoF ($\tilde\Phi$) set to zero at all times (blue dot-dashed line).  For both panels, $m=$ 1 GeV and $\tau_{\rm eq}$ = 0.5 fm/c is used.  In the top panels I fixed the initial spheroidal anisotropy parameter $\xi_0=0$ and in the bottom panels i chose $\xi_0 = 100$.  For the bulk initial condition I take $\tilde\Phi_0 = 0$ since this is consistent with the spheroidal initial condition assumed in the exact solution.

 Considering first Fig.~\ref{fig:aniso1}, one sees that allowing for the bulk DoF significantly improves agreement between aHydro and the exact solution.  The equations derived previously \cite{Florkowski:2014bba} correspond to the assumption that $\tilde\Phi=0$ at all proper times.  As can be seen from this figure, if this assumption is made (blue dot-dashed line) the agreement with the exact solution is quite poor.  Alternatively, one could assume that the ellipsoidal bulk parameter $\Phi=0$ at all times.  This case is not shown, because it is vastly inferior and does not even reproduce the late-time dynamics of the system.  Turning our attention to Fig.~\ref{fig:aniso1}, one sees from the top panel that for a system which is initially isotropic, the different prescriptions seem to give nearly identical results for $P_L/P_T$.  However, if the initial pressure anisotropy is large (bottom panel), then solution of the full aHydro equations including the bulk DoF seems to be further away from the exact solution.  It seems that, within the framework advocated here, it not possible to improve the agreement with the exact solutions for the bulk pressure without causing some discrepancy in the pressure anisotropy.  Since the number of parameters used to describe the system is quite small, this may not be surprising, but it is still worrisome that the uniform convergence towards the exact result in all bulk observables is not observed. 
\begin{figure*}[t]
      \includegraphics[width=.57\linewidth]{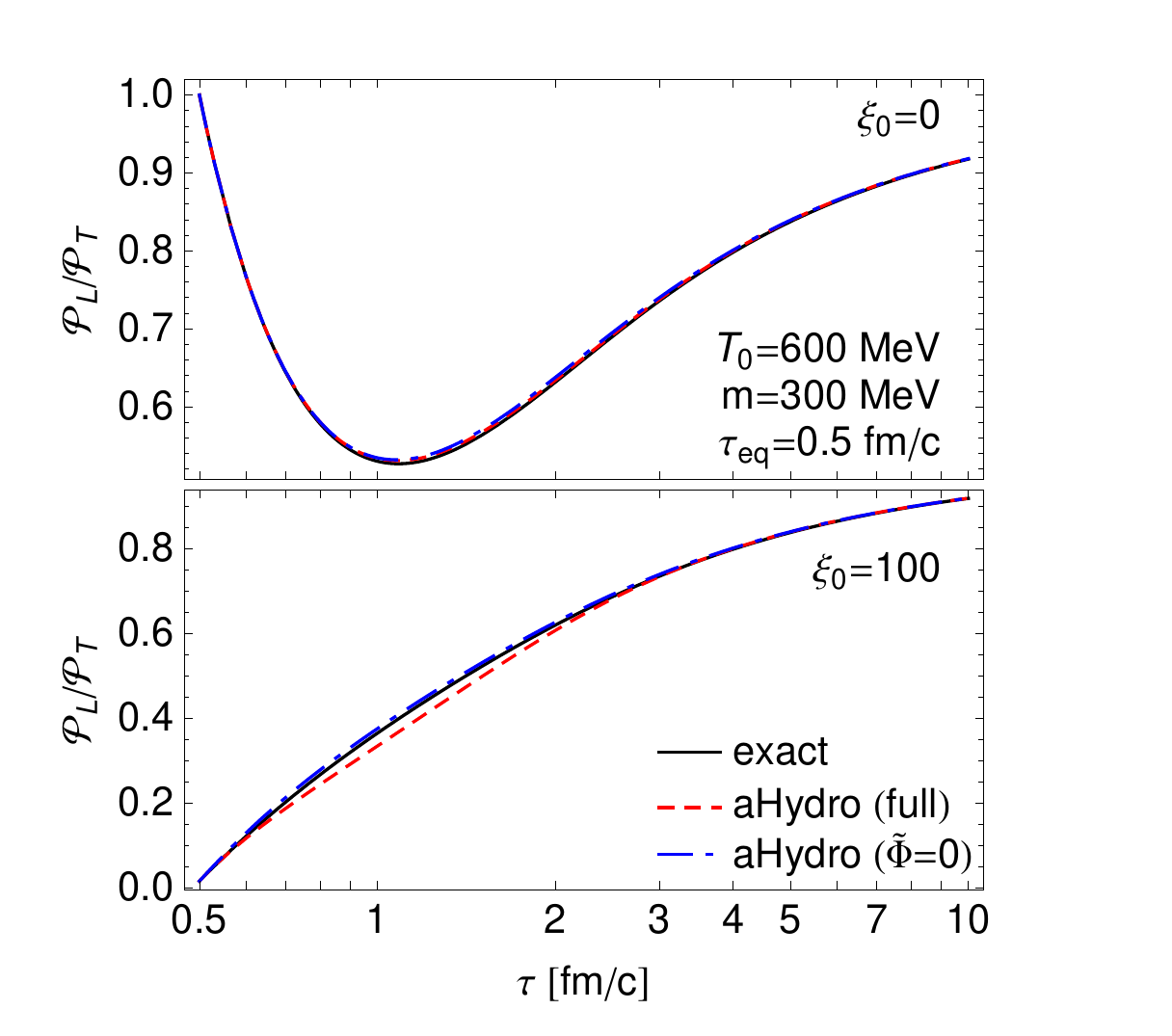}\hspace{-1cm}
    \includegraphics[width=.57\linewidth]{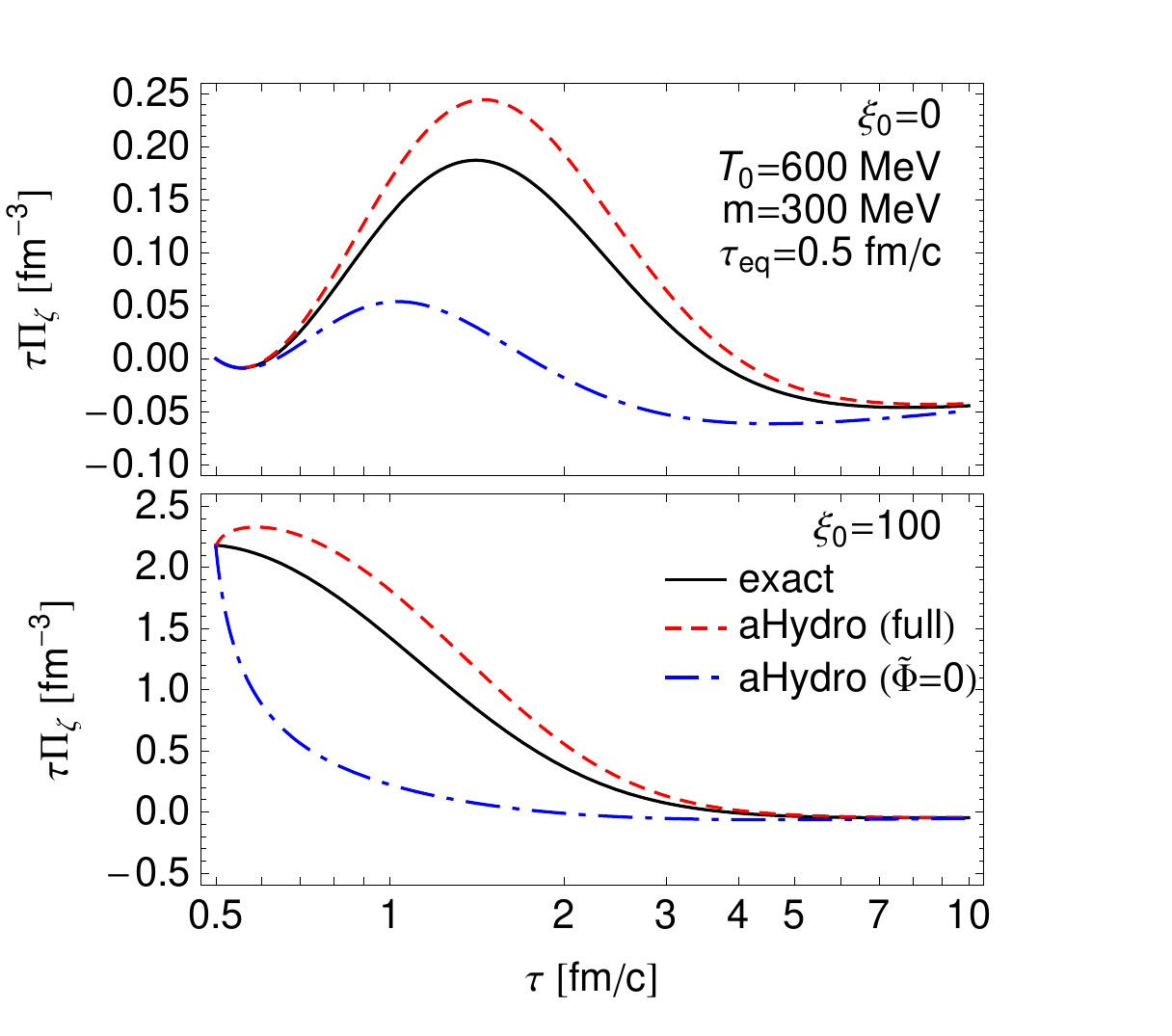}
    \caption{(Color online) Proper-time evolution of $P_L/P_T$. Parameters and descriptions are the same as in Fig.~\ref{fig:aniso1} except here I take $m=300$ MeV. (Color online) Proper-time evolution of the bulk pressure. }
    \label{fig:aniso2}
\end{figure*}
Now consider a somewhat lower mass as an additional check of the performance of the aHydro equations obtained herein.  In Figs.~\ref{fig:aniso2} I plot the proper-time evolution of $P_L/P_T$ and the bulk pressure $\Pi_\zeta$, respectively.  The parameters and descriptions are the same as Figs.~\ref{fig:aniso1}, except for these figures I take $m =$ 300 MeV.  These figures once again show that including the bulk DoF improves agreement between aHydro and the exact solution for the bulk pressure; however, including the bulk DoF seems to cause a somewhat poorer agreement with the pressure anisotropy when the system has a large initial momentum-space anisotropy.

\begin{figure}[t]
\centerline{\includegraphics[angle=0,width=0.7\textwidth]{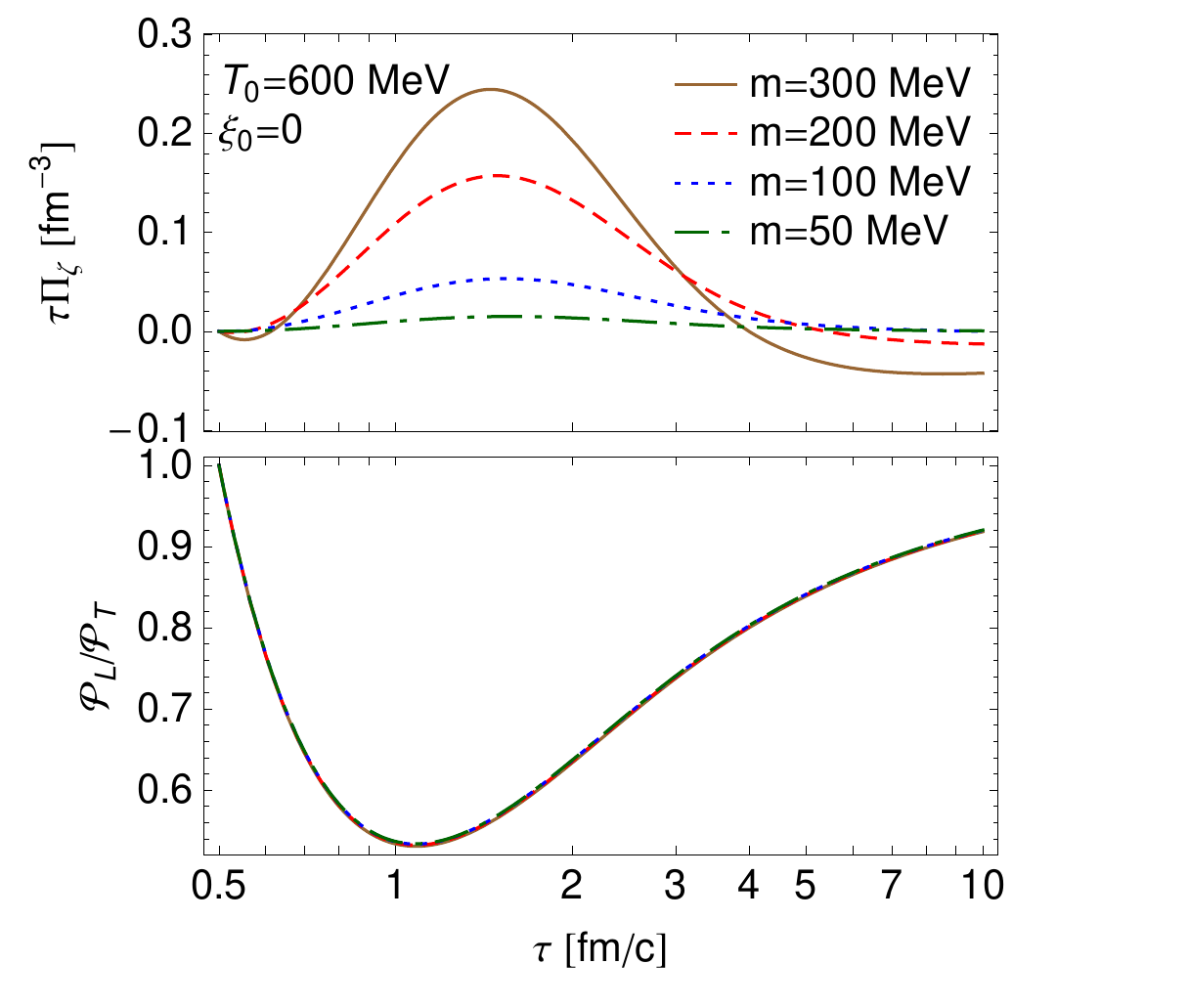}}
\caption{(Color online) Bulk pressure and pressure anisotropy as a function of proper time.  All curves correspond to the aHydro evolution including the bulk DoF.  I took $\xi_0=0$ with all other the initial conditions and parameters being the same as in previous figures.}
\label{fig:bulk_varm}
\end{figure}

As our final numerical result, in Fig.~\ref{fig:bulk_varm} I plot the bulk pressure (\ref{eq:PIkz}) as a function of proper time for different assumed particle masses ranging from $m=300$ MeV down to $m=50$ MeV.  Except for the masses, the parameters, descriptions, and initial conditions are the same as in the preceding figures.  As can be seen from the top panel of Fig.~\ref{fig:bulk_varm}, as one lowers the mass, the bulk pressure goes to zero as it should.  From the bottom panel one learns that there is very little dependence of the pressure anisotropy on the assumed mass of the particles.

\section{Conclusions}

In this chapter I extended the treatment of Tinti and Florkowski \cite{Tinti:2013vba} to include an explicit bulk DoF.  This was done by introducing a general form for the anisotropy tensor $\Xi^{\mu\nu}$ and then decomposing it into components parallel to and orthogonal to the fluid four-velocity.  I then further decomposed the orthogonal piece into traceless and traceful components analogously to how the viscous tensor is decomposed in standard relativistic viscous hydrodynamics.  Using this as a starting point, I then derived explicit expressions for the number density, energy density, and pressures for a massive anisotropic gas.  I then took moments of the Boltzmann equation in the RTA.  Restricting to boost-invariant and cylindrically symmetric systems, the full set of dynamical equations necessary to evolve the effective temperature, momentum-space anisotropies, and the bulk DoF is obtained.  

In order to test the efficacy of the approach,  a transversally homogeneous system which reduces the system to 0+1d is considered.  For such a boost-invariant and transversally homogeneous system of massive particles it is possible to solve the relaxation time approximation Boltzmann equation exactly \cite{Florkowski:2014sfa}.  Our comparisons of aHydro with the exact solution showed that adding the bulk DoF improves agreement between aHydro and the exact results for the bulk pressure.  However, including this DoF seems to cause some small early-time discrepancy with the pressure anisotropy evolution when the system is assumed to have a large initial momentum-space anisotropy.

On the formal side, an important result of the work presented in this chapter concerns the question of how to select which moments of the Boltzmann equation to use for the evolution of the microscopic parameters.  Even in the case of a 0+1d system, with the addition of the bulk DoF, it is not obvious a priori which moment, either zeroth moment or the $uu$-projection of the second moment, should be used as the additional equation of motion.  I demonstrated herein that in general, for a system of massive particles, the zeroth moment and $uu$-projections of the second moment of the Boltzmann equation give the same dynamical equation.  As a consequence, there is less ambiguity about how to proceed in the case of a 0+1d system.  If one considers a 1+1d boost-invariant cylindrically symmetric system there are two more equations than the number of unknowns.  The prescription of Tinti and Florkowski \cite{Tinti:2013vba} is to disregard the equation generated by the $ux$-projection of the second moment and the sum of the ${\cal I}_i$ equations, which seems to work in practice; however, it would be nice to have a more firm physics justification for this procedure.  To us, the mismatch in number of equations and parameters suggests that for a 1+1d system one can introduce an additional parameter in the ansatz for the one-particle distribution; however, it is unclear at this moment in time what additional physics parameters are required/well-motivated.

Looking forward, despite the progress reported here, there are still important open questions to be addressed.  The first and foremost question in our minds concerns the massless limit of the equations obtained herein.  In this limit the bulk DoF should be irrelevant, which is evidenced by our numerical results (in that the bulk pressure goes to zero negating the need for this DoF).  However, in terms of the microscopic parameters $\xi_i$ and $\Phi$ it is not obvious to us at this moment that the system of equations obtained herein reduces smoothly to the system of equations obtained by Tinti and Florkowski.  It is straightforward to show that the first and second moment equations become the same as those obtained by Tinti and Florkowski, however, the zeroth moment equation remains part of the system of equations even in the massless limit and one then has an overdetermined system.  In their approach Tinti and Florkowski disregarded the zeroth moment equation, so it is not clear to us how our equations and theirs can be smoothly connected.  

Another important open question raised by this work concerns how to simultaneously improve the description of the pressure anisotropy and bulk pressure.  In order to get better agreement with the exact solutions for the pressure anisotropy and bulk pressure, particularly at early times, it seems that one needs to account for non-ellipsoidal components of the one-particle distribution function.  This can be done using methods similar to Ref.~\cite{Bazow:2013ifa}; however, the anisotropic background would now be much more complicated.  It may be more efficient in the end to linearize around the spheroidal background and include the ellipsoidal and bulk corrections perturbatively since in this case many of the integrals (analogs of the ${\cal H}_3$ functions used herein) can be evaluated analytically.  

Finally, note that in this work, a system of particles with fixed masses is considered which means that the equation of state is fixed.  Looking forward, within the kinetic approach one would like to have a way to implement a realistic equation of state that can reasonably reproduce the lattice equation of state.  One possibility is to use a quasiparticle approach as proposed by Romatschke~\cite{Romatschke:2011qp}.  
   
\chapter{\bf Anisotropic hydrodynamics under the conformal Gubser flow}

\label{chap:gubser}
\setcounter{figure}{0}
\setcounter{table}{0}
\setcounter{equation}{0}
\section{Introduction}
In the previous chapter, I discussed the inclusion of the bulk degree of freedom in non-conformal aHydro formalism. Such a formalism is useful in studying a non-conformal QGP.
As discussed before, dissipative hydrodynamics has been a successful tool for studying the spatio-temporal evolution of the medium. For phenomenological applications, second-order viscous hydrodynamics is the most often-used hydrodynamical framework. On the other hand, the aHydro studies the early-time strong momentum-space anisotropy more systematically. One of the most important challenges with regard to aHydro is whether or not it is able to describe the dynamics better than standard vHydro models. In this chapter, I want to present compelling evidence that aHydro describes the non-equilibrium behavior of the QGP far better than contemporary vHydro models for a system presenting both transverse and longitudinal expansions. The work presented in this chapter is crucially important because it demonstrates that incorporating non-equilibrium effects using the aHydro framework is mathematically more accurate. This sets the stage for aHydro to become the most accurate hydrodynamics framework for phenomenological study of the QGP. 

In this chapter, I derive the anisotropic hydrodynamics dynamical equations for a system subject to {\it conformal Gubser flow} \cite{Gubser:2010ze,Gubser:2010ui} and compare the results to recently obtained analytical solutions to the Boltzmann equation in the relaxation-time approximation subject to the same flow \cite{Denicol:2014xca,Denicol:2014tha}.  I  also compare to recently obtained solutions using the Israel-Stewart second-order viscous hydrodynamics framework \cite{Marrochio:2013wla} and a complete second-order Grad 14-moment approximation (DNMR) \cite{Denicol:2014tha}.  Since Gubser flow includes both cylindrically-symmetric transverse and boost-invariant longitudinal (1+1d) expansion, this will allow us to test the efficacy of anisotropic hydrodynamic model in a more realistic set up. 
In order to implement the anisotropic hydrodynamics framework, I begin by assuming that, to leading order, the one-particle distribution function is ellipsoidally symmetric in the momenta conjugate to the {\it de Sitter coordinates} used to parameterize the Gubser flow and that the argument of the distribution function only depends quadratically on the de Sitter-space momenta.  I then demonstrate that the $SO(3)_q$ symmetry in de Sitter space further constrains the anisotropy tensor to be of spheroidal form.  The resulting system of coupled ordinary differential equations for the de Sitter-space momentum scale $\hat\lambda$ and anisotropy parameter $\hat\alpha_\varsigma$ are solved numerically. I show that aHydro describes the spatio-temporal evolution of the system better than all currently known dissipative hydrodynamics approaches.  In addition, I prove that anisotropic hydrodynamics simplifies to the exact solution of the relaxation-time approximation Boltzmann equation in the ideal, $\eta/s \rightarrow 0$, and free-streaming, $\eta/s \rightarrow \infty$, limits.
\section{Conventions and setup}
In this chapter, in order to distinguish among three coordinate systems I use special symbols. Cartesian Milne coordinates are defined by $\check{x}^\mu=(\tau,x,y,\varsigma)$. The polar Milne coordinates is defined as $\tilde{x}^\mu = (\tau,r,\phi,\varsigma)$.   In all cases, the flow velocity $u^\mu$ is normalized as $u_{\mu }u^{\mu }=-1$. Finally, all variables connected to the de Sitter coordinate are going to hold ``hat'', i.e. $\hat{x}^\mu=(\rho,\theta,\phi,\varsigma)$.
Herein I assume that the system is boost invariant and cylindrically symmetric with respect to the beamline at all times.  With this assumption, one can construct a flow with $SO(3)_q{\otimes}SO(1,1){\otimes}Z_2$ symmetry (``Gubser symmetry'') \cite{Gubser:2010ui,Gubser:2010ze}.  In this case, one can show that all dynamical variables depend on $\tau$ and $r$ through the dimensionless combination
\ba 
G(\tau,r)=\frac{1-q^2{\tau}^2+q^2r^2}{2 q{\tau}}\, ,
\label{eq:G}
\ea
where $q$ is an arbitrary energy scale.\footnote{The final results presented herein are expressed in de Sitter space and hold for arbitrary $q$.}  In order to study the dynamics, I start by specifying a basis appropriate for treating a boost-invariant and cylindrically-symmetric system and then simplify the equations of motion by introducing de Sitter coordinates.
In this chapter, the metric is taken to be ``mostly plus'', i.e. $\eta^{\mu\nu}={\rm diag}(-1,\mathds{1})$, such that in Minkowski space with $x^{\mu }=(t,x,y,z)$, the line element is
\be
ds^2 = \eta_{\mu\nu} dx^\mu dx^\nu = -dt^2 + dx^2 + dy^2 + dz^2 \,.
\ee
%
\subsection{Weyl transformation}
\label{sec:weyl}

In order for a system to be conformally invariant, the dynamics should be invariant under {\it Weyl rescaling} \cite{Gubser:2010ui}. A $(m,n)$ tensor of the form $Q^{\mu_1 ...\mu_m}_{\nu_1 ...\nu_n}(x)$ with canonical dimension $\Delta$ transforms under Weyl rescaling as
\ba
 Q^{\mu_1 ...\mu_m}_{\nu_1 ...\nu_n}(x)\,\rightarrow\,\Omega^{\Delta+m-n}Q^{\mu_1 ...\mu_m}_{\nu_1 ...\nu_n}(x)\, ,
\label{eq:weyl-rescaling}
\ea
where $\Omega(x) = \exp[\omega(x)]$ with $\omega(x)$ being a function of space and time.
For example, the metric tensor $\eta_{\mu\nu}$ is a dimensionless tensor of rank 2. Using the relation above with $m=0$, $n=2$, and $\Delta=0$, one finds $[\eta_{\mu\nu}]=-2$. This means that  $\eta_{\mu\nu}$ has a conformal weight of $-2$ and transforms under Weyl rescaling as \cite{Gubser:2010ui}
\ba 
\eta_{\mu\nu} \rightarrow \Omega^{-2}\,\eta_{\mu\nu}\, .
\label{eq:g-weyl}
\ea

\subsection{Gubser flow and de Sitter coordinates}

The Gubser flow is completely determined by symmetry constraints to be \cite{Gubser:2010ze,Gubser:2010ui}
\ba 
\tilde{u}^\tau &=&\cosh\left[\tanh^{-1} \left(\frac{2q^2\tau r}{1+q^2\tau^2+q^2r^2}\right)\right]\, ,\notag \\
\tilde{u}^r &=&\sinh\left[\tanh^{-1} \left(\frac{2q^2 \tau r}{1+q^2\tau^2+q^2r^2}\right)\right]\, , \notag \\
\tilde{u}^\phi &=&\tilde{u}^\varsigma= 0\, .
\label{eq:gubser-flow}
\ea
Comparing basis vectors defined in Eqs.~ (\ref{eq:basis1+1d}) and (\ref{eq:gubser-flow}), one has
\be 
\theta_\perp=\tanh^{-1} \left(\frac{2q^2\tau r}{1+q^2\tau^2+q^2r^2}\right).
\label{eq:thetaperp}
\ee
In what follows, a Weyl rescaling and a change of variables to de Sitter coordinates are performed.  I begin by introducing the de Sitter ``time'' $\rho$ and polar angle $\theta$ \cite{Gubser:2010ui}
\ba 
\sinh{\rho} =  - \frac{1-q^2{\tau^2}+q^2r^2}{2q{\tau}}\,; \qquad
\tan{\theta} = \frac{2qr}{1+q^2{\tau}^2-q^2r^2}\, .
\label{eq:desitter2}
\ea
Note that, for fixed $r$, the limits $\tau \rightarrow 0^+$ and $\tau\rightarrow\infty$ correspond to $\rho \rightarrow -\infty$ and $\rho \rightarrow \infty$, respectively.  This means that the de Sitter map covers the future (forward) light cone.

In order to map the flow (\ref{eq:gubser-flow}) to a static one, I follow the prescription of Gubser \cite{Gubser:2010ui,Gubser:2010ze} and make a coordinate transformation combined with a  Weyl rescaling to pass from ${\bf R}^{3,1}$ to $dS_3 \times {\bf R}$ (de Sitter space).  Quantities defined in de Sitter space will be indicated with a hat.  Using Eq.~(\ref{eq:g-weyl}) and the rules for general coordinate transformations of tensors, one can relate the de Sitter-space metric with the Minkowski space metric via
\ba
\hat{g}_{\mu\nu}=\frac{1}{\tau^2}\frac{\partial x^\alpha}{\partial \hat{x}^\mu}\frac{\partial x^\beta}{\partial\hat{x}^\nu}g_{\alpha\beta}\, .
\ea
The de Sitter-space metric tensor in matrix form is
\be
\hat{g}_{\mu\nu} = {\rm diag}(-1,\, \cosh^2\!\rho,\, \cosh^2\!\rho  \sin^2\!\theta ,\, 1) \, ,
\label{eq:desitter-metric}
\ee
and $\hat{g}^{\mu\nu} = \hat{g}^{-1}_{\mu\nu}$, which can also be expressed in terms of tetrads (see (\ref{eq:trans}) below) as $\hat{g}_{\mu\nu}=\hat{u}_\mu\hat{u}_\nu+\hat{\Theta}_\mu\hat{\Theta}_\nu+\hat{\Phi}_\mu\hat{\Phi}_\nu+\hat{\varsigma}_\mu\hat{\varsigma}_\nu$.  The determinant of $\hat{g}_{\mu\nu}$ is
\ba
\hat{\mathrm{g}} \equiv \det{\hat{g}_{\mu\nu}} = -\cosh^4\!\rho\,\sin^2\!\theta\, .
\label{eq:g}
\ea

In order to proceed, I need to establish relations between the Minkowski-space basis vectors and the de Sitter-space basis vectors.  To do this, I first need to know the conformal weights of the Minkowski-space basis vectors.  Knowing that $[\eta_{\mu\nu}]=-2$ and using $\eta_{\mu\nu} X^\mu X^\nu=1$ (where $X^\mu$ generally stands for the spacelike Minkowski basis vectors), one concludes that $[X^\mu]=1$ and $[X_\mu]=-1$. Also, using $\eta_{\mu\nu} u^\mu u^\nu=-1$, one obtains $[u^\mu]=1$ and $[u_\mu]=-1$. The tensor transformation to relate 4-vectors in de Sitter coordinates to 4-vectors in Minkowski coordinates can be written as follows
\be
\begin{aligned}
\hat{u}^\mu &=& \tau\,\frac{\partial \hat{x}^\mu}{\partial x^\nu}\,u^\nu,\nonumber \\
\hat{\Theta}^\mu &=& \tau\,\frac{\partial \hat{x}^\mu}{\partial x^\nu}\,X^\nu,
\end{aligned}
\hspace{1.5cm}
\begin{aligned}
\hat{\Phi}^\mu &=& \tau\,\frac{\partial \hat{x}^\mu}{\partial x^\nu}\,Y^\nu,\nonumber \\
\hat{\varsigma}^\mu &=& \tau\,\frac{\partial \hat{x}^\mu}{\partial x^\nu}\,Z^\nu\,.
\end{aligned}
\label{eq:trans}
\ee
Starting with the Minkowski basis vectors in the boost-invariant cylindrically-symmetric case (\ref{eq:basis1+1d}), one can use Eq.~(\ref{eq:trans}) and the de Sitter-space identities listed in App.~\ref{app:desitterids} to obtain
\ba 
\hat{u}^\mu &=& (1,\,0,\,0,\,0)\, , \notag \\
\hat{\Theta}^\mu &=& (0,\,(\cosh\rho)^{-1},\,0,\,0)\, , \notag \\
\hat{\Phi}^\mu &=& (0,\,0,\,(\cosh\rho \sin\theta )^{-1},\,0)\, , \notag \\
\hat{\varsigma}^\mu &=& (0,\,0,\,0,\,1)\, .
\label{eq:desitter-4vectors}
\ea
One can check explicitly that the orthonormality conditions for the basis vectors are satisfied:
\ba
\hat{u}\cdot\hat{u}&\equiv & \hat{u}^\mu\hat{u}_\mu=-1\, , \notag \\
\hat{\Theta}\cdot\hat{\Theta}&\equiv & \hat{\Theta}^\mu\hat{\Theta}_\mu=1\, , \notag \\
\hat{\Phi}\cdot\hat{\Phi}&\equiv & \hat{\Phi}^\mu\hat{\Phi}_\mu=1\, , \notag \\
\hat{\varsigma}\cdot\hat{\varsigma}&\equiv & \hat{\varsigma}^\mu\hat{\varsigma}_\mu=1\, ,
\label{eq:u.u}
\ea 
and all other dot products vanish. In de Sitter coordinates, $\theta$ and $\phi$ are transverse coordinates and $\varsigma$ is the longitudinal one.

\subsection{Ellipsoidal form for the distribution function}
Following Chap.~\ref{chap:ahydro}, one can introduce an ansatz for the conformal anisotropic distribution function in de Sitter space using (\ref{eq:xi}) where all quantities are defined in de Sitter space
\ba 
\hat{\Xi}^{\mu\nu}=\hat{u}^\mu \hat{u}^\nu+\hat{\xi}^{\mu\nu}\,.
\label{eq:aniso-tensor1}
\ea
Herein, the anisotropy tensor is expanded over the de Sitter basis vectors \footnote{Note that I assume the anisotropy tensor to be diagonal in the local rest frame with respect to the de Sitter coordinate. In App.~\ref{app:expansion} I have shown that this leads to diagonal anisotropy tensor respect to polar Milne coordinate, as well.} 
\ba
\hat{\xi}^{\mu\nu} = \hat{\xi}_\theta \hat{\Theta}^\mu \hat{\Theta}^\nu+\hat{\xi}_\phi \hat{\Phi}^\mu \hat{\Phi}^\nu+\hat{\xi}_\varsigma \hat{\varsigma}^\mu \hat{\varsigma}^\nu\, .
\label{eq:aniso-tensor2}
\ea
The above quantities obey the following identities 
\be
\begin{aligned}
\hat{\xi}^{\mu}_{\ \mu} &= 0 \, , 
\label{eq:tracelessness}
\\
\hat{u}_\mu \hat{\xi}^{\mu\nu} &= 0 \, ,
\end{aligned}
\hspace{2.5cm}
\begin{aligned}
{\hat{\Xi}^\mu}_{\ \mu} &= -1 \, ,
\\
\hat{u}_\mu \hat{\Xi}^{\mu\nu} &= -\hat{u}^\nu \, .
\end{aligned}
\ee
Using tensor $\hat{\Xi}^{\mu\nu}$, one can construct an anisotropic distribution function following Ref.~\cite{Nopoush:2014qba}\footnote{I assume herein that the chemical potential is zero.} 
\ba
f(\hat{x},\hat{p})=f_{\rm eq}\left(\frac{1}{\hat\lambda}\sqrt{\hat{p}_\mu\hat{\Xi}^{\mu\nu} \hat{p}_\nu}\right)\, ,
\label{eq:pdf}
\ea
where $\hat\lambda$ can be identified with the de Sitter-space temperature, $\hat{T}$, only when $\hat\xi^{\mu\nu}=0$.

\subsection{Dynamical variables}
In the conformal case, $\hat\xi^{\mu\nu}$ is traceless and diagonal
\ba
\hat{\xi}_\theta+\hat{\xi}_\phi+\hat{\xi}_\varsigma=0 \, ,
\label{eq:xi-trace}
\ea
which can be verified using Eqs.~(\ref{eq:desitter-metric}) and (\ref{eq:desitter-4vectors}). In order to satisfy $SO(3)_q$ invariance, the distribution function can only depend on $\hat{p}_\Omega^2\equiv\hat{p}_\theta^2+\hat{p}_\phi^2/\!\sin^2\!\theta$ \cite{Denicol:2014tha}. As a result, one must have $\hat{\xi}_\theta=\hat{\xi}_\phi$. Using this, the condition (\ref{eq:xi-trace}) implies
\ba
\hat{\xi}_\theta=-\frac{\hat{\xi}_\varsigma}{2} \, .
\label{eq:xi-trace2}
\ea
Again for convenience, one can define new parameters $\hat\alpha_i$ as $\hat\alpha_i \equiv (1+\hat{\xi}_i)^{-1/2}$ where $i \in \{\theta,\,\phi,\,\varsigma\}$.  Using Eqs.~(\ref{eq:desitter-metric}),$\,$(\ref{eq:aniso-tensor1}) one can simplify the distribution function to 
\ba 
f(\hat{x},\hat{p})=f_{\rm iso}\left(\frac{1}{\hat\lambda}\sqrt{\sum_i\frac{\hat{p}^i\hat{p}_i}{\hat\alpha_i^2}}\right)\, ,
\ea
where $\hat{p}_i$ and $\hat{p}^i$ are related through the metric (\ref{eq:desitter-metric}) as before. Note that $\hat{\xi}_\theta=\hat{\xi}_\phi$ implies
\be
\hat\alpha_\theta = \hat\alpha_\phi \ .
\label{eq:at=af}
\ee

\section{Bulk variables in de Sitter coordinates}
\label{sec:bulkvars}
In order to extract the energy density and pressures from the energy-momentum tensor, one can expand it in a tensor basis (\ref{eq:desitter-4vectors}) in de Sitter coordinates.  Since the distribution function is of ellipsoidal form, the energy-momentum tensor is diagonal in de Sitter space
\ba 
\hat{T}^{\mu\nu} &=& \hat{\varepsilon}\hat{u}^\mu\hat{u}^{\nu}+
\hat{P}_\theta\hat{\Theta}^\mu\hat{\Theta}^{\nu}
+\hat{P}_\phi\hat{\Phi}^\mu\hat{\Phi}^{\nu}+
\hat{P}_\varsigma\hat{\varsigma}^\mu\hat{\varsigma}^{\nu}\, ,
\label{eq:energy-mom}
\ea
where $\hat{\varepsilon}$, $\hat{P}_\theta$, $\hat{P}_\phi$, and $\hat{P}_\varsigma$ are de Sitter energy density and pressures. In the kinetic theory framework, one can use the integral form of $\hat{T}^{\mu\nu}$ to evaluate these quantities. Following Chap.~\ref{chap:ahydro} one can define the $n^{\rm th}$-moment of the distribution function as
\ba
&& \hat{\mathcal{I}}^{\mu_1...\mu_n} \equiv \int d\hat{P} \, \hat{p}^{\mu_1}...\,\hat{p}^{\mu_n} f(\hat{x},\hat{p})\, .
\label{eq:nth moments}
\ea
The integral measure has the following form in de Sitter coordinates
\ba
\int d\hat{P}\equiv\tilde{N}\int \frac{ d\hat{p}_\theta\, d\hat{p}_\phi\, d\hat{p}_\varsigma}{E\,\cosh^2\!\rho
\,\sin\!\theta} \, , 
\ea
where I have used $\sqrt{-\hat{\mathrm{g}}}=\cosh^2\!\rho\,\sin\theta$ defined in Eq.~(\ref{eq:g}).
Taking $n=2$ in Eq.~(\ref{eq:nth moments}), the integral form of the energy-momentum tensor is obtained
\ba
&& \hat{T}^{\mu\nu} \equiv \int d\hat{P}\,\hat{p}^\mu \hat{p}^\nu f(\hat{x},\hat{p})\, .
\label{eq:energy-mom-int}
\ea
Taking projections of $\hat{T}^{\mu\nu}$ with the de Sitter-space basis vectors (\ref{eq:desitter-4vectors}), one finds
\ba 
\hat{\varepsilon} &\equiv & \hat{u}_\mu \hat{T}^{\mu\nu} \hat{u}_\nu = \int d\hat{P} \, \hat{p}^\rho \hat{p}^\rho f_{\rm iso}\left(\frac{1}{\hat\lambda}\sqrt{\sum_i\frac{\hat{p}_i\hat{p}^i}{\hat\alpha^2_i}}\right)\, ,
\label{eq:e-def} \\
\hat{P_\theta} &\equiv & \hat{\Theta}_\mu \hat{T}^{\mu\nu} \hat{\Theta}_\nu = \int d\hat{P}\cosh^2\!\rho\, \hat{p}^\theta \hat{p}^\theta f_{\rm iso}\left(\frac{1}{\hat\lambda}\sqrt{\sum_i\frac{\hat{p}_i\hat{p}^i}{\hat\alpha^2_i}}\right)\, ,
\label{eq:ept-def} \\
\hat{P_\phi} &\equiv & \hat{\Phi}_\mu \hat{T}^{\mu\nu} \hat{\Phi}_\nu = \int d\hat{P} \cosh^2\!\rho\sin^2\!\theta\, \hat{p}^\phi \hat{p}^\phi f_{\rm iso}\left(\frac{1}{\hat\lambda}\sqrt{\sum_i\frac{\hat{p}_i\hat{p}^i}{\hat\alpha^2_i}}\right)\, ,
\label{eq:pf-def} \\
\hat{P_\varsigma} &\equiv & \hat{\varsigma}_\mu \hat{T}^{\mu\nu} \hat{\varsigma}_\nu = \int d\hat{P} \, \hat{p}^\varsigma \hat{p}^\varsigma f_{\rm iso}\left(\frac{1}{\hat\lambda}\sqrt{\sum_i\frac{\hat{p}_i\hat{p}^i}{\hat\alpha^2_i}}\right)\, ,
\label{eq:pv-def}
\ea
where I have used $\hat{p}^0 = \hat{p}^\rho$. To obtain these results, the following identities were used
\ba 
\hat{u}_\mu \hat{p}^\mu &=& -\hat{p}^\rho\, , \notag \\
\hat{\Theta}_\mu \hat{p}^\mu &=& \hat{p}^\theta\cosh\rho\, , \notag \\
\hat{\Phi}_\mu \hat{p}^\mu &=& \hat{p}^\phi\cosh\rho\sin\theta\, , \notag \\
\hat{\varsigma}_\mu \hat{p}^\mu &=& \hat{p}^\varsigma \, .
\ea
Computing the integrals, the bulk variables in de Sitter coordinates are
\ba 
\hat{\varepsilon}&=&\frac{6\hat\alpha_\theta\hat\alpha_\phi}{(2\pi)^3}\hat\lambda^4 \int_0^{2\pi}d\phi\,\hat\alpha_\perp^2 {\cal H}_2(y)\, ,
\label{eq:e} \\
\hat{P}_\theta &=&\frac{6\hat\alpha_\theta^3\hat\alpha_\phi}{(2\pi)^3}\hat\lambda^4 \int_0^{2\pi}d\phi\cos^2\!\phi {\cal H}_{2T}(y)\, , 
\label{eq:pt}\\
\hat{P}_\phi &=&\frac{6\hat\alpha_\theta\hat\alpha_\phi^3}{(2\pi)^3}\hat\lambda^4 \int_0^{2\pi}d\phi\sin^2\!\phi {\cal H}_{2T}(y)\, , 
\label{eq:pf}\\
\hat{P}_\varsigma &=&\frac{6\hat\alpha_\theta\hat\alpha_\phi}{(2\pi)^3}\hat\lambda^4 \int_0^{2\pi}d\phi\,\hat\alpha_\perp^2 {\cal H}_{2L}(y)\, .
\label{eq:pv}
\ea
where $\hat\alpha_\perp\equiv\sqrt{\hat\alpha^2_\theta\cos^2\phi+\hat\alpha^2_\phi\sin^2\phi}$, $y \equiv\hat\alpha_\varsigma/\hat\alpha_\perp$, and the ${\cal H}$-functions are defined in (\ref{eq:H2})-(\ref{eq:H2L}).
As discussed earlier, requiring $SO(3)_q$ invariance in de Sitter space implies (\ref{eq:at=af}).  Together with Eq.~(\ref{eq:xi-trace}), this condition implies that one can write $\hat\alpha_\theta$ in terms of $\hat\alpha_\varsigma$
\ba 
\hat\alpha_\theta = \sqrt{\frac{2\hat\alpha_\varsigma^2}{3\hat\alpha_\varsigma^2-1}} \, .
\label{eq:at}
\ea
Additionally, $\hat\alpha_\phi = \hat\alpha_\theta$ implies that $\hat\alpha_\perp=\hat\alpha_\theta$.  Therefore, one can simplify Eqs.~(\ref{eq:e}-\ref{eq:pv}) to
\ba
\hat{\varepsilon}&=&\frac{3\,\hat\alpha_\theta^4\hat\lambda^4}{2\pi^2} {\cal H}_2(\bar{y})\, ,
\label{eq:e2} \\
\hat{P}_\theta &=&\hat{P}_\phi=\frac{3\,\hat\alpha_\theta^4\hat\lambda^4}{4\pi^2} {\cal H}_{2T}(\bar{y})\, , 
\label{eq:pt2} \\
\hat{P}_\varsigma &=&\frac{3\,\hat\alpha_\theta^4\hat\lambda^4}{2\pi^2} {\cal H}_{2L}(\bar{y})\, ,
\label{eq:pv2}
\ea
where $\bar{y} \equiv \hat\alpha_\varsigma/\hat\alpha_\theta$ is
\ba
\bar{y}=\sqrt{\frac{3\hat\alpha_\varsigma^2-1}{2}}\, .
\label{eq:ybar}
\ea 

\section{Moments of the Boltzmann equation}
\label{sec:moments}

The Boltzmann equation in the relaxation-time approximation in de Sitter space is	
\ba 
\hat{p} \cdot {\mathfrak D} f &=& \frac{\hat{p}\cdot \hat{u}}{\hat\tau_{\rm eq}}(f-f_{\rm eq})\, ,
\label{eq:boltzmanneq}
\ea
where ${\mathfrak D}_\mu$ is the covariant derivative defined in App. \ref{app:covderiv}, $f_{\rm eq}$ denotes the isotropic equilibrium distribution function, and $\hat\tau_{\rm eq}$ is the relaxation time. Conformal invariance requires that $\hat\tau_{\rm eq}$ is inversely proportional to the temperature, i.e. $\hat\tau_{\rm eq} \propto 1/\hat{T}$. Since I work in the RTA, the exact relation is $\hat\tau_{\rm eq} = 5\hat{\bar\eta}/\hat{T}$, where $\hat{\bar\eta}=\hat\eta/\hat{s}=\eta/s$ with $\hat\eta$ being the Weyl-rescaled shear viscosity and $\hat{s}$ being the  Weyl-rescaled entropy density.  As before, dynamical equations are derived using the moments of  the Boltzmann equation in de Sitter coordinates.

\subsection{First moment}

Taking the first moment of the Boltzmann equation (\ref{eq:boltzmanneq}) gives
\ba 
{\mathfrak D}_\mu \hat{T}^{\mu\nu} &=& 0\, ,
\label{eq:boltzmann1}
\ea
where $\hat{T}^{\mu\nu}$ is the energy-momentum tensor. To obtain (\ref{eq:boltzmann1}) I require that the first moment of the right-hand side of the Boltzmann equation vanishes, so that the energy and momenta are conserved. This results in the so-called dynamical Landau matching condition, which allows us to express the effective temperature $\hat{T}$ in terms of the microscopic parameters
\ba
\hat{T}=\frac{\hat\alpha_\varsigma}{\bar{y}}\left(\frac{{\cal H}_2(\bar{y})}{2}\right)^{1/4} \hat\lambda\, .
\label{eq:matching-final}
\ea
Using Eqs.~(\ref{eq:energy-mom}) and (\ref{eq:covariant-derivative}) in Appendix (\ref{app:covderiv}), one can expand Eq.~(\ref{eq:boltzmann1}) to obtain 
\ba
\Gamma^\nu_{\lambda\mu} \hat{T}^{\lambda\mu} &+& \hat{T}^{\mu\nu} \frac{\partial_\mu\sqrt{-\hat{g}\,}}{\sqrt{-\hat{g}\,}} + \hat{u}^{\nu}(\hat{u}^\mu\partial_\mu)\hat{\varepsilon}+\hat{u}^\nu(\partial_\mu\hat{u}^\mu)\hat{\varepsilon}+\hat{\varepsilon}(\hat{u}^\mu\partial_\mu)\hat{u}^\nu \notag \\
&+& \hat{\Theta}^{\nu}(\hat{\Theta}^\mu\partial_\mu)\hat{P}_\theta+\hat{\Theta}^\nu(\partial_\mu\hat{\Theta}^\mu)\hat{P}_\theta+\hat{P}_\theta(\hat{\Theta}^\mu\partial_\mu)\hat{\Theta}^\nu  \notag \\ 
&+& \hat{\Phi}^{\nu}(\hat{\Phi}^\mu\partial_\mu)\hat{P}_\phi+\hat{\Phi}^\nu(\partial_\mu\hat{\Phi}^\mu)\hat{P}_\phi+\hat{P}_\phi(\hat{\Phi}^\mu\partial_\mu)\hat{\Phi}^\nu  \notag \\
&+& \hat{\varsigma}^{\nu}(\hat{\varsigma}^\mu\partial_\mu)\hat{P}_\varsigma+\hat{\varsigma}^\nu(\partial_\mu\hat{\varsigma}^\mu)\hat{P}_\varsigma+\hat{P}_\varsigma(\hat{\varsigma}^\mu\partial_\mu)\hat{\varsigma}^\nu  = 0\, .
\label{eq:boltzman-expand}
\ea
Using of the de Sitter 4-vectors (\ref{eq:desitter-4vectors}), one can take different projections of Eq.~(\ref{eq:boltzman-expand}) as
\ba
\partial_\rho\hat{\varepsilon} + \tanh\!\rho \,(2\hat{\varepsilon} + \hat{P}_\theta + \hat{P}_\phi) &=& 0\, ,
\label{eq:1th-mom-e}\\
\partial_\theta\hat{P}_\theta +(\hat{P}_\theta-\hat{P}_\phi)\cot\theta &=& 0\, , 
\label{eq:1th-mom-pt}\\
\partial_\phi\hat{P}_\phi =\partial_\varsigma\hat{P}_\varsigma &=& 0 \, .
\label{eq:1th-mom-pv}
\ea
Using the $SO(3)_q$ symmetry and Eq.~(\ref{eq:at=af}), one can simplify the equations above to
\ba
\partial_\rho\hat{\varepsilon} + 2\tanh\!\rho \,(\hat{\varepsilon} + \hat{P}_\theta) &=& 0\, ,
\label{eq:1th-mom-e2}\\
\partial_\theta\hat{P}_\theta=\partial_\phi\hat{P}_\phi =\partial_\varsigma\hat{P}_\varsigma &=& 0\, .
\label{eq:1th-mom-pv2}
\ea 
The set of equations above demonstrates that, subject to $SO(3)_q$ symmetry, all fields and physical quantities are functions of $\rho$ exclusively. In other words, the differential equations describing the system reduce to coupled first-order ordinary differential equations, which can be solved by providing initial conditions in de Sitter space. Having the final expressions for $\hat{\varepsilon}$ and $\hat{P}_\theta$, Eqs.~(\ref{eq:e2}) and (\ref{eq:pt2}), one finds the first moment of the Boltzmann equation in de Sitter space
\ba
4\frac{d\log\hat\lambda}{d\rho}+\frac{1}{3\hat\alpha_\varsigma^2-1}\bigg[3 \hat\alpha_\varsigma^2\left(\frac{{\cal H}_{2
   L}(\bar{y})}{{\cal H}_2(\bar{y})}+1\right)-4\bigg]
   \, \frac{d\log\hat\alpha_\varsigma}{d\rho}+ \tanh\rho\left(\frac{{\cal H}_{2T}(\bar{y})}{{\cal H}_2(\bar{y})}+2\right)=0\, .
   \label{eq:1st-mom-final}
\ea

\subsection{Equivalence to second-order viscous hydrodynamics}
As a check that our starting point given by Eqs.~(\ref{eq:1th-mom-e})-(\ref{eq:1th-mom-pv}) is consistent with the results obtained previously in the context of second-order viscous hydrodynamics, I can rewrite them in terms of the shear tensor.  To do this, I begin by expanding the shear viscous tensor in terms of the de Sitter-space basis vectors (\ref{eq:desitter-4vectors})
\ba
\hat{\pi}_{\mu\nu} =\hat{\pi}_\theta^\theta\hat{\Theta}_\mu\hat{\Theta}_\nu
+\hat{\pi}_\phi^\phi\hat{\Phi}_\mu\hat{\Phi}_\nu
+\hat{\pi}_\varsigma^\varsigma\hat{\varsigma}_\mu\hat{\varsigma}_\nu\, ,
\label{eq:pi-exp}
\ea
where the different components obey
\ba
\hat{\pi}_\theta^\theta+\hat{\pi}_\phi^\phi+\hat{\pi}_\varsigma^\varsigma=0\, .
\label{eq:pi-condition}
\ea
To proceed, I can use the definition of the shear viscous stress tensor as the correction to the isotropic equilibrium pressures
\be
\hat{P}_i = \hat{P}_{\rm eq}+\hat{\pi}_i^i\, ,\\
\label{eq:pi-def}
\ee
where $i \in \{\theta,\phi,\varsigma\}$.
Using $\hat{P}_{\rm eq}=\hat{\varepsilon}/3$, one obtains
\ba
\hat{P}_\theta+\hat{P}_\phi=\frac{2}{3}\hat{\varepsilon}-\hat{\pi}_\varsigma^\varsigma\, .
\label{eq:pt+pf}
\ea
Substituting Eq.~(\ref{eq:pt+pf}) into Eq.~(\ref{eq:1th-mom-e}) gives
\ba
\partial_\rho\hat{\varepsilon} + \tanh\!\rho \,\left(\frac{8}{3}\hat{\varepsilon}-\hat{\pi}_\varsigma^\varsigma\right) = 0\, .
\ea
Using the thermodynamic relation $\hat{\varepsilon}+\hat{P}_{\rm eq}=\hat{T}\hat{s}$, where $\hat{s}$ is Weyl-rescaled entropy density, one finds $\hat{T}\hat{s}=4\hat{\varepsilon}/3$. In conformal field theory I have $\hat{\varepsilon}\propto \hat{T}^4$. Defining $\bar{\pi}^\varsigma_\varsigma \equiv \hat{\pi}^\varsigma_\varsigma/(\hat{T}\hat{s})$, one obtains the following equation
\ba
\frac{\partial_\rho\hat{T}}{\hat{T}} + \frac{2}{3}\tanh\!\rho = \frac{1}{3}\,\bar{\pi}^\varsigma_\varsigma\tanh\!\rho\, .
\label{eq:denicol}
\ea
This is precisely the same as the first-moment equation obtained originally in Ref.~\cite{Marrochio:2013wla}.

\subsection{Second moment}
Computing the second moment of Boltzmann equation (\ref{eq:boltzmanneq}) gives
\ba 
{\mathfrak D}_\lambda\hat{\mathcal{I}}^{\lambda\mu\nu}=-\frac{1}{\hat\tau_{\rm eq}} \left(\hat{u}_\lambda \hat{\mathcal{I}}^{\lambda\mu\nu}_{\rm eq}-\hat{u}_\lambda \hat{\mathcal{I}}^{\lambda\mu\nu}\right) ,
\label{eq:boltzmann2}
\ea
where $\hat{\mathcal{I}}^{\lambda\mu\nu}$ and $\hat{\mathcal{I}}^{\lambda\mu\nu}_{\rm eq}$ can be obtained by taking $n=3$ in Eq.~(\ref{eq:nth moments})
\ba 
\hat{\mathcal{I}}^{\lambda\mu\nu} &=& \int d\hat{P} \, \hat{p}^\lambda \hat{p}^\mu \hat{p}^\nu f(\hat{x},\hat{p})\, , \\
\hat{\mathcal{I}}^{\lambda\mu\nu}_{\rm eq} &=& \int d\hat{P} \, \hat{p}^\lambda \hat{p}^\mu\hat{p}^\nu f_{\rm eq}(\hat{x},\hat{p})\, .
\label{eq:2th-mom-int}
\ea

From the symmetry of the integrands in the definition of $\hat{\mathcal{I}}^{\lambda\mu\nu}$ above, one concludes that $\hat{\mathcal{I}}^{\lambda\mu\nu}$ only contains terms which have an even number of spatial indices. Using the de Sitter-space basis (\ref{eq:desitter-4vectors}), one can expand $\hat{\mathcal{I}}^{\lambda\mu\nu}$ in covariant form as
\ba 
\hat{\mathcal{I}} \equiv  \hat{\mathcal{I}}_\rho\Big[\hat{u}\otimes \hat{u} \otimes \hat{u} \Big] \notag \!&\!+\!&\! \hat{\mathcal{I}}_\theta\Big[\hat{u}\otimes \hat{\Theta} \otimes \hat{\Theta}\,+\,\hat{\Theta}\otimes \hat{u} \otimes \hat{\Theta}\,+\,\hat{\Theta}\otimes \hat{\Theta} \otimes \hat{u} \Big] \notag  \\ &+& \hat{\mathcal{I}}_\phi\Big[\hat{u}\otimes \hat{\Phi} \otimes \hat{\Phi}\,+\,\hat{\Phi}\otimes \hat{u} \otimes \hat{\Phi}\,+\,\hat{\Phi}\otimes \hat{\Phi} \otimes \hat{u} \Big] \notag \\ &+&\hat{\mathcal{I}}_\varsigma\Big[\hat{u}\otimes \hat{\varsigma} \otimes \hat{\varsigma}\,+\,\hat{\varsigma}\otimes \hat{u}\otimes \hat{\varsigma}\,+\,\hat{\varsigma}\otimes \hat{\varsigma} \otimes \hat{u} \Big]\, .
\label{2nd-mom-exp}
\ea
For a massless system, one has $\hat{p}^\mu\hat{p}_\mu=0$, which gives the following useful identity
\ba 
\hat{p}^\rho=-\hat{p}_\rho=\sqrt{\frac{\hat{p}_\theta^2}{\cosh^2\!\rho}+\frac{\hat{p}_\phi^2}{\cosh^2\!\rho\sin^2\!\theta}+\hat{p}_\varsigma^2} \; .
\ea
Using Eq.~(\ref{eq:u.u}), one can take different projections of $\hat{\mathcal{I}}^{\lambda\mu\nu}$ to obtain the following expressions
\ba \notag
\hat{\mathcal{I}}_\rho &\equiv & -\hat{u}_\lambda\hat{u}_\mu\hat{u}_\nu\hat{\mathcal{I}}^{\lambda\mu\nu} =\int d\hat{P}\hat{p}^3_\rho f_{\rm eq}\!\left(\frac{1}{\hat\lambda}\sqrt{\frac{\hat{p}^2_\theta}{\hat\alpha^2_\theta\cosh^2\!\rho}+\frac{\hat{p}^2_\phi}{\hat\alpha^2_\phi\cosh^2\!\rho\sin^2\!\theta}+\frac{\hat{p}^2_\varsigma}{\hat\alpha^2_\varsigma}}\right) \notag\\
\hat{\mathcal{I}}_\theta &\equiv & -\hat{u}_\lambda\hat{\Theta}_\mu\hat{\Theta}_\nu\hat{\mathcal{I}}^{\lambda\mu\nu} =\int d\hat{P}\frac{\hat{p}_\rho\hat{p}_\theta^2}{\cosh^2\!\rho}   f_{\rm eq}\!\left(\frac{1}{\hat\lambda}\sqrt{\frac{\hat{p}^2_\theta}{\hat\alpha^2_\theta\cosh^2\!\rho}+\frac{\hat{p}^2_\phi}{\hat\alpha^2_\phi\cosh^2\!\rho\sin^2\!\theta}+\frac{\hat{p}^2_\varsigma}{\hat\alpha^2_\varsigma}}\right)\notag \\ 
\hat{\mathcal{I}}_\phi &\equiv & -\hat{u}_\lambda\hat{\Phi}_\mu\hat{\Phi}_\nu\hat{\mathcal{I}}^{\lambda\mu\nu} = \int d\hat{P}\frac{\hat{p}_\rho\hat{p}_\phi^2}{\cosh^2\!\rho\sin^2\!\theta} f_{\rm eq}\!\left(\frac{1}{\hat\lambda}\sqrt{\frac{\hat{p}^2_\theta}{\hat\alpha^2_\theta\cosh^2\!\rho}+\frac{\hat{p}^2_\phi}{\hat\alpha^2_\phi\cosh^2\!\rho\sin^2\!\theta}+\frac{\hat{p}^2_\varsigma}{\hat\alpha^2_\varsigma}}\right) , \notag\\
\hat{\mathcal{I}}_\varsigma &\equiv & -\hat{u}_\lambda\hat{\varsigma}_\mu\hat{\varsigma}_\nu\hat{\mathcal{I}}^{\lambda\mu\nu} = \int d\hat{P} \hat{p}_\rho\hat{p}_\varsigma^2 f_{\rm eq}\!\left(\frac{1}{\hat\lambda}\sqrt{\frac{\hat{p}^2_\theta}{\hat\alpha^2_\theta\cosh^2\!\rho}+\frac{\hat{p}^2_\phi}{\hat\alpha^2_\phi\cosh^2\!\rho\sin^2\!\theta}+\frac{\hat{p}^2_\varsigma}{\hat\alpha^2_\varsigma}}\right) ,
\ea
The results of the integrals above can be compactly written as
\ba 
&&\hat{\mathcal{I}}_\rho=\hat\alpha\left[\sum_{i=\theta,\phi,\varsigma}\hat\alpha_i^2\right]\hat{\mathcal{I}}_{\rm eq} \, , \label{eq:I-rho}\\
&&\hat{\mathcal{I}}_i=\hat\alpha\hat\alpha_i^2\hat{\mathcal{I}}_{\rm eq} \, ,\label{eq:I-i}
\ea
where $\hat\alpha \equiv \hat\alpha_\theta\hat\alpha_\phi\hat\alpha_\varsigma$ and $\hat{\mathcal{I}}_{\rm eq}\equiv4\hat\lambda^5/\pi^2$.  The coefficients above clearly obey
\ba
\hat{\mathcal{I}}_\rho=\sum_{i=\theta,\phi,\varsigma}\hat{\mathcal{I}}_i \, ,
\label{eq:I-sum}
\ea
which follows from $\hat{g}_{\mu\nu}\hat{\mathcal{I}}^{\mu\nu\lambda}=\hat{g}_{\mu\lambda}\hat{\mathcal{I}}^{\mu\nu\lambda}=\hat{g}_{\nu\lambda}\hat{\mathcal{I}}^{\mu\nu\lambda}=0$ since $\hat{p}^\mu \hat{p}_\mu=0$.  If the system possesses $SO(3)_q$ symmetry, one has $\hat{\mathcal{I}}_\theta=\hat{\mathcal{I}}_\phi$
\ba 
\hat{\mathcal{I}}_\theta=\hat{\mathcal{I}}_\phi\, . \label{eq:It=If}
\ea

\subsection{Dynamical equations}

Using Eq.~(\ref{2nd-mom-exp}), one can expand Eq.~(\ref{eq:boltzmann2}) as
\ba 
{\mathfrak D}_\lambda\hat{\mathcal{I}}^{\lambda\mu\nu}&=&\hat{u}^\mu\hat{u}^\nu(\hat{u}^\lambda {\mathfrak D}_\lambda)\hat{\mathcal{I}}_\rho+
\hat{u}^\mu\hat{u}^\nu({\mathfrak D}_\lambda\hat{u}^\lambda)\hat{\mathcal{I}}_\rho +
\hat{\mathcal{I}}_\rho(\hat{u}^\lambda {\mathfrak D}_\lambda)\hat{u}^\mu\hat{u}^\nu \notag \\
&&+\hat{\Theta}^\mu\hat{\Theta}^\nu (\hat{u}^\lambda {\mathfrak D}_\lambda)\hat{\mathcal{I}}_\theta
+\hat{\Theta}^\mu\hat{\Theta}^\nu ({\mathfrak D}_\lambda\hat{u}^\lambda) \hat{\mathcal{I}}_\theta
+\hat{\mathcal{I}}_\theta(\hat{u}^\lambda {\mathfrak D}_\lambda)\hat{\Theta}^\mu\hat{\Theta}^\nu \notag \\
&&+\hat{u}^\mu\hat{\Theta}^\nu (\hat{\Theta}^\lambda {\mathfrak D}_\lambda) \hat{\mathcal{I}}_\theta+
\hat{u}^\mu\hat{\Theta}^\nu ({\mathfrak D}_\lambda\hat{\Theta}^\lambda)\hat{\mathcal{I}}_\theta +
\hat{\mathcal{I}}_\theta(\hat{\Theta}^\lambda {\mathfrak D}_\lambda)\hat{u}^\mu\hat{\Theta}^\nu \notag \\
&&+\hat{\Theta}^\mu\hat{u}^\nu (\hat{\Theta}^\lambda {\mathfrak D}_\lambda)\hat{\mathcal{I}}_\theta
+\hat{\Theta}^\mu\hat{u}^\nu ({\mathfrak D}_\lambda\hat{\Theta}^\lambda)\hat{\mathcal{I}}_\theta
+\hat{\mathcal{I}}_\theta(\hat{\Theta}^\lambda {\mathfrak D}_\lambda)\hat{\Theta}^\mu\hat{u}^\nu
\notag \\
&&+(\hat{\Theta}\rightarrow\hat{\Phi})+(\hat{\Theta}\rightarrow\hat{\varsigma}) =-\frac{1}{\hat\tau_{\rm eq}}\left[\hat{u}_\lambda\hat{\mathcal{I}}^{\lambda\mu\nu}_{\rm eq}-\hat{u}_\lambda\hat{\mathcal{I}}^{\lambda\mu\nu}\right] .
\label{eq:Dynamical-equations}
\ea
Using the identities in Appendices \ref{app:covderiv}, one can simplify Eq.~(\ref{eq:Dynamical-equations}) to
\ba 
{\mathfrak D}_\lambda\hat{\mathcal{I}}^{\lambda\mu\nu}&=&\hat{u}^\mu\hat{u}^\nu\left[\partial_\rho\hat{\mathcal{I}}_\rho
+2\tanh\rho\,(\hat{\mathcal{I}}_\rho+\hat{\mathcal{I}}_\theta+\hat{\mathcal{I}}_\phi)\right] \notag \\
&+&\hat{\Theta}^\mu \hat{\Theta}^\nu\left[\partial_\rho\hat{\mathcal{I}}_\theta
+4\tanh\rho\,\hat{\mathcal{I}}_\theta\right] +\frac{\hat{u}^\mu\hat{\Theta}^\nu+\hat{\Theta}^\mu\hat{u}^\nu}{\cosh\rho}\left[\partial_\theta\hat{\mathcal{I}}_\theta+\cot\theta (\hat{\mathcal{I}}_\theta-\hat{\mathcal{I}}_\phi)\right]\notag \\
&+&\hat{\Phi}^\mu \hat{\Phi}^\nu\left[\partial_\rho\hat{\mathcal{I}}_\phi
+4\tanh\rho\,\hat{\mathcal{I}}_\phi\right]+\frac{\hat{u}^\mu\hat{\Phi}^\nu+\hat{\Phi}^\mu\hat{u}^\nu}{\cosh\rho\sin\theta}\left[\partial_\phi\hat{\mathcal{I}}_\phi\right] \notag \\
&+&\hat{\varsigma}^\mu \hat{\varsigma}^\nu\left[\partial_\rho\hat{\mathcal{I}}_\varsigma
+2\tanh\rho\,\hat{\mathcal{I}}_\varsigma\right] +\left(\hat{u}^\mu\hat{\varsigma}^\nu+\hat{\varsigma}^\mu\hat{u}^\nu\right)\left[\partial_\varsigma\hat{\mathcal{I}}_\varsigma\right]\notag \\
&=&-\frac{1}{\hat\tau_{\rm eq}}\left[\hat{u}_\lambda\hat{\mathcal{I}}^{\lambda\mu\nu}_{\rm eq}
-\hat{u}_\lambda\hat{\mathcal{I}}^{\lambda\mu\nu}\right]\, .
\ea
From the expression above, I can obtain various scalar projections.  The diagonal projections, $\hat{u}\hat{u}$, $\hat{\Theta}\hat{\Theta}$, $\hat{\Phi}\hat{\Phi}$, and $\hat{\varsigma}\hat{\varsigma}$, are:
\ba
\partial_\rho\hat{\mathcal{I}}_\rho
+2\tanh\rho\,(\hat{\mathcal{I}}_\rho+\hat{\mathcal{I}}_\theta+\hat{\mathcal{I}}_\phi)
= \frac{1}{\hat\tau_{\rm eq}}\left[\hat{\mathcal{I}}_{\rho,\rm eq}-\hat{\mathcal{I}}_\rho\right]\, ,
\label{eq:uu-projection} \\
\partial_\rho\hat{\mathcal{I}}_\theta + 4\tanh\rho\,\hat{\mathcal{I}}_\theta 
=\frac{1}{\hat\tau_{\rm eq}}\left[\hat{\mathcal{I}}_{\theta,\rm eq}-\hat{\mathcal{I}}_\theta\right]\, ,
\label{eq:tt-projection}
\\
\partial_\rho\hat{\mathcal{I}}_\phi +4\tanh\rho\,\hat{\mathcal{I}}_\phi 
=\frac{1}{\hat\tau_{\rm eq}}\left[\hat{\mathcal{I}}_{\phi,\rm eq}-\hat{\mathcal{I}}_\phi\right]\, ,
\label{eq:ff-projection}
\\
\partial_\rho\hat{\mathcal{I}}_\varsigma
+2\tanh\rho\,\hat{\mathcal{I}}_\varsigma
=\frac{1}{\hat\tau_{\rm eq}}\left[\hat{\mathcal{I}}_{\varsigma,\rm eq}-\hat{\mathcal{I}}_\varsigma\right]\, .
\label{eq:vv-projection}
\ea
Using (\ref{eq:I-sum}), one can verify that the equations above are not independent, i.e. Eqs.~(\ref{eq:tt-projection})-(\ref{eq:vv-projection}) imply that Eq.~(\ref{eq:uu-projection}) is automatically satisfied.  The non-trivial off-diagonal projections, i.e. $\hat{u}\hat{\Theta}$, $\hat{u}\hat{\Phi}$, and $\hat{u}\hat{\varsigma}$, are:
\ba 
\partial_\theta\hat{\mathcal{I}}_\theta+\cot\theta(\hat{\mathcal{I}}_\theta-\hat{\mathcal{I}}_\phi)=0\, ,
\label{eq:ut-projection}
\\
\partial_\phi\hat{\mathcal{I}}_\phi=\partial_\varsigma\hat{\mathcal{I}}_\varsigma=0\, .
\ea
All other projections give equations that are trivially satisfied. Using $SO(3)_q$ symmetry, one can use Eq.~(\ref{eq:It=If}) to find the set of independent second-moment equations 
\ba
\partial_\rho\hat{\mathcal{I}}_\theta + 4\tanh\rho\,\hat{\mathcal{I}}_\theta 
&=&\frac{1}{\hat\tau_{\rm eq}}\left[\hat{\mathcal{I}}_{\theta,\rm eq}-\hat{\mathcal{I}}_\theta\right]\, ,
\label{eq:tt-projection2} \\
\partial_\rho\hat{\mathcal{I}}_\varsigma + 2\tanh\rho\,\hat{\mathcal{I}}_\varsigma 
&=&\frac{1}{\hat\tau_{\rm eq}}\left[\hat{\mathcal{I}}_{\varsigma,\rm eq}-\hat{\mathcal{I}}_\varsigma\right]\, ,
\label{eq:vv-projection2} \\
\partial_\theta\hat{\mathcal{I}}_\theta &=& \partial_\phi\hat{\mathcal{I}}_\phi = \partial_\varsigma\hat{\mathcal{I}}_\varsigma = 0\, , \\
\hat{\mathcal{I}}_\rho &=&2\hat{\mathcal{I}}_\theta+\hat{\mathcal{I}}_\varsigma\, , \\
\hat{\mathcal{I}}_{\rho,\rm eq} &=&2\hat{\mathcal{I}}_{\theta,\rm eq}+\hat{\mathcal{I}}_{\varsigma,\rm eq}\, .
\ea
Using Eqs.~(\ref{eq:I-i}), (\ref{eq:tt-projection2}), and (\ref{eq:vv-projection2}), one finds
\ba
\frac{6\hat\alpha_{\varsigma }}{1-3 \hat\alpha _\varsigma ^2}\frac{d \hat\alpha_\varsigma}{d\rho}-\frac{3 \left(3 \hat\alpha_\varsigma^4-4\hat\alpha_\varsigma^2+1\right)}{4\hat\tau_{\rm eq} \hat\alpha _{\varsigma }^5} \left(\frac{\hat{T}}{\hat\lambda}\right)^5+2\tanh\rho=0\, .
\label{eq:2nd-mom-final}
\ea
For the effective temperature appearing above, one uses Eq.~(\ref{eq:matching-final}), which was obtained by requiring energy conservation.

\section{Final anisotropic hydrodynamics equations}

Equations~(\ref{eq:1st-mom-final}), (\ref{eq:2nd-mom-final}), and (\ref{eq:matching-final}) form the complete set of equations required in order to describe the de Sitter-space evolution using anisotropic hydrodynamics.  I list them again here in order to provide easier access in the forthcoming discussion
\ba
4\frac{d\log\hat\lambda}{d\rho}+\frac{3 \hat\alpha_\varsigma^2\left(\frac{{\cal H}_{2
   L}(\bar{y})}{{\cal H}_2(\bar{y})}+1\right)-4}{3\hat\alpha_\varsigma^2-1} \, \frac{d\log\hat\alpha_\varsigma}{d\rho}+ \tanh\rho\left(\frac{{\cal H}_{2T}(\bar{y})}{{\cal H}_2(\bar{y})}+2\right) &=& 0\, ,
\label{eq:1st-mom-final2} \\
\frac{6\hat\alpha_{\varsigma }}{1-3 \hat\alpha _\varsigma ^2} \frac{d \hat\alpha_\varsigma}{d\rho} -\frac{3 \left(3 \hat\alpha_\varsigma^4-4\hat\alpha_\varsigma^2+1\right)}{4\hat\tau_{\rm eq} \hat\alpha _{\varsigma }^5} \left(\frac{\hat{T}}{\hat\lambda}\right)^5+2\tanh\rho &=&0 \, ,
\label{eq:2nd-mom-final2}
\ea
where $\bar{y} \equiv \hat\alpha_\varsigma/\hat\alpha_\theta = \sqrt{(3\hat\alpha_\varsigma^2-1)/2}$. The ${\cal H}$-functions are defined in Eqs.~(\ref{eq:H2})-(\ref{eq:H2L}).  The set of equations can be closed by using the dynamical Landau matching condition
\ba
\hat{T}=\frac{\hat\alpha_\varsigma}{\bar{y}}\left(\frac{{\cal H}_2(\bar{y})}{2}\right)^{1/4} \hat\lambda .
\label{eq:matching-final2}
\ea

\section{Limiting cases}
\label{sec:limitingcases}

In this section, I consider two limiting cases of Eqs.~(\ref{eq:1st-mom-final2})-(\ref{eq:matching-final2}).  The cases I consider are the ideal ($\hat\tau_{\rm eq}\rightarrow0$) and free-streaming ($\hat\tau_{\rm eq}\rightarrow\infty$) limits. In these two cases, one can dramatically simplify the equations and solve them analytically as first-order ordinary differential equations.  As I will see below, this will allow us to compare our results with the exact solution of Boltzmann equation in these limits, which one can also obtain analytically.

\subsection{Ideal hydrodynamics limit}

In order to take the ideal limit of Eqs.~(\ref{eq:1st-mom-final2})-(\ref{eq:2nd-mom-final2}), one has to impose the following conditions which require the system to be isotropic and remains so for all de Sitter time $\rho$
\ba
\hat\alpha_\varsigma\rightarrow1\,;\quad
\partial_\rho \hat\alpha_\varsigma\rightarrow0\,;\quad
\hat\tau_{\rm eq}\rightarrow0\, .
\ea
With these assumptions, $\bar{y}\rightarrow1$ and $\hat\lambda(\rho)\rightarrow\hat{T}(\rho)$. Using these relations, one finds that Eq.~(\ref{eq:2nd-mom-final2}) is trivially satisfied.  Eq.~(\ref{eq:1st-mom-final2}) simplifies to
\ba 
\hat{T}(\rho)= \hat{T}_0 \left(\frac{\cosh\rho_0}{\cosh\rho}\right)^{2/3} ,
\ea
with $\hat{T}_0 = \hat{T}(\rho_0)$.  This is the solution obtained originally by Gubser and Yarom \cite{Gubser:2010ui}.\footnote{I have generalized the solution to allow the boundary condition to be specified at an arbitrary $\rho_0$.  The form of the Gubser and Yarom solution is recovered when $\rho_0=0$.}

\subsection{Free-streaming limit}
In order to take the free-streaming (FS) limit, one has to take the limit $\hat\tau_{\rm eq}\rightarrow\infty$ of Eqs.~(\ref{eq:1st-mom-final2}) and (\ref{eq:2nd-mom-final2}).  As it turns out, it is also possible to solve the anisotropic hydrodynamics equations analytically in this case.  In this limit, solving Eq.~(\ref{eq:2nd-mom-final2}) gives
\ba
\hat\alpha_\varsigma^2(\rho)&=&\frac{1}{3}+\left(\hat\alpha_{\varsigma,0}^2-\frac{1}{3}\right)\frac{\cosh^2\rho}{\cosh^2\rho_0}\, ,
\ea
where I have specified the boundary condition at $\rho=\rho_0$ and required that $\hat\alpha_\varsigma(\rho_0)=\hat\alpha_{\varsigma,0}$.  With this result, one can obtain an expression for $\bar{y}_{\rm FS}$ using Eq.~(\ref{eq:ybar}) 
\ba
\bar{y}_{\rm FS}=\sqrt{\frac{3\hat\alpha_{\varsigma,0}^2-1}{2}}\frac{\cosh\rho}{\cosh\rho_0}\,.
\ea
Substituting the previous two results into Eq.~(\ref{eq:1st-mom-final2}) and solving it analytically gives
\ba
\hat\lambda(\rho)=\frac{\hat\lambda_0\hat\alpha_{\varsigma,0}}{\hat\alpha_\varsigma(\rho)}\,,
\ea
where I have required $\hat\lambda(\rho_0)=\hat\lambda_0$.  Finally, one can use Eq.~(\ref{eq:matching-final2}) to find the free-streaming limit for the (effective) temperature 
\ba
\hat{T}(\rho)=\hat\lambda_0\,\hat\alpha_{\varsigma,0}\mathcal{H}_\varepsilon^{1/4}(\mathcal{C}_{\rho_0,\rho}) \, ,
\label{eq:T-hat}
\ea 
where 
\ba 
\mathcal{H}_\varepsilon(x) &\equiv& \frac{x^2}{2}+\frac{x^4}{2}\frac{{\rm tanh}^{-1}\sqrt{1-x^2}}{\sqrt{1-x^2}} \, ,\\
\mathcal{C}_{\rho_0,\rho} &\equiv& \frac{1}{\bar{y}_{\rm FS}} = \frac{\hat\alpha_{\theta,0}\cosh\rho_0}{\hat\alpha_{\varsigma,0}\cosh\rho}\, .
\ea
The free-streaming limit of the energy density can be obtained using (\ref{eq:T-hat}) 
\ba
\hat{\varepsilon}_{\rm FS}=\frac{3\hat\lambda_0^4\hat\alpha_{\varsigma,0}^4}{\pi^2}\mathcal{H}_\varepsilon(\mathcal{C}_{\rho_0,\rho}) .
\label{eq:eFS}
\ea
In addition, one can use Eq.~(\ref{eq:at}) to find $\hat\alpha_\theta(\rho)$. 
Finally, using Eqs.~(\ref{eq:pv2}) and (\ref{eq:pi-def}) one can determine the $\varsigma\varsigma$-component of the viscous stress tensor in the free-streaming limit
\ba 
(\hat{\pi}^\varsigma_\varsigma)_{\rm FS} 
&=& \frac{\hat\lambda_0^4\hat\alpha^4_{\varsigma,0}}{\pi^2}\mathcal{H}_\pi\!\left(\mathcal{C}_{\rho_0,\rho}^{-1}\right),
\label{eq:piFS}
\ea
where 
\ba 
\mathcal{H}_\pi(x)&\equiv &\frac{x\sqrt{x^2-1}(1+2x^2)+(1-4x^2){\rm coth}^{-1}\!\left(x/\sqrt{x^2-1}\right)}{2x^3(x^2-1)^{3/2}}\, .
\ea
I note that the functions $\mathcal{H}_\varepsilon$ and $\mathcal{H}_\pi$ introduced above are closely related to the \mbox{$H$-functions} previously defined in Eq.~(\ref{eq:H2})-(\ref{eq:H2L}) as
\ba
\mathcal{H}_\varepsilon(x)&=&\frac{1}{2}x^4 {\cal H}_2\!\left(x^{-1}\right)\, , \\
\mathcal{H}_\pi(x)&=&\frac{3}{2x^4}\left(\!{\cal H}_{2L}(x)-\frac{{\cal H}_2(x)}{3}\!\right)\, .
\ea

\section{Exact solution}
\label{sec:exactsolution}

Denicol et al. obtained an exact solution to the Boltzmann equation subject to Gubser flow in the RTA~\cite{Denicol:2014xca,Denicol:2014tha}.  In order to assess the efficacy of the anisotropic hydrodynamics equations, one can compare the results against this exact solution.  Note that one limitation of the exact solution obtained in Refs.~\cite{Denicol:2014xca,Denicol:2014tha} is that the distribution function was assumed to be isotropic at $\rho_0$.  As I will show below, if one assumes that the initial distribution function is of spheroidal form in de Sitter space, then it is possible to allow for an arbitrary pressure anisotropy at $\rho_0$.  This will allow us to compare anisotropic hydrodynamics with the exact solution subject to a variety of different de Sitter-space initial conditions.
In general, the exact solution can be expressed in the form~\cite{Denicol:2014xca,Denicol:2014tha}
\ba
\hat{\varepsilon}(\rho)&=& D(\rho,\rho_0)\hat{\varepsilon}_{\rm FS}+\frac{3}{\pi^2c}\int_{\rho_0}^\rho\!d\rho'D(\rho,\rho')\mathcal{H}_\varepsilon\!\left(\frac{\cosh\rho'}{\cosh\rho}\right)\hat{T}^5(\rho')\, ,
\label{eq:e-exact} \\
\hat{\pi}^\varsigma_\varsigma(\rho)&=& D(\rho,\rho_0)(\hat{\pi}^\varsigma_\varsigma)_{\rm FS}
+ \frac{1}{\pi^2c}\!\int_{\rho_0}^\rho\!d\rho'D(\rho,\rho')\mathcal{H}_\pi\!\left(\frac{\cosh\rho}{\cosh\rho'}\right)\hat{T}^5(\rho')\, ,
\label{eq:pv-exact}
\ea
where 
\begin{equation}
D(\rho_2,\rho_1)=\exp\!\left(-\int_{\rho_1}^{\rho_2} d\rho''\,
\frac{\hat T(\rho'')}{c} \right) .
\label{defineD}
\end{equation}
Above $\hat{T}(\rho) = (\pi^2\hat\varepsilon(\rho)/3)^{1/4}$ is the effective temperature and $c\equiv5\hat\eta/\hat{s}$.  Using Eqs.~(\ref{eq:e-def})-(\ref{eq:pv-def}), $\hat{\varepsilon}_{\rm FS}$ and $(\hat{\pi}^\varsigma_\varsigma)_{\rm FS}$ can be obtained 
\ba
\hat{\varepsilon}_{\rm FS}&\equiv&\int d\hat{P}(\hat{p}^\rho)^2 f_{\rm iso}\!\left(\frac{1}{\hat\lambda_0}\sqrt{\frac{\hat{p}^2_\theta}{\hat\alpha^2_{\theta,0}\cosh^2\!\rho_0}+\frac{\hat{p}^2_\phi}{\hat\alpha^2_{\theta,0}\cosh^2\!\rho_0\sin^2\!\theta}+\frac{\hat{p}^2_\varsigma}{\hat\alpha^2_{\varsigma,0}}}\right) \notag \\
&=& \frac{3\hat\lambda_0^4\hat\alpha_{\varsigma,0}^4}{\pi^2}\mathcal{H}_\varepsilon\!\left(\mathcal{C}_{\rho_0,\rho}\right), 
\label{eq:e-exact-f}
\ea
\ba
(\hat{\pi}^\varsigma_\varsigma)_{\rm FS} &\equiv & \int d\hat{P}\left(\hat{p}_\varsigma^2-\frac{(\hat{p}^\rho)^2}{3}\right) f_{\rm eq}\!\left(\frac{1}{\hat\lambda_0}\sqrt{\frac{\hat{p}^2_\theta}{\hat\alpha^2_{\theta,0}\cosh^2\!\rho_0}+\frac{\hat{p}^2_\phi}{\hat\alpha^2_{\theta,0}\cosh^2\!\rho_0\sin^2\!\theta}+\frac{\hat{p}^2_\varsigma}{\hat\alpha^2_{\varsigma,0}}}\right) \notag \\
&=& \frac{\hat\lambda_0^4\hat\alpha^4_{\varsigma,0}}{\pi^2}\mathcal{H}_\pi\left(\mathcal{C}_{\rho_0,\rho}^{-1}\right). 
\label{eq:eq:pv-exact-f}
\ea
By using the results above, the integral equations (\ref{eq:e-exact}) and (\ref{eq:pv-exact}) allow for an arbitrary momentum-space anisotropy at $\rho=\rho_0$ with $\hat\alpha_\varsigma(\rho_0)=\hat\alpha_{\varsigma,0}$ and $1/3<\hat{\alpha}_{\varsigma,0}^2<\infty$. In the original work \cite{Denicol:2014xca,Denicol:2014tha}, the solutions obtained were restricted to the case $\hat\alpha_{\varsigma,0}=1$.  If one takes $\hat\alpha_{\varsigma,0}=1$, the expressions above reduce to the ones obtained in \cite{Denicol:2014xca,Denicol:2014tha}.  Importantly, I find that Eqs.~(\ref{eq:eFS})-(\ref{eq:piFS}) correspond precisely to the exact free-streaming limits obtained above.  This means that, if the initial distribution function at $\rho_0$ is of spheroidal form in de Sitter space, anisotropic hydrodynamics gives the exact solution in the free-streaming limit.

\section{Numerical results}

In the general case, it is necessary to solve Eqs.~(\ref{eq:1st-mom-final2}), (\ref{eq:2nd-mom-final2}), and (\ref{eq:matching-final2}) numerically.  Since they are ordinary first-order differential equations, this task is rather straightforward.  In order to complete the solution, however, I need to specify a boundary condition.  
This might be non-trivial task since not all choices lead to physical results.  As shown in Appendix (B) of Ref.~\cite{Denicol:2014tha}, in the exact solution, some initial conditions can result in complex-valued temperatures, etc.  While such solutions may be mathematically sound, they are clearly not physical.  However, as discussed in Appendix (B) of Ref.~\cite{Denicol:2014tha}, if one fixes the boundary condition on the ``left'' ($\rho \rightarrow -\infty$), which corresponds to the ``distant past'' in de Sitter time, one has freedom to choose the initial condition.  In addition, with this boundary condition, one can smoothly take the limit $\eta/s \rightarrow 0$ in order to obtain the ideal hydrodynamics result (see Fig. 8 of Ref.~\cite{Denicol:2014tha}).  This limit is not guaranteed for other choices of $\rho_0$.  
\begin{figure}[!t]
\hspace{-7mm}\includegraphics[width=1.06\linewidth]{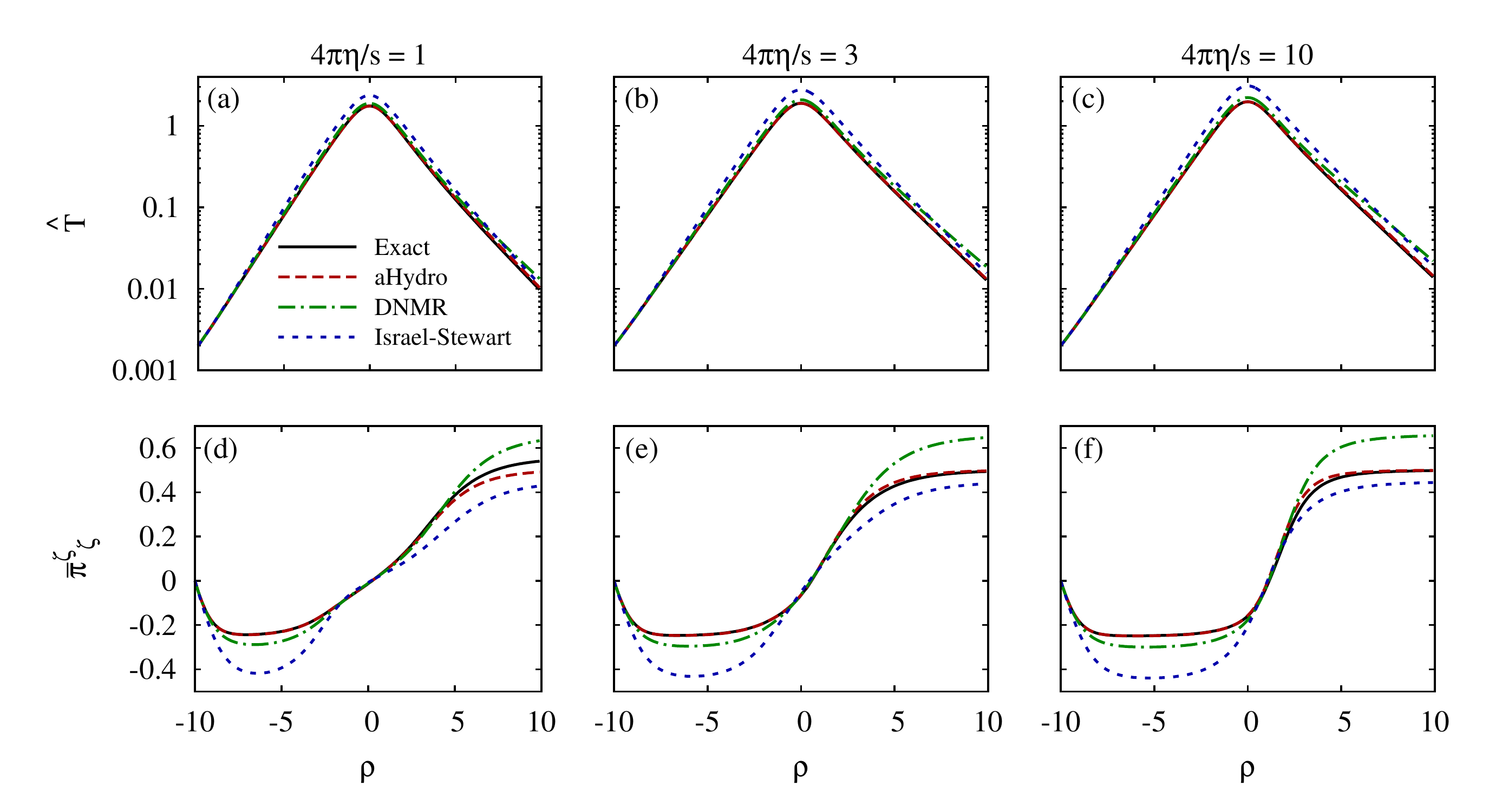}
\vspace{-16mm}
\caption{In the top row, I compare the de Sitter-space effective temperature $\hat{T}$ obtained from the exact solution (black solid line), the anisotropic hydrodynamics equations obtained in this chapter (red dashed line), the DNMR second-order approach (green dot-dashed line), and the Israel-Stewart second-order approximation (blue dotted line) for three values of shear viscosity to entropy density ratio with $4 \pi \eta/s \in \{1,3,10\}$, respectively.  In the bottom row, I compare results for the scaled shear $\bar{\pi}_{\varsigma}^{\varsigma} \equiv \hat{\pi}_\varsigma^\varsigma/(\hat{T}\hat{s})$.  The labeling and values of $4 \pi \eta/s$ in the bottom row are the same as in the top row.  In all cases, at $\rho=\rho_0=-10$, I fixed the initial effective temperature to be $\hat{T}_0 = 0.002$ and the initial anisotropy to be  $\hat\alpha_{\varsigma,0}=1$, which corresponds to an isotropic initial condition in de Sitter space.
}
\label{fig:fig-rho-10-T0p002-az1}
\end{figure}
More importantly, I want to specify a set of initial conditions on a fixed proper-time surface $\tau=\tau_0$ and then take the limit $\tau_0 \rightarrow 0^+$ so that I can describe the system's evolution in the entire forward light cone.  For this reason, in what follows I will always fix the boundary condition on the left.  As discussed above, these boundary conditions will also allow us to smoothly go from the ideal to free-streaming limits unambiguously.  In practice, specifying numerical boundary conditions at extremely large negative $\rho$ and obtaining the full solution also for positive $\rho$ is time-consuming, particularly for the exact solution that I intend to compare with.  For this reason, I will present solutions in which the boundary condition is fixed at a large, but finite, negative $\rho$.  In all plots shown, I fix the boundary condition at $\rho_0 = -10$.

\begin{figure}[!t]
\hspace{-7mm}\includegraphics[width=1.06\linewidth]{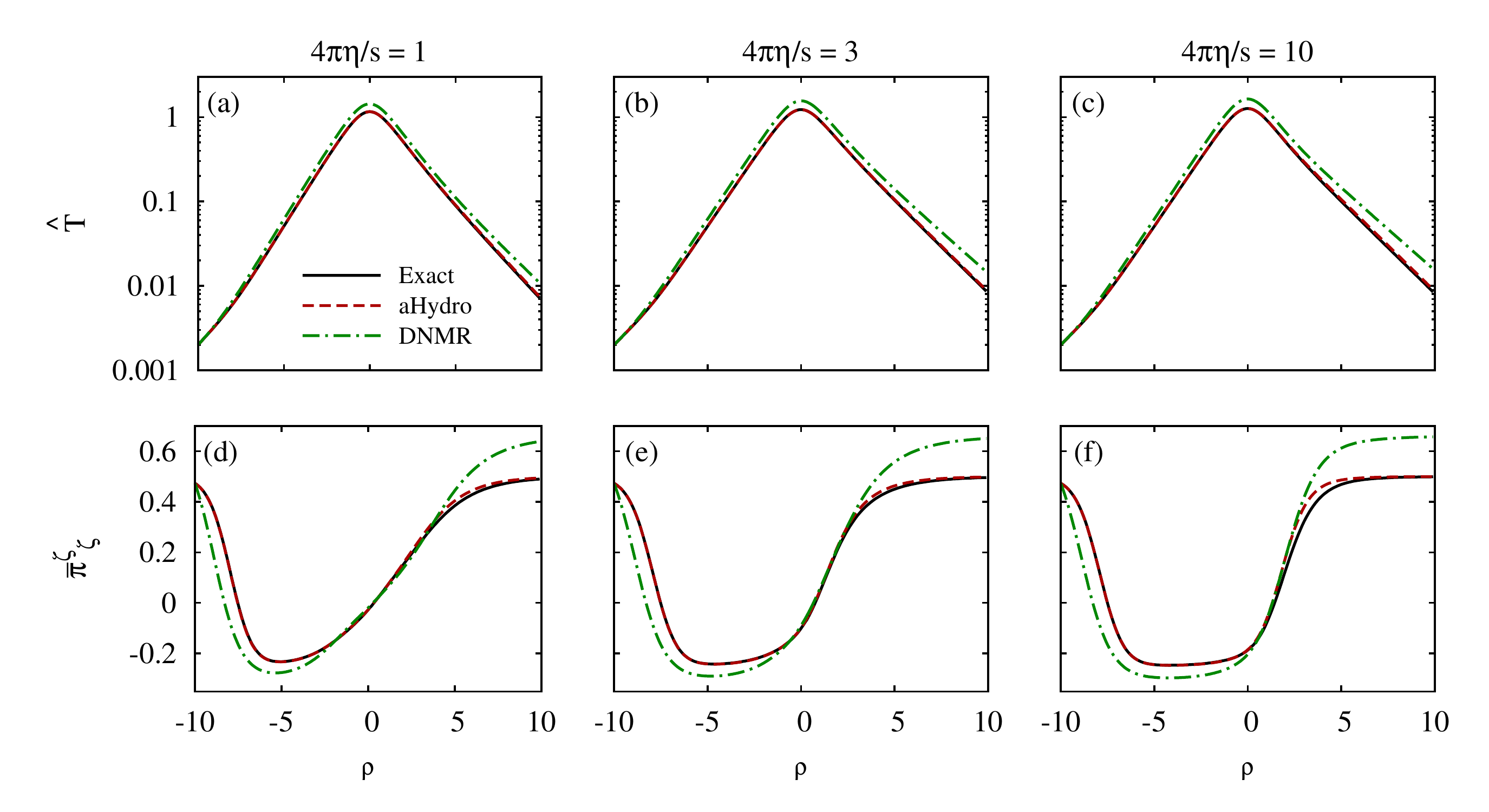}
\vspace{-16mm}
\caption{Same as Fig.~\ref{fig:fig-rho-10-T0p002-az1} except here I take $\hat{\alpha}_{\varsigma,0}=10$, which corresponds to a prolate initial condition in de Sitter space.  The Israel-Stewart approximation result is not included, because, for this boundary condition, they are unstable and diverge in the negative-$\rho$ region.}
\label{fig:fig-rho-10-T0p002-az10}
\end{figure}

\begin{figure}[t]
\hspace{-7mm}\includegraphics[width=1.06\linewidth]{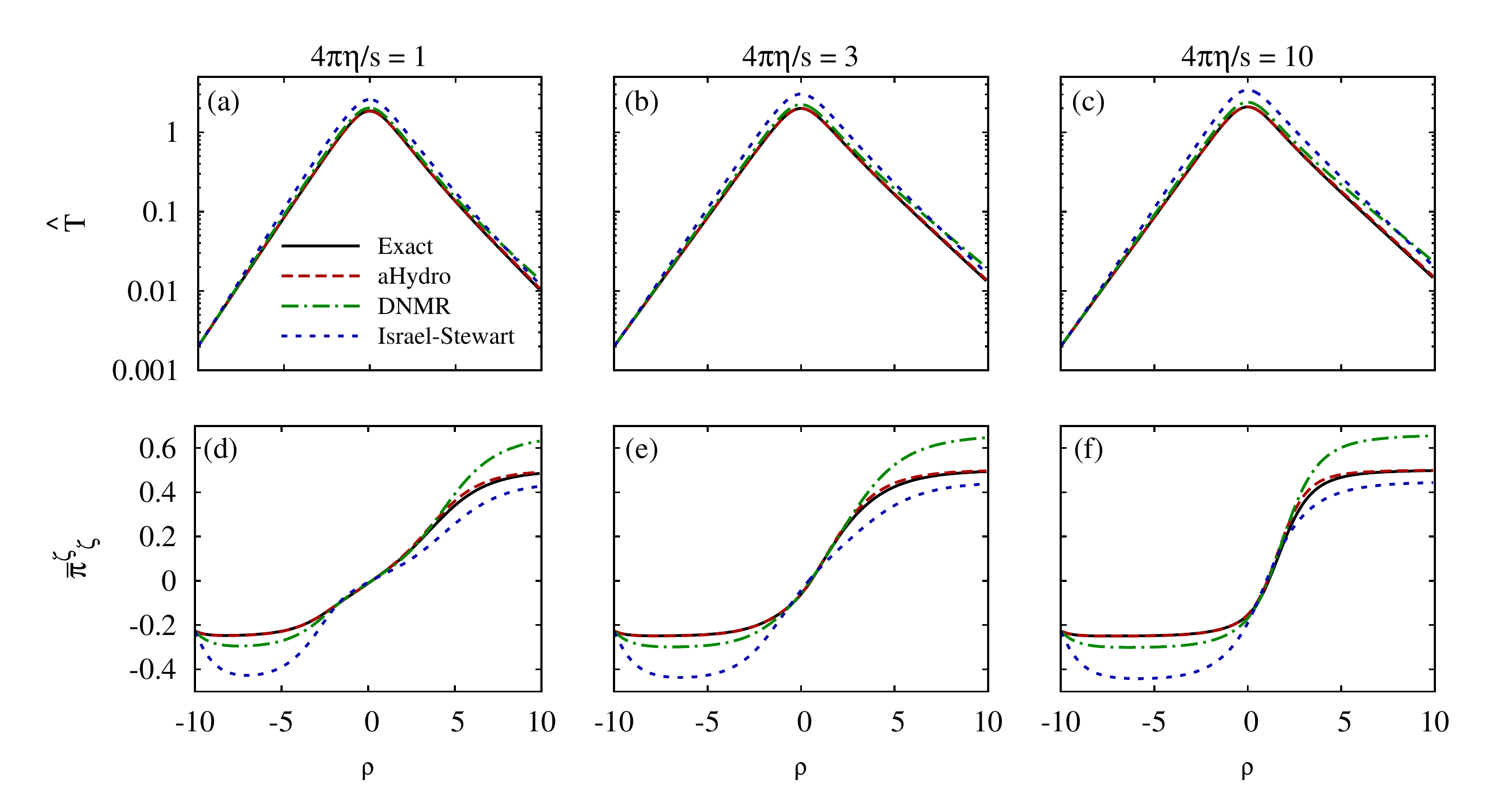}
\vspace{-16mm}
\caption{Same as Fig.~\ref{fig:fig-rho-10-T0p002-az1} except here I take $\hat{\alpha}_{\varsigma,0}=0.6$, which corresponds to an oblate initial condition in de Sitter space.}
\label{fig:fig-rho-10-T0p002-az0p6}
\end{figure}

In addition to comparing to the generalization of the exact result of Refs.~\cite{Denicol:2014xca,Denicol:2014tha}, I will also compare with results obtained using the Israel-Stewart second-order viscous hydrodynamics approximation \cite{Marrochio:2013wla} and a complete second-order Grad 14-moment approximation \cite{Denicol:2014tha}.  For the second-order hydrodynamic approximations, one has to 
solve two coupled ordinary differential equations subject to a boundary condition at $\rho = \rho_0$. For the Israel-Stewart (IS) case, the necessary equations are~\cite{Marrochio:2013wla}
\ba
&&\frac{1}{\hat{T}}\frac{d\hat{T}}{d\rho }+\frac{2}{3}\tanh \rho =
\frac{1}{3}\bar{\pi}_{\varsigma }^{\varsigma }(\rho )\,\tanh \rho \,, 
\label{eq:istemp}
\\
&&\frac{d\bar{\pi}_{\varsigma }^{\varsigma }}{d\rho }
+\frac{4}{3}\left( \bar{\pi}_{\varsigma }^{\varsigma }\right)^{2}\tanh \rho
+\frac{\bar{\pi}_\varsigma^\varsigma}{\hat{\tau}_\pi} =\frac{4}{15}\tanh\rho \,,
\label{eq:isshear}
\ea
where $\bar{\pi}_{\varsigma}^{\varsigma} \equiv \hat{\pi}_\varsigma^\varsigma/(\hat{T}\hat{s})$ and $\hat{\tau}_\pi = 5\hat\eta/(\hat{s}\hat{T})$. As mentioned above, one can go beyond the IS approximation presented in Ref.~\cite{Marrochio:2013wla} and also include the complete second-order contribution (see Appendix (A) of Ref.~\cite{Denicol:2014tha} for further details). In this case, the second equation above should be replaced by
\begin{equation}
\frac{d\bar{\pi}_{\varsigma }^{\varsigma }}{d\rho }
+\frac{4}{3}\left(\bar{\pi}_{\varsigma }^{\varsigma }\right) ^{2}\tanh \rho 
+\frac{\bar{\pi}_\varsigma^\varsigma}{\hat{\tau}_\pi} 
= \frac{4}{15} \tanh \rho 
+ \frac{10}{21}\bar{\pi}_\varsigma^\varsigma \tanh\rho \,.
\label{eq:dnmrshear}
\end{equation}
If Eq.~(\ref{eq:dnmrshear}) is used, the result is labeled as DNMR.

In Figs.~\ref{fig:fig-rho-10-T0p002-az1}-\ref{fig:fig-rho-10-T0p002-az0p6} I present our numerical solutions of Eqs.~(\ref{eq:1st-mom-final2}), (\ref{eq:2nd-mom-final2}), and (\ref{eq:matching-final2}) and compare the results to the exact solution and the two second-order viscous hydrodynamics approximations.  In these three figures I take $\hat\alpha_{\varsigma,0}=1, 10,$ and $0.6$, which correspond to an initially isotropic, prolate ($\hat{P}_\theta = \hat{P}_\phi < \hat{P}_\varsigma$), and oblate  ($\hat{P}_\theta = \hat{P}_\phi > \hat{P}_\varsigma$) initial condition, respectively.  In all cases, at $\rho=\rho_0=-10$,  the initial effective temperature is fixed to be $\hat{T}_0 = 0.002$.  In the top row of all three figures, I compare the de Sitter-space effective temperature $\hat{T}$ obtained from the exact solution (black solid line), the anisotropic hydrodynamics equations obtained herein (red dashed line), the DNMR second-order approach (green dot-dashed line), and the Israel-Stewart second-order approach (blue dotted line).  The columns from left to right correspond to three different choices of the shear viscosity to entropy density ratio with $4 \pi \eta/s \in \{1,3,10\}$, respectively.  In the bottom row of all three figures, I compare results for the scaled shear $\bar{\pi}_{\varsigma}^{\varsigma} \equiv \hat{\pi}_\varsigma^\varsigma/(\hat{T}\hat{s})$.  The labeling and values of $4 \pi \eta/s$ in the bottom row are the same as in the top row.  

As can be seen from Figs.~\ref{fig:fig-rho-10-T0p002-az1}-\ref{fig:fig-rho-10-T0p002-az0p6}, the anisotropic hydrodynamics equations obtained in this chapter provide the best approximation to the exact result in all cases.  For the temperature, it is very difficult to distinguish the anisotropic hydrodynamics result from the exact result.  For the scaled shear $\bar{\pi}_{\varsigma}^{\varsigma} \equiv \hat{\pi}_\varsigma^\varsigma/(\hat{T}\hat{s})$, there are visible differences between the aHydro solutions and the exact solution in the region between $\rho \gtrsim  0$ for small $\eta/s$, but at large $\rho$ one sees that anisotropic hydrodynamics has the correct asymptotic behavior.\footnote{Using Eqs.~(\ref{eq:1st-mom-final2})-(\ref{eq:matching-final2}), one finds that in the limit $\rho \rightarrow \infty$, $\hat\alpha_\varsigma \sim \exp(\rho/3)$ and $\hat\lambda \sim \exp(-2\rho/3)$.  As a consequence, one finds that the anisotropic hydrodynamics equations give $\bar{\pi}_{\varsigma}^{\varsigma} = 0.5$ in the limit $\rho \rightarrow \infty$ independent of the value of $\hat\tau_{\rm eq}$.}

Between the two hydrodynamic approximations, one finds that, for negative $\rho$, the DNMR solutions better reproduces the exact solution for $\bar{\pi}_{\varsigma}^{\varsigma} \equiv \hat{\pi}_\varsigma^\varsigma/(\hat{T}\hat{s})$, whereas for positive $\rho$ the IS solution seems to perform better overall.  That being said, one finds that in the range of de Sitter times considered, the DNMR solution better reproduces the exact solution for the effective temperature.

\section{Conclusions}

In this chapter I used the framework of anisotropic hydrodynamics to derive two coupled ordinary differential equations that describe the evolution of the de Sitter-space scale parameter $\hat\lambda$ and anisotropy parameter $\hat\alpha_{\varsigma}$.  My final analytic results are listed in Eqs.~(\ref{eq:1st-mom-final2}), (\ref{eq:2nd-mom-final2}), and (\ref{eq:matching-final2}).  Using these equations one could find the evolution of the effective temperature $\hat{T}$ and shear correction $\bar{\pi}_{\varsigma}^{\varsigma}$ in de Sitter time.  I demonstrated that these equations reproduce both the ideal ($\eta/s=0$) and free-streaming ($\eta/s \rightarrow \infty$) limits of the exact solution obtained in Ref.~\cite{Denicol:2014tha}.  In order to make a more general comparison, I extended the exact solution of Ref.~\cite{Denicol:2014tha} to allow for arbitrary momentum-space anisotropy in the de Sitter-space initial condition.  Our numerical results indicate that Eqs.~(\ref{eq:1st-mom-final2}), (\ref{eq:2nd-mom-final2}), and (\ref{eq:matching-final2}) provide an excellent approximation to the exact solution and, hence, this work provides further evidence that the anisotropic hydrodynamics approximation   might provide a superior approximation even when including transverse expansion.  That being said, the transverse flow pattern considered herein (``Gubser'' flow) is rather special, and I cannot generalize beyond the specific case studied herein to a general transverse flow at the moment.

In the results section, I presented solutions for the de Sitter-space evolution of the effective temperature and scaled shear.  The solutions obtained herein can be mapped back to Milne space, giving the full spatio-temporal evolution for a boost-invariant and cylindrically-symmetric system for arbitrary values of parameter $q$, which sets the spatial extent of the solution.  Using this, one can obtain the radial temperature profile at any given proper time.  This can be used as an initial condition for subsequent evolution in Milne space.  

\chapter{\bf Phenomenological predictions of quasiparticle anisotropic hydrodynamics}    
\label{chap:qp}
\setcounter{figure}{0}
\setcounter{table}{0}
\setcounter{equation}{0}

\section{Introduction}

In previous chapters, after a quick review of conformal aHydro, I have discussed the crucial the role of bulk degree of freedom when dealing with non-conformal, e.g. massive, anisotropic systems. I also discussed the comparison of aHydro and the standard vHydro models against the exact solution of Boltzmann equation for a 1+1d system under conformal gubser flow. The results indicated that aHydro describes the QGP far more accurately than any other standard dissipative hydrodynamics model. 
Despite this promise, turning anisotropic hydrodynamics into a practical phenomenological tool for use in modeling heavy-ion collisions requires two additional fundamental components to be implemented: (1) a realistic equation of state (EoS) based on lattice QCD and (2) self-consistent anisotropic freeze-out to hadronic degrees of freedom. 
With this in mind, in this chapter I am going to discuss systematic way of introducing realistic equation of state in the framework of aHydro using a quasiparticle model. I will also explain how the standard freezeout schemes is extended to aHydro, considering anisotropic distributions for hadrons at freezeout. Having implemented these two components enables us to study the phenomenology of the QGP using the aHydro model. I will wrap up the discussion of this chapter by comparing aHydro phenomenological predictions with URHIC experimental results.

The inclusion of the equation of state in aHydro is challenging. That is because the equation of state, by definition, relates the isotropic equilibrium pressure and equilibrium energy density. However, in aHydro one deals with anisotropic pressures and a non-equilibrium energy density and it is not a priori obvious how to deal with this issue. 
 Early attempts at implementing a realistic EoS were based on exploiting the conformal multiplicative factorization of the components of the energy-momentum tensor~\cite{Martinez:2010sc,Florkowski:2010cf}.  With this method, one relies on the assumption of factorization even in the non-conformal (massive) case. Such an approach is justified by the smallness of the corrections to factorization in the massive case in the near-equilibrium limit.  For details concerning this method, I refer the reader to Refs.~\cite{Nopoush:2015yga,Ryblewski:2012rr}. Although this method is relatively straightforward to implement, it is only approximate since for non-conformal systems there is no longer exact multiplicative factorization of the components of the energy-momentum tensor.  This introduces a theoretical uncertainty which is difficult to quantitatively estimate.
 
 Extending massless aHydro model to a massive one and introducing a bulk degree of freedom (as explained in Chap.~\ref{chap:bulk}) cannot be used since it relies on the assumption of a constant mass for particles in the plasma.  As we know, plasma particles acquire a thermal mass which is the result of their QCD interactions with the plasma constituents with $m\sim gT$ at high temperatures. 
 
From the perspective of quantum field theory, the study of the QGP near and above the critical temperature is of fundamental interest.  One can gain some insight into the physics of the QGP using perturbation theory since the asymptotic freedom of quantum chromodynamics (QCD) ensures that, for the high temperatures, $T \gg \Lambda_{\rm QCD}$, the QGP can be thought of as a weakly-coupled many-body system. In this regime, perturbative methods, such as hard thermal loop (HTL) resummation, can be used \cite{Andersen:2011sf,Haque:2014rua,Braaten:1989mz,Andersen:1999fw,Andersen:2003zk,Andersen:2004fp}.\footnote{HTL-resummed calculations of the thermodynamic potential at finite temperature and quark chemical potential(s) describe the lattice data well for $T \gtrsim 300$ MeV with no free parameters~\cite{Haque:2014rua,Bellwied:2015lba,Ding:2015fca}.}  In the HTL framework, the quarks and gluons can be thought of as quasiparticles having temperature-dependent (thermal) masses with $m_{q,\bar{q},g} \sim g T$, where $g$ is the strong coupling.

Such a picture provides motivation to try to model the QGP as a gas of massive quasiparticles for the purposes of obtaining self-consistent hydrodynamic equations.  However, perturbation theory needs to be supplemented since, for temperatures $T \lesssim 2 T_c$, first-principles perturbative calculations based on deconfined quarks and gluons break down.  In order to proceed, one can use non-perturbative lattice calculations of QCD (LQCD) thermodynamics to determine information about the necessary quasiparticle mass(es).  In practice, one can perform this procedure at all temperatures and determine a non-perturbative temperature-dependent quasiparticle mass, $m(T)$.  Once $m(T)$ is determined, one can use this to enforce the target equation of state (EoS) in an effective kinetic field theory framework.  One complication is that, in order to guarantee thermodynamic consistency in equilibrium and related out-of-equilibrium constraints, it is necessary to introduce a background (vacuum energy) contribution to the energy-momentum tensor~\cite{Romatschke:2011qp,Gorenstein:1995vm,Jeon:1995zm}.  The resulting EoS, together with a self-consistent non-equilibrium energy-momentum tensor and modified Boltzmann equation, can be used to derive relativistic hydrodynamic equations for such a quasiparticle gas.
The new method above will be referred to herein as the ``quasiparticle EoS''.

\section{Setup}
\label{sec:setup}
The goal of this chapter is to derive the general 3+1d non-conformal anisotropic hydrodynamics equations for a system of quasiparticles with a temperature-dependent mass. To accomplish this goal, an effective Boltzmann equation for thermal quasiparticles is obtained.  I then take moments of the resulting kinetic equation to obtain the leading-order 3+1d anisotropic hydrodynamics equations. Using a general set of basis vectors defined in (\ref{eq:basis-gen}), the equations are expanded explicitly. Then, using various simplifying assumptions (e.g. boost-invariance, etc.) I specialize to 0+1d aHyrdoQP and 0+1d (massless) standard aHydro cases, for investigating the evolution of bulk variables and discussing the two available methods of implementing realistic equation of state. Later on, in order to study the phenomenology of QGP, a modified version of Cooper-Frye freezeout model is used which is adjusted to take care of non-conformal mechanisms in aHydro framework \cite{Alqahtani:2016rth,Alqahtani:2017mhy}. In the last step, I will compare the phenomenological predictions of 3+1d aHydroQP model with experimental data.
 The anisotropic distribution function in the non-conformal case including the bulk degree of freedom is also defined at section \ref{sec:massivedf}. 

\section{Quasiparticle equation of state}
As is well-known from the literature \cite{Gorenstein:1995vm}, one cannot simply substitute temperature-dependent masses into the thermodynamic functions obtained for constant masses because this would violate thermodynamic consistency.  For an equilibrium system, one can ensure thermodynamic consistency by adding a background contribution to the energy-momentum tensor, i.e.
\be
T^{\mu\nu}_{\rm eq} = T^{\mu\nu}_{\rm kinetic,eq} + g^{\mu\nu} B_{\rm eq}  \, ,
\ee
with $B_{\rm eq}\equiv B_{\rm eq}(T)$ being the additional background contribution.  The kinetic contribution to the energy momentum tensor is given by
\be
T^{\mu\nu}_{\rm kinetic,eq} = \intdP \, p^\mu p^\nu f_{\rm eq}(x,p) \, .
\ee

For an equilibrium Boltzmann gas, the number and entropy densities are unchanged, while, due to the additional background contribution, the energy density and pressure are shifted by $+B_{\rm eq}$ and $-B_{\rm eq}$, respectively, giving
\ba
n_{\rm eq}(T,m) &=& 4 \pi \tilde{N} T^3 \, \hat{m}_{\rm eq}^2 K_2\left( \hat{m}_{\rm eq}\right) , \label{eq:neq} \\
{\cal S}_{\rm eq}(T,m) &=&4 \pi \tilde{N} T^3 \, \hat{m}_{\rm eq}^2 \Big[4K_2\left( \hat{m}_{\rm eq}\right)+\hat{m}_{\rm eq}K_1\left( \hat{m}_{\rm eq}\right)\Big] ,
\label{eq:Seq} \\
\epsilon_{\rm eq}(T,m) &=& 4 \pi \tilde{N} T^4 \, \hat{m}_{\rm eq}^2
 \Big[ 3 K_{2}\left( \hat{m}_{\rm eq} \right) + \hat{m}_{\rm eq} K_{1} \left( \hat{m}_{\rm eq} \right) \Big]+B_{\rm eq} \, , 
\label{eq:Eeq} \\
 P_{\rm eq}(T,m) &=& 4 \pi \tilde{N} T^4 \, \hat{m}_{\rm eq}^2 K_2\left( \hat{m}_{\rm eq}\right)-B_{\rm eq} \, ,
\label{eq:Peq}
\ea
where $\hat{m}_{\rm eq} = m/T$ with $m$ implicitly depending on the temperature from here on.  In order to fix $B_{\rm eq}$, one can require, for example, the thermodynamic identity
\be
T {\cal S}_{\rm eq} = \epsilon_{\rm eq} + P_{\rm eq} = T \frac{\partial P_{\rm eq}}{\partial T} \, ,
\label{eq:thermoid}
\ee
be satisfied.  Using Eqs.~(\ref{eq:Eeq}), (\ref{eq:Peq}), and (\ref{eq:thermoid}) one obtains 
\ba
\frac{dB_{\rm eq}}{dT} = - \frac{1}{2} \frac{dm^2}{dT} \intdP \, f_{\rm eq}(x,p)
= -4\pi \tilde{N}m^2 T K_1(\hat{m}_{\rm eq}) \frac{dm}{dT} \, .
\label{eq:BM-matching-eq-1}
\ea
If the temperature dependence of $m$ is known, then Eq.~(\ref{eq:BM-matching-eq-1}) can be used to determine $B_{\rm eq}(T)$.
\begin{figure}[t]
\hspace{-6mm}
\includegraphics[width=0.46\linewidth]{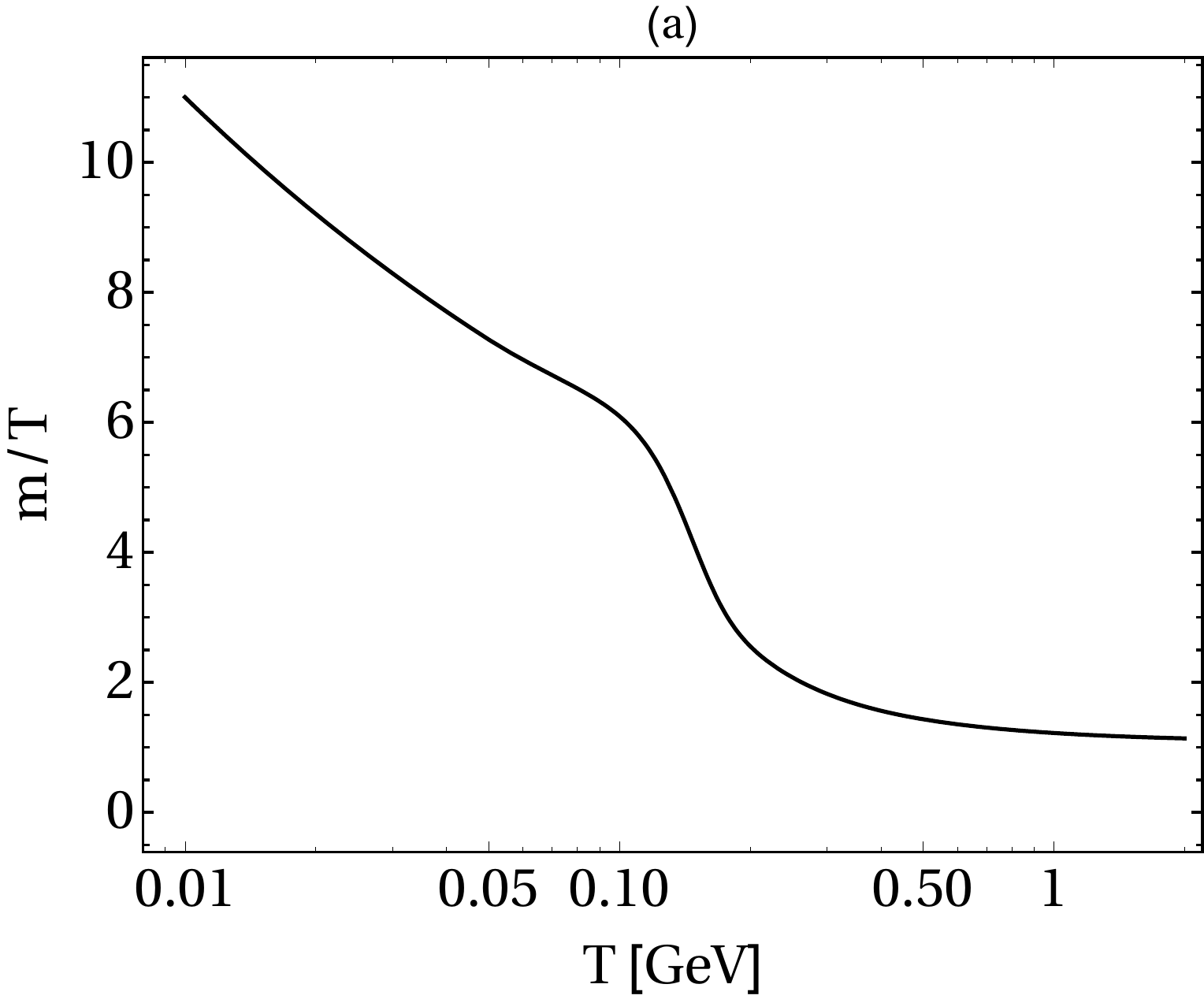}
$\;\;$
\includegraphics[width=0.48\linewidth]{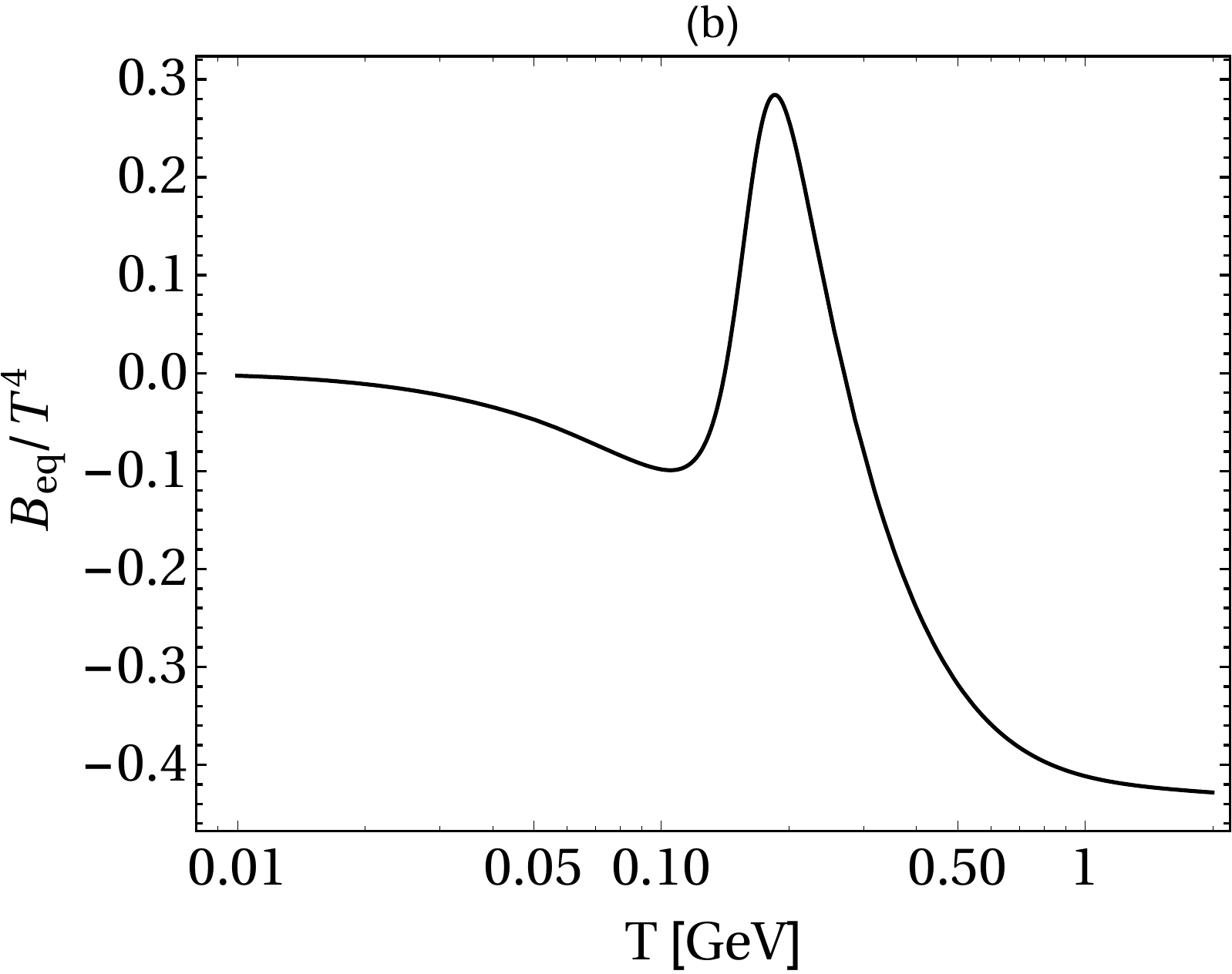}
\caption{In panel (a) the temperature dependence of the quasiparticle mass scaled by the temperature obtained using Eq.~(\ref{eq:meq}) is plotted.  In panel (b) the temperature dependence of the background term $B_{\rm eq}$ scaled by the temperature obtained using (\ref{eq:BM-matching-eq-1}) is plotted.}
\label{fig:eos2}
\end{figure}
In order to determine $m$, one can use the thermodynamic identity
\be
\epsilon_{\rm eq}+P_{\rm eq}=T{\cal S}_{\rm eq} = 4 \pi \tilde{N} T^4 \, \hat{m}_{\rm eq}^3 K_3\left( \hat{m}_{\rm eq}\right) .
\label{eq:meq}
\ee
Using the lattice parameterization (\ref{eq:I-func}) to compute the equilibrium energy density and pressure, one can numerically solve for $m(T)$.  In Fig.~\ref{fig:eos2}a, I plot the resulting solution for $m/T$ as a function of the temperature.  Once $m$ is determined using Eq.~(\ref{eq:meq}), one can solve Eq.~(\ref{eq:BM-matching-eq-1}) subject to the boundary condition $B_{\rm eq}(T=0) = 0$ to find $B_{\rm eq}(T)$. Note that, using this method, one can exactly reproduce the lattice results for energy density, pressure, and entropy density.  In Fig.~\ref{fig:eos2}b, I plot the resulting solution for the normalized quantity $B_{\rm eq}(T)/T^4$  as a function of the temperature.

\section{Effective Boltzmann equation}
\label{sec:thermal-boltzmann}
If the quasiparticles have a temperature-dependent mass, one has to generalize the Boltzmann equation in order to take into account gradients in the mass. Generally, the Boltzmann equation for on-shell quasiparticles can be written as~\cite{Jeon:1995zm}
\be
p^\mu \partial_\mu f+\frac{1}{2}\partial_i m^2\partial^i_{(p)} f=-\mathcal{C}[f]\,,
\label{eq:boltz2}
\ee
where $i$ is a spatial coordinate index, and
$\partial^i_{(p)}\equiv -\partial/\partial p^i$.  Note that the extra term, $(\partial_i m^2/2)\partial^i_{(p)} f$, corresponds precisely to the result obtained from derivation of the Boltzmann equation using quantum field theoretical methods~\cite{Berges:2005md}. In the constant mass limit, the Eq.~(\ref{eq:boltz2}) simplifies to Eq.~(\ref{eq:boltzmann}). 

To obtain a realistic model for $\tau_{\rm eq}$, which is valid for massive systems, one can relate $\tau_{\rm eq}$ to the shear viscosity to entropy density ratio.  For a massive system, one has \cite{anderson1974relativistic,Czyz:1986mr}
\be
\eta(T)=\frac{\tau_{\rm eq}(T) P_{\rm eq}(T)}{15}\kappa(\hat{m}_{\rm eq})\,.
\ee
In this formula the function $\kappa(x)$ is defined as
\be 
\kappa(x)\equiv x^3 \bigg[\frac{3}{x^2}\frac{K_3(x)}{K_2(x)}-\frac{1}{x}+\frac{K_1(x)}{K_2(x)}-\frac{\pi}{2}\frac{1-xK_0(x)L_{-1}(x)-xK_1(x)L_0(x)}{K_2(x)}\bigg] ,
\ee
where $K_n(x)$ are modified Bessel functions of second kind and $L_n(x)$ are modified Struve functions. Assuming that  $\eta/{\cal S}_{\rm eq}\equiv \bar{\eta}$ is held fixed during the evolution and using the thermodynamic relation $\epsilon_{\rm eq}+P_{\rm eq}=T{\cal S}_{\rm eq}$ one obtains
\be 
\tau_{\rm eq}(T)=\frac{15 \bar{\eta}}{\kappa(\hat{m}_{\rm eq})T}\bigg(1+\frac{\epsilon_{\rm eq}(T)}{P_{\rm eq}(T)}\bigg) .
\label{eq:teq}
\ee
Note that, in the massless limit, $m\rightarrow 0$, one has $\kappa(\hat{m}_{\rm eq})\rightarrow12$, giving 
\be
\tau_{\rm eq}(T)= \frac{5 \eta}{4P_{\rm eq}(T)} \, .
\label{eq:teq0}
\ee

\section{Moments of Boltzmann equation}
\label{sec:boltzmann-moments}
In order to study the evolution of the bulk properties of a system, one can take the moments of Eq.~(\ref{eq:boltz2}). 
The zeroth, first, and second moments yields, respectively 
\ba
\partial_\mu J^\mu&=&-\intdP \, {\cal C}[f]\, , \label{eq:J-conservation} \\
\partial_\mu T^{\mu\nu}&=&-\intdP \, p^\nu {\cal C}[f]\, , \label{eq:T-conservation} \\
\partial_\mu {\cal I}^{\mu\nu\lambda}- J^{(\nu} \partial^{\lambda)} m^2 &=&-\intdP \, p^\nu p^\lambda{\cal C}[f]\, \label{eq:I-conservation},   
\ea 
the functions are given by
\ba
J^\mu &\equiv& \intdP \, p^\mu f(x,p)\, , \label{eq:J-int} \\
T^{\mu\nu}&\equiv& \intdP \, p^\mu p^\nu f(x,p)+B g^{\mu\nu}, \label{eq:T-int}\\
{\cal I}^{\mu\nu\lambda} &\equiv& \intdP \, p^\mu p^\nu p^\lambda  f(x,p) \, .
\label{eq:I-int}
\ea
Note that I have introduced the non-equilibrium background field $B\equiv B({\boldsymbol\alpha},\lambda)$, which is the analogue of the equilibrium background $B_{\rm eq}$ in order to guarantee the correct equilibrium limit of $T^{\mu\nu}$.  For obtaining Eq.~(\ref{eq:T-conservation}), one finds a constraint equation for $B$ as 
\be
\partial_\mu B = -\frac{1}{2} \partial_\mu m^2 \intdP  f(x,p)\,.
\label{eq:BM-matching}
\ee
In practice, one can use (\ref{eq:BM-matching}) to write the derivative of $B$ with respect to any variable in terms of the derivative of the thermal mass times the $E^{-1}$ moment of the non-equilibrium distribution function.

\section{Quasiparticle bulk variables}
\label{sec:bulk-var}
In this section, the necessary dynamical equations for bulk variables are computed. The quantities $J^\mu$, $T^{\mu\nu}$, and ${\cal I}^{\mu\nu\lambda}$ defined in Eqs.~(\ref{eq:J-int}), (\ref{eq:T-int}), and (\ref{eq:I-int}) follow the tensor expansions introduced at (\ref{eq:expand}). 

The conservation of energy and momentum is enforced by $\partial_\mu T^{\mu\nu}=0$. Using Eq.~(\ref{eq:T-int}) one has $\epsilon _{\rm kinetic} = \epsilon _{\rm kinetic,eq}$, or more explicitly
\be
\tilde{\cal H}_3 \lambda^4 = \tilde{\cal H}_{3,\rm eq} T^4.
 \label{eq:matching}
\ee

Considering Eq.~(\ref{eq:T-conservation}) and taking $U$-, $X$-, $Y$-, and $Z$-projections, one obtains
\ba
D_u\epsilon +\epsilon \theta_u+ P_x u_\mu D_xX^\mu+ P_y u_\mu D_yY^\mu +P_z u_\mu D_zZ^\mu &=&0\, , \nonumber\\
D_x P_x+P_x\theta_x -\epsilon X_\mu D_uu^\mu -P_y X_\mu D_yY^\mu - P_z X_\mu D_zZ^\mu &=& 0\,, \nonumber\\
D_y P_y+P_y \theta_y-\epsilon Y_\mu D_uu^\mu -P_x Y_\mu D_xX^\mu - P_z Y_\mu D_zZ^\mu  &=& 0\,, \nonumber\\
D_z P_z+P_z \theta_z-\epsilon Z_\mu D_uu^\mu- P_x Z_\mu D_xX^\mu - P_y Z_\mu D_yY^\mu &=& 0\,.
\label{eq:1st-mom-gen}
\ea
Using Eqs.~(\ref{eq:fform}) and (\ref{eq:T-int}) to take projections of $T^{\mu\nu}$, one can obtain its components as
\ba
\epsilon  &=& {\cal H}_3({\boldsymbol\alpha},\hat{m}) \, \lambda^4+B \, ,\nonumber \\
P_x &=& {\cal H}_{3x}({\boldsymbol\alpha},\hat{m}) \, \lambda^4-B \, ,\nonumber \\
P_y &=& {\cal H}_{3y}({\boldsymbol\alpha},\hat{m}) \, \lambda^4-B \, ,\nonumber \\
P_z &=& {\cal H}_{3L}({\boldsymbol\alpha},\hat{m}) \, \lambda^4-B \, ,
\ea
The various ${\cal H}$-functions appearing above are defined in App.~\ref{subapp:h-functions-1}. 

 The second moment of Boltzmann equation in the RTA is
\be
\partial_\mu {\cal I}^{\mu\nu\lambda}-J^{(\nu} \partial^{\lambda)} m^2= \frac{1}{\tau_{\rm eq}}(u_\mu {\cal I}^{\mu\nu\lambda}_{\rm eq}-u_\mu {\cal I}^{\mu\nu\lambda})\,,
\label{eq:2moment}
\ee
where ${\cal I}$ and ${\cal I}_{\rm eq}$ are defined in (\ref{eq:I-funcs}) and (\ref{eq:Ieq}). 
Taking its $uu$-, $XX$-, $YY$-, and $ZZ$-projections to obtain 
\ba
D_u {\cal I}_u + {\cal I}_u \theta_u + 2 {\cal I}_x u_\mu D_x X^\mu+ 2 {\cal I}_y u_\mu D_y Y^\mu+ 2 {\cal I}_z u_\mu D_z Z^\mu
-nD_u m^2
&=& \frac{{\cal I}_{u,\rm eq} - {\cal I}_u}{\tau_{\rm eq}}  , \label{eq:uu} \\
D_u {\cal I}_x + {\cal I}_x (\theta_u + 2 u_\mu D_x X^\mu)
&=& \frac{1}{\tau_{\rm eq}} ( {\cal I}_{\rm eq} - {\cal I}_x ),  \hspace{1cm} \label{eq:xx} \\
D_u {\cal I}_y + {\cal I}_y (\theta_u + 2 u_\mu D_y Y^\mu)
&=& \frac{1}{\tau_{\rm eq}} ( {\cal I}_{\rm eq} - {\cal I}_y ) , \label{eq:yy}\\
D_u {\cal I}_z + {\cal I}_z (\theta_u + 2 u_\mu D_z Z^\mu)
&=& \frac{1}{\tau_{\rm eq}} ( {\cal I}_{\rm eq} - {\cal I}_z )  . \label{eq:zz} 
\ea
Also, taking $uX$-, $uY$-, and $uZ$-projections one can find
\ba
D_x {\cal I}_x+{\cal I}_x \theta_x+({\cal I}_x+{\cal I}_u) u_\mu D_u X^\mu -{\cal I}_y X_\mu D_y Y^\mu-{\cal I}_z X_\mu D_z Z^\mu-\frac{1}{2}nD_xm^2=0\,, \label{eq:ux} \\
D_y{\cal I}_y+{\cal I}_y \theta_y+({\cal I}_y+{\cal I}_u) u_\mu D_uY^\mu -{\cal I}_x Y_\mu D_x X^\mu-{\cal I}_z Y_\mu D_z Z^\mu-\frac{1}{2}nD_ym^2=0\,, \label{eq:uy}\\
D_z{\cal I}_z+{\cal I}_z \theta_z+({\cal I}_z+{\cal I}_u) u_\mu D_u Z^\mu -{\cal I}_x Z_\mu D_x X^\mu-{\cal I}_y Z_\mu D_y Y^\mu-\frac{1}{2}nD_zm^2=0\, , \label{eq:uz}
\ea
and finally projecting with $XY$, $XZ$, and $YZ$ gives
\ba
{\cal I}_x(Y_\mu D_u X^\mu+Y_\mu D_x u^\mu)+ {\cal I}_y (X_\mu D_u Y^\mu+X_\mu D_y u^\mu)&=&0\,, \label{eq:xy}\\
{\cal I}_x (Z_\mu D_u X^\mu+Z_\mu D_x u^\mu)+ {\cal I}_z (X_\mu D_u Z^\mu+X_\mu D_z u^\mu)&=&0\,, \label{eq:xz}\\
{\cal I}_y (Z_\mu D_u Y^\mu+Z_\mu D_y u^\mu)+ {\cal I}_z (Y_\mu D_u Z^\mu+Y_\mu D_z u^\mu)&=&0\,.  \label{eq:yz}
\ea
Note that (\ref{eq:uu}) is not independent and can be obtained from the the equations (\ref{eq:xx})-(\ref{eq:zz}). In brief, the dynamical equations are constructed by Eqs.~(\ref{eq:matching}), (\ref{eq:1st-mom-gen}), and (\ref{eq:xx})-(\ref{eq:zz}).

\section{Numerical results}

In this section, the comparisons of our aHydroQP model results with $\sqrt{s_{NN}}$ = 2.76 TeV Pb-Pb collision data available from the ALICE collaboration are presented.  To set the initial conditions, the system is assumed to be initially isotropic in momentum space ($ \alpha_{i}(\tau_0)=1 $), with zero transverse flow ($ {\bf u}_{\perp}(\tau_0) =0$), and Bjorken flow in the longitudinal direction ($ \vartheta(\tau_0) = \eta $). The initial energy density distribution in the transverse plane is computed from a ``tilted'' profile \cite{Bozek:2010bi}.  The distribution used is a linear combination of smooth Glauber wounded-nucleon and binary-collision density profiles, with a binary-collision mixing factor of $\chi = 0.15$.  In the longitudinal direction, I used a profile with a central plateau and Gaussian ``tails'', resulting in a longitudinal profile function of the form 
\be
\rho(\varsigma) \equiv \exp \left[ - (\varsigma - \Delta \varsigma)^2/(2 \sigma_\varsigma^2) \, \Theta (|\varsigma| - \Delta \varsigma) \right] \, .
\label{eq:rhofunc}
\ee
The parameters entering (\ref{eq:rhofunc}) were fitted to the pseudorapidity distribution of charged hadrons with the results being $\Delta\varsigma = 2.3$ and $\sigma_{\varsigma} = 1.6$.  The first quantity sets the width of the central plateau and the second sets the width of the Gaussian ``tails''.  

The resulting initial energy density at a given transverse position ${\bf x}_\perp$ and spatial rapidity $\varsigma$ was computed using 
\be 
{\cal E}({\bf x}_\perp,\varsigma) \propto (1-\chi) \rho(\varsigma) \Big[ W_A({\bf x}_\perp) g(\varsigma) + W_B({\bf x}_\perp) g(-\varsigma)\Big] + \chi \rho(\varsigma) C({\bf x}_\perp) \, ,
\ee
 where $W_{A,B}({\bf x}_\perp)$ is the wounded nucleon density for nucleus $A$ or $B$, $C({\bf x}_\perp)$ is the binary collision density, and $g(\varsigma)$ is the ``tilt function''.  The tilt function  is defined through
\ba 
g(\varsigma) =
\left\{ \begin{array}{lcccc}
0  & \,\,\,\,\,\,\,\,\,\,\,\,\,\,\,\ & \mbox{if} & \,\,\,
& \varsigma < -y_N \, ,
 \\ (\varsigma+y_N)/(2y_N) & & \mbox{if} &
& -y_N \leq \varsigma \leq y_N \, , \\
1 & & \mbox{if} & 
& \varsigma > y_N\, ,
\end{array}\right. \,\,\,\,\,\,\,\,\,\,\,\,\,
\ea 
where $y_N = \log(2\sqrt{s_{NN}}/(m_p + m_n))$ is the nucleon momentum rapidity \cite{Bozek:2010bi}.

\begin{figure}[t!]
\includegraphics[width=0.99\linewidth]{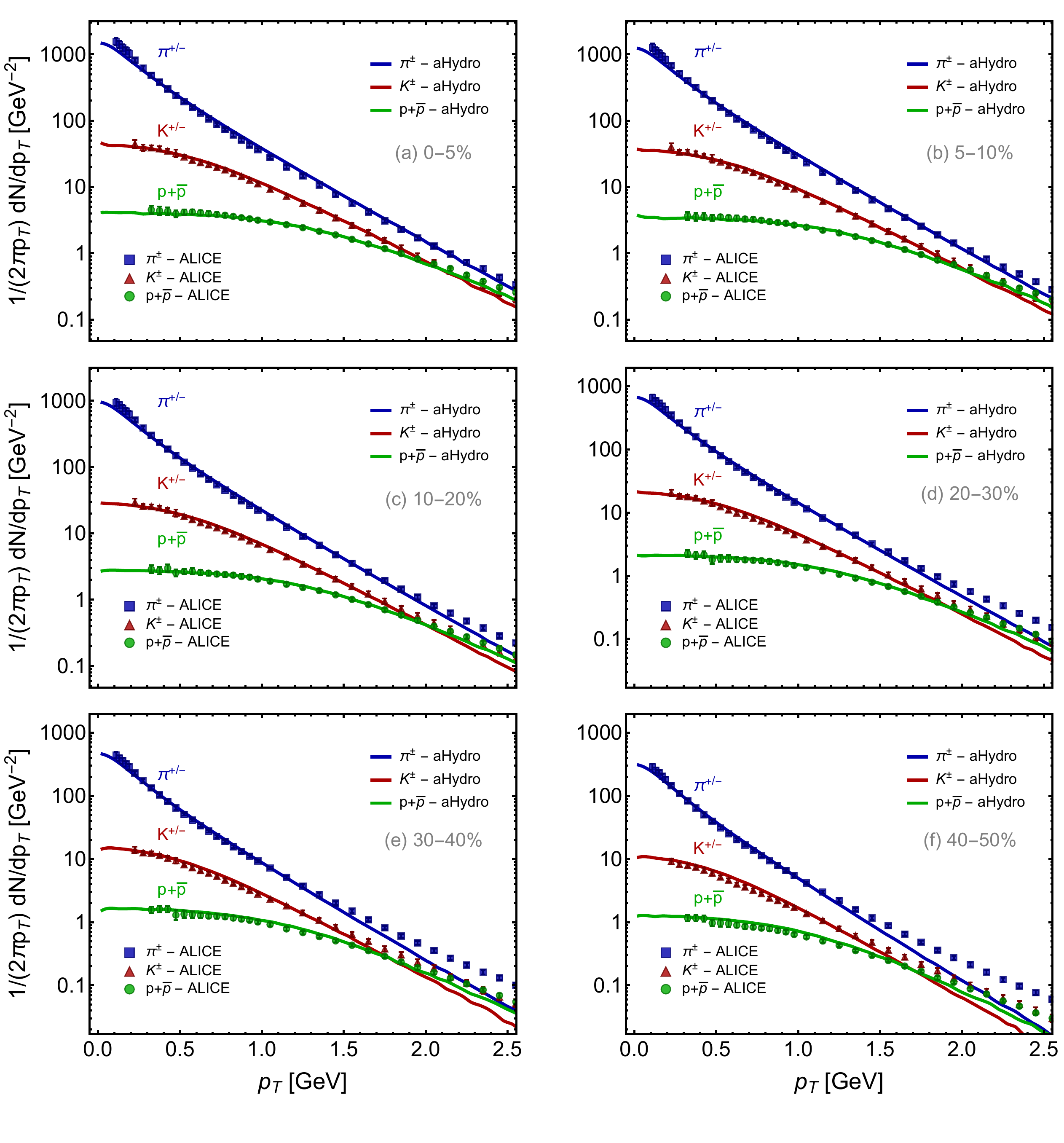}
\caption{Transverse momentum spectra of $\pi^\pm$, $K^\pm$, and $p+\bar{p}$ for six centrality classes.  All results are for 2.76 TeV Pb-Pb collisions and  data are from the ALICE collaboration \cite{Abelev:2013vea}.}
\label{fig:spectra}
\end{figure}

\begin{figure*}[t!]
\centerline{
\hspace{-1.5mm}
\includegraphics[width=1\linewidth]{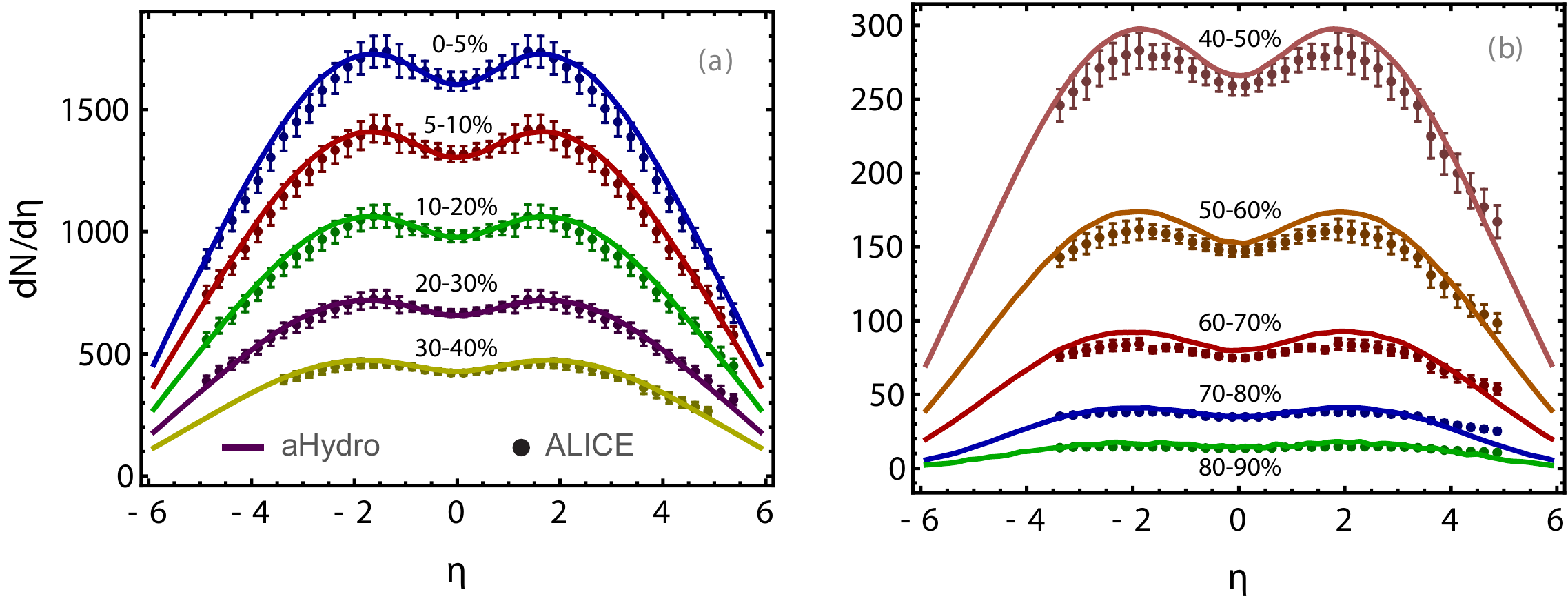}}
\caption{The charged-hadron multiplicity in different centrality classes as a function of pseudorapidity. Results are for 2.76 TeV Pb-Pb collisions and data  are from the ALICE collaboration \cite{Abbas:2013bpa,Adam:2015kda}. }
\label{fig:multiplicity}
\end{figure*}
 I solved the aHydroQP dynamical equations on a $64^3$ lattice with lattice spacings $\Delta x = \Delta y = 0.5$ fm and \mbox{$\Delta \varsigma$ = 0.375}. To compute spatial derivatives a fourth-order centered-differences is used and, for temporal updates, I used fourth-order Runge-Kutta with step size of $\Delta\tau = 0.02$ fm/c. To regulate potential numerical instabilities associated with the centered-differences scheme,  a weighted-LAX smoother~\cite{Martinez:2012tu} is used. In most cases, I set the weighted-LAX fraction to be $0.005$, however, for large impact parameters I used $0.02$.\footnote{This does not affect the evolution considerably since, for high impact parameters, the system reaches $T_{\rm FO}$ at times $\lesssim 4$ fm/c.} The aHydroQP evolution was started at $\tau_0 = 0.25$ fm/c and stopped when the highest effective temperature in the entire volume was sufficiently below $T_{\rm FO}$.

\begin{figure*}[t!]
\centerline{
\includegraphics[width=.5\linewidth]{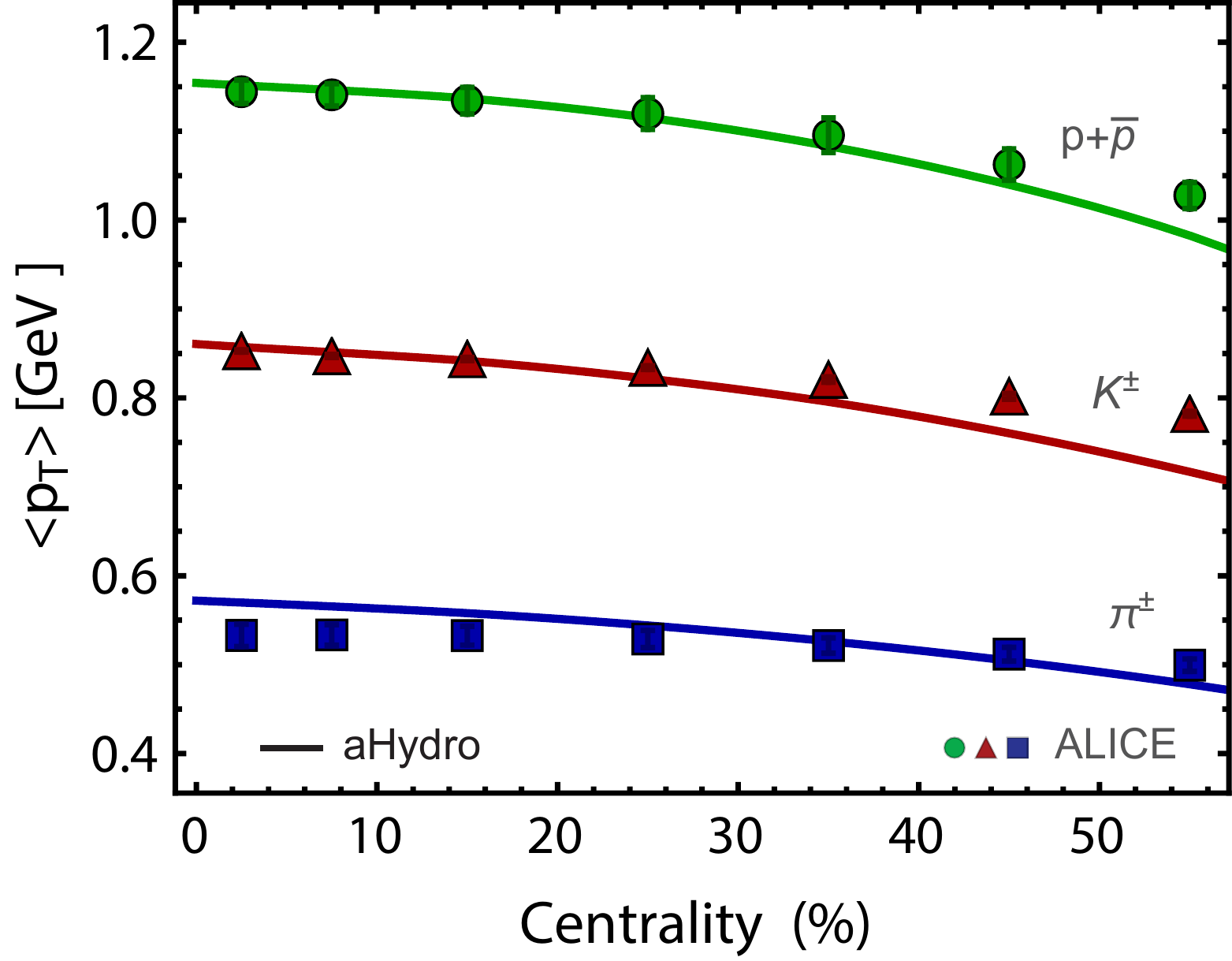}
}
\caption{The average $p_T$ of $\pi^\pm$, $K^\pm$, and $p+\bar{p}$ as a function of centrality for 2.76 TeV Pb-Pb collisions. Data taken from the ALICE collaboration \cite{Abelev:2013vea}. }
\label{fig:ptavg}
\end{figure*}

\begin{figure*}[t!]
\centerline{\includegraphics[width=.5\linewidth]{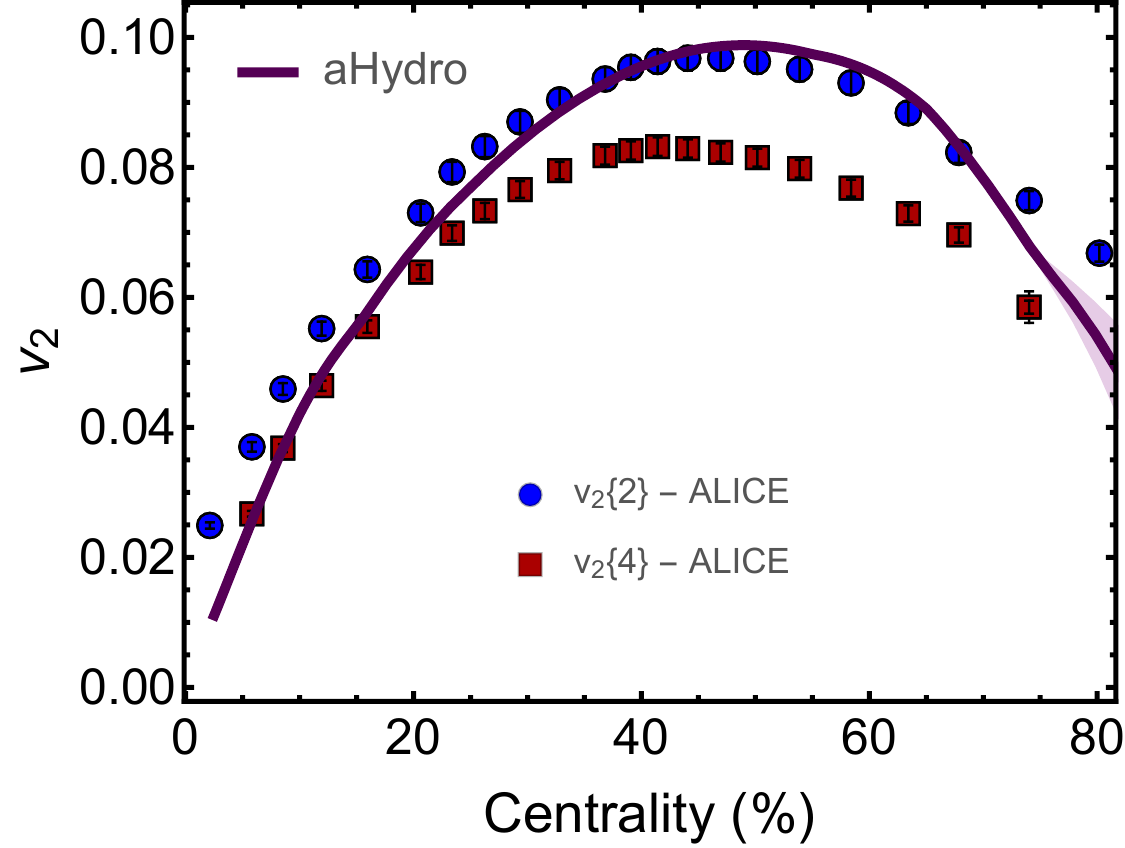}}
\caption{The integrated $v_2$ for charged hadrons as a function of centrality ($0.2 < p_T < 3$ GeV, $\eta < 0.8$).  All data takenfrom 2.76 TeV Pb-Pb ALICE collaboration \cite{Abelev:2014mda}. }
\label{fig:v2integrated}
\end{figure*}

\begin{figure}[t!]
\includegraphics[width=0.99\linewidth]{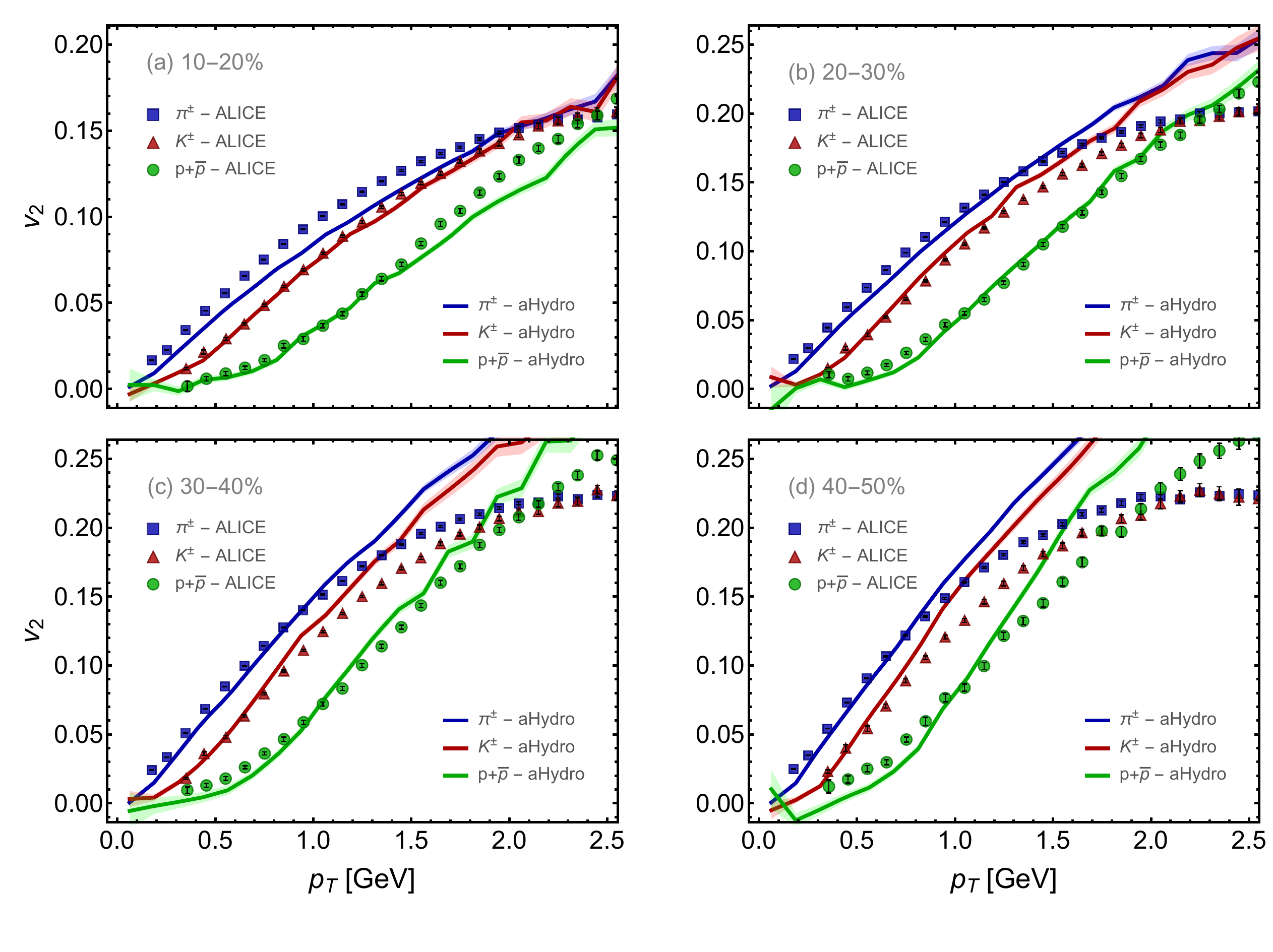}
\caption{The elliptic flow coefficient for identified hadrons as a function of  $p_T$ for four centrality classes. All results and data are for 2.76 TeV Pb-Pb collisions.  Data shown are from the ALICE collaboration and were extracted using the scalar product method~\cite{Abelev:2014pua}.}
\label{fig:v2}
\end{figure}

Using aHydroQP, the full 3+1d evolution of the system is ran first and, then a freeze-out hypersurface based on the effective temperature is extracted. I assumed that all hadronic species were in chemical equilibrium and had the same fluid anisotropy tensor ($\Xi_{\mu\nu}$) and scale parameter ($ \lambda$). The distribution function parameters on the freeze-out hypersurface were fed  into a customized version of THERMINATOR 2 which allows for an ellipsoidal distribution function. THERMINATOR 2 performs sampled event-by-event hadronic production from the exported freeze-out hypersurface  using Monte-Carlo sampling. It then performs hadronic feed down (resonance decays) for each sampled event. Depending on the observables under consideration and the centrality class considered, one may need to generate more hadronic events for the purposes of improved statistics.  For all plots shown herein, between 7,400 and 36,200 hadronic events per centrality class is used. I indicate the statistical uncertainty of our model results associated with the hadronic Monte-Carlo sampling by a shaded band surrounding the hadronic event-averaged value (the central line).

In this model, there are three remaining free parameters: (1) the initial central temperature $T_0$ obtained in a perfectly central collision at ${\bf x}_\perp=0$ and $\varsigma=0$, (2) the freeze-out temperature $T_{\rm FO}$, and (3) $\eta/s$ which is assumed to be a (temperature-independent) constant. In order to fix these parameters a scan over them is performed and I compared the theoretical predictions resulting from this scan with experimental data from the ALICE collaboration for the differential spectra of pions, kaons, and protons in both the 0-5\% and 30-40\% centrality classes. The fitting error was minimized across species, with equal weighting for the three particle types.  The parameters obtained from this procedure are $T_0 = 600$ MeV, $\eta/s = 0.159$, and \mbox{$T_{\rm FO} = 130$ MeV}. 
   
\begin{figure*}[t!]
\centerline{
\hspace{-1.5mm}
\includegraphics[width=.99\linewidth]{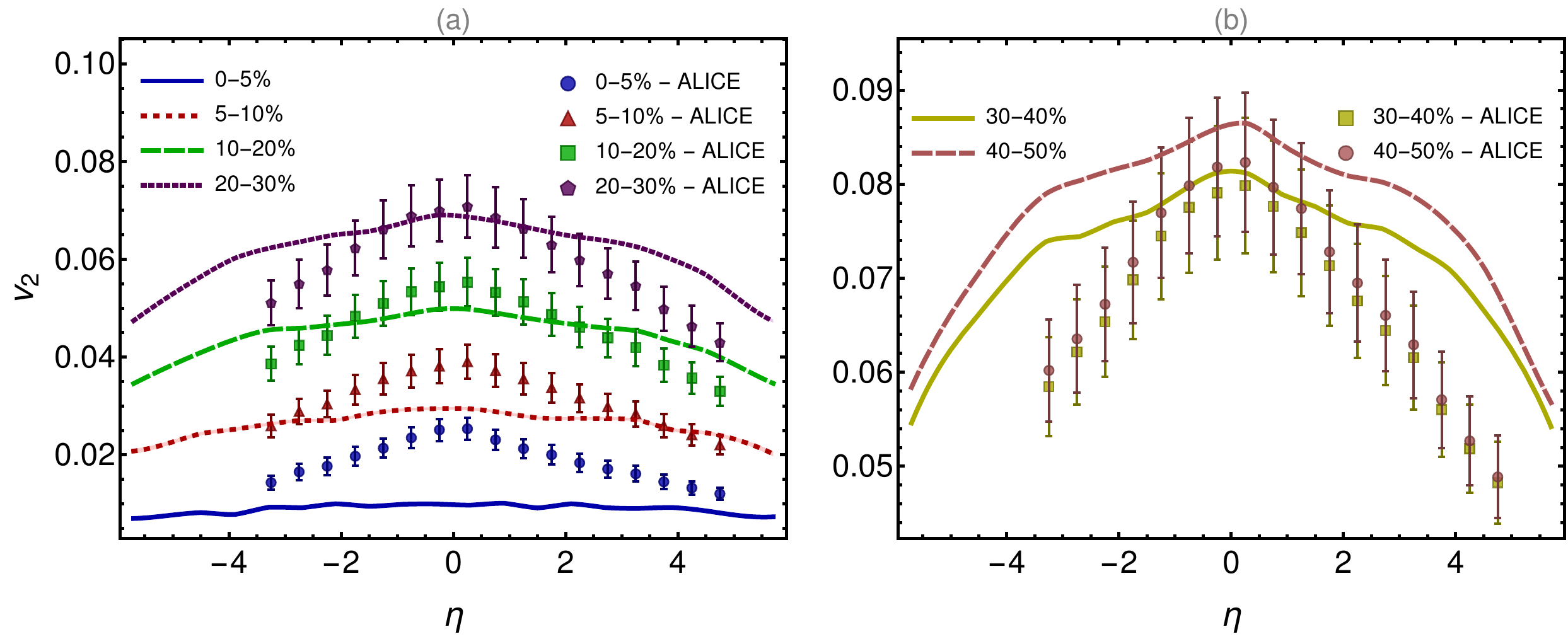}
}
\caption{The pseudorapidity dependence of the elliptic  flow $v_2$ for charged hadrons in different centrality classes where I take $0 < p_T < 100$ GeV.  All results are for 2.76 TeV Pb-Pb collisions and  data are from the ALICE collaboration \cite{Adam:2016ows}. }
\label{fig:v2_eta}
\end{figure*}

First the comparisons of the  transverse momentum spectra of $\pi^\pm$, $K^\pm$, and $p+\bar{p}$ in six centrality classes  0-5$\%$, 5-10$\%$, 10-20$\%$, 20-30$\%$, 30-40$\%$, and 40-50$\%$ in Fig.~\ref{fig:spectra} is presented. These comparisons show that our model provides a very good simultaneous description of the ALICE data for the pion, kaon, and proton spectra \cite{Abelev:2013vea}, with largest differences at $p_T \gtrsim 1.5$ GeV and relatively high centrality classes 30-40$\%$, and 40-50$\%$. Note that our model slightly underpredicts the pion spectrum at low transverse momentum which is similar to what is observed in other hydrodynamic models (see e.g. Ref.~\cite{Ryu:2015vwa}). One possible explanation for this discrepancy that has been suggested is pion condensation \cite{Begun:2013nga}.

In Fig.~\ref{fig:multiplicity},  the charged-hadron multiplicity in different centrality classes as a function of pseudorapidity, $\eta$ is shown. In panel (a), the 0-5$\%$, 5-10$\%$, 10-20$\%$, 20-30$\%$, and 30-40$\%$ centrality classes is shown, and in panel (b) I show the 40-50$\%$, 50-60$\%$, 70-80$\%$, 80-90$\%$, and  90-100$\%$ centrality classes.  As can be seen from both panels, our model is able to describe the charged hadron multiplicity as a function of pseudorapidity \cite{Abbas:2013bpa,Adam:2015kda} quite well in all centrality classes. Another observable to consider is  the average transverse momentum of pions, kaons, and protons as a function of centrality. This is shown in Fig.~\ref{fig:ptavg}, where our model is again able to reproduce the data reasonably well. 

Next, in Fig.~\ref{fig:v2integrated}, the integrated elliptic flow coefficient $v_2$ for charged hadrons as a function of centrality is shown.  The model predictions were computed using the geometrical definition of the elliptic flow coefficient, $v_2 \sim \langle \cos(2\phi) \rangle$, for all charged hadrons.  The experimental data were obtained using second- and fourth-order cumulants $v_2\{2\}$ and $v_2\{4\}$ \cite{Abelev:2014mda}. From this figure one sees that this model agrees well with $v_2\{4\}$ measurements at low centrality, but agrees better with $v_2\{2\}$ at higher centrality.  One would expect better agreement with $v_2\{4\}$ than $v_2\{2\}$, since the former has non-flow effects subtracted.  The fact that the agreement is better with $v_2\{2\}$ at high centrality could be due to the fact that our smooth initial condition is too simple or that the off-diagonal components of the anisotropy tensor in the evolution and freeze-out is not included.

In Fig.~\ref{fig:v2},  the comparisons of the identified-particle $v_2$ as a function of $p_T$ obtained using our model with experimental data reported by ALICE collaboration~\cite{Abelev:2014pua} is presented. Our model provides a quite reasonable description of the identified-particle elliptic flow as can be seen in panels (b) and (c), 20-30\% and 30-40\% centrality classes, respectively.  In panel (b), the 20-30\% centrality class, one sees that our model reproduces the data very well for the pion, kaon, and proton data out to $p_T \sim$ 1.5, 1.5, and 2.5 GeV, respectively.  A very similar agreement is seen in panel (c), the 30-40\% centrality class, where the model is in good  agreement with the pion, kaon, and proton data out to $p_T \sim$ 1, 1, and 2 GeV, respectively. However, in panels (a) and (d), 10-20\% and 40-50\% centrality classes, respectively, one sees poorer agreement than panels (b) and (c). For example, my results underpredict the pion elliptic flow in the 10-20$\%$ centrality class as can be seen from panel (a).  Again, as in the case of Fig.~\ref{fig:v2integrated}, this is related to our use of smooth Glauber initial conditions.

In order to further  examine how well our model describes various observables, I look at the pseudorapidity dependence of $v_2$ for different centrality classes in Fig.~\ref{fig:v2_eta}. As can be seen from Fig.~\ref{fig:v2_eta} our model results do not fall fast enough at large pseudorapidity compared to the experimental data~\cite{Adam:2016ows}.  One possible remedy for this may be including temperature-dependent $\eta/s$, since this has been shown to improve agreement with this observable in the context of viscous hydrodynamics \cite{Denicol:2015nhu}.  

\section{Conclusions}

In this chapter I presented phenomenological comparisons of aHydroQP with LHC experimental data collected in 2.76 TeV Pb-Pb collisions. This work was originally reported in Ref.~\cite{Alqahtani:2017jwl}.  Herein, I gave more details about the formalism used and presented a more thorough comparison between our model and LHC data for a variety of observables. In  aHydroQP, three momentum anisotropy parameters in the underlying distribution function is included, both in the dissipative hydrodynamic stage and at freeze-out. Also a quasiparticle implementation of the LQCD EoS is included to take into account the non-conformality of the system.  At freeze-out, a customized version of THERMINATOR 2 which was modified to accept anisotropic distribution functions of generalized Romatschke-Strickland form, is used.  As a first test, in this chapter, I used smooth Glauber initial conditions which were obtained from  a linear combination of wounded-nucleon and binary-collision profiles. I additionally assumed the system to be initially isotropic in momentum space with no initial transverse flow. 

To fix the remaining phenomenological parameters, a parameter scan is performed where I compared my results with experimentally observed identified-particle spectra in the 5-10\% and 30-40\% centrality classes. The resulting set of best fit parameters was $T_0 = 600$ MeV, $\eta/s = 0.159$, and \mbox{$T_{\rm FO} = 130$ MeV}.  After this fitting was complete, I computed an array of different heavy-ion observables, finding quite good agreement between our model and experimental data despite our simple smooth initial condition. I looked at particle multiplicity and spectra, average transverse momentum, $v_2$.  Compared to Ref.~\cite{Alqahtani:2017jwl}, I have added additional centrality classes in some cases and increased the statistics associated with the hadronic Monte-Carlo sampling where necessary \cite{Alqahtani:2017tnq, Alqahtani:2017diss}. 

The phenomenological results presented in this chapter represent the first aHydro results to include three separate anisotropy parameters together with the quasiparticle method for imposing the EoS and self-consistent anisotropic freeze-out.  Compared to prior results which used a single anisotropy parameter and/or an approximate conformal-factorization implementation of the equation of state \cite{Ryblewski:2012rr,Nopoush:2016qas,Strickland:2016ezq} much better agreement with the pion, kaon, and proton spectra and, relatedly, the total multiplicity as a function of pseudorapidity is observed.  Prior studies which used the approximate conformal-factorization implementation of the equation of state dramatically underestimated the low $p_T$ spectra \cite{Nopoush:2016qas,Strickland:2016ezq}, making this the first phenomenological study within the context of aHydro which is able to reproduce both the experimentally observed spectra and elliptic flow.

\part{Applications of anisotropic hydrodynamics}

\chapter{\bf Gluon self-energy in anisotropic hydrodynamics}  
\label{chap:gluon}
\setcounter{figure}{0}
\setcounter{table}{0}
\setcounter{equation}{0}

\section{introduction}
At the beginning of the 20th century, two extremely useful theories were born: quantum mechanics and special relativity. Later on, people tried to construct a formalism for studying the fundamental properties of all elementary particles. Reconciling quantum mechanics and special relativity led to quantum field theory (QFT). Despite being astonishingly successful in studying the basic interactions among particles, i.e. particle scattering cross sections in accelerators, vacuum QFT is unable to describe a system of particles that has finite temperature/density. In fact, conventional QFT is formulated at zero temperature or when the effect of temperature/density is negligible. However, our real world systems certainly evolve at non-zero temperatures/density limit and one needs to extend the dynamics to the finite temperature/density in order to study the phenomenology. In fact, when the temperature is comparable to (or larger than) the energy scale of the theory\footnote{This can be introduced as inverse of characteristic length scale of the system which can be mean inter-particle distance or Debye length, etc.}, thermodynamic effects become important. To be specific, there are mainly two important contexts in physics that thermal QFT effects play  inevitable roles and are widely applied: cosmology and URHIC experiments. To clarify, one needs to see what happens when the system of particles is heated up or when it is compressed. For instance, at extremely high temperature the matter experiences a transition to QGP which is the topic of URHIC experiments. Also, in the astronomical studies one encounters a whole set of important phenomena all of which are studied based on finite density/temperature effects in QFT, e.g. formation and evolution of neutron stars, black holes radiation, and cosmological inflation. For the purpose of discussion, there have been some approaches developed in the mid 20th century. The {\it imaginary time formalism} (ITF), developed based on quantum statistical mechanics, is a relevant framework for equilibrium finite-temperature field theory calculations. In this framework, starting from the quantum mechanical transition amplitude based on the Feynman path integral, one can replace the (real) time variable with an imaginary time variable $\tau=it$ with an appropriate prescription for the boundary conditions for bosonic and fermionic states. Through this, the transition amplitude looks the same as a quantum mechanical ensemble average. By analytic continuation, one can extend this formalism to the real time formalism (RTF). However, there is an alternative method to ITF, which is called {\it Schwinger-Keldysh formalism} (SK). The real-time SK formalism is suitable for studying finite-temperature systems both in- and out of equilibrium.  It is been proven that both approaches are consistent in the scopes of their mutual validity. In this chapter, I am going to discuss the SK formalism briefly and then use it to calculate the heavy-quark potential in an anisotorpic QGP.

As discussed briefly in Chap.~\ref{chap:intro}, studying heavy-quarkonia states is of crucial importance in the phenomenology of QGP. For instance, heavy-quark bound states interact with QGP and through QCD Debye screening experience in-medium suppression which depends on QGP temperature; therefore they can be considered as internal probes for measuring the temperature of the QGP. Also, due to the fact that heavy quarks explicitly break the chiral symmetry they remain unchanged during the QCD phase transition and they can carry a lot of useful information from that region. On the other hand, studying the dynamics of heavy-quarkonia without a precise calculation of heavy-quark potential essentially based on QGP phenomenology is ill-advised. Following the discussion in previous chapters, the large early stage momentum-space anisotropy is a phenomena which has to be considered carefully in the dynamics and suppression of heavy-quarks bound states. For instance, potential non-equilibrium corrections could have a large impact on the heavy-quark potential. One can attempt to include this deviation from equilibrium ``perturbatively'' as is done in the framework of viscous hydrodynamics, however, at very early times after the impact or near the transverse/longitudinal edges of the system this method becomes unreliable since such approaches rely on an explicit assumption that the system is near equilibrium.

One of the shortcomings of all previous calculations of the heavy-quark potential in a momentum space anisotropic QGP is that, although the real part of the potential was obtained to all orders in the plasma momentum-space anisotropy parameter $\xi$ in Ref. \cite{Dumitru:2007hy}, calculations of the imaginary part of the heavy-quark potential have relied on a Taylor expansion around $\xi=0$ to linear order. In this chapter we explain the steps necessary to calculate the heavy-quark potential based on (semi-)static hard-loop (HL) dressed propagators using the real time formalism (SK formalism) in a momentum-space anisotropic quark-gluon plasma. For this purpose, we need to calculate the dressed retarded, advanced, and Feynman propagators and the related self-energies using the leading order anisotropic distribution function. Once these are determined we can compute the potential using
\be
V({\bf r},\xi) = - \frac{g^2 C_F}{2} \int \frac{d^3{\bf p}}{(2\pi)^3} \left(e^{i {\bf p}\cdot {\bf r}} - 1 \right) 
\left[ \tilde{\cal D}^{00}_R + \tilde{\cal D}^{00}_A + \tilde{\cal D}^{00}_F \right]_{\omega \rightarrow 0} ,
\ee
where $g$ is the strong coupling constant, $C_F = (N_c^2-1)/(2 N_c)$ is the quadratic Casimir in the fundamental representation of $SU(N_c)$.

\section{Setup and notation}
\label{sec:seutp}
In this chapter, the metric is taken to be ``mostly minus'', i.e. in Minkowski space $\eta^{\mu\nu}\equiv(1,-\mathds{1})$. Lower-case letters denote four-vectors and bold lower-case letters denote three vectors. All Greek-letter indices stand for the components of four-vectors while Latin indices indicate spatial components of four-vectors. For any two four-vectors $x^\mu\equiv (x_0,{\bf x})$ and $y^\mu\equiv (y_0,{\bf y})$ the inner product is defined as $x\cdot y\equiv x^\mu y_\mu=x_0 y_0-{\bf x}\cdot{\bf y}$. The subscripts `R', `A', and `F' for propagators and self-energies stand for retarded, advanced, and Feynman propagators, respectively. According to Feynman slash notation, for any four-vector $x$, we have $\slashed{x} \equiv \gamma\cdot x$, with $\gamma^\mu$ being Dirac matrices. Herein, the anisotropic distribution function is taken to be of spheroidal form \ref{eq:anisodist3}. Anisotropic Fermi-Dirac $f_F$, ($a=1$), or Bose-Einstein $f_B$, ($a=-1$) distribution functions are
\be
f_{F\!/\!B}({\bf p})=\left[\exp\!\left(\frac{1}{\lambda}\sqrt{{\bf p}^2+\xi ({\bf p\cdot n})^2}\right)+a\right]^{-1}\,. 
\label{eq:FB-dist}
\ee
 We will perform the calculations initially in QED and then, in the end, we will generalize our results to QCD.

\section{The real-time formalism for non-equilibrium field theories}

In this section, we present the basic formalism used to obtain our results in a concise and self-contained manner.  We will use the real-time SK \cite{Dumitru:2009fy, Carrington:1998jj, Mrowczynski:2000ed, Mrowczynski:2016etf}. The SK formalism is based on contour Green's functions. For a spinor field $\psi$ and a vector field $A^\mu$, we can define the fermionic and bosonic QED Green's functions, respectively
\ba
i(S(x,y))_{\alpha\beta}\equiv \langle \hat{T}[\psi_\alpha(x)\bar{\psi}_\beta(y)]\rangle \,,\\
i({\cal D}(x,y))_{\mu\nu}\equiv \langle \hat{T}[A_\mu(x) A_\nu(y)]\rangle\,,
\label{eq:gfuncs1}
\ea
where $\{\alpha,\beta\} \in \{1,2,3,4\}$ are spinor indices, $\{\mu,\nu\} \in \{0,1,2,3\}$ are Lorentz indices. The angle brackets denote the quantum expectation value and $\hat{T}$ is the time-ordering operator
\ba 
\hat{T}[X(x)Y(y)]\equiv \Theta(x_0-y_0)X(x)Y(y)\pm \Theta(y_0-x_0)Y(y)X(x)\,,
\ea
where $\Theta$ is the Heaviside step function. Corresponding to different ways of propagation with respect to the contour, one can define four functions based on the contour propagator:
\ba 
i(S^>(x,y))_{\alpha\beta} &\equiv&  \langle \psi_\alpha(x)\bar{\psi}_\beta(y)\rangle\,, \nonumber \\
i(S^<(x,y))_{\alpha\beta}&\equiv& -\langle \bar{\psi}_\beta(y)\psi_\alpha(x)\rangle\,, \nonumber \\
i(S^c(x,y))_{\alpha\beta}&\equiv& \langle \hat{T}^c[\psi_\alpha(x)\bar{\psi}_\beta(y)]\rangle\,, \nonumber \\
i(S^a(x,y))_{\alpha\beta}&\equiv& \langle \hat{T}^a[\psi_\alpha(x)\bar{\psi}_\beta(y)]\rangle\,. 
\label{eq:gfuncs2}
\ea
Likewise, for bosons, we have  
\ba
i({\cal D}^>(x,y))_{\mu\nu} &\equiv&  \langle A_\mu(x)A_\nu(y)\rangle\,, \nonumber \\
i({\cal D}^<(x,y))_{\mu\nu}&\equiv& \langle A_\nu(y)A_\mu(x)\rangle\,, \nonumber \\
i({\cal D}^c(x,y))_{\mu\nu}&\equiv& \langle \hat{T}^c[A_\mu(x)A_\nu(y)]\rangle\,, \nonumber \\
i({\cal D}^a(x,y))_{\mu\nu}&\equiv& \langle \hat{T}^a[A_\mu(x)A_\nu(y)]\rangle\,. 
\label{eq:gfuncs3}
\ea
 In the relation above, $\hat{T}^c$ and $\hat{T}^a$ are time-ordering and anti-time-ordering operators, respectively, which are defined as
\ba 
\hat{T}^c[X(x)Y(y)]&\equiv& \Theta(x_0-y_0)X(x)Y(y)\pm \Theta(y_0-x_0)Y(y)X(x) \,,\\
\hat{T}^a[X(x)Y(y)]&\equiv& \Theta(y_0-x_0)X(x)Y(y)\pm \Theta(x_0-y_0)Y(y)X(x)\,.
\label{eq:timeorder2}
\ea
The upper and lower signs above correspond to bosonic and fermionic cases, respectively. In practice, one can define four different Green's functions with the following meanings:
\ba
S^c(x,y)&\equiv& S(x,y)\quad \text{\small with both $x_0$ and $y_0$ on the upper branch}\,, \nonumber \\
S^a(x,y)&\equiv& S(x,y)\quad \text{\small with both $x_0$ and $y_0$ on the lower branch}\,, \nonumber  \\
S^<(x,y)&\equiv& S(x,y)\quad \text{\small with $x_0$ on the upper and $y_0$ on the lower branch}\,, \nonumber  \\
S^>(x,y)&\equiv& S(x,y)\quad \text{\small with $x_0$ on the lower and $y_0$ on the upper branch}\,.\nonumber 
\label{eq:gfunction3}
\ea
In the SK formulation it is useful to introduce the following $2\times 2$ matrix 
\ba
S=
  \begin{pmatrix}
     S_{11} & S_{12} \\ S_{21} & S_{22}
     \end{pmatrix} 
     =  \begin{pmatrix}
     S^c & S^< \\ S^> & S^a
     \end{pmatrix}\,.
     \label{eq:keldysh}
\ea
Since, the system under consideration is assumed to be translationally invariant, the two-point function only depends on $x-y$. Therefore, we can safely set $y=0$ and study the two point function function of a single variable ($x$). The components of the electron propagator $S$ satisfy the following relations
\ba
S^{c/a}(x)&=&\Theta(x_0)S^\gtrless(x)+\Theta(-x_0) S^\lessgtr(x)\,, \\
S^c(x)+S^a(x)&=&S^<(x)+S^>(x) \,.
\ea
The retarded, advanced, and Feynman propagators are defined as 
\ba
S_{R/A} (x) &\equiv& \pm \Theta(\pm x_0)(S^>(x)-S^<(x))\,,\\
S_{R/A} (-x) &=& \pm \Theta(\mp x_0)(S^>(-x)-S^<(-x))\,,\\
S_F(x)&=& S^>(x)+S^<(x)\,.
\label{eq:SF<>}
\ea
Using the above relations one finds some useful identities such as
\ba
S_R(\pm x)-S_A(\pm x)&=&S^>(\pm x)-S^<(\pm x)\,, \label{eq:SRA<>}\\
 S_{R/A}(x)&=&\pm \Theta(\pm x_0)[S_R(x)-S_A(x)]\,,
 \label{eq:SRA} \\
  S_{R/A}(-x)&=&\pm \Theta(\mp x_0)[S_R(- x)-S_A(- x)]\,.
 \label{eq:SRA2}
\ea
In QED, the one-loop photon self-energy is 
\be
\Pi^{\mu\nu}(x)=-ie^2 {\rm Tr}[\gamma^\mu S(x) \gamma^\nu S(-x)]\,.
\ee
The retarded, advanced, and Feynman self-energies are defined as following
\ba
\Pi_{R/A}^{\mu\nu}(x)&=&\pm \Theta(\pm x_0)(\Pi^>(x)-\Pi^<(x))\,,
\label{eq:PiRA<>}\\
\Pi_F^{\mu\nu}(x)&=&\Pi^>(x)+\Pi^<(x)\,,
\label{eq:PiF<>}
\ea
with
\be
\left(\Pi^\lessgtr(x)\right)^{\mu\nu} =-i e^2{\rm Tr} [\gamma^\mu S^\lessgtr(x)\gamma^\nu S^\gtrless(-x)]\,.
\label{eq:Pi<>}
\ee
Substituting (\ref{eq:Pi<>}) into (\ref{eq:PiRA<>}), the retarded/advanced self-energy can be written as
\be
\Pi_{R/A}^{\mu\nu}(x)= \mp ie^2 \Theta(\pm x_0) {\rm Tr}[\gamma^\mu S^>(x) \gamma^\nu S^<(-x)-\gamma^\mu S^<(x) \gamma^\nu S^>(-x)]\,.
\label{eq:PiR<>}
\ee
Using (\ref{eq:SF<>}) and (\ref{eq:SRA<>}) we have
\ba
S^ \lessgtr(x)&=&\frac{\mp S_R(x)\pm S_A(x)+S_F(x)}{2}\,.
\label{eq:S<>}
\ea
Thus, Eq.~(\ref{eq:PiR<>}) gives
\ba
\Pi_{R/A}^{\mu\nu}(x) &=& \mp \frac{i}{2}e^2 \Theta(\pm x_0) {\rm Tr}[\gamma^\mu S_F(x) \gamma^\nu S_A(-x)-\gamma^\mu S_F(x) \gamma^\nu S_R(-x)\nonumber \\
&& \hspace{3cm}
+\gamma^\mu S_R(x) \gamma^\nu S_F(-x)-\gamma^\mu S_A(x) \gamma^\nu S_F(-x)] \nonumber \\
&=& -\frac{i}{2}e^2 {\rm Tr}[\gamma^\mu S_F(x) \gamma^\nu S_{A/R}(-x)
+\gamma^\mu S_{R/A}(x) \gamma^\nu S_F(-x)]\,.
\ea
where in the last line we have used (\ref{eq:SRA}). Performing the Fourier transform of both sides
\be
\Pi_{R/A}^{\mu\nu}(p)=-i\frac{e^2}{2}\int \frac{d^4k}{(2\pi)^4}{\rm Tr}[\gamma^\mu S_{R/A}(k)\gamma^\nu S_F(q)+\gamma^\mu S_F(k)\gamma^\nu S_{A/R}(q)]\,,
\label{eq:PiR0}
\ee
with $q\equiv k-p$. The Feynman self-energy can be obtained by substituting (\ref{eq:Pi<>}) in (\ref{eq:PiF<>}) and then using (\ref{eq:S<>})
\ba
 \Pi_F^{\mu\nu}(x) &=&-ie^2{\rm Tr}[\gamma^\mu S^>(x)\gamma^\nu S^<(-x)+\gamma^\mu S^<(x)\gamma^\nu S^>(-x)] \\ 
 &=&  -i\frac{e^2}{2}{\rm Tr}\Big[\gamma^\mu S_F(x)\gamma^\nu S_F(-x)-\gamma^\mu \left[S_R(x)-S_A(x)\right]\gamma^\nu [S_R(-x)-S_A(-x)]\Big]\nonumber \,.
\ea
After performing the Fourier transform of both sides one has
\be 
 \Pi_F^{\mu\nu}(p) =  -i\frac{e^2}{2}\int \frac{d^4k}{(2\pi)^4}{\rm Tr}  \Big[\gamma^\mu S_F(k)\gamma^\nu S_F(q)-\gamma^\mu \left[S_R(k)-S_A(k)\right]\gamma^\nu \left[S_R(q)-S_A(q)\right]\Big]\,.
 \label{eq:PiF0}
\ee
I calculate the hard loop self-energies and propagators using SK formalism in the limit of vanishing chemical potential~\cite{Carrington:1997sq}. The  ``bare'' propagators are then $2 \times 2 $ matrices such as
\ba
S(k)=  \left (\begin{array}{cc}
\frac{\slashed{k}}{k^2+i\epsilon} & 0\\
0 & \frac{-\slashed{k}}{k^2-i\epsilon}\\
                          \end{array} \right )
 +2\pi i\, \slashed{k}\,\delta (k^2)\>\left (\begin{array}{cc}
f_F({\bf k}) & -\Theta (-k_0)+f_F({\bf k})\\
-\Theta (k_0)+f_F({\bf k}) & f_F({\bf k}) \\ \end{array} \right ),
\label{2a5}
\ea
for a massless Dirac field and 
\ba
 \label{2a4}
  {\cal D}(k)  =  \left (\begin{array}{cc} \frac{1}{k^2+i\epsilon} & 0\\
                             0 & \frac{-1}{k^2-i\epsilon}\\
            \end{array} \right ) -  2\pi i\, \delta (k^2)
 \left (\begin{array}{cc}
f_B({\bf k}) & \Theta (-k_0)+f_B({\bf k})\\
\Theta (k_0)+f_B({\bf k}) & f_B({\bf k}) \\ \end{array} \right )\,,
\ea
for a massless scalar field. In the above relations, $\epsilon$ is a small positive number which is sent to zero only at the end of calculation. It should be noted that since Eqs.~(\ref{2a5}) and (\ref{2a4}) are bare propagators, the hard-loop resummation has yet to be performed. The retarded, advanced, and Feynman propagators can be obtained from the SK representation (which satisfies ${\cal D}_{11}-{\cal D}_{12}-{\cal D}_{21}+{\cal D}_{22}=0$) via 
\ba
\label{2a6}
   {\cal D}_R = {\cal D}_{11} - {\cal D}_{12} ,\qquad {\cal D}_A = {\cal D}_{11} - {\cal D}_{21} ,\qquad
   {\cal D}_F = {\cal D}_{11} + {\cal D}_{22}  ~,
\ea
with analogous expressions holding for the fermionic propagators. In momentum space, the explicit expressions for the bare propagators as a function of a general momentum $k$ are
\ba
\begin{aligned}
S_R(k) & =  \frac{\slashed{k}}{k^2+i\, \mbox{sgn}(k_0) \epsilon}, \\
S_A(k) & =  \frac{\slashed{k}}{k^2-i\, \mbox{sgn}(k_0) \epsilon},\\
S_F(k) & =  -2\pi i\, \slashed{k} \, [1-2f_F({\bf k})]\, \delta (k^2)\,,
\label{eq:S&D}
\end{aligned}
\hspace{1.5cm}
\begin{aligned}
{\cal D}_R(k) & =  \frac{1}{k^2+i\, \mbox{sgn}(k_0) \epsilon}, \\
{\cal D}_A(k) & =  \frac{1}{k^2-i\, \mbox{sgn}(k_0) \epsilon}, \\
{\cal D}_F(k) & =  -2\pi i\, [1+2f_B({\bf k})]\, \delta (k^2)\,, 
\end{aligned}
\ea
for fermions and bosons, respectively. 
In the real-time formalism, the following relations hold for the self
energies:
\ba
\Pi_{11}+\Pi_{12}+\Pi_{21}+\Pi_{22}=0 \label{2a12}\,,
\ea
and
\ba
\Pi_R  =  \Pi_{11}+\Pi_{12}, \qquad
\Pi_A  =  \Pi_{11}+\Pi_{21} , \qquad
\Pi_F  =  \Pi_{11}+\Pi_{22} ~. \label{2a13}
\ea
Note that, for vector fields, one must add the appropriate Lorentz indices to the propagators and self-energies.
\section{The dressed propagator}
 The dressed propagator can be obtained from the Dyson-Schwinger equation 
\be
i\tilde{\cal D}=i{\cal D}+i{\cal D}(-i\Pi)i\tilde{\cal D} \,,
\ee
where, in the SK formalism, both the propagators and self-energies are $2\times2$ matrices, ${\cal D}$ and $\tilde{\cal D}$ are bare and dressed propagators. For the dressed retarded propagator, on has 
\be 
\tilde{\cal D}_{R/A}={\cal D}_{R/A}+{\cal D}_{R/A}\Pi_{R/A} \tilde{\cal D}_{R/A}\,.
\label{eq:ds-PiR}
\ee
The dressed Feynman propagator satisfies
\be
\tilde{\cal D}_F={\cal D}_{F}+{\cal D}_{R}\Pi_R \tilde{\cal D}_F+{\cal D}_{F}\Pi_A \tilde{\cal D}_A+{\cal D}_{R}\Pi_F \tilde{\cal D}_A \,,
\ee 
which, upon using $p = (\omega,{\bf p})$ and ${\cal D}_F(p)=[1+2f_B({\bf p})]{\rm sgn}(\omega)[{\cal D}_R(p)-{\cal D}_A(p)]$, becomes
\ba
\tilde{\cal D}_{F}(p)&= & (1+2f_B({\bf p}))\, \mbox{sgn}(\omega)\,
[\tilde{\cal D}_{R}(p)-\tilde{\cal D}_{A}(p)]
\nonumber \\
& +&\tilde{\cal D}_{R}(p)\,\{\Pi _F(p)-(1+2f_B({\bf p}))\, \mbox{sgn}(\omega)\, [\Pi
_R(p)-\Pi _A(p)]\} \,  \tilde{\cal D}_A(p)~. \label{eq:2b8}
\ea
Note that in the relations introduced so far the distribution functions are general. We will specify the precise forms in the forthcoming sections.

\section{Tensor decomposition in a momentum-space anisotropic plasma}
Since propagators and self energies are tensor quantities, we must find a suitable tensor basis and construct the corresponding scalar coefficient functions.  For anisotropic systems there are more independent projectors than for the standard equilibrium case due to the fact that there is an additional spacelike vector ${\bf n}$ which defines the anisotropy direction~\cite{Romatschke:2003ms}. Here, we use a four-tensor basis which is appropriate for systems with one anisotropy direction~\cite{Dumitru:2007hy}. Specifically, we introduce four tensors
\ba
A^{\mu \nu}&=& -\eta^{\mu\nu}+\frac{p^\mu
p^\nu}{p^2}+\frac{\tilde{m}^\mu \tilde{m}^\nu}{\tilde{m}^2}\,,\nonumber \\
B^{\mu \nu}&=& -\frac{p^2}{(m\cdot p) ^2}\frac{\tilde{m}^\mu
\tilde{m}^\nu}{\tilde{m}^2}\,,\nonumber \\
C^{\mu \nu}&=& \frac{\tilde{m}^2p^2}{\tilde{m}^2p^2+(n\cdot
p)^2}[\tilde{n}^\mu
\tilde{n}^\nu-\frac{\tilde{m}\cdot\tilde{n}}{\tilde{m}^2}(\tilde{m}^\mu
\tilde{n}^\nu+\tilde{m}^\nu
\tilde{n}^\mu)+\frac{(\tilde{m}\cdot\tilde{n})^2}{\tilde{m}^4}\tilde{m}^\mu
\tilde{m}^\nu]\,,\nonumber \\
D^{\mu \nu}&=& \frac{p^2}{m\cdot p}
\left[ 2\frac{\tilde{m}\cdot\tilde{n}}{\tilde{m}^2}\tilde{m}^\mu
\tilde{m}^\nu-
\left(\tilde{n}^\mu \tilde{m}^\nu+\tilde{m}^\mu
\tilde{n}^\nu\right) \right]\,.
\label{eq:ABCD}
\ea
Here, $m^{\mu}$ is the heat-bath four-velocity, which in the LRF is given by $m^{\mu}=(1,0,0,0)$, and
\begin{equation}
\tilde{m}^\mu=m^{\mu}-\frac{m\cdot p}{p^2} \,p^\mu\,,
\label{eq:mt}
\end{equation}
is the component of $m^\mu$ orthogonal to $p^\mu$. The direction of anisotropy in momentum space is determined by the vector
\begin{equation}
n^{\mu}=(0,{\bf n})\,,
\label{eq:n}
\end{equation}
where ${\bf n}$ is a three-dimensional unit vector. Likewise, $\tilde{n}^\mu$ is the component of $n^\mu$ orthogonal to $p^\mu$.
The self-energies and dressed propagators can be expanded in terms of the tensor basis (\ref{eq:ABCD}) as
\ba
\Pi^{\mu\nu}_{R,A,F} &=&\alpha_{R,A,F} A^{\mu\nu}+\beta_{R,A,F} B^{\mu\nu} + \gamma_{R,A,F}
C^{\mu\nu} + \delta_{R,A,F} D^{\mu\nu}\,,\label{eq:Piexp}\\
\tilde{\cal D}^{\mu\nu}_{R,A,F}&=&\alpha'_{R,A,F} A^{\mu\nu}+\beta'_{R,A,F} B^{\mu\nu} + \gamma'_{R,A,F}
C^{\mu\nu} + \delta'_{R,A,F} D^{\mu\nu}\,.
\label{eq:Dexp}
\ea
Note that, due to the transversality of the self-energy $p_\mu \Pi^{\mu\nu}=0$, not all components of $\Pi^{\mu\nu}$ are independent.  One has four equations which can be used, for example, to write the timelike rows/columns of the self-energy tensor in terms of the space-like components
\ba 
\nu=0 \; &\Rightarrow& \;  \omega \Pi^{00}+p_i \Pi^{i0}=0\,,\\
\nu=i \; &\Rightarrow& \; \omega \Pi^{0i}+p_j \Pi^{ji}=0\,.
\ea
Using the symmetry of $\Pi^{\mu\nu} = \Pi^{\nu\mu}$, one finds \mbox{$\omega^2\Pi^{00}=p_i\Pi^{ij}p_j$}. These relations show that having $\Pi^{ii}$ and $\Pi^{xy}$, $\Pi^{xz}$, $\Pi^{yz}$ (6 components overall), one can obtain all components of $\Pi^{\mu\nu}$. 

Restricting our attention to the spatial block of $\Pi^{\mu\nu}$, $\Pi^{ij}$, one can obtain the expansion tensor coefficients of self-energy using the following projections
\ba
p^i \Pi^{ij}p^j &=&{\bf p}^2 \beta\,, \nonumber \\
A^{il}n^l \Pi^{ij}p^j &=&({\bf p}^2-(n\cdot p)^2)\delta\,, \nonumber \\
A^{il}n^l \Pi^{ij}A^{jk}n^k &=&\frac{{\bf p}^2-(n\cdot p)^2}{{\bf p}^2}(\alpha+\gamma)\,, \nonumber \\
{\rm Tr}\,\Pi^{ij} &=&2\alpha+\beta+\gamma\,.
\label{eq:abgd}
\ea
An alternative method for extracting the coefficient functions, which is based on the four-tensor form of $\Pi^{\mu\nu}$, is presented in App.~\ref{app:alternative}.

The dressed retarded/advanced propagators satisfy (\ref{eq:ds-PiR}), which can be solved to give
\be
\tilde{\cal D}_{R,A}^{-1}= ({\cal D}_{R,A})^{-1}-\Pi_{R,A}\,.
\ee
Using the definition of bare propagator and (\ref{eq:Piexp}) one has
\ba
(\tilde{\cal D}_{R,A}^{-1})^{\mu\nu}&=&-p^2 \eta^{\mu\nu} +p^\mu p^\nu-\Pi_{R,A}^{\mu\nu}-\frac{1}{\zeta}p^\mu p^\nu \\
&=&(p^2-\alpha_{R,A})A^{\mu\nu}+(\omega^2-\beta_{R,A})B^{\mu\nu}-\gamma_{R,A} C^{\mu\nu}-\delta_{R,A} D^{\mu\nu}-\frac{1}{\zeta}p^\mu p^\nu\,.
\ea
where $\zeta$ is the gauge fixing parameter.  One can obtain $\tilde{\cal D}$ by inverting the above relation~\cite{Dumitru:2007hy}
\ba
\tilde{\cal D}_{R,A}^{\mu\nu}&=&\Delta_A\left[A^{\mu\nu}-C^{\mu\nu}\right]-\frac{\zeta}{p^4}p^\mu p^\nu\nonumber \\
&+&\Delta_G\left[(p^2-\alpha_{R,A}-\gamma_{R,A})\frac{\omega^4}{p^4}B^{\mu\nu}+(\omega^2-\beta_{R,A})C^{\mu\nu}+\delta_{R,A}\frac{\omega^2}{p^2}D^{\mu\nu}\right]\,,
\label{eq:tildeDRA}
\ea
with 
\ba
\Delta_A^{-1} &=& p^2 - \alpha_{R,A} \label{eq:deltaa} \,,\\
\Delta_G^{-1} &=& (p^2-\alpha_{R,A}-\gamma_{R,A})(\omega^2-\beta_{R,A})-\delta_{R,A}^2\left[{\bf p}^2-(n\cdot p)^2\right]\,.
\label{eq:deltag}
\ea
Comparing to Eq.~(\ref{eq:Dexp}), the expansion coefficients are
\ba
\alpha'_{R,A}&=&\Delta_A\,,\nonumber\\
\beta'_{R,A}&=&\Delta_G(p^2-\alpha_{R,A}-\gamma_{R,A})\frac{\omega^4}{p^4}\,,\nonumber\\
\gamma'_{R,A}&=&\Delta_G(\omega^2-\beta_{R,A})-\Delta_A\,,\nonumber\\
\delta' _{R,A}&=&\Delta_G \frac{\omega^2}{p^2} \delta_{R,A}\,.
\label{eq:coefPrime}
\ea
Herein, we take the gauge parameter to be zero, $\zeta=0$, which is allowed since the static limit of the gauge propagator is gauge invariant.

\section{The HL retarded and advanced photon self-energies}
On of the quantities that we are interested to calculate in this chapter is the HL limit of gluon self-energy. Since in HL limit the photon and gluon self-energies are the same up to definition of Debye mass we start with photon self energy which is less challenging. Starting from Eq.~(\ref{eq:PiR0}), notice that $\Pi_R$ and $\Pi_A$ are complex conjugates of each other based on Eq.~(\ref{eq:S&D}) and we only need to find one of them. To proceed, we start with the retarded photon self-energy
\be
\Pi_R^{\mu\nu}(p,\xi)=-i\frac{e^2}{2}\int \frac{d^4k}{(2\pi)^4}{\rm Tr}[\gamma^\mu S_R(k)\gamma^\nu S_F(q)+\gamma^\mu S_F(k)\gamma^\nu S_A(q)]\,,
\ee
where, specializing to anisotropic distribution function (\ref{eq:FB-dist}) in this section, we made the dependence on the anisotropy parameter, $\xi$,  explicit. Using $S_{R,A,F}(k)=\slashed{k}\Delta_{R,A,F}(k)$ with
\ba
\Delta_R(k) & = & \frac{1}{k^2+i\, \mbox{sgn}(k_0) \epsilon}, \\
\Delta_A(k) & = & \frac{1}{k^2-i\, \mbox{sgn}(k_0) \epsilon}, \\
\Delta_F(k) & = & -2\pi i\, \big[1-2f_F({\bf k})\big]\, \delta (k^2)\,,
\label{eq:propag}
\ea
and ${\rm Tr}[\gamma^\mu\gamma^\alpha\gamma^\nu\gamma^\beta]=
4(\eta^{\mu\alpha}\eta^{\nu\beta}-\eta^{\mu\nu}\eta^{\alpha\beta}+\eta^{\mu\beta}\eta^{\alpha\nu})$, we have
\ba
\Pi^{\mu\nu}_R(p,\xi)&=&-2ie^2\int \frac{d^4k}{(2\pi)^4}(q^\mu k^\nu+ q^\nu k^\mu-\eta^{\mu\nu}q\cdot k) \Big[\Delta_R(k)\Delta_F(q)+\Delta_F(k)\Delta_A(q)\Big]\nonumber \\
&=&-4ie^2\int \frac{d^4k}{(2\pi)^4}(q^\mu k^\nu+ q^\nu k^\mu-\eta^{\mu\nu}q\cdot k)\Delta_F(k)\Delta_A(q)\,,
\label{eq:PiR2}
\ea
where, in going from the first to the second line, we have used the fact that two terms in the integrand are equal under the transformation $k\rightarrow -k+p\,(=-q)$.
Using (\ref{eq:propag}), one has
\ba
  \Pi^{\mu\nu}_R(p,\xi)&=&16\pi e^2 \int \frac{d^4k}{(2\pi )^4}
f_F({\bf k}) \left[
\frac{q^\mu k^\nu+ q^\nu k^\mu-\eta^{\mu\nu}q\cdot k}{k^2+p^2-2k \cdot p-i\epsilon\,{\rm sgn}(k_0-\omega)}
\right]
\delta(k^2)\,.
\label{eq:PiR3}
\ea
Note that the first term in $\Delta_F(k)$ in Eq.~(\ref{eq:propag}) which corresponds to a divergent vacuum contribution, is subtracted to obtain the in-medium photon self-energy. Now one can take the HL approximation, that is, taking all internal momenta $k$ to be of order $\lambda$ (hard) and the external momenta $p$ to be of order $e\lambda$ (soft) and Taylor-expand the integrand around $e=0$. At the LO of the HL approximation, the quantity in square brackets in (\ref{eq:PiR3}) is
\be
\left[ \; \cdots \; \right] \longrightarrow \frac{2k^\mu k^\nu}{-2 k\cdot p-i \epsilon\,{\rm sgn}(k_0)}\,.
\ee
Note that the terms containing $k^2$ are effectively zero due to the delta function which enforces the mass shell condition. Substituting this into the integral (\ref{eq:PiR3}), using 
\be
\delta(k^2)=\delta(k_0^2-{\bf k}^2)=\frac{1}{2|{\bf k}|} \Big(\delta(k_0-|{\bf k}|)+\delta(k_0+|{\bf k}|)\Big)\,,
\label{eq:delta-f}
\ee
integrating over $k_0$, and finally setting ${\bf k}\rightarrow -{\bf k}$ in negative-energy contribution, one finds
\be
4e^2\int \frac{d^3{\bf k}}{(2\pi )^3}
\frac{f_F({\bf k})}{|{\bf k}|}
\bigg[\frac{-2k^\mu k^\nu}{2|{\bf k}|\omega-2{\bf k\cdot p}+i \epsilon\,}+\frac{2k^\mu k^\nu}{2|{\bf k}|\omega-2{\bf k\cdot p}+i \epsilon\,}\bigg]
\Bigg|_{k_0 = |{\bf k}|}
=0\,.
\ee
At next-to-leading order we have 
\be
\left[ \; \cdots \; \right] \longrightarrow  -\frac{\eta^{\mu\nu}}{2}+\frac{k^\mu p^\nu+k^\nu p^\mu}{2k\cdot p+i\epsilon\,{\rm sgn}(k_0)}-\frac{2k^\mu k^\nu p^2}{(2k\cdot p+i\epsilon\, {\rm sgn}(k_0))^2}\,,
\ee
where by substituting into the integral one obtains the retarded photon self-energy in the HL limit  
\ba
  \Pi^{\mu\nu}_R(p,\xi)=8\pi e^2 \!\int \frac{d^4k}{(2\pi )^4}
f_F({\bf k})
\bigg[\!-\!\eta^{\mu\nu}+\frac{k^\mu p^\nu+k^\nu p^\mu}{k\cdot p+i\epsilon\, {\rm sgn}(k_0)\,}-\frac{k^\mu k^\nu p^2}{(k\cdot p+i\epsilon\, {\rm sgn}(k_0))^2}\bigg]\delta(k^2).\hspace{.2cm}
\label{eq:PiR4}
\ea
Performing the integral over $k_0$ and again setting ${\bf k}\rightarrow -{\bf k}$ in the negative energy contribution, one has
\be
  \Pi^{\mu\nu}_R(p,\xi)=4 e^2 \int \frac{d^3{\bf k}}{(2\pi )^3}
\frac{f_F({\bf k})}{|{\bf k}|}
\bigg[-\eta^{\mu\nu}+\frac{k^\mu p^\nu + k^\nu p^\mu}{ k\cdot p+i\epsilon}-\frac{k^\mu k^\nu p^2}{(k\cdot p+i\epsilon)^2}\bigg] \Bigg|_{k_0 = |{\bf k}|}
\,.
\label{eq:PiR5}
\ee
One can show that, for on-shell momentum $k^\mu$, 
\be
 |{\bf k}|\frac{\partial}{\partial k^l}\bigg[\frac{k^\mu k^\nu p^l}{|{\bf k}|(p\cdot k+i \epsilon)}-\frac{k^\mu \eta^{\nu l}}{|{\bf k}|}\bigg]=\frac{k^\mu p^\nu+k^\nu p^\mu}{p\cdot k+i \epsilon}-\frac{k^\mu k^\nu p^2}{(p\cdot k+i\epsilon)^2}-\eta^{\mu\nu}  \,,
\ee
so, we after integrating by parts one obtains
\ba
  \Pi^{\mu\nu}_R(p,\xi)&=&- 4 e^2 \int \frac{d^3{\bf k}}{(2\pi )^3}
\frac{\partial f_F({\bf k})}{\partial k^l} 
\bigg[\frac{k^\mu k^\nu p^l}{|{\bf k}|(p\cdot k+i \epsilon)}-\frac{k^\mu \eta^{\nu l}}{|{\bf k}|}\bigg] \Bigg|_{k_0 = |{\bf k}|}
. \label{eq:PiR6} 
\ea
By specializing to the RS form for the anisotropic distribution function, we can simplify $\Pi_R^{\mu\nu}$ a bit more. 
Using the anisotropic Fermi-Dirac distribution (\ref{eq:FB-dist}), one has
\ba
\frac{\partial f_F({\bf k})}{\partial k^l} &=&\frac{v^l+\xi ({\bf v\cdot n}) n^l}{1+\xi({\bf v\cdot n})^2}\frac{\partial f_F({\bf k})}{\partial |{\bf k}|}\,, \label{eq:dist-identity1} \\
\frac{2e^2}{\pi^2} \int d|{\bf k}|\,{\bf k}^2\frac{\partial f_F({\bf k})}{\partial|{\bf k}|}&=&  \frac{1}{1+\xi({\bf v\cdot n})^2} \frac{2 e^2}{\pi^2}  \int d|{\bf k}|\,{\bf k}^2\frac{\partial f_F^{\rm iso}({\bf k})}{\partial|{\bf k}|}=-\frac{m_D^2}{1+\xi({\bf v\cdot n})^2}\,,
\label{eq:dist-identity2}
\ea
where $v^\mu\equiv k^\mu/|{\bf k}|=(1,{\bf k}/|{\bf k}|)$ and the QED Debye mass is defined as
\be 
m_D^2\equiv -\frac{2e^2}{\pi^2}\int d|{\bf k}|{\bf k}^2 \frac{\partial f_F^{\rm iso}({\bf k})}{\partial |{\bf k}|}=\frac{e^2\lambda^2}{3}\,.
\label{eq:mDQED}
\ee
Substituting (\ref{eq:dist-identity1}) and (\ref{eq:dist-identity2}) into (\ref{eq:PiR6}), one obtains
\be
  \Pi^{\mu\nu}_R(p,\xi)= m_D^2 \int \frac{d\Omega}{4\pi} \,
v^\mu \frac{v^l+\xi ({\bf v\cdot n}) n^l}{(1+\xi({\bf v\cdot n})^2)^2} \bigg[-\eta^{\nu l} +\frac{v^\nu  p^l}{p\cdot v+i\epsilon}\bigg]\,.
\label{eq:PiR8}
\ee
This precisely corresponds to the results obtained in \cite{Romatschke:2003ms,Mrowczynski:2000ed} using relativistic kinetic theory.

\section{The HL retarded and advanced gluon self-energies }
\label{sec:retarded-PiR}
As mentioned previously, in the hard-loop limit, the photon and gluon self-energies are the same up to the definition of the Debye mass. The effective QCD distribution function includes contributions from quarks, anti-quarks, and gluons, including the degeneracy factors for the number of quark flavors and gluon color-charge states
\be
f({\bf k})=2N_c f_g({\bf k})+N_f (f_q({\bf k}) +f_{\bar q}({\bf k}))\,,
\label{eq:qcd-dist}
\ee
where $f_g({\bf k})$ is gluonic distribution function, $f_q({\bf k})$ and $f_{\bar q}({\bf k})$ are quarks and anti-quarks distribution functions.
Eq.~(\ref{eq:PiR8}) is valid for the gluon self-energy  provided that QED Debye mass is replaced with its QCD counterpart, defined as
\be 
m_D^2\equiv -\frac{g^2}{2\pi^2}\int d|{\bf k}|{\bf k}^2 \frac{\partial f_{\rm iso}({\bf k})}{\partial |{\bf k}|}=\frac{(2N_c+N_f)g^2\lambda^2}{6}\,.
\label{eq:qcd-debye}
\ee
 To calculate the expansion coefficients (\ref{eq:Piexp}) we only need the spatial block of $\Pi_R^{\mu\nu}$ which is
\be
  \Pi^{ij}_R(p,\xi)= m_D^2 \int \frac{d\Omega}{4\pi} \,
v^i \frac{v^l+\xi ({\bf v\cdot n}) n^l}{(1+\xi({\bf v\cdot n})^2)^2} \Bigg[\delta^{lj} +\frac{v^j  p^l}{p\cdot v+i\epsilon}\Bigg] .
\label{eq:PiRij}
\ee
In order to obtain the tensor expansion coefficients for the retarded self-energy $\Pi_R^{\mu\nu}$ (\ref{eq:Piexp}) we can make use of Eqs.~(\ref{eq:abgd}). For this purpose, we need to choose a frame to be able to define the vectors ${\bf k}$, ${\bf p}$, and ${\bf n}$, and subsequently, the tensor basis matrices. The trivial choice is to take ${\bf n}$ along $z$-axis and ${\bf p}$ in the $x$-$z$ plane making an angle $\theta_n\equiv \arctan(p_x/p_z)$ with $z$-axis. Based on this coordinate we have 
\ba
{\bf n}&=&(0,0,1);\:\quad {\bf v}=(\sin\theta\cos\phi,\sin\theta\sin\phi,\cos\theta);\:\quad \hat{\bf p}=\left(\frac{p_x}{|{\bf p}|},0,\frac{p_z}{|{\bf p}|}\right)\,.
\label{eq:def1}
\ea
Up to next-to-leading order in $\omega$, we have 
\ba
\alpha_{R/A}(p,\xi)\!&=&\!-\frac{m_D^2}{2p_x^2} \Bigg[\frac{p_z^2\arctan\sqrt{\xi}}{\sqrt{\xi}}-\frac{{\bf p}^2 p_z\arctan\Big[\frac{p_z\sqrt{\xi}}{\sqrt{{\bf p}^2+\xi p_x^2}}\Big]}{\sqrt{\xi}\sqrt{{\bf p}^2+\xi p_x^2}}\pm i\frac{{\bf p}^2p_x^2 \pi(1+\xi)}{2({\bf p}^2+\xi p_x^2)^{3/2}}\omega\Bigg]+{\cal O}(\omega^2)\,,\nonumber \\
\beta_{R/A}(p,\xi)&=&-\frac{m_D^2\omega^2}{2{\bf p}^2}\Bigg[\frac{{\bf p}^2}{{\bf p}^2+\xi p_x^2}+\frac{\arctan\sqrt{\xi}}{\sqrt{\xi}}+\frac{{\bf p}^2 p_z\sqrt{\xi}\arctan\Big[\frac{p_z\sqrt{\xi}}{\sqrt{{\bf p}^2+\xi p_x^2}}\Big]}{({\bf p}^2+\xi p_x^2)^{3/2}}\nonumber \\&& \hspace{30mm}\pm i\frac{2{\bf p}^4+{\bf p}^2(2{\bf p}^2+p_x^2)\xi+p_x^2(p_x^2-p_z^2)\xi^2}{2({\bf p}^2+\xi p_x^2)^{5/2}}\pi\omega\Bigg]+{\cal O}(\omega^4)\,, \nonumber \\
\gamma_{R/A}(p,\xi)\!&\!=\!&\!-\frac{m_D^2}{2}\Bigg[\frac{{\bf p}^2}{{\bf p}^2+\xi p_x^2}-\frac{({\bf p}^2+p_z^2)\arctan\sqrt{\xi}}{\sqrt{\xi}p_x^2}+\frac{{\bf p}^2 p_z(2{\bf p}^2+3\xi p_x^2)\arctan\Big[\frac{p_z\sqrt{\xi}}{\sqrt{{\bf p}^2+\xi p_x^2}}\Big]}{p_x^2\sqrt{\xi}({\bf p}^2+\xi p_x^2)^{3/2}}\nonumber \\ &&\hspace{63mm}\mp i\frac{\xi\pi p_x^2({\bf p}^2+\xi({\bf p}^2+2p_z^2))}{2({\bf p}^2+\xi p_x^2)^{5/2}}\omega\Bigg]+{\cal O}(\omega^2)\,,\nonumber \\
\delta_{R/A}(p,\xi)&=&\frac{m_D^2\omega}{2{\bf p}^2p_x^2}\Bigg[\mp i \frac{{\bf p}^2p_x^2p_z\pi\xi}{2({\bf p}^2+\xi p_x^2)^{3/2}}+\frac{3{\bf p}^2p_x^2p_z\xi\omega}{({\bf p}^2+\xi p_x^2)^2}+\frac{p_z\omega}{\sqrt{\xi}}\arctan\sqrt{\xi}\nonumber \\ &&\hspace{5mm}- \frac{{\bf p}^2({\bf p}^4+\xi {\bf p}^2p_x^2-3\xi^2p_x^2p_z^2)\omega}{\sqrt{\xi}({\bf p}^2+\xi p_x^2)^{5/2}}\arctan\bigg[\frac{p_z\sqrt{\xi}}{\sqrt{{\bf p}^2+\xi p_x^2}}\bigg]\Bigg]+{\cal O}(\omega^3)\,.
\label{eq:PiRcoef}
\ea
\subsubsection*{Static limit}

One finds that, in the static limit, $\Delta_A$ (\ref{eq:deltaa}) and $\Delta_G$ (\ref{eq:deltag}) become
\ba
\lim_{\omega\rightarrow 0}\Delta_A^{-1} &=& -( {\bf p}^2 + m_\alpha^2) \, , \\
\lim_{\omega\rightarrow 0}\Delta_G^{-1} &=& -\frac{\omega^2}{{\bf p}^2}\left[({\bf p}^2+m_\alpha^2+m_\gamma^2)({\bf p}^2+m_\beta^2)
  - m_\delta^4 \right] . 
\ea
with \cite{Romatschke:2003ms,Dumitru:2007hy}  
\ba
m_\alpha^2 &\equiv& -\frac{m_D^2}{2 p_x^2 \sqrt{\xi}}%
\left(p_z^2 {\rm{arctan}}{\sqrt{\xi}}-\frac{p_z {\bf{p}}^2}{\sqrt{{\bf{p}}^2+\xi p_x^2}}%
{\rm{arctan}}\bigg[\frac{\sqrt{\xi} p_{z}}{\sqrt{{\bf{p}}^2+\xi p_x^2}}\bigg]\right) , \nonumber\\
m_\beta^2&\equiv&m_{D}^2
\frac{(\sqrt{\xi}+(1+\xi){\rm{arctan}}{\sqrt{\xi}})({\bf{p}}^2+\xi p_x^2)+\xi p_z\left(%
p_z \sqrt{\xi} + \frac{{\bf{p}}^2(1+\xi)}{\sqrt{{\bf{p}}^2+\xi p_x^2}} %
{\rm{arctan}}\Big[\frac{\sqrt{\xi} p_{z}}{\sqrt{{\bf{p}}^2+\xi p_x^2}}\Big]\right)}{%
2  \sqrt{\xi} (1+\xi) ({\bf{p}}^2+ \xi p_x^2)},  \nonumber \\
m_\gamma^2&\equiv&-\frac{m_D^2}{2}\left(\frac{{\bf{p}}^2}{{\bf{p}}^2 + \xi p_x ^2}%
-\frac{{\bf p}^2+p_z^2}{\sqrt{\xi}p_x^2}{\rm{arctan}}{\sqrt{\xi}}+\frac{
p_z{\bf{p}}^2(2{\bf{p}}^2+3\xi p_x^2)}{\sqrt{\xi}(\xi
p_x^2+{\bf{p}}^2)^{{3/2}}
p_x^2}{\rm{arctan}}\bigg[\frac{\sqrt{\xi}
p_{z}}{\sqrt{{\bf{p}}^2+\xi p_x^2}}\bigg]\right)\!, \nonumber \\
m_\delta^4&\equiv&\frac{\pi^2 m_D^4 \xi^2 p_z^2 p_x^2 {\bf{p}}^2}{16({\bf{p}}^2 + \xi p_x ^2)^{3}}\, . \label{eq:m}
\ea
The above expressions apply when ${\bf n}=(0,0,1)$ points along the $z$-axis and ${\bf p}$ lies in the $x-z$ plane; in the general case, $p_z$ and $p_x$ should be replaced by $\bf{p\cdot n}$ and $|\bf{p- (p\cdot n)n}|$, respectively.  With this, we can write down an expression for the real part of the potential which is valid to all orders in $\xi$ \cite{Dumitru:2007hy,Strickland:2011aa}
\ba
V({\bf{r}},\xi) &=& -g^2 C_F\int \frac{d^3{\bf{p}}}{(2\pi)^3} \,
\left( e^{i{\bf{p \cdot r}}} -1 \right) \tilde{\cal D}_R^{00}(\omega=0, \bf{p},\xi) \nonumber\\
&=& -g^2 C_F\int \frac{d^3{\bf{p}}}{(2\pi)^3} \,
\left( e^{i{\bf{p \cdot r}}} -1 \right) \frac{{\bf{p}}^2+m_\alpha^2+m_\gamma^2}
 {({\bf{p}}^2 + m_\alpha^2 +
     m_\gamma^2)({\bf{p}}^2+m_\beta^2)-m_\delta^4} \, ,
     \label{eq:repot}
\ea
where we have used the fact that, in this frame, $A^{00} = C^{00} = D^{00}=0$ and $B^{00} = {\bf p}^2/\omega^2$.  Unfortunately, from this point forward one must compute this integral numerically except in some special limiting cases \cite{Dumitru:2007hy,Strickland:2011aa}.

\section{The HL Feynman photon self-energy}
Starting from the relation (\ref{eq:PiF0}) and using $S_{R,F,A}(k)=\slashed{k}\Delta_{R,A,F}(k)$  we have
\ba
\Pi^{\mu\nu}_F(p,\xi)=-2i e^2\int &&\!\!\!\frac{d^4k}{(2\pi)^4} \Big(q^\mu k^\nu+q^\nu k^\mu-\eta^{\mu\nu}q\cdot k\Big)\times\nonumber \\ &&\Big[\Delta_F(k)\Delta_F(q)
-\Big(\Delta_R(k)-\Delta_A(k)\Big)
\Big(\Delta_R(q)-
\Delta_A(q)\Big)\Big] \,.
\label{eq:PiF2}
\ea
Using (\ref{eq:propag}) and
\be
\Delta_R(q)-\Delta_A(q)=-2\pi i\, \mbox{sgn}(q_0)\delta(q^2) \,,
\ee
one obtains the term inside [...] in (\ref{eq:PiF2}) as
\be
[...]= (-2\pi i)^2 \delta(k^2)\delta(q^2)\Big[(1-2f_F({\bf k}))(1-2f_F({\bf q}))-{\rm sgn}(k_0){\rm sgn}(q_0)\Big]\,.
\ee
Also by definition
\ba
\delta(k^2)=\frac{1}{2|{\bf k}|} \Big(\delta(k_0-|{\bf k}|)+\delta(k_0+|{\bf k}|)\Big)\,, \\
\delta(q^2)=\delta(k_0^2-{\bf k}^2+p^2-2k_0\omega+2{\bf k}\cdot{\bf p})\,.
\ea
Using the relations above and performing the integral over $k_0$, and setting ${\bf k}\rightarrow -{\bf k}$ in the negative-energy contribution one has
\ba
\frac{\Pi^{\mu\nu}_F(p,\xi)}{2i\pi e^2}\!&=&\! \int_\text{on shell} \frac{d^3{\bf k}}{(2\pi)^3|{\bf k}|} (q^\mu k^\nu+q^\nu k^\mu-\eta^{\mu\nu}q\cdot k)\delta(p^2-2|{\bf k}| \omega+2{\bf k\cdot p})\times \nonumber \\ &&\hspace{3cm}\Big[(1-2f_F({\bf k}))(1-2f_F({\bf q}))-{\rm sgn}(|{\bf k}|){\rm sgn}(|{\bf k}|-\omega) \Big]\nonumber \\
&+& \int_\text{on shell} \frac{d^3{\bf k}}{(2\pi)^3|{\bf k}|} (\bar{q}^\mu k^\nu+\bar{q}^\nu k^\mu-\eta^{\mu\nu}\bar{q}\cdot k)\delta(-p^2-2|{\bf k}| \omega+2{\bf k\cdot p})\times \nonumber \\ &&\hspace{3cm}\Big[(1-2f_F({\bf k}))(1-2f_F({\bf q}))-{\rm sgn}(|{\bf k}|){\rm sgn}(|{\bf k}|+\omega) \Big]\!, \;\;
\ea
where   $\bar{q}\equiv k+p$.
The Feynman photon self-energy in the HL limit is
\be 
\Pi^{\mu\nu}_F(p,\xi)= \frac{16i\pi e^2}{{\bf |p|}}\int \frac{d^3{\bf k}}{(2\pi)^3}
v^\mu v^\nu f_F({\bf k})(f_F({\bf k})-1)  \delta\left({\bf v\cdot {\hat p}}-\frac{\omega}{|{\bf p}|}\right)\,.
\label{eq:PiF3}
\ee
By specializing to the anisotropic distribution function, we can simplify $\Pi_F^{\mu\nu}$ a bit more. Using an anisotropic Fermi-Dirac distribution (\ref{eq:FB-dist}) and Eq.~(\ref{eq:dist-identity2})  
\ba
e^2\int_0^\infty \!\!d|{\bf k}|{\bf k}^2 f_F({\bf k})(f_F({\bf k})-1)=
\frac{\lambda e^2}{\sqrt{1+\xi ({\bf v\cdot n})^2}}\int_0^\infty \!\!d|{\bf k}|{\bf k}^2\frac{\partial f_F({\bf k})}{\partial|{\bf k}|}= \frac{-\lambda \pi^2m_D^2}{2(1+\xi ({\bf v\cdot n})^2)^{3/2}}  . \nonumber
\ea
Using this identity, (\ref{eq:PiF3}) becomes
\be
\Pi^{\mu\nu}_F(p,\xi)=-\frac{i\lambda  m_D^2}{|{\bf p}|} \int d\Omega \frac{v^\mu v^\nu}{(1+\xi ({\bf v\cdot n})^2)^{3/2}} \delta\Big({\bf v\cdot \hat{p}}-\frac{\omega}{|\bf p|}\Big) \,.
\label{eq:PiF4}
\ee
%

\section{The HL Feynman gluon self-energy}
Once more, the expression for the HL Feynman photon self-energy (\ref{eq:PiF4}) can be used to obtain the HL Feynman gluon self-energy provided that the Debye mass is replaced by (\ref{eq:qcd-debye}). The spatial block of Feynman gluon self-energy is
\be
\Pi^{ij}_F(p,\xi)=-\frac{i\lambda  m_D^2}{|{\bf p}|} \int d\Omega \frac{v^i v^j}{(1+\xi ({\bf v\cdot n})^2)^{3/2}} \delta\Big({\bf v\cdot \hat{p}}-\frac{\omega}{|\bf p|}\Big) \,.
\label{eq:PiFij}
\ee
 The integral above can be solved analytically as a function of ${\bf p}$, $\omega$, and $\xi$. For this purpose, it is more convenient to take ${\bf \hat{p}}$ along $z$-axis and ${\bf n}$ in the $x-z$ plane characterized by \mbox{$(\theta_n,\phi_n)=(\arctan(p_x/p_z),\pi)$}, where $p_x \equiv |{\bf p} - ({\bf p}\cdot{\bf n}) {\bf n}|$ and $p_z \equiv {\bf p} \cdot {\bf n}$ are the perpendicular and parallel components of ${\bf p}$ respect to ${\bf n}$ . Using this setup one has
\ba
{\bf n}&=&\left(-\frac{p_x}{|{\bf p}|},0,\frac{p_z}{|{\bf p}|}\right);\:\quad {\bf v}=(\sin\theta\cos\phi,\sin\theta\sin\phi,\cos\theta);\:\quad \hat{\bf p}=(0,0,1)\,,
\label{eq:def2}
\ea
which gives
\ba
 {\bf v\cdot n}&=&-\frac{p_x}{|{\bf p}|}\sin\theta\cos\phi+\frac{p_z}{|{\bf p}|}\cos\theta\,, \\
 {\bf v\cdot \hat{p}}&=&\cos\theta\,.
\ea
Now by defining 
\be
{\bf u}=\left(\sqrt{1-\frac{\omega^2}{{\bf p}^2}}\cos\phi,\sqrt{1-\frac{\omega^2}{{\bf p}^2}}\sin\phi,\frac{\omega}{|{\bf p}|}\right)\,,
\ee
and taking the integral over $\theta$ in (\ref{eq:PiFij}) we have
\be
\Pi^{ij}_F(p,\xi)=-\frac{i\lambda  m_D^2}{|{\bf p}|} \Theta(|{\bf p}|^2-\omega^2) \int d\phi \frac{{\bf u}_i {\bf u}_ j}{[1+\xi ({\bf u\cdot n})^2]^{3/2}} \,.
\ee
Note that, for ease of calculation, the integral form of Feynman self-energy is performed in a coordinate system that is different from the one used for retarded and advanced self-energies (\ref{eq:def1}). This is mathematically sound, because we only use the new coordinate to calculate the scalars, i.e. expansion coefficients of $\Pi^{\mu\nu}_F$, which are coordinate independent (\ref{eq:abgd}). Finally, one can use the matrices  $A$, $B$, $C$, and $D$ calculated in original coordinate (\ref{eq:def1}) to construct $\Pi_F^{\mu\nu}$. Note that the tensor $A$ as appears in (\ref{eq:abgd}) should be defined in new frame (\ref{eq:def2}).
\subsection*{static limit}
The HL tensor expansion coefficients for Feynman self-energy in the static limit are
\ba
\lim_{\omega \rightarrow 0} \alpha_F&=& -\frac{4 i \lambda m_D^2}{|{\bf p}| \varsigma} \Big[ E(-\varsigma) - K(-\varsigma) \Big] ,\nonumber \\
\lim_{\omega \rightarrow 0} \beta_F&=& -\frac{4 i \lambda m_D^2 \omega^2}{{\bf p}^3 (1+\varsigma)} E(-\varsigma) \, ,\nonumber \\
\lim_{\omega \rightarrow 0} \gamma_F&=& -\frac{4 i \lambda m_D^2}{|{\bf p}|} \bigg[ \frac{2}{\varsigma} K(-\varsigma) - \frac{2+\varsigma}{\varsigma(1+\varsigma)} E(-\varsigma) \bigg] ,\nonumber \\
\lim_{\omega \rightarrow 0} \delta_F&=& -\frac{4 i \lambda m_D^2 \omega^2 p_z }{{\bf p}^3 p_x^2 (1+\varsigma)} \bigg[ \frac{1-\varsigma}{1+\varsigma} E(-\varsigma) - K(-\varsigma) \bigg] ,
\ea
where $\varsigma \equiv \xi p_x^2/{\bf p}^2$, and $K$ and $E$ are complete elliptic integrals of the first and second kind, respectively, defined by 
\ba
K(x)&\equiv& \int _0^{\pi/2}\frac{1}{\sqrt{1-x\sin^2\phi}} \,d\phi\,,\nonumber \\
E(x)&\equiv& \int _0^{\pi/2}\sqrt{1-x\sin^2\phi}\,\,d\phi\,.
\ea
In the isotropic case, the relations above simplify to
\ba
\lim_{\xi\rightarrow 0}  \lim_{\omega\rightarrow 0} \alpha_F &\rightarrow & -\frac{ i \pi \lambda m_D^2}{|{\bf p}|} \,, \nonumber \\
\lim_{\xi\rightarrow 0}  \lim_{\omega\rightarrow 0} \beta_F &\rightarrow & -\frac{2 i \pi \lambda m_D^2 \omega^2}{{\bf p}^3}\,, \nonumber \\
\lim_{\xi\rightarrow 0}  \lim_{\omega\rightarrow 0} \gamma_F &\rightarrow & 0\,, \nonumber \\
\lim_{\xi\rightarrow 0}  \lim_{\omega\rightarrow 0} \delta_F &\rightarrow & 0\,.
\ea
The first two agree with the isotropic results given by Eqs.~(20) and (19) of Ref.~\cite{Carrington:1997sq} for $\Pi^T_F$ and $\Pi^L_F$, respectively, upon using $\beta_F = (\omega^2/{\bf p}^2) \Pi^L_F$ \cite{Romatschke:2003ms} and $m^2_D$ is replaced by its QED definition, i.e. Eq.~(\ref{eq:mDQED}).

\section{The hard-loop Feynman gluon propagator}

In this section, we obtain the static limit ($\omega\rightarrow 0$) of $\tilde{\cal D}^{00}_F$ using Eq.~(\ref{eq:2b8})
\ba
\tilde{\cal D}_{F}(p)&=& (1+2f_B({\bf p}))\, \mbox{sgn}(\omega)\,
[\tilde{\cal D}_{R}(p)-\tilde{\cal D}_{A}(p)]
\nonumber \\
&+&\tilde{\cal D}_{R}(p)\,\{\Pi _F(p)-(1+2f_B({\bf p}))\, \mbox{sgn}(\omega)\, [\Pi
_R(p)-\Pi _A(p)]\} \,  \tilde{\cal D}_A(p)~. \label{eq:2b88}
\ea
Taking the $00$ component using (\ref{eq:ABCD}) and (\ref{eq:Dexp}), the first term becomes
\be
 \left[1+2f_B({\bf p})\right]\, \mbox{sgn}(\omega) (\tilde{\cal D}_R^{00}-\tilde{\cal D}_A^{00})=\left[1+2f_B({\bf p})\right]\, \mbox{sgn}(\omega) (\beta'_R-\beta'_A)\frac{{\bf p}^2}{\omega^2}\,,
 \label{eq:trm1}
\ee
where considering (\ref{eq:ABCD}) we have used $A^{00}=C^{00}=D^{00}=0$ and $B^{00} = {\bf p}^2/\omega^2$.
Using (\ref{eq:PiRcoef}), one can write the tensor basis coefficients of $\Pi_{R/A}$ as
\ba
\alpha_{R/A}&=&\alpha_0 \pm  i \omega \alpha_1\,, \nonumber\\
\beta_{R/A}&=&(\beta_0 \pm  i \omega \beta_1)\omega^2\,,\nonumber \\
\gamma_{R/A}&=&\gamma_0 \pm  i \omega \gamma_1\,, \nonumber \\
\delta_{R/A}&=&(\pm i\delta_0 + \omega \delta_1)\omega\,,
\label{eq:PiRcoef2}
\ea
where all new coefficients with subscripts `0' and `1' are independent of $\omega$ and can be easily read off from (\ref{eq:PiRcoef}). Specializing to the anisotropic distribution function, Eq.~(\ref{eq:FB-dist}), yields
\be
\lim_{\omega \rightarrow 0} (1+2f_B({\bf p}))\, \mbox{sgn}(\omega) \approx \frac{2 \lambda}{\sqrt{1+\xi {\bf (v\cdot n)}^2}}\frac{1}{\omega}+{\cal O}(\omega^0) \, .
\ee
Using relations (\ref{eq:coefPrime}) for $\beta'_{R/A}$, one finds  $ (1+2f_B)\, \mbox{sgn}(\omega)(\tilde{\cal D}_R^{00}-\tilde{\cal D}_A^{00})$ in the static limit 
\ba
 &\lim_{\omega \rightarrow 0} (1+2f_B({\bf p}))\mbox{sgn}(\omega)(\tilde{\cal D}_R^{00}-\tilde{\cal D}_A^{00})= \nonumber \\
 &\hspace{10mm}\frac{4i\lambda}{{\bf p}^2\sqrt{1+\xi {\bf (v\cdot n)}^2}}\frac{-p_x^2(\alpha_1+\gamma_1)\delta_0^2+({\bf p}^2+\alpha_0+\gamma_0)
 \left(\beta_1({\bf p}^2+\alpha_0+\gamma_0)-2p_x^2\delta_0\delta_1\right)}{\left((\beta_0-1)({\bf p}^2+\alpha_0+\gamma_0)+p_x^2\delta_0^2\right)^2}\,. \label{eq:term1}
\ea
Now, we turn to the next term, $(\tilde{\cal D}_R\Pi_F \tilde{\cal D}_A)^{00}$ which, in the static limit, becomes 
\ba
&\lim_{\omega\rightarrow 0}(\tilde{\cal D}_R \Pi_F \tilde{\cal D}_A)^{00}= \nonumber\\&
\frac{-4i \lambda m_D^2}{[(\beta_0-1)({\bf p}^2+\alpha_0+\gamma_0)+p_x^2 \delta_0^2]^2}\!\!
\left[ \frac{ \xi ( {\bf p}^2+\alpha_0+\gamma_0)^2 - \delta_0^2 {\bf p}^4 }{\xi {\bf p}^3 ({\bf p}^2 + \xi p_x^2)}  E\!\left(\frac{-\xi p_x^2}{{\bf p}^2} \right) +
\frac{\delta_0^2}{|{\bf p}|\xi} K\!\left(\frac{-\xi p_x^2}{{\bf p}^2} \right) 
\right] \!. \label{eq:term2}
\ea
The last term can be calculated similarly using the setup above and relations listed in App.~\ref{app:alternative} 
\ba
&&\lim_{\omega\rightarrow 0}(1+2f_B({\bf p}))\mbox{sgn}(\omega)[\tilde{\cal D}_R(\Pi_R-\Pi_A)\tilde{\cal D}_A]^{00}= \\
&&\hspace{10mm}\frac{-4i\lambda}{{\bf p}^2\sqrt{1+\xi {\bf (v\cdot n)}^2}}\frac{-p_x^2(\alpha_1+\gamma_1)\delta_0^2+({\bf p}^2+\alpha_0+\gamma_0)
 \Big(\beta_1({\bf p}^2+\alpha_0+\gamma_0)-2p_x^2\delta_0\delta_1\Big)}{\Big((\beta_0-1)({\bf p}^2+\alpha_0+\gamma_0)+p_x^2\delta_0^2\Big)^2}\,.
\nonumber  \label{eq:term3}
\ea
From the relations above (\ref{eq:term1}) and (\ref{eq:term3}), one can see that the first and the last terms of Eq.~(\ref{eq:2b88}) cancel each other leaving the second term which, using the parametrizations (\ref{eq:m}),  gives
\ba
&&\lim_{\omega\rightarrow 0}\tilde{\cal D}^{00}_{F}(p,\xi)= \\
&&\hspace{2mm} \frac{4i \lambda m_D^2}{\varsigma{|\bf p}|[({\bf p}^2+m_\beta^2)({\bf p}^2 + m_\alpha^2 + m_\gamma^2)-m_\delta^4]^2}
\left[ \frac{m_\delta^4- \varsigma ( {\bf p}^2+m_\alpha^2 + m_\gamma^2)^2 }{1+\varsigma}  E\!\left(-\varsigma \right) -
m_\delta^4 K\!\left(-\varsigma \right) 
\right]\!. \nonumber \label{eq:term2alt}
\ea
This is our final result for the static Feynman propagator.  Expanding our final result in terms of powers of $\xi$
\be
\lim_{\omega\rightarrow 0} \tilde{\cal D}^{00}_{F}(p,\xi) = - \frac{2 \pi i m_D^2 \lambda}{|{\bf p}| ({\bf p}^2 + m_D^2)^2}+\frac{i\pi m_D^2 \lambda \xi}{6 {\bf p}^3({\bf p}^2+m_D^2)^3}\Big[9{\bf p}^2p_x^2+m_D^2(8{\bf p}^2-15p_x^2)\Big]+{\cal O}(\xi^2) \, ,
\ee
which is in agreement with earlier results obtained in the small-$\xi$ limit \cite{Dumitru:2009fy,Laine:2006ns,Burnier:2009yu}.  Note that, if one expands (\ref{eq:term2alt}) to higher order in $\xi$, one finds increasingly negative powers of $|{\bf p}|$ which result in infrared divergences in the corresponding corrections to the imaginary part of the static heavy quark potential.  The full result (\ref{eq:term2alt}) is, however, infrared safe. 

\section{Pinch singularity}
As mentioned in the introduction, the imaginary part of the heavy-quark potential can be obtained from the Fourier transform of the static limit of the Feynman propagator
\be
V_I({\bf r},\xi) \equiv - \frac{g^2 C_F}{2} \int \frac{d^3{\bf p}}{(2\pi)^3} \left(e^{i {\bf p}\cdot {\bf r}} - 1 \right) 
\tilde{\cal D}^{00}_F \Big|_{\omega \rightarrow 0} \; .
\ee
However, (\ref{eq:term2alt}) contains a pinch singularity which is related to the (chromo-)Weibel instability in momentum-space anisotropic plasmas \cite{Romatschke:2003ms,Mrowczynski:2000ed}.  This pinch singularity causes the imaginary-part of the potential to be ill-defined.  To see that this is the case, we point out that (\ref{eq:term2alt}) can be written more compactly using
\be
({\bf p}^2+m_\beta^2)({\bf p}^2 + m_\alpha^2 + m_\gamma^2)-m_\delta^4 = ({\bf  p}^2 + m_+^2)({\bf  p}^2 + m_-^2) \, ,
\ee
where
\be
2 m_{\pm}^2 = M^2 \pm \sqrt{M^4-4[m_\beta^2(m_\alpha^2+m_\gamma^2)-m_\delta^4]} \; ,
\label{mpm}
\ee
with $M^2 = m_\alpha^2+m_\beta^2+m_\gamma^2$ \cite{Romatschke:2003ms}.  One can show that $m_+^2$ is positive for all $\xi$ and angles of propagation; however, $m_-^2$ can be negative for some propagation angles.  This is illustrated in Fig.~2 of Ref.~\cite{Romatschke:2003ms} and discussed in the surrounding text.  As a result, in unstable regions of phase space, ${\bf p}^2 + m_-^2$ can go to zero.  This occurs already in the integral necessary to obtain the real part of the potential (\ref{eq:repot}); however, in this case there is only one power of ${\bf p}^2 + m_-^2$ in the denominator, which results in a simple pole that can be integrated using a principle part prescription, e.g. ${\bf p}^2 + m_-^2 \rightarrow (|{\bf p}| + |m_-|)(|{\bf p}| - |m_-|)$.  In the case of $V_I$, however, the denominator of the integrand contains $({\bf  p}^2 + m_-^2)^2$, which results in a double pole in the Fourier transform.   To see that this is, in fact, a pinch singularity we note that the prefactor of (\ref{eq:term2alt}) which causes the trouble comes from the product of retarded and advanced propagators, $\Delta_{G}^R \Delta_{G}^A$.  Keeping track of the $i \epsilon$'s, one finds two simple poles shifted by $\pm i\epsilon$ which collapse onto the real axis as $\epsilon \rightarrow 0$, forming a double pole.

\section{Conclusions}

In this chapter, I presented a calculation of the hard-loop resummed retarded, advanced, and Feynman (symmetric) gluon propagators in a momentum-space anisotropic plasma with a single anisotropy direction, ${\bf n}$.  We used the real-time formalism throughout and, when available, we compared to previously obtained results.  Our main new result is an expression for the Feynman gluon propagator which is accurate to all orders in the anisotropy parameter $\xi$ (\ref{eq:term2alt}).  Unlike results obtained using Taylor expansion in $\xi$, (\ref{eq:term2alt}) is infrared finite, however, it possesses a pinch singularity which formally renders the imaginary part of the heavy-quark potential infinite.   The existence of this pinch singularity can be traced back to the existence of unstable modes in a momentum-space anisotropic quark-gluon plasma \cite{Romatschke:2003ms,Mrowczynski:2000ed,Mrowczynski:2016etf}.  

A pinch-singularity emerges because one presumes that the collective modes, which are determined through a linearized treatment, apply at all times.  In an equilibrium (stable) situation the field amplitudes are bounded (and small) and such a treatment makes some sense.  However, in our case, the system is unstable and some subset of the linearized collective modes grow exponentially for all times, which upon taking the static limit ($\omega \rightarrow 0$ or  $t \rightarrow \infty$) results in an infinite effect.  As a result, in the presence of unstable modes this scheme is ill-defined and it seems necessary to impose an upper time limit for unstable mode growth.  At the most conservative, the upper time limit for unstable mode growth would be set by the lifetime of the QGP, however, in practice one finds that plasma instabilities may saturate on a shorter timescale.  

In terms of the calculation presented herein, one could attempt to implement the physics of instability saturation or finite plasma lifetime by imposing an infrared cutoff on the frequency $\omega_0 \sim {\rm max}(\tau^{-1}_{\rm instability},\tau^{-1}_{\rm QGP})$ where $\tau_{\rm instability}$ is the expected timescale for the saturation of unstable field growth and $\tau_{\rm QGP} \sim $ 10 fm/c $\sim$ \mbox{1/(20 MeV)} is the typical lifetime of the quark-gluon plasma.
Detailed simulations of anisotropic non-abelian plasmas in fixed boxes show that unstable exponential growth terminates when the gauge field amplitude reaches the soft scale and the subsequent gauge field dynamics transform into a much slower turbulent cascade of energy from soft scales to hard scales \cite{Arnold:2005ef,Arnold:2005qs,Rebhan:2005re,Strickland:2007fm,
Dumitru:2006pz}.  More recent studies of chromo-Weibel dynamics in an expanding non-Abelian plasmas found that unstable modes saturate on a time scale of 3-4 fm/c at LHC energies \cite{Rebhan:2008uj,Attems:2012js}.  Combined with the QGP lifetime estimate, one has $\omega_0 \sim 20 - 70 $ MeV.  In practice, an infrared cutoff such as this will lift the poles off the real axis, even in the limit $\epsilon \rightarrow 0$ due to the finite imaginary linear correction in $\omega$ to the structure functions $\alpha_{R/A}$, $\beta_{R/A}$, $\gamma_{R/A}$, and $\delta_{R/A}$ listed in (\ref{eq:PiRcoef}).  

While such a phenomenological prescription may work in practice, it introduces a fundamental problem, since the 00-component gluon propagator which is used to define the potential is not gauge invariant for finite $\omega$ (see Eq.~(\ref{eq:tildeDRA})).  For ``reasonable gauges'' the dependence may not be large, but nevertheless this is an unsatisfactory resolution of this problem on general grounds.  For this reason, one should simultaneously pursue the possibility to measure the potential numerically using classical gauge theory simulations similar to those used to measure the imaginary part of the heavy quark potential in the equilibrium limit \cite{Laine:2007qy}.  In this method, one determines the imaginary part of the potential by measuring the classical Wilson loop which amounts to a two-point correlation function of two spatial Wilson lines.  With this method one would be able to obtain a gauge-invariant imaginary part of the potential, however, one still would not be able to take the $t \rightarrow \infty$ limit due to finite computational resources, break-down of the classical hard loop limit, etc.  

Finally, as another path forward, one might consider adding the effect of collisions in the computation of the anisotropic structure functions.  Previous studies \cite{Schenke:2006xu} have shown that at fixed $\xi$, if the collision rate exceeds a certain threshold, then unstable modes are eliminated from the spectrum.  This would provide another way to regulate/eliminate the ill-defined effect of unstable modes in the heavy-quark potential.

\chapter{\bf Quark self-energy in anisotropic hydrodynamics}   
\label{chap:quark}
\setcounter{figure}{0}
\setcounter{table}{0}
\setcounter{equation}{0}
\section{Introduction}
As discussed briefly in the Chap.~\ref{chap:intro}, one useful formalism for analyzing the thermal properties of a hot QGP is finite temperature (thermal) field theory (TFT) \cite{kapusta2006finite}. The real-time formalisms of TFT are relevant for studying the non-equilibrium dynamics \cite{1997ftft.book}.  The  imaginary  part  of  the  self-energy  is  related  to  inverse  decay  rates  which  provides the information  about emission/absorption (enhancement/suppression of production) of particles \cite{weldon1983simple,Bodeker:2015exa}.  For high-temperature plasmas where the medium is thermalized, the hard-thermal-loop (HTL) approximation
has been widely used in order to simplify the analysis of thermodynamics, transport, and collective
behaviour of the QGP \cite{Blaizot:2001nr}.  For non-thermal systems, one can use scale separation to define the
so-called hard-loop (HL) approximation which relaxes the need for thermal equilibrium \cite{Romatschke:2016hle,Romatschke:2003ms,Schenke:2006fz}. 

In this chapter, the momentum-anisotropy of the quark self-energy in an anisotropic QGP will be studied. In prior works, this effect was studied using a spheroidal anisotropic distribution function, with one anisotropy parameter along the longitudinal direction \cite{schenke2006fermionic}. Herein, I extend the formalism to include an ellipsoidal distribution function, with three distinct anisotropy parameters corresponding to deformation of distribution function both in the longitudinal and transverse directions in momentum space. This is important, since an ellipsoidal distribution gives more general and realistic quark distribution function. 
 This work sets the stage for a fully self-consistent calculation of photon production and collective flow from an anisotropic QGP.

\section{Anisotropic quark self-energy}
\label{quarkse}

The general expression for the gauge-independent retarded quark self-energy in a momentum anisotropic system in the hard-loop (HL) approximation is \cite{Mrowczynski:2000ed} 
\ba 
 \Sigma(k) = \frac{C_F}{4} g^2 \int_{\bf p} \frac{f ({\bf p})}{|{\bf p}|} \frac{p \cdot \gamma}{p\cdot k} \, ,\label{retself}
\ea
where $p = (\omega_p,{\bf p})$ and $k = (\omega,{\bf k})$ are the Minkowski-space partonic momentum four-vectors, $C_F \equiv (N_c^2 -1)/2N_c$, $\int_{\bf p} \equiv \int d^3 {\bf p}/(2 \pi)^3 $, $g$ is the QCD coupling, and the distribution function $f ({\bf p}) $ is the sum of the momentum distributions for quark and gluon partons $f ({\bf p}) \equiv 2 \left( n({\bf p}) + \bar n ({\bf p})
\right) + 4 n_g({\bf p})$.  

\section{Ellipsoidal self-energy setup}

Generalizing the setup used in Refs.~\cite{Romatschke:2003ms, Schenke:2006fz}, herein, the local rest frame distribution function $f({\bf p})$ is required to be parametrized by
\be
f({\bf p})=f_{\boldsymbol\xi}({\bf p}) = f_{\rm iso}\left(\frac{1}{\lambda}\sqrt{{\bf p}^2+\xi_x({\bf p}\cdot{\bf \hat x})^2+\xi_y({\bf p}\cdot{\bf \hat y})^2+\xi_z({\bf p}\cdot{\bf \hat z})^2}\right)%
,
\label{squashing}
\ee
where $\hat{x}$, $\hat{y}$, and $\hat{z}$ are Cartesian unit vectors in the local rest frame of the matter, $\pmb{\xi}\equiv(\xi_x,\xi_y,\xi_z)$ are anisotropy parameters corresponding to three spatial dimensions, and $\lambda$ is a temperature-like scale. In this parametrization, $f_{\rm iso}$ is a general isotropic distribution function which reduces to the appropriate equilibrium distribution function in the isotropic equilibrium limit ($\pmb{\xi}=0$).  The anisotropy parameters $\xi_x$ and $\xi_y$ characterize the strength of anisotropy in transverse plane and $\xi_z$ characterizes the strength of anisotropy in the longitudinal direction. In other words, the spherical equal occupation number surfaces (isosurfaces) in momentum-space for the isotropic case transform to ellipsoidal isosurfaces in the anisotropic case. Using Eq.~(\ref{squashing}) one obtains
\ba \label{q-self2}
\Sigma(k) = \frac{m_q^2}{4\pi} \int d\Omega \, \Big(1+\xi_x({\hat {\bf p}}\cdot{\bf \hat x})^2+\xi_y({\hat {\bf p}}\cdot{\bf \hat y})^2+\xi_z({\hat {\bf p}}\cdot{\bf \hat z})^2 \Big)^{-1}
\frac{p \cdot \gamma}{p\cdot k} \, ,
\ea
where
\ba
m_{\rm q}^2 = \frac{g^2 C_F}{8 \pi^2} \int_0^\infty d{|\bf p|} \,
   |{\bf p}| \, f_{\rm iso}\Big(\frac{{\bf p}}{\lambda}\Big) \, .
\ea
As a result, all dependence on the form of the underlying isotropic distribution function is subsumed into the numerical value of $m_q$.

\section{Dirac decomposition and collective modes}

The self-energy (\ref{q-self2}) can be expanded as
\ba \Sigma(k) = \gamma^0 \Sigma_0 + {\boldsymbol\gamma}\cdot{\mathbf
\Sigma}\,,
\ea
where $\gamma^\mu$ are Dirac matrices. The quark collective modes are determined by finding all four-momenta $k$ for which the determinant of the inverse propagator $S$ vanishes
\ba
{\rm det}\,S^{-1} = 0 \; ,
\ea
where
\ba
i S^{-1}(k) &=& \gamma^\mu k_\mu - \Sigma(k) \, , \nonumber \\
&\equiv& \gamma^\mu \Delta_\mu \,,
\ea
with $\Delta(k)\equiv(\omega - \Sigma_0,{\bf k} - {\bf \Sigma})$.
Using the fact that ${\rm det}(\gamma^\mu \Delta_\mu) = (\Delta^\mu \Delta_\mu)^2$ and defining $\Delta_s^2 = {\bf \Delta}\cdot{\bf \Delta}$, the dispersion relations for the quark collective modes becomes 
\ba
\Delta_0 = \pm \Delta_s \, .
\label{fermiondisp}
\ea

\section{Calculation of the ellipsoidal quark self-energy}

I now turn to the explicit calculation of the self-energy (\ref{q-self2}) for an ellipsoidally anisotropic distribution function.   In the high-energy limit, to good approximation, one can ignore the quark bare masses and, as a result, the system is approximately conformal.  

Our method is based on three anisotropy parameters corresponding to two transverse and one longitudinal directions. Expanding the relation (\ref{q-self2}), one finds the following relation
\ba 
\Sigma^i(k) = \frac{m_q^2}{4\pi |{\bf k}|}\!\int_{-1}^{1}\!dx\!\int_0^{2\pi}\!\frac{d\phi}{c_x \cos^2\phi+c_y \sin^2\phi+c_z} \frac{v^i}{a-b\cos\phi-c\sin\phi} \,,
\ea
where the variables $a$, $b$, $c$, $c_x$, $c_y$, and $c_z$, are defined as
\be
\begin{aligned}
a&=\frac{\omega}{|{\bf k}|}-x \cos\theta_k\,,\nonumber\\
b&=\sin\theta_k \cos\phi_k\sqrt{1-x^2}\,,\nonumber \\
c&=\sin\theta_k \sin\phi_k\sqrt{1-x^2}\,,
\end{aligned}
\hspace{1.5cm}
\begin{aligned}
c_x&\equiv \xi_x(1-x^2)\, , \\
c_y&\equiv \xi_y(1-x^2)\, , \\
c_z&\equiv 1+\xi_z x^2\, .
\end{aligned}
\ee
Using partial-fraction decomposition, one can transform the integral over $\phi$ for each component of $\Sigma$ into four non-trivial simpler ones:
\ba
\Sigma^i=\frac{m_{\rm q}^2}{|{\bf k}|}\sum_{j=1}^4 \int_{-1}^1 dx\,n_{j i}(x) {\cal I}_j(x) \ ; \ \ \ \ \ \ \ (i=t,x,y,z)\,.
\label{eq:sigma}
\ea
The ${\cal I}$-functions used here are defined as
\ba
{\cal I}_1(x)&\equiv&\frac{2}{a+r}\sqrt{\frac{a+r}{a-r}}\, ,\\
{\cal I}_2(x)&\equiv&1-\frac{a}{2}{\cal I}_1(x)\, , \\
{\cal I}_3(x)&\equiv&\frac{1}{\sqrt{c_2^2-c_1^2}}\, , \\
{\cal I}_4(x)&\equiv&-c_2 {\cal I}_3(x)+1\, ,
\ea
and 
\ba
r &\equiv&\sqrt{1-x^2}\sin\theta_k\,, \\
c_1&\equiv&c_x-c_y=\xi_a (1-x^2)\,,\\
c_2 &\equiv&c_x+c_y+2c_z=-\xi_b x^2+\xi_x+\xi_y+2\, ,
\ea
with $\xi_a\equiv \xi_x-\xi_y$ and $\xi_b\equiv \xi_x+\xi_y-2\xi_z$.
By defining $e\equiv -re^{i\phi_k}/2$, and $f\equiv -re^{-i\phi_k}/2$, and the following functions,
\ba
{\cal S} &\equiv& a^4 c_1^2 +{\cal R}^2(c_2,c_1)-2a^2c_1{\cal R}(c_1,c_2)\,,\\
{\cal R}(x_1,x_2) &\equiv& 2 x_1 e f -x_2 (e^2+f^2)\, .
\ea
The coefficients $n_{ji}$ used in Eq.~(\ref{eq:sigma}) are defined as
\ba
n_{1t}&=&\frac{e}{{\cal S}}\Big[a^2 c_1 e+f {\cal R}(c_2,c_1)\Big]=n_{1z}/x\, ,\\
n_{2t}&=&\frac{a c_1}{{\cal S}}\Big[e^2-f^2\Big]=n_{2z}/x\, ,\\
n_{3t}&=&\frac{a c_1}{{\cal S}}\Big[a^2c_1 +2f(-c_1 e +c_2 f)\Big]=n_{3z}/x\,, \\
n_{4t}&=&-n_{2t}=n_{4z}/x\,, \\
n_{1x}&=&-\frac{a e}{2{\cal S}}\Big[a^2 c_1+(c_1-c_2)(e^2-f^2)-{\cal R}(c_1,c_2)\Big]\, ,\\
n_{2x}&=&-\frac{(e-f)}{2{\cal S}}\Big[a^2 c_1-{\cal R}(c_2,c_1)\Big]\, ,\\
n_{3x}&=&\frac{1}{2{\cal S}}\bigg[a^2 c_1\Big(2c_2 e -c_1(e+f)\Big)+\Big(2c_2 f-c_1(e+f)\Big){\cal R}(c_2,c_1)\bigg]\,, \\
n_{4x}&=&-n_{2x}\,, \\
n_{1y}&=&-\frac{ia e}{2{\cal S}}\Big[a^2 c_1-(c_1+c_2)(e^2-f^2)-{\cal R}(c_1,c_2)\Big]\,, \\
n_{2y}&=&\frac{-i(e+f)}{2{\cal S}}\Big[a^2 c_1+{\cal R}(c_2,c_1)\Big]\, ,\\
n_{3y}&=&\frac{i}{2{\cal S}}\bigg[a^2 c_1\Big(2c_2 e +c_1(e-f)\Big)+\Big(2c_2 f-c_1(e-f)\Big){\cal R}(c_2,c_1)\bigg]\, ,\\
n_{4y}&=&-n_{2y}\,.
\ea

\section{Results}
\label{resultssec}
In this section, the results for the components of the quark self-energy as a function of phase velocity $\omega/|\bf  k|$ are presented. In what follows, the real and imaginary parts of the four components of the quark self-energy are normalized by the quantity $m_{\rm q}^2/|\bf k|$. Then, for presentation purposes, each individual component of the quark self-energy is scaled by a trivial geometrical factor which depends on the particular component being considered.  Following this scaling procedure, let's consider the following quantities
\ba
\bar{\Sigma}_0&\equiv&\frac{|{\bf k}|\Sigma_0}{m_{\rm q}^2}\,,\\
\bar{\Sigma}_x&\equiv&\frac{1}{\sin\theta_k\cos\phi_k}\frac{|{\bf k}|\Sigma_x}{m_{\rm q}^2}\,,\\
\bar{\Sigma}_y&\equiv&\frac{1}{\sin\theta_k\sin\phi_k}\frac{|{\bf k}|\Sigma_y}{m_{\rm q}^2}\,,\\
\bar{\Sigma}_z&\equiv&\frac{1}{\cos\theta_k}\frac{|{\bf k}|\Sigma_z}{m_{\rm q}^2}\,.
\ea 

\begin{figure}[H]
\centerline{
\includegraphics[width=0.92\linewidth]{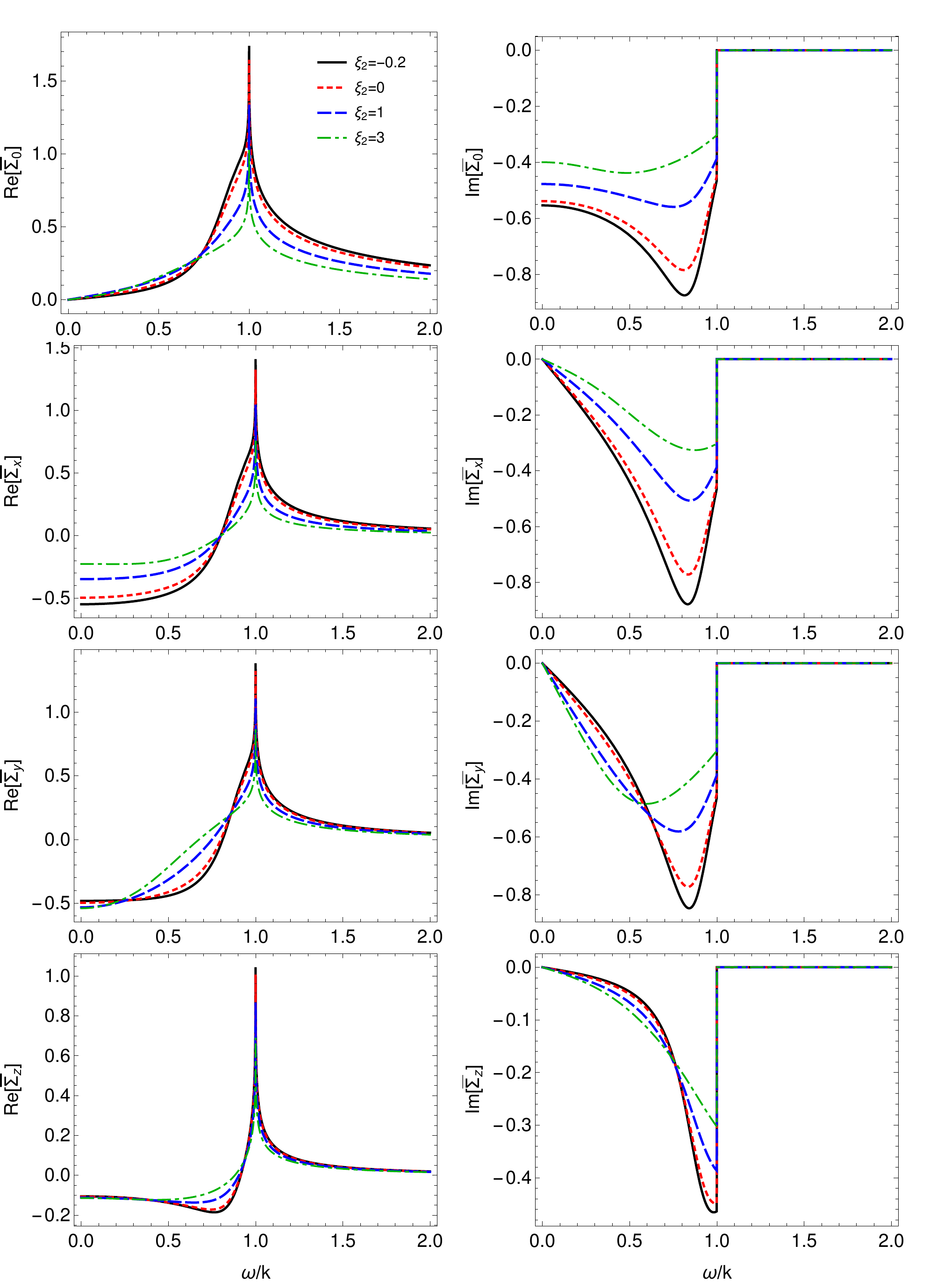}
}
\caption{The real and imaginary parts of $\bar{\Sigma}_0$, $\bar{\Sigma}_x$, $\bar{\Sigma}_y$, and $\bar{\Sigma}_z$ as a function of $\omega/|{\bf k}|$ for $\xi_1=10$, $\theta_k=\pi/3$, $\phi_k=\pi/6$, and $\xi_2=\{-0.2,0,1,3\}$.} 
\label{plot:plot1}
\end{figure}

Generally speaking, one finds that the analytic structure of fermion self-energy is the same as in the anisotropic case, namely that for time-like momenta, $\omega/|{\bf k}| > 1$,  the self-energy is real-valued and for space-like momenta, $\omega/|{\bf  k}|<1$, there is a cut in the complex plane which spans the line $\Im[\omega/|{\bf k}|]=0$. In Fig.~\ref{plot:plot1}, the components of the scaled quark self-energy for $\xi_1=10$, $\theta_k=\pi/3$, $\phi_k=\pi/6$ is presented, while varying the transverse anisotropy parameter with $\xi_2=\{-0.2,0,1,3\}$. As can be seen in this plot, the real part of the components of the quark self-energy tend to zero for large $\omega/|\bf k|$, while the imaginary parts drop to zero abruptly for $\omega/|{\bf k}| > 1$ due to the absence of the Landau cut for time-like momenta. The plots also show that the magnitude of self-energy components depend on the magnitude of the transverse anisotropy, as one can expect on general grounds.

\begin{figure}[H]
\centerline{
\includegraphics[width=0.92\linewidth]{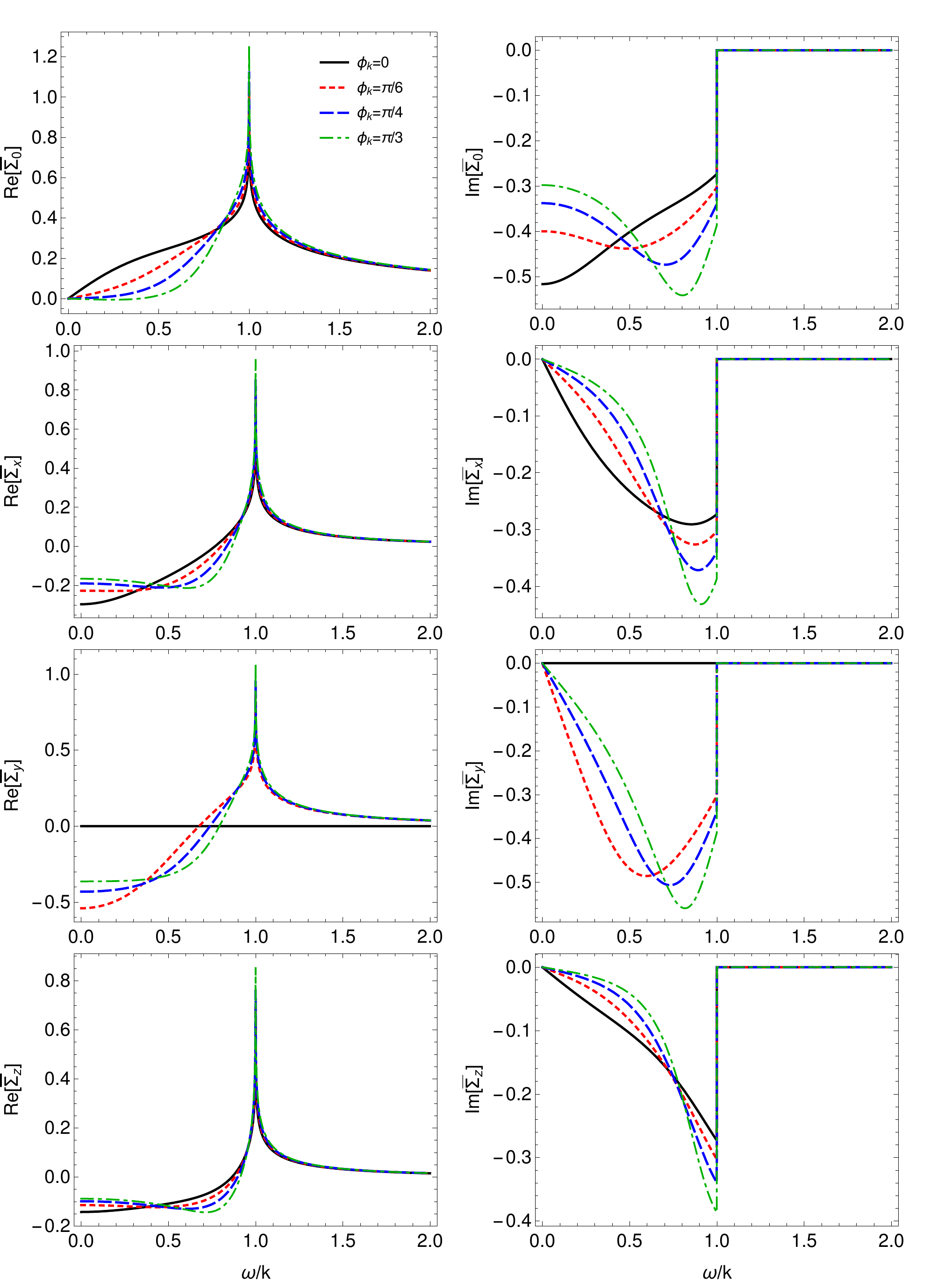}
}
\caption{The real and imaginary parts of $\bar{\Sigma}_0$, $\bar{\Sigma}_x$, $\bar{\Sigma}_y$, and $\bar{\Sigma}_z$ as a function of $\omega/|{\bf k}|$ for $\xi_1=10$, $\xi_2=3$, $\theta_k=\pi/3$, and $\phi_k=\{0,\pi/6,\pi/4,\pi/3\}$.}
\label{plot:plot2}
\end{figure}
 This dependence would be reflected in a photon production rate that possesses explicit azimuthal anisotropies which are independent from those generated solely due to QGP collective flow.  To demonstrate this feature more explicitly, in Fig.~\ref{plot:plot2} I use $\xi_1=10$, $\xi_2=3$, $\theta_k=\pi/3$ and vary the azimuthal scattering angle as $\phi_k=\{0,\pi/6,\pi/4,\pi/3\}$. 

\section{Conclusions}
In this chapter, I present the steps of calculation of the quarks self-energy in an ellipsoidally-anisotropic QGP by using the method of partial-fraction decomposition together with numerical evaluation of the resulting one-dimensional integrals. Previous results for the hard-loop self-energy of quarks in a spheroidally-anisotropic QGP were extended by generalizing the parametrization of the momentum distribution functions to incorporate anisotropies in transverse momentum-space directions. 

 With the introduction of the additional anisotropies in the transverse plane, the calculations become a bit more tedious compared to the case of a spheroidal momentum anisotropy, however, the final results can be expressed as modifications of the previously considered case.  The results show that anisotropies in transverse momentum directions affect the quark self-energy, as can be expected on general grounds and herein I demonstrated how to evaluate the effects quantitatively.  I have shown that the self-energy modifications due to transverse anisotropies induce additional angular dependence of the self-energy in transverse-momentum plane. As a result, there might be observable effects of an ellipsoidal momentum anisotropy in heavy-ion collision experiments.   In particular, the transverse anisotropies can introduce azimuthal angular dependence in the photon production rate, which would result in explicit azimuthal anisotropies in photon production, e.g. elliptic flow, triangle flow, etc.  This source of azimuthal anisotropy is distinct from that induced solely by the collective flow of the QGP itself and is, instead, directly related to viscous effects.

As a demonstration of the underlying source of the effect, I presented the variation of both the real and imaginary parts of the quark self-energy for different combinations of the anisotropy parameters and azimuthal angles. Comparing to previous results obtained in the spheroidal case, in an ellipsoidally anisotropic system one observes modifications to the real part of self-energy which are related to the effective mass of quasi-particles.  As a result, quarks obtain effective masses which depend on their full 3d direction of propagation. I found that the effect on the imaginary part of self-energy, which is related to the decay or production rates of particles, is larger than the effect on the real part. These modifications will affect QGP differential photon production rates.

Looking to the future, the results obtained herein form the basis of a self-consistent calculation of photon production from a QGP as created in relativistic heavy-ion collisions.  The underlying anisotropic formalism guarantees that the photon production rate is positive-definite at all momenta, which is not guaranteed using typical viscous hydrodynamics approaches.  Anisotropic hydrodynamics codes which take into account ellipsoidal anisotropies already exist and the output of the space-time evolution of the momentum-space anisotropies $\pmb{\xi}$, hard-momentum scale $\lambda$, and the collective flow generated during QGP evolution can now be folded together to obtain the final photon spectra including the effect of explicit azimuthal anisotropies in the rate.  This will extend previous works \cite{Schenke:2006yp,Bhattacharya:2015ada} which employed a spheroidal approximation.  The computation of the integrated photon spectra is left to future work.  Finally, I also note that the method of partial-fraction decomposition presented in this paper can also be applied to the gluon polarization tensor in an ellipsoidally anisotropic QGP.

\chapter{\bf Summary and outlook}
\label{chap:summary}
\setcounter{figure}{0}
\setcounter{table}{0}
\setcounter{equation}{0}
\section{Summary}
As discussed previously, the QGP created at heavy-ion collisions experiences significant longitudinal expansion right after strong coherent gluonic fields melt into thermalized QGP. This causes the QGP to have a strong momentum-space anisotropy in the LRF at early times and near the transverse edges of the system, before partonic interactions derive the system toward isotropy. Studying non-equilibrium hydrodynamics  is quite challenging for hydrodynamics models based on standard vHydro approaches, where a perturbative expansion about isotropic state is performed.
My dissertation concerned the formulation and applications of anisotropic hydrodynamics in order to create a more quantitatively reliable non-equilibrium hydrodynamics framework for studying the QGP produced in heavy-ion collisions. Anisotropic hydrodynamics integrates the main anisotropy effects into the leading order term while guaranteeing positivity of the distribution function. As a result, large momentum-space anisotropy resulted from  hydrodynamics expansion will not push the framework out of its range of applicability. In our model, a realistic equation of state based on lattice QCD measurements is included. Anisotropic hydrodynamics using a consistent approach for the equation of states based on quasiparticle model has been shown to be a very successful in describing both experimental data and available exact solutions of the Boltzmann equation.
AHydro can serve as a useful tool even for extreme hydrodynamics systems, i.e. small systems like p-A and systems with strong anisotropy like cold atom systems. 
 My dissertation has two main parts: 
 
 In the first part, in a sequence of chapters I introduced the basic conformal anisotropic hydrodynamics formalism and then explained the ways to include various realistic features, i.e. bulk degree of freedom \cite{Nopoush:2014pfa}, quasiparticle implementation of realistic equation of state \cite{Borsanyi:2010cj}, more realistic collisional kernel \cite{Alqahtani:2015qja}, to make aHydro more reliable to study the hydrodynamics of the QGP generated at heavy-ion collisions. For verification of our model the evolution of the model parameters predicted by aHydro and vHydro, with the exact analytical solution of the conformal Boltzmann equation are compared \cite{Denicol:2014xca}. However, the QGP is not conformal and aHydro needed to be improved to include a prescription for implementing a realistic equation of state which takes care of non-ideal effect in the dynamics. To deal with this problem, a novel method for implementing a realistic equation of state (provided by lattice QCD) in the aHydro formalism is introduced \cite{Alqahtani:2016rth,Alqahtani:2017jwl}. This model, called the quasiparticle aHydro model, self-consistently integrates non-conformal effects in the aHydro model. The non-conformal effects are due to strong interactions of plasma constituents which leads to temperature-dependence of the particles' effective mass in the system. Based on the quasiparticle picture, the quasiparticle aHydro (aHydroQP) model is developed which has all necessary components for studying the phenomenology of the QGP created at heavy-ion collisions. This part of my dissertation is concluded with comparing phenomenological predictions of the aHydroQP model with experimental observations as another benchmark. Comparisons illustrate a high level of consistency between our model and the experimental data \cite{Alqahtani:2017tnq}. 
 
 In the second part, I presented two important applications of  aHydro in field-theoretical measurables in the QGP. In this part of the dissertation, 
I presented the calculation of gluon self-energy in hard loop approximation in an anisotropic QGP \cite{Nopoush:2017zbu}. This gives a more realistic picture for assessing the suppression/enhancement of heavy-quarkonia bound states in an anisotropic QGP. Heavy quarkonia serves as an important probe in the QGP and provides useful information about the temperature of the medium, among other things. The calculations were performed in the real-time formalism framework of the finite temperature field theory using the hard loop approximation. For the hydrodynamics set up spheroidal anisotropic distribution function with one longitudinal anisotropy parameter is used. In the last chapter, I have calculated the quark self-energy in an ellipsoidally anisotropic QGP \cite{Kasmaei:2016apv}. The hydrodynamics set up is based on generalized anisotropic distribution function with two anisotropy parameters. 

The overall conclusion is that anisotropic hydrodynamics has been established as a cutting edge non-equilibrium hydrodynamics model for studying the hydrodynamics systems with strong momentum space anisotropy. Such an anisotropic behavior can be observed in a large varieties of systems with anisotropic distribution function which results from strong directional expansion, existence of external magnetic fields, etc. The aHydro framework integrates the important of anisotropic deformation of distribution function at the leading order. Therefore, it treats the realistic (non-ellipsoidal) deformations as a small perturbation around the leading order (aka NLO aHydro). 

\section{Outlook}
As discussed in this thesis, in reproducing the experimental results and exact solution of Boltzmann equation in non-equilibrium regimes, aHydro has demonstrated that it is more qualitatively reliable than other hydrodynamics frameworks. Nevertheless, heading to the future, one can always improve the model by including more realistic components and relaxing the assumed symmetries.  

One of the missing features in aHydro is off-diagonal elements of anisotropic tensor. Inclusion of these contributions  helps to study systems with ellipsoidal anisotropic distribution function where the ellipse is not oriented along the local rest frame's principal axes. In an ongoing project we are working on this, trying to find a coordinate transformation in which the anisotropy tensor is diagonal. 

Another important missing component is the realistic collisional kernel in the Boltzmann equation. This is crucial to the dynamics since it provides a more accurate picture of the system's out-of-equilibrium behavior, e.g. transport properties, thermal conductivity, viscosity, and etc. In fact, the current prescription for collisional kernel based on linearizion around an equilibrium state (relaxation-time approximation) is perhaps too simple.
One needs to provide aHydro with a collisional kernel based on realistic interactions among particle inside the plasma. Such a kernel should be calculated based on quantum chromodynamics calculations for possible interactions. For instance, for gluon scattering  the possible interactions are 2 $\leftrightarrow$ 2, 2 $\leftrightarrow$ 3, 3 $\leftrightarrow$ 3 and etc. Currently, my colleagues are actively working on this component \cite{Almaalol:2018ynx}.  

One more important component which needs to be added in aHydro phenomenologically is the initial states quantum fluctuations. This effects originates from the randomness of internal quantum structure of nucleons prior to the collisions which strongly affect the evolution of the QGP at later times. Our current model includes the fluctuations only due to random configuration of the targets respect to one another, which results in a distribution of impact parameters for colliding nuclei. 

 In order to simulate the initial state of the target in HICs more realistically, one needs to consider the fact that in real experiments millions of ions from opposite sides collide with a distribution of impact parameters. Our current model is based on one event where two nuclei collide with a single-valued impact parameter. However, the data collection in real URHIC experiments is performed over multiple events with random configurations. This corresponds to simulating multiple events with randomly selected impact parameters.  

As another interesting component to be added to the model, is inclusion of a temperature-dependent shear viscosity to entropy density ratio since in this thesis it was assumed to be constant. Based on prior studies in the context of viscous hydrodynamics~\cite{Denicol:2015nhu}, there is some hope that this will improve the agreement between our model and the experimental data, in particular, with regards to the pseudorapidity dependence of $v_2$.

Additionally, the future plan is to look at different collision energies, e.g.  RHIC 200 GeV collisions \cite{Almaalol:2018gjh} and LHC 5.023 TeV collisions, and different colliding systems, e.g. pA and pp, in the near future.  The application of aHydro in pA and pp is of particular interest, since in these systems viscous hydrodynamics is being pushed to its limits, especially at freeze-out \cite{Nopoush:2015yga}.  

Finally, the gluon self-energy to all order in the anisotropy parameter calculated in Chap. \ref{chap:gluon} ended up with a pinch singularity. As discussed at the end of that chapter, there are some ways to resolve this issue which can be the topic of the future projects. Analyzing the gluon-self energy for an ellipsoidally anisotropic QGP, helps to improve our understanding of heavy-quarkonia bound states in the QGP.

\appendix
													

 \chapter{\bf Units}
 \label{app:units}
In high-energy physics, we adopt the system of natural units, where the velocities and actions are measured in terms of $c$ and $\hbar$, respectively. The logic behind this, is that we will have very simple expressions for relating the variables. In this system of units, energy, momentum, temperature, and mass are measured in terms of GeV (or MeV), length and time are measured in terms of GeV$^{-1}$. 
Using 
\ba
 1\,{\rm fm} &=& 10^{-15}\,{\rm m}\,, \nonumber \\
 1\hspace{.1cm}{\rm eV}&=& 1.602\times 10^{-19} \,\,{\rm J}\,,
 \label{eq:fmev}
 \ea
we have in SI units
 \ba
 h&=&6.626070150(81)\times 10^{-34} \,{\rm Js}\,,\\
 c&=& 2.99792458\times 10^8 \, \,{\rm m/s}\,.
 \ea
 One can calculate the key quantities that frequently appear in the equations
 \ba
 \hbar  &=& 6.5821220(20) \times 10^{-25}\, {\rm GeVs}\,,\\
 \hbar c  &=& 0.197327053(59)\, {\rm GeV\, fm}\,.
 \ea
 In natural units, one sets $\hbar=c=1$ and then obtains some useful conversion relations
 \ba
  1\, {\rm GeV}^{-1} \equiv 0.197\,{\rm fm} \equiv 6.582\times 10^{-25} \,{\rm s}\,.
  \label{eq:mevfm}
 \ea
Using the relations (\ref{eq:fmev}) and (\ref{eq:mevfm}), it is straightforward to convert the energies and spatial/temporal lengths in natural units to SI units. Also, it is interesting to obtain an estimation of the temperature of the QGP in terms of Kelvin. Using the general form of equipartition theorem (which is actually an extension of the virial theorem) for a system of particles whose Hamiltonian has $n$ degrees of freedom is
\ba 
\left< x_m\frac{\partial H}{\partial x_n}\right>=\delta_{mn}\,k_B T\,,
\ea
where the notation $\langle\dots\rangle$ indicates ensemble average, $\delta_{mn}$ is  Kronecker delta, and $k_B$ is Boltzmann's constant.  For an ultra-relativistic gas, e.g. QGP, one has $H_{\rm kin}={\bf p}c=c\sqrt{p_x^2+p_y^2+p_z^2}$ which specifies three degrees of freedom. One obtains 
\ba
\langle H_{\rm kin}\rangle=\sum_{i=x,y,z}\left<p_i\frac{\partial H}{\partial p_i}\right>=3 \,k_B T\,.
\ea
For a typical temperature of QGP, i.e. $T_c \simeq 150$ MeV, and using Boltzmann constant $k_B=1.38\times 10^{-23}$ it yields
 \ba
 T_c \simeq 1.74 \times 10^{12} \,{\rm  K} \,.
 \ea

\chapter{\bf Basic definitions}
\label{app:basicdef}
In this section, some basic definitions and experimental measurable used throughout this dissertation are defined. In the context of heavy-ion experiments, the collision energy is expressed in terms of $\sqrt{s_{NN}}$ which refers to the center of momentum energy per nucleons (inside the nuclei) right before collision. To clarify how it is calculated lets start from mandelstam variables
\ba
s=(p_1+p_2)^2 \,,\hspace{1cm}
t=(p_1-p_3)^2 \,,\hspace{1cm}
u=(p_1-p_4)^2 \,,
\ea
with $p$ being the momentum four-vectors for  $p_1+p_2\rightarrow p_3+p_4$. Because the Mandelstam variables are Lorentz invariant, the choice of coordinate is arbitrary. For a general AB collision (A or B can be heavy nuclei, deuterium, or proton) which collide with energies ($E_A,E_B$), one can find energy per nucleon $E_N$ of the pair as
\ba
E_N=\left(\frac{E_A}{A},\frac{E_B}{B}\right)\,,
\ea
where herein A is the mass number of the collider, i.e. number of nucleons. Typically we deal with ultrarelativistic regimes where $E\approx |{\bf p}|$
\ba
\sqrt{s_{NN}}=\sqrt{\left(E_{Nx}+E_{Ny}\right)^2 -\left({\bf p}_{Nx}-{\bf p}_{Ny}\right)^2}\approx \sqrt{4E_{Nx} E_{Ny}}\,,
\ea
where $E_{Nx}$ indicates the energy per nucleon in collider x. Also, the momentum of particle is projected along and transverse to the beamline direction, i.e. $p_z$ and $p_T$. Using the total energy, momentum, and rest mass, one can obtain the speed of the particle and the relativistic factor $\gamma=(1-v^2)^{-1/2}$. In the natural unit one has
\ba
\frac{\bf p}{E}&=&\frac{\gamma m_0 v}{\gamma m_0}=v\,, \nonumber \\
\frac{E}{m_0}&=&\gamma\,.
\ea 
The spacetime rapidity in high-energy physics is a bit different than in special relativity 
\ba
y=\frac{1}{2}\ln\left(\frac{E+p_z}{E-p_z}\right) \,.
\ea
The pseudorapidity is defined as following
\ba
\eta=\frac{1}{2}\ln\left(\frac{|{\bf p}|+p_z}{|{\bf p}|-p_z}\right)=-\ln\Big[\tan(\theta/2)\Big]\,,
\ea
where $\theta$ is the angle that particle makes with the longitudinal direction (beamline direction). So, by measuring $\theta$ one can calculate $\eta$. Pseudorapidity ranges $(-\infty,\infty)$ for the particle along the beamline moving backward and forward, respectively. For the massless particles one has $\eta=y$.
\chapter{\bf Explicit formulas for derivatives}
\label{app:identities}

In this section, first I introduce the notations used in derivation of the general moment-based hydrodynamics equations and then, by taking the appropriate limits, I simplify them for the transversally-homogeneous 0+1d case. Using the definitions
\ba
D_1&\equiv&\cosh(\vartheta-\varsigma)\partial_\tau+\frac{1}{\tau}\sinh(\vartheta-\varsigma)\partial_\varsigma\nonumber\,, \\
D_2&\equiv&\sinh(\vartheta-\varsigma)\partial_\tau+\frac{1}{\tau}\cosh(\vartheta-\varsigma)\partial_\varsigma\,, \\
\nabla_\perp\cdot{\bf u}_\perp &\equiv&\partial_x u_x+\partial_y u_y \nonumber\,, \\
{\bf u}_\perp\cdot\nabla_\perp &\equiv& u_x\partial_x+u_y\partial_y \nonumber\,,\\
{\bf u}_\perp\times\nabla_\perp &\equiv& u_x \partial_y-u_y\partial_x\,,
\label{eq:identities-gen}
\ea
and four-vectors defined in Eq.~(\ref{eq:4vectors}) one has
\ba
D_u&\equiv&u^\mu \partial_\mu=u_0D_1+{\bf u}_\perp\cdot\nabla_\perp \nonumber\,,\\
D_x&\equiv&X^\mu \partial_\mu=u_\perp D_1+\frac{u_0}{u_\perp}({\bf u}_\perp\cdot\nabla_\perp)\nonumber\,,\\
D_y&\equiv&Y^\mu \partial_\mu=\frac{1}{u_\perp}({\bf u}_\perp\times\nabla_\perp)\nonumber\,,\\
D_z&\equiv&Z^\mu \partial_\mu= D_2\,.
\label{eq:deriv-gen}
\ea
The divergences are defined as
\ba 
\theta_u&\equiv&\partial_\mu u^\mu=D_1 u_0+u_0D_2\vartheta+\nabla_\perp\cdot{\bf u}_\perp\nonumber\,,\\
\theta_x&\equiv&\partial_\mu X^\mu=D_1 u_\perp+u_\perp D_2\vartheta+\frac{u_0}{u_\perp}(\nabla_\perp\cdot{\bf u}_\perp)-\frac{1}{u_0 u_\perp^2}({\bf u}_\perp\cdot\nabla_\perp) u_\perp\nonumber\,,\\
\theta_y&\equiv&\partial_\mu Y^\mu=-\frac{1}{u_\perp}({\bf u}_\perp\cdot\nabla_\perp) \varphi\nonumber\,, \\
\theta_z&\equiv&\partial_\mu Z^\mu=D_1 \vartheta\,,
\label{eq:diverg-gen}
\ea
where $\varphi=\tan^{-1}(u_y/u_x)$. Also one has
\be
\begin{aligned}
u_\mu D_\alpha X^\mu &=\frac{1}{u_0}D_\alpha u_\perp \,, \\
u_\mu D_\alpha Y^\mu &= u_\perp D_\alpha\varphi  \,, \\
u_\mu D_\alpha Z^\mu &= u_0 D_\alpha\vartheta  \,,
\end{aligned}
\hspace{2.5cm}
\begin{aligned}
X_\mu D_\alpha Y^\mu &= u_0 D_\alpha\varphi \,,\\
X_\mu D_\alpha Z^\mu &= u_\perp D_\alpha\vartheta \,, \\
Y_\mu D_\alpha Z^\mu &= 0\,,
\end{aligned}
\label{eq:contractions-basis}
\ee
where $\alpha\in\{u,x,y,z\}$.  Note that the contractions such as $X^\mu D_\alpha u_\mu$ are also non-vanishing, however, such terms can be written in terms of the expressions above by using the orthogonality of the basis vectors, i.e. $D_\alpha(X^\mu u_\mu) = 0$ implies that $X^\mu D_\alpha u_\mu = -u_\mu D_\alpha X^\mu$.

\section{Simplification for 1+1d}
In the case of boost-invariant and cylindrically-symmetric flow one has $\varphi \rightarrow \phi$ and
$\vartheta \rightarrow \varsigma$, where $\varsigma$ is the spatial rapidity. Using $u_\perp\equiv\sinh\theta_\perp$, identities (\ref{eq:deriv-gen}) and (\ref{eq:diverg-gen}) simplify to
\ba
D_u&=&\cosh\theta_\perp \partial_\tau+\sinh\theta_\perp\partial_r\,,\hspace{1cm} D_y=\frac{1}{r}\partial_\phi\,,\nonumber\\
D_x&=&\sinh\theta_\perp\partial_\tau+\cosh\theta_\perp\partial_r\,, \hspace{1cm}D_z=\frac{1}{\tau}\partial_\varsigma\,,\nonumber\\
\theta_u&=&\cosh\theta_\perp\Big(\frac{1}{\tau}+\partial_r\theta_\perp\Big)+\sinh\theta_\perp\Big(\frac{1}{r}+\partial_\tau\theta_\perp\Big) ,\nonumber\\
\theta_x&=&\sinh\theta_\perp\Big(\frac{1}{\tau}+\partial_r\theta_\perp\Big)+\cosh\theta_\perp\Big(\frac{1}{r}+\partial_\tau\theta_\perp\Big) ,\nonumber\\
\theta_y&=&\theta_z=0\,.
\ea
In this limit, the only non-vanishing terms in (\ref{eq:contractions-basis}) are
\be
\begin{aligned}
u_\mu D_u X^\mu &= D_u\theta_\perp \nonumber\,, \\
u_\mu D_x X^\mu &=D_x\theta_\perp \nonumber \,,\\
u_\mu D_y Y^\mu&=\frac{1}{r}\sinh\theta_\perp \nonumber\,,
\end{aligned}
\hspace{2.5cm}
\begin{aligned}
u_\mu D_z Z^\mu&=\frac{1}{\tau}\cosh\theta_\perp \nonumber\,, \\
X_\mu D_y Y^\mu&=\frac{1}{r}\cosh\theta_\perp \nonumber\,, \\
X_\mu D_z Z^\mu&=\frac{1}{\tau} \sinh\theta_\perp\,.
\end{aligned}
\ee
%

\section{Simplification for 0+1d}
For this case, one has $\theta_\perp=0$ and the surviving terms are
\be
\begin{aligned}
D_u&=\partial_\tau\,,\\
D_x&=\partial_r\,,
\end{aligned}
\hspace{1.3cm}
\begin{aligned}
D_y&=\frac{\partial_\phi}{r}\,, \\
D_z&=\frac{\partial_\varsigma}{\tau}\,,
\end{aligned}
\hspace{1.3cm}
\begin{aligned}
&\theta_u=\frac{1}{\tau}\,,\\
&\theta_x=\frac{1}{r}\,,
\end{aligned}
\hspace{1.3cm}
\begin{aligned}
u_\mu D_z Z^\mu&=&\frac{1}{\tau} \nonumber\,, \\
X_\mu D_y Y^\mu&=&\frac{1}{r} \nonumber \,.
\end{aligned}
\ee

\chapter{\bf Special functions}
\label{app:h-functions}

In this section, I provide definitions of the special functions appearing in the body of the text. Starting by introducing $\mathds{C}_1\equiv y^2-1$ and $\mathds{C}_2\equiv y^2+z^2$ and $\mathds{C}^\pm\equiv \mathds{C}_2\pm \mathds{C}_1$
\ba
 {\cal H}_2(y,z) &\equiv&y \int_{-1}^1 dx  \; \sqrt{ \mathds{C}_1x^2 +  \mathds{C}^-}
= \frac{y}{\sqrt{ \mathds{C}_1}} \left[  \mathds{C}^-\tanh^{-1} \sqrt{\frac{ \mathds{C}_1}{ \mathds{C}_2}} + \sqrt{ \mathds{C}_1 \mathds{C}_2} \, \right] ,
\label{eq:H2}
\ea
\ba
{\cal H}_{2T}(y,z) &\equiv& y \, \int\limits_{-1}^1 \frac{dx (1-x^2) }{\, \sqrt{ \mathds{C}_1x^2 +  \mathds{C}^-}} 
= \frac{y}{ \mathds{C}_1^{3/2}}\left[ \mathds{C}^+ \tanh^{-1}\sqrt{\frac{ \mathds{C}_1}{ \mathds{C}_2}}
-\sqrt{ \mathds{C}_1 \mathds{C}_2} \right] , \hspace{1cm}
\label{eq:H2T}
\ea
\ba
{\cal H}_{2L}(y,z) &\equiv& y^3 \, 
 \int\limits_{-1}^1 \frac{ dx\,x^2 }{\, \sqrt{ \mathds{C}_1x^2 +  \mathds{C}^-}}
= \frac{y^3}{ \mathds{C}_1^{3/2}}
\left[
\sqrt{ \mathds{C}_1 \mathds{C}_2}- \mathds{C}^-
\tanh^{-1}\sqrt{\frac{ \mathds{C}_1}{ \mathds{C}_2}} \,\,\right]. 
\label{eq:H2L}
\ea
Derivatives of these functions satisfy the following relations
\ba
\frac{\partial {\cal H}_2(y,z)}{\partial y}&=&\frac{1}{y}\Big[{\cal H}_2(y,z)+{\cal H}_{2L}(y,z)\Big] , \\
\frac{\partial {\cal H}_2(y,z)}{\partial z}&=&\frac{1}{z}\Big[{\cal H}_2(y,z)-{\cal H}_{2L}(y,z)-{\cal H}_{2T}(y,z)\Big] .
\ea
 
\section{Massive Case} 
\label{subapp:h-functions-1}

The ${\cal H}$-functions appearing in the body of the text are  
\ba
{\cal H}_3({\boldsymbol\alpha},\hat{m}) &\equiv &  \tilde{N} \alpha_x \alpha_y
\int_0^{2\pi} d\phi \, \alpha_\perp^2 \int_0^\infty d\hat{p} \, \hat{p}^3  f_{\rm eq}\!\left(\!\sqrt{\hat{p}^2 + \hat{m}^2}\right) {\cal H}_2\!\left(\frac{\alpha_z}{\alpha_\perp},\frac{\hat{m}}{\alpha_\perp \hat{p}} \right),
\label{eq:h3gen}
\\
{\cal H}_{3x}({\boldsymbol\alpha},\hat{m}) &\equiv &  \tilde{N}\alpha_x^3 \alpha_y
\int_0^{2\pi} d\phi \, \cos^2\phi \int_0^\infty d\hat{p} \, \hat{p}^3  f_{\rm eq}\!\left(\!\sqrt{\hat{p}^2 + \hat{m}^2}\right) {\cal H}_{2T}\!\left(\frac{\alpha_z}{\alpha_\perp},\frac{\hat{m}}{\alpha_\perp \hat{p}} \right),
\;\;\;\;
\label{eq:h3xgen}
\\
{\cal H}_{3y}({\boldsymbol\alpha},\hat{m}) &\equiv &  \tilde{N}\alpha_x \alpha_y^3
\int_0^{2\pi} d\phi \, \sin^2\phi \int_0^\infty d\hat{p} \, \hat{p}^3  f_{\rm eq}\!\left(\!\sqrt{\hat{p}^2 + \hat{m}^2}\right) {\cal H}_{2T}\!\left(\frac{\alpha_z}{\alpha_\perp},\frac{\hat{m}}{\alpha_\perp \hat{p}} \right) ,
\label{eq:h3ygen}
\\
{\cal H}_{3T}({\boldsymbol\alpha},\hat{m}) &\equiv &  \frac{1}{2} \Big[ {\cal H}_{3x}({\boldsymbol\alpha},\hat{m}) + {\cal H}_{3y}({\boldsymbol\alpha},\hat{m}) \Big] ,
\label{eq:h3tgen}
\ea 
\ba
{\cal H}_{3L}({\boldsymbol\alpha},\hat{m}) &\equiv &  \tilde{N} \alpha_x \alpha_y
\int_0^{2\pi} d\phi \, \alpha_\perp^2 \int_0^\infty d\hat{p} \, \hat{p}^3  f_{\rm eq}\!\left(\!\sqrt{\hat{p}^2 + \hat{m}^2}\right) {\cal H}_{2L}\!\left(\frac{\alpha_z}{\alpha_\perp},\frac{\hat{m}}{\alpha_\perp \hat{p}} \right) ,
\label{eq:h3lgen} 
\\
{\cal H}_{3m}({\boldsymbol\alpha},\hat{m}) &\equiv &  \tilde{N}\alpha_x\alpha_y\hat{m}^2
\int_0^{2\pi} d\phi \, \alpha_\perp^2\int_0^\infty d\hat{p} \,\hat{p}^3 \frac{f_{\rm eq}\!\left(\!\sqrt{\hat{p}^2 + \hat{m}^2}\right)}{\sqrt{\hat{p}^2+\hat{m}^2}}   {\cal H}_2\!\left(\frac{\alpha_z}{\alpha_\perp},\frac{\hat{m}}{\alpha_\perp \hat{p}} \right)\!,
\label{eq:h3mgen} 
\\
{\cal H}_{3B}({\boldsymbol\alpha},\hat{m}) &\equiv &  \tilde{N}\alpha_x\alpha_y
\int_0^{2\pi} d\phi\int_0^\infty d\hat{p} \, \hat{p} f_{\rm eq}\!\left(\!\sqrt{\hat{p}^2 + \hat{m}^2}\right) {\cal H}_{\rm 2B}\!\left(\frac{\alpha_z}{\alpha_\perp},\frac{\hat{m}}{\alpha_\perp \hat{p}} \right) ,
\label{eq:h3Bgen} 
\\
\Omega_T({\boldsymbol\alpha},\hat{m}) &\equiv& {\cal H}_3+{\cal H}_{3T}\,,  
\label{eq:omgt}\\
\Omega_L({\boldsymbol\alpha},\hat{m}) &\equiv& {\cal H}_3+{\cal H}_{3L}\,,  
\label{eq:omgl}\\
\Omega_m({\boldsymbol\alpha},\hat{m}) &\equiv& {\cal H}_3-{\cal H}_{3L}-2{\cal H}_{3T}-{\cal H}_{3m}\,,
\label{eq:omgm}
\ea
where $\alpha_\perp^2 \equiv \alpha_x^2 \cos^2\phi  + \alpha_y^2 \sin^2\phi $ and
\be
{\cal H}_{2B}(y,z)\equiv {\cal H}_{2T}(y,z)+ \frac{{\cal H}_{2L}(y,z)}{y^2}=\frac{2}{\sqrt{y^2-1}}\tanh^{-1} \sqrt{\frac{y^2-1}{y^2+z^2}} \, .
\ee
In 0+1d case one has $\alpha_x = \alpha_y$ such that $\alpha_\perp = \alpha_x$ and $\tilde{\cal H}_{3T} \equiv  \tilde{\cal H}_{3x}  = \tilde{\cal H}_{3y}$, so one obtains
\ba
\tilde{\cal H}_3({\boldsymbol\alpha},\hat{m}) &\equiv&  2 \pi \tilde{N} \alpha_x^4
\int_0^\infty d\hat{p} \, \hat{p}^3  f_{\rm eq}\!\left(\!\sqrt{\hat{p}^2 + \hat{m}^2}\right) {\cal H}_2\!\left(\frac{\alpha_z}{\alpha_x},\frac{\hat{m}}{\alpha_x\hat{p}} \right) ,
\label{eq:h3tilde}
\\
\tilde{\cal H}_{3T}({\boldsymbol\alpha},\hat{m}) &\equiv&  \pi \tilde{N} \alpha_x^4
\int_0^\infty d\hat{p} \, \hat{p}^3  f_{\rm eq}\!\left(\!\sqrt{\hat{p}^2 + \hat{m}^2}\right) {\cal H}_{2T}\!\left(\frac{\alpha_z}{\alpha_x},\frac{\hat{m}}{\alpha_x\hat{p}} \right) ,
\label{eq:h3ttilde}
\\
\tilde{\cal H}_{3L}({\boldsymbol\alpha},\hat{m}) &\equiv&  2 \pi \tilde{N} \alpha_x^4
\int_0^\infty d\hat{p} \, \hat{p}^3  f_{\rm eq}\!\left(\!\sqrt{\hat{p}^2 + \hat{m}^2}\right) {\cal H}_{2L}\!\left(\frac{\alpha_z}{\alpha_x},\frac{\hat{m}}{\alpha_x\hat{p}} \right) ,
\label{eq:h3ltilde}
\\
\tilde{\cal H}_{3m}({\boldsymbol\alpha},\hat{m}) &\equiv&  2 \pi \tilde{N} \alpha_x^4 \hat{m}^2
\int_0^\infty d\hat{p} \, \hat{p}^3  \frac{f_{\rm eq}\!\left(\!\sqrt{\hat{p}^2 + \hat{m}^2}\right)}{\sqrt{\hat{p}^2 + \hat{m}^2}} {\cal H}_2\!\left(\frac{\alpha_z}{\alpha_x},\frac{\hat{m}}{\alpha_x\hat{p}} \right) ,
\label{eq:h3mtilde}
\\
\tilde{\cal H}_{3B}({\boldsymbol\alpha},\hat{m}) &\equiv &  2\pi \tilde{N}\alpha_x^2
\int_0^\infty d\hat{p} \, \hat{p} f_{\rm eq}\!\left(\!\sqrt{\hat{p}^2 + \hat{m}^2}\right) {\cal H}_{\rm 2B}\!\left(\frac{\alpha_z}{\alpha_x},\frac{\hat{m}}{\alpha_x \hat{p}} \right) .
\label{eq:h3Btilde} 
\ea
For $\tilde{\Omega}_T$, $\tilde{\Omega}_L$, and $\tilde{\Omega}_m$ one should only replace ${\cal H}$-functions with $\tilde{\cal H}$-functions in (\ref{eq:omgt})-(\ref{eq:omgm}). Also, derivatives of $\tilde{\cal H}_3$ satisfy
\ba
\frac{\partial \tilde{\cal H}_3}{\partial\alpha_x}=\frac{2}{\alpha_x}\tilde\Omega_T\,,\hspace{1.5cm}
\frac{\partial \tilde{\cal H}_3}{\partial\alpha_z}=\frac{1}{\alpha_z}\tilde\Omega_L\,,\hspace{1.5cm}
\frac{\partial \tilde{\cal H}_3}{\partial\hat{m}}=\frac{1}{\hat{m}}\tilde\Omega_m\,.
\ea
For the isotropic equilibrium case, one has $\alpha_i\rightarrow 1$, $\lambda\rightarrow T$, and $\hat{m}\rightarrow\hat{m}_{\rm eq}$
\ba
\tilde{\cal H}_{3,\rm eq}(\hat{m}_{\rm eq}) &=&  4 \pi \tilde{N} \hat{m}^2_{\rm eq}\Big[\hat{m}_{\rm eq}K_1(\hat{m}_{\rm eq})+3K_2(\hat{m}_{\rm eq})\Big] \,,
\label{eq:h3iso}
\\
\tilde{\cal H}_{3T,\rm eq}(\hat{m}_{\rm eq}) &=& \tilde{\cal H}_{3L,\rm eq}(\hat{m}_{\rm eq}) =4 \pi \tilde{N} \hat{m}^2_{\rm eq}K_2(\hat{m}_{\rm eq}) \,,
\label{eq:h3tiso}
\\
\tilde{\cal H}_{3m,\rm eq}(\hat{m}_{\rm eq}) &=& 4 \pi \tilde{N} \hat{m}^4_{\rm eq}K_2(\hat{m}_{\rm eq}) \,.
\label{eq:h3miso}
\ea
%

\section{Massless Case}
\label{subapp:h-functions-2}
Taking the massless limit of Eqs.~(\ref{eq:h3gen}) - (\ref{eq:h3Bgen}) one obtains
\ba
\hat{{\cal H}}_3 ({\boldsymbol\alpha})&\equiv &\lim_{m\rightarrow 0}{\cal H}_3({\boldsymbol\alpha},\hat{m}) =  6\tilde{N} \alpha_x \alpha_y
\int_0^{2\pi} d\phi \, \alpha_\perp^2 \bar{{\cal H}}_2\Big(\frac{\alpha_z}{\alpha_\perp} \Big) ,
\label{eq:h30}
\\
\hat{{\cal H}}_{3x} ({\boldsymbol\alpha})&\equiv &\lim_{m\rightarrow 0}{\cal H}_{3x}({\boldsymbol\alpha},\hat{m}) =  6 \tilde{N}\alpha_x^3 \alpha_y
\int_0^{2\pi} d\phi \, \cos^2\phi \, \bar{{\cal H}}_{2T}\Big(\frac{\alpha_z}{\alpha_\perp} \Big) ,
\label{eq:h3x0}
\\
\hat{{\cal H}}_{3y} ({\boldsymbol\alpha})&\equiv &\lim_{m\rightarrow 0}{\cal H}_{3y}({\boldsymbol\alpha},\hat{m}) =  6 \tilde{N}\alpha_x \alpha_y^3
\int_0^{2\pi} d\phi \, \sin^2\phi \, \bar{{\cal H}}_{2T}\Big(\frac{\alpha_z}{\alpha_\perp} \Big) ,
\label{eq:h3y0}
\\
\hat{{\cal H}}_{3L} ({\boldsymbol\alpha})&\equiv &\lim_{m\rightarrow 0}{\cal H}_{3L}({\boldsymbol\alpha},\hat{m}) =  6\tilde{N} \alpha_x \alpha_y
\int_0^{2\pi} d\phi \, \alpha_\perp^2 \bar{{\cal H}}_{2L}\Big(\frac{\alpha_z}{\alpha_\perp} \Big) ,
\label{eq:h3l0}
\\
\hat{{\cal H}}_{3m} ({\boldsymbol\alpha})&\equiv &\lim_{m\rightarrow 0}{\cal H}_{3m}({\boldsymbol\alpha},\hat{m})=0\,,
\label{eq:h3m0}
\ea
where $\bar{{\cal H}}_{2,2T,2L}(y) \equiv {\cal H}_{2,2T,2L}(y,0)$.
In the transversally-symmetric case, $\alpha_x=\alpha_y$ and $\bar{\cal H}_{3T} \equiv  \bar{\cal H}_{3x}  = \bar{\cal H}_{3y}$, and the functions above simplify to
\ba
\bar{{\cal H}}_3 ({\boldsymbol\alpha}) &=&  12\pi\tilde{N} \alpha_x^4
 \bar{{\cal H}}_2\Big(\frac{\alpha_z}{\alpha_x} \Big)\xrightarrow[]{\rm eq\,\,}24\pi\tilde{N} ,
\label{eq:h30-trans}
\\
\bar{{\cal H}}_{3T} ({\boldsymbol\alpha})&=&  6\pi \tilde{N}\alpha_x^4  \bar{{\cal H}}_{2T}\Big(\frac{\alpha_z}{\alpha_x} \Big)\xrightarrow{\rm eq\,\,}8\pi\tilde{N} ,
\label{eq:h3t0-trans}
\\
\bar{{\cal H}}_{3L} ({\boldsymbol\alpha}) &=&  12\pi\tilde{N} \alpha_x^4  \bar{{\cal H}}_{2L}\Big(\frac{\alpha_z}{\alpha_x} \Big) \xrightarrow{\rm eq\,\,}8\pi\tilde{N},
\label{eq:h3l0-trans}
\ea
where arrows indicate the equilibrium limit of the functions where $\alpha_i\rightarrow 1$ and $\lambda\rightarrow T$.

\chapter{\bf De Sitter coordinates identities}
\label{app:desitterids}

In this appendix, I present some useful identities and derivatives of the de Sitter coordinates which are used in our calculations in Chap.~\ref{chap:gubser}. As mentioned in Chap.~\ref{chap:gubser}, de Sitter coordinates are defined as 
\ba 
{\rho}(\tau,r) &=&  \arcsinh\left(-\frac{1-q^2{\tau^2}+q^2r^2}{2q{\tau}}\right) ,
 \\
{\theta}(\tau,r) &=& \arctan\left(\frac{2qr}{1+q^2{\tau}^2-q^2r^2}\right) .
\ea
Taking partial derivatives and using Eq.~(\ref{eq:thetaperp}) for $\theta_\perp$, one can obtain the necessary derivatives of $(\rho,\theta)$ with respect to $(\tau,r)$
\ba
\frac{\partial\rho}{\partial\tau}&=& q \, (\cosh\rho-\sinh\rho\cos\theta)\, , \qquad 
 \frac{\partial\rho}{\partial r}=- q \sin\theta \, , \nonumber \\
 \frac{\partial\theta}{\partial\tau}&=&-q \frac{\sin\theta}{\cosh\rho} \, , \hspace{1.3in}
 \frac{\partial\theta}{\partial r}= q\,(1-\cos\theta\tanh\rho)\, ,
\ea
and inversely
\be
\begin{aligned}
\frac{\partial\tau}{\partial\rho}&=\tau\cosh\theta_\perp\, , \\
\frac{\partial\tau}{\partial\theta}&=q\tau r\, , 
\end{aligned}
\hspace{2.5cm}
\begin{aligned}
\frac{\partial r}{\partial\rho}&=\tau\sinh\theta_\perp\, , \\
\frac{\partial r}{\partial\theta}&=q\tau r\coth\theta_\perp\, .
\end{aligned}
\ee
Note that the variables above are also related through the following useful relations
\ba
\tau&=&\frac{1}{q}\frac{\sinh\theta_\perp}{\sin\theta}\, , \\
r&=&\frac{1}{q}\cosh\rho\sinh\theta_\perp\, . 
\ea

\chapter{\bf The covariant derivative}
\label{app:covderiv}

The covariant derivative is the generalization of the directional derivative of a vector field which acts as a derivative along tangent vectors of a manifold.  Its action on an arbitrary scalar $\varphi$ and rank-1 and rank-2 tensors (indicated by $V$ below) is 
\be
\begin{aligned}
{\mathfrak D}_\mu \varphi &=\partial_\mu \varphi\, , \\
{\mathfrak D}_\mu V_\nu &=\partial_\mu V_\nu-\Gamma^\lambda_{\beta\mu}V_\lambda\, , \\
{\mathfrak D}_\mu V_{\alpha\beta}&=\partial_\mu V_{\alpha\beta}-\Gamma^\lambda_{\alpha\mu}V_{\lambda\beta} \label{eq:covariant-derivative}
\end{aligned}
\qquad
\begin{aligned}
{\mathfrak D}_\mu V^\nu &=\partial_\mu V^\nu+\Gamma^\nu_{\mu\lambda}V^\lambda\, , \\
{\mathfrak D}_\mu V^{\mu\nu}&=\frac{1}{\sqrt{-\mathrm{g}\,}}\partial_\mu\left(\sqrt{-{\mathrm{g}}\,}V^{\mu\nu}\right)+\Gamma^\nu_{\lambda\mu} V^{\lambda\mu}\, ,\\
{\mathfrak D}_\lambda V^{\mu\nu}&=\partial_\lambda V^{\mu\nu}+\Gamma^\mu_{\lambda\eta} V^{\eta\nu}+\Gamma^\nu_{\lambda\eta} V^{\mu\eta}\, ,
\end{aligned}
\ee
where $\Gamma^\nu_{\mu\lambda}$ are Christoffel symbols, which are 
\ba
\Gamma^\nu_{\mu\lambda}=\frac{1}{2}g^{\nu\sigma}(\partial_\mu g_{\sigma\lambda}+\partial_\lambda g_{\sigma\mu}-\partial_\sigma g_{\mu\lambda})\, .
\label{eq:christoffel}
\ea
Starting from Eq.~(\ref{eq:christoffel}) and using the de Sitter metric (\ref{eq:desitter-metric}),  one obtains the following non-vanishing Christoffel symbols for de-Sitter coordinates
\be
\begin{aligned}
\Gamma^\rho_{\theta\theta}&=\sinh\rho\cosh\rho\, , \\
\Gamma^\rho_{\phi\phi}&=\sin^2\theta\sinh\rho\cosh\rho\, , \\
\Gamma^\theta_{\rho\theta}&=\Gamma^\theta_{\theta\rho}=\tanh\rho\, ,
\end{aligned}
\qquad
\begin{aligned}
\Gamma^\theta_{\phi\phi}&=-\sin\theta\cos\theta\, , \\
\Gamma^\phi_{\rho\phi}&=\Gamma^\phi_{\phi\rho}=\tanh\rho\, , \\
\Gamma^\phi_{\theta\phi}&=\Gamma^\phi_{\phi\theta}=\cot\theta\, .
\end{aligned}
\ee
%

\chapter{\bf The anisotropy tensor in different coordinate systems}
\label{app:xitrans}

In this section, I present the transformation of the anisotropy tensor from de Sitter to Milne and polar Milne coordinates.  The tensors in the different cases are indicated by $\hat{\xi}^\mu_\nu$, $\tilde{\xi}^\mu_\nu$, and $\check\xi^\mu_\nu$ in de Sitter, polar Milne, and Milne coordinates, respectively. According to Sec.~\ref{sec:weyl}, since $\hat{\xi}^\mu_\nu$ is a dimensionless tensor of rank 2 with one up and one down index, it has a conformal weight of 0. Therefore, 
\ba
\tilde{\xi}^\mu_\nu=\frac{\partial \tilde{x}^\mu}{\partial\hat{x}^\alpha}\frac{\partial\hat{x}^\beta}{\partial\tilde{x}^\nu}\hat{\xi}^\alpha_\beta\, .
\ea
The anisotropy tensor in de Sitter space is expanded using Eqs.~(\ref{eq:aniso-tensor2}). Using Eq.~(\ref{eq:desitter-4vectors}), one can expand it in matrix form as 
\ba
&&\hat{\xi}^\mu_\nu={\rm diag}\left(0,\hat{\xi}_\theta,\hat{\xi}_\phi,\hat{\xi}_\varsigma\right) .
\ea
Using the derivative relations in App.~\ref{app:desitterids}, one can find the matrix forms of $\tilde{\xi}^\mu_\nu$ and $\check{\xi}^\mu_\nu$
\ba
&&\tilde{\xi}^\mu_\nu=
  \begin{pmatrix}
     -\hat{\xi}_\theta\sinh^2\theta_\perp & \hat{\xi}_\theta\sinh\theta_\perp\cosh\theta_\perp & 0 & 0 \\-\hat{\xi}_\theta\sinh\theta_\perp\cosh\theta_\perp & \hat{\xi}_\theta\cosh^2\theta_\perp & 0 & 0\\  0 & 0 & \hat{\xi}_\phi
 & 0 \\ 0 & 0 & 0 & \hat{\xi}_\varsigma \end{pmatrix} ,
\ea
\be
\!\check\xi^\mu_\nu\!=\! 
  \begin{pmatrix}
     \!-\hat{\xi}_\theta\sinh^2\theta_\perp & \hat{\xi}_\theta\frac{\sinh(2\theta_\perp\!)}{2}\cos\phi & \hat{\xi}_\theta\frac{\sinh(2\theta_\perp\!)}{2}\sin\phi & 0 \\\!-\hat{\xi}_\theta\frac{\sinh(2\theta_\perp\!)}{2}\cos\phi &\, \hat{\xi}_\theta\cosh^2\!\theta_\perp\cos^2\!\phi+\hat{\xi}_\phi\sin^2\!\phi & (\hat{\xi}_\theta\cosh^2\!\theta_\perp\!- \hat{\xi}_\phi)\sin\!\phi\cos\!\phi & 0\!\\ \!-\hat{\xi}_\theta\frac{\sinh (2\theta_\perp\!)}{2}\sin\!\phi &  (\hat{\xi}_\theta\cosh^2\!\theta_\perp\!-\!\hat{\xi}_\phi)\sin\!\phi\cos\!\phi & \hat{\xi}_\theta\cosh^2\!\theta_\perp\!\sin^2\!\phi\!+\!\hat{\xi}_\phi\cos^2\!\phi\!
 & 0 \\ 0 & 0 & 0 & \hat{\xi}_\varsigma \end{pmatrix} \! ,
\ee
where, in  $\tilde{\xi}^\mu_\nu$ the indices are taken from $(\mu,\nu) \in \{\tau,r,\phi,\varsigma\}$, and in $\check\xi^\mu_\nu$ they are taken from $(\mu,\nu) \in \{\tau,x,y,\varsigma\}$.  Since I started with the basis vectors in Minkowski space LF, one needs to boost them to find their form in the LRF.\footnote{In both cases, only a transverse boost is required.  For the case of polar Milne coordinates, one can make a pure radial boost.}  Constructing the necessary boost from the fluid velocity 4-vector appropriate to each coordinate system one finds
\ba
\left(\tilde{\xi}^\mu_\nu\right)_{\rm LRF}&=&{\rm diag}\left(0,\hat{\xi}_\theta,\hat{\xi}_\phi,\hat{\xi}_\varsigma\right)\,,\\
\left(\check\xi^\mu_\nu\right)_{\rm LRF}&=&
  \begin{pmatrix}
     0 & 0 & 0 & 0 \\ 0 & \frac{\hat{\xi}_++\hat{\xi}_-\cos(2\phi)}{2} & \hat{\xi}_-\sin\phi\cos\phi & 0 \\ 0 & \hat{\xi}_-\sin\phi\cos\phi & \frac{\hat{\xi}_+-\hat{\xi}_-\cos(2\phi)}{2} & 0 \\ 0 & 0 & 0 & \hat{\xi}_\varsigma \end{pmatrix} ,
\ea
where $\hat{\xi}_+\equiv\hat{\xi}_\theta+\hat{\xi}_\phi$ and $\hat{\xi}_-\equiv\hat{\xi}_\theta-\hat{\xi}_\phi$. Using $SO(3)_q$ symmetry, one finds $\hat{\xi}_+\rightarrow 2\hat{\xi}_\theta$ and $\hat{\xi}_-\rightarrow0$ and, therefore,
\ba
&&\left(\tilde{\xi}^\mu_\nu\right)_{\rm LRF}=\left(\check\xi^\mu_\nu\right)_{\rm LRF}= {\rm diag}\left(0,\hat{\xi}_\theta,\hat{\xi}_\theta,\hat{\xi}_\varsigma\right)\,.
\ea
From the results above, one concludes that the LRF anisotropy tensor in polar Milne coordinates is diagonal, irrespective of whether the system is $SO(3)_q$-symmetric or not, however, in Milne coordinates, the anisotropy is only diagonal if the system is $SO(3)_q$-symmetric.

\chapter{\bf More on Gluon self-energy}
\label{app:alternative}
In this section, I review the useful identities for the tensor basis used in Chap.~\ref{chap:gluon}. I also  present an alternative way to calculate the expansion coefficients. 
In this subsection, I present various tensor identities obeyed by our basis tensors (\ref{eq:ABCD}). Useful identities for the contraction of any two basis tensors are as following
\be
\begin{aligned}[c]
A\cdot A &= -A,\\
A\cdot C&=C\cdot A=C\cdot C= -C,\\
A\cdot B&= B\cdot A=B\cdot C= C \cdot B = 0,\\
A\cdot D&=C\cdot D=-D\cdot C-D,\\
D\cdot A&= D\cdot C=-A\cdot D-D\,,
\end{aligned}
\qquad\qquad
\begin{aligned}[c]
B\cdot B &= -Z B\,,\\
D\cdot D &=-p_\perp^2 (B+Z C)\,,\\
D\cdot B&=Z A\cdot D,\\
B\cdot D&=Z D\cdot C=-D\cdot B-ZD\,,\\
\end{aligned}
\ee
with $Z\equiv p^2/\omega^2=1- {\bf p}^2/\omega^2$. Out of 64 possible contractions of any three tensors of $A$, $B$, $C$, and $D$ the non-trivial ones are
\ba 
&A\cdot A \cdot A=A \hspace{1.1cm}; \hspace{1.1cm} D\cdot D\cdot D=Zp_\perp^2D\,,  \nonumber\\
&\frac{B\cdot B\cdot B}{Z^2}=\frac{B\cdot D\cdot D}{Zp_\perp^2}=\frac{D\cdot D\cdot B}{Z p_\perp^2}=\frac{D\cdot A\cdot D}{p_\perp^2}=\frac{D\cdot C\cdot D}{p_\perp^2}= B\,,  \nonumber\\
&A\cdot A\cdot C=A\cdot C\cdot C=A\cdot C\cdot A=C\cdot A\cdot A=C\cdot C\cdot C=C\cdot A\cdot C=C\cdot C\cdot A=C\,,   \nonumber\\
&A\cdot D\cdot D=C\cdot D\cdot D=D\cdot D\cdot A=D\cdot D\cdot C=\frac{D\cdot B\cdot D}{Z}= Zp_\perp^2 C\,,   \nonumber\\
&A\cdot A\cdot D=C\cdot C\cdot D=C\cdot A\cdot D=A\cdot C\cdot D=\frac{D\cdot B\cdot B}{Z^2}=\frac{A\cdot D\cdot B}{Z}=\frac{C\cdot D\cdot B}{Z}\,, \label{eq:AAD} \nonumber\\
&D\cdot A\cdot A=D\cdot C\cdot C=D\cdot A\cdot C=D\cdot C\cdot A=\frac{B\cdot B\cdot D}{Z^2}=\frac{B\cdot D\cdot A}{Z}=\frac{B\cdot D\cdot C}{Z}\,, \label{eq:DAA}  \nonumber\\
&B \cdot  D \cdot B = A \cdot D \cdot A = C \cdot D \cdot C = A \cdot D \cdot C = C \cdot D \cdot A = 0 \, ,
\ea
where (\ref{eq:AAD}) and (\ref{eq:DAA}) contain the contractions that cannot be expressed in terms of any single basis tensors. In all relations above $p_\perp\equiv |{\bf p}-({\bf p}\cdot {\bf n}){\bf n}|$ is the component of ${\bf p}$ perpendicular to ${\bf n}$.
Using the identities listed above, one can calculate the contraction of any two dressed propagators with a self-energy in between, $\tilde{\cal D}_1\Pi_2 \tilde{\cal D}_3$, as 
\ba
\tilde{\cal D}_1\Pi_2 \tilde{\cal D}_3&=&\alpha'_1\alpha_2\alpha'_3 A \\
&+&\Big[\beta'_1\beta_2\beta'_3 Z^2+\beta'_1\delta_2\delta'_3p_\perp^2Z+\delta'_1\delta_2\beta'_3p_\perp^2Z+
\delta'_1\alpha_2\delta'_3p_\perp^2+ \delta'_1\gamma_2\delta'_3p_\perp^2\Big]B\nonumber\\
&+&\Big[\delta'_1\beta_2\delta'_3p_\perp^2Z^2+\delta'_1\delta_2\gamma'_3 p_\perp^2Z+\gamma'_1\delta_2\delta'_3 p_\perp^2Z+\alpha'_1\delta_2\delta'_3p_\perp^2Z
+\delta'_1\delta_2\alpha'_3p_\perp^2Z
+\alpha'_1\alpha_2\gamma'_3\nonumber \\&+&\gamma'_1\alpha_2\alpha'_3
+\alpha'_1\gamma_2\alpha'_3+\gamma'_1
(\gamma_2+\alpha_2)\gamma'_3+\alpha'_1\gamma_2\gamma'_3+\gamma'_1\gamma_2\alpha'_3\Big]C +\delta'_1\delta_2\delta'_3Z p_\perp^2 D\nonumber \\
&-& \Big[\beta'_1\beta_2\delta'_3 Z^2+\beta'_1\delta_2\alpha'_3 Z+\beta'_1\delta_2\gamma'_3 Z+\delta'_1\alpha_2\alpha'_3+
\delta'_1\alpha_2\gamma'_3 +\delta'_1\gamma_2\alpha'_3+\delta'_1\gamma_2\gamma'_3\Big]D\cdot A \nonumber \\
&-&\Big[\delta'_1\beta_2\beta'_3Z^2+\alpha'_1\delta_2\beta'_3Z
+\gamma'_1\delta_2\beta'_3Z+\alpha'_1\alpha_2\delta'_3+\gamma'_1\gamma_2\delta'_3
+\alpha'_1\gamma_2\delta'_3
+\gamma'_1\alpha_2\delta'_3\Big]A\cdot D \, .\nonumber \ea

\chapter{\bf Small anisotropy expansion of quark self-energies}
\label{app:expansion}
In this section, I simplify the quark self-energy discussed in Chap.~\ref{chap:quark} in small anisotropy limit. By Taylor expanding the quark self-energy around $\boldsymbol{\xi}=0$, the integrals can be calculated analytically. To leading order in the anisotropy parameters, one finds
\ba
\Sigma_0=\Sigma^{\rm iso}_0+\Big[{\cal F}_{0,1}+\Sigma^{\rm iso}_0{\cal F}_{0,2}\Big]\, , \\
\frac{\Sigma_x}{\sin\theta_k\cos\phi_k}=\Sigma^{\rm iso}_s+\Big[{\cal F}_{x,1}+\Sigma^{\rm iso}_s{\cal F}_{x,2}\Big]\, , \\
\frac{\Sigma_y}{\sin\theta_k\sin\phi_k}=\Sigma^{\rm iso}_s+\Big[{\cal F}_{y,1}+\Sigma^{\rm iso}_s{\cal F}_{y,2}\Big]\, , \\
\frac{\Sigma_z}{\cos\theta_k}=\Sigma^{\rm iso}_s+\Big[{\cal F}_{z,1}+\Sigma^{\rm iso}_s{\cal F}_{z,2}\Big]\, ,
\ea
with
\ba
\Sigma^{\rm iso}_0&=&\frac{m_{\rm q}^2}{2\bf k} \log\frac{\omega+\bf k}{\omega-\bf k}\, , \\
\Sigma^{\rm iso}_s&=&\frac{m_{\rm q}^2}{\bf k} \Big(\frac{\omega}{2\bf k}\log\frac{\omega+\bf k}{\omega-\bf k}-1\Big)\,. 
\ea
The various functions are
\ba
{\cal F}_{0,1}&=&\frac{zm_{\rm q}^2}{8\bf k}\Big[6\xi_a\cos2\phi_k\sin^2\theta_k-\xi_b(3\cos2\theta_k+1)\Big]\, ,\\
{\cal F}_{0,2}&=& \frac{1}{8}\Big[2\xi_a \cos2\phi_k\sin^2\theta_k-\xi_b\big(\cos2\theta_k+3\big)-8\xi_z\Big] -\frac{z\bf k}{m_{\rm q}^2} {\cal F}_{0,1}\, , \\
{\cal F}_{x,1}&=&\frac{m_{\rm q}^2}{24\bf k}\Big[\xi_a\big(10\cos2\phi_k\sin^2\theta_k-4\big)-\xi_b\big(5\cos2\theta_k+3\big)\Big]\,,\\
{\cal F}_{x,2}&=&\frac{1}{8}\Big[6\xi_a\cos2\phi_k\sin^2\theta_k-\xi_b(3\cos2\theta_k+1)-8\xi_x\Big]-3\frac{{\bf k} z^2}{m_{\rm q}^2} {\cal F}_{x,1}\,,\\
{\cal F}_{y,1}&=&\frac{m_{\rm q}^2}{24\bf k}\Big[\xi_a\big(10\cos2\phi_k\sin^2\theta_k+4\big)-\xi_b\big(5\cos2\theta_k+3\big)\Big]\,,
\ea
\ba
{\cal F}_{y,2}&=&\frac{1}{8}\Big[6\xi_a\cos2\phi_k\sin^2\theta_k-\xi_b(3\cos2\theta_k+1)-8\xi_y\Big]-3\frac{{\bf k} z^2}{m_{\rm q}^2} {\cal F}_{y,1}
\,,\\
{\cal F}_{z,1}&=&\frac{m_{\rm q}^2}{24\bf k}\Big[10\xi_a\cos2\phi_k\sin^2\theta_k-\xi_b(5\cos2\theta_k-1)\Big]\,,\\
{\cal F}_{z,2}&=&\frac{1}{8}\Big[6\xi_a\cos2\phi_k\sin^2\theta_k-\xi_b(3\cos2\theta_k+1)-8\xi_z\Big] -3\frac{{\bf k} z^2}{m_{\rm q}^2} {\cal F}_{z,1}\,.
\ea
%



\bibliographystyle{unsrt}

\bibliography{mydiss}


\end{document}